\documentclass[10pt, conference, letterpaper]{IEEEtran}
\usepackage{algpseudocode}
\usepackage{algorithmicx}
\usepackage[ruled,vlined,linesnumbered]{algorithm2e}
\usepackage{amssymb}
\usepackage{booktabs}
\usepackage{graphicx, cite}
\usepackage{subfigure}
\usepackage{amsmath}
\usepackage{amsthm}
\usepackage{float}
\usepackage{stfloats}
\usepackage{url}
\usepackage{multirow}
\usepackage{xcolor}
\usepackage{CJK}
\usepackage{enumitem}
\usepackage{textcomp}

\newcommand{\hytt}[1]{\texttt{\hyphenchar\font=\defaulthyphenchar #1}}

\def\BibTeX{{\rm B\kern-.05em{\sc i\kern-.025em b}\kern-.08em T\kern-.1667em\lower.7ex\hbox{E}\kern-.125emX}}
\graphicspath{{figures/}}

\IEEEoverridecommandlockouts
\begin{document}
\title{\LARGE Energy Drain of the Object Detection Processing Pipeline for Mobile Devices: Analysis and Implications}
\author{\IEEEauthorblockN{Haoxin Wang\IEEEauthorrefmark{1}, BaekGyu Kim\IEEEauthorrefmark{2}, Jiang Xie\IEEEauthorrefmark{1}, and Zhu Han\IEEEauthorrefmark{3}\\
\IEEEauthorblockA{\IEEEauthorrefmark{1}University of North Carolina at Charlotte, Charlotte, NC 28223, U.S.A.}
\IEEEauthorblockA{\IEEEauthorrefmark{2}Toyota Motor North America (TMNA) R\&D InfoTech Labs, U.S.A.}
\IEEEauthorblockA{\IEEEauthorrefmark{3}University of Houston, Houston, TX 77004, U.S.A.}}
\thanks{This is a personal copy of the authors. Not for redistribution. The final version of this paper was accepted by IEEE Transactions on Green Communications and Networking.}
}
\maketitle

\begin{abstract}
Applying deep learning to object detection provides the capability to accurately detect and classify complex objects in the real world. However, currently, few mobile applications use deep learning because such technology is computation-intensive and energy-consuming. This paper, to the best of our knowledge, presents the first detailed experimental study of a mobile augmented reality (AR) client's energy consumption and the detection latency of executing Convolutional Neural Networks (CNN) based object detection, either locally on the smartphone or remotely on an edge server. In order to accurately measure the energy consumption on the smartphone and obtain the breakdown of energy consumed by each phase of the object detection processing pipeline, we propose a new measurement strategy. Our detailed measurements refine the energy analysis of mobile AR clients and reveal several interesting perspectives regarding the energy consumption of executing CNN-based object detection. Furthermore, several insights and research opportunities are proposed based on our experimental results. These findings from our experimental study will guide the design of energy-efficient processing pipeline of CNN-based object detection.
\end{abstract}
\IEEEpeerreviewmaketitle

\section{Introduction}
\label{sc:introduction}
With the advancement in \emph{Deep Learning} in the past few years, we are able to create complex machine learning models for detecting objects in real-time video frames. This advancement has the potential to make Augmented Reality (AR) devices highly intelligent and enable industries to favor machine learning models with superior performance. For example, AR automotive applications (e.g., deep learning-based AR head-up-displays (HUDs)) are promised to help increase road safety, bring intuitive activities to driving, and enhance driving experience in the future. Meanwhile, as people nowadays are using their smartphones to a larger extent and also expect increasingly advanced performance from their mobile applications, the industry needs to adopt more advanced technologies to meet such expectations. One such adoption can be the use of deep learning-based AR applications.

However, few mobile AR applications use deep learning today because of inadequate infrastructure support (e.g., limited computation capacity and battery resource of smartphones). Deep learning algorithms are computation-intensive, and executed locally in ill-equipped smartphones may not provide acceptable latency for end users. For instance, in Deepmon \cite{huynh2017deepmon}, it takes approximately $600$ ms for small and medium \emph{convolutional neural network} (CNN\footnote{A CNN is a deep learning algorithm which has demonstrated great success on image recognition, image classifications, object detection, etc.}) models and almost $3$ seconds for large CNN models to process one frame, which is obviously not acceptable for real-time processing \cite{chatzopoulos2017mobile}. 

Two research directions have emerged to address this challenge. The first direction is to tailor the computation-intensive deep learning algorithms to be executed on smartphones. For instance, Tiny-YOLO \cite{yolov3} that has only $9$ convolutional layers ($24$ convolutional layers in a full YOLO network) is developed and optimised for use on embedded and mobile devices. TensorFlow Lite \cite{Lite} is TensorFlow's lightweight solution for embedded and mobile devices. It enables low-latency inference of on-device machine learning models with a small binary size and fast performance supporting hardware acceleration. However, the reduction of the inference latency is at the cost of the precision degradation of the detection.
The other research direction that is widely used in running deep learning in smartphones is to transfer all the computation data to more powerful infrastructures (e.g., the remote cloud and edge servers) and execute deep learning algorithms there \cite{chen2015glimpse,jain2016low,wang2020architectural}. Such offloading-based solutions can reduce the inference latency and extend smartphones' battery life only when the network access is reliable and sufficiently fast. 

\textbf{Our Motivation.} Although the complexity and capabilities of smartphones continue to grow at an amazing pace, smartphones are expected to continually become lighter and slimmer. When combined with energy-hungry deep learning-based applications, the limited battery capacity allowed by these expectations now motivates significant investment into smartphone power management research. In order to better investigate and understand the relationship between the energy consumption and the performance of deep learning-based applications such as CNN-based object detection, we propose the following questions:

\textbf{RQ 1.} How is energy consumed when a CNN-based object detection application is executed locally on a mobile AR client? In order to help a mobile AR device to extend its battery life, conducting a comprehensive measurement study is significantly important.

\textbf{RQ 2.} Does offloading the object detection tasks to a powerful infrastructure significantly decrease both the energy consumption and latency? When a CNN-based object detection application is executed remotely, communication latency is non-negligible and unstable, especially in wireless networks. Previous work \cite{carroll2010analysis} shows that smartphone's radio interfaces account for up to $50$\% of the total power budget. In addition, improved communication speeds generally come at the cost of higher power consumption \cite{saha2015power}.


\textbf{RQ 3.} Besides the network condition, what else impacts the energy consumption and latency when executed remotely, and how? Executing object detection on a remote edge server is one of the most commonly used approaches to assist resource-constrained smartphones in improving their energy efficiency and performance \cite{liu2019edge}. Therefore, to further improve the efficiency for executing object detection remotely, understanding the factors that may impact the detection performance is critical.

\textbf{Our Contributions.} In this paper, we conduct, to the best of our knowledge, the first comprehensive experimental study that investigates how a mobile AR client's energy efficiency, latency, and detection accuracy are influenced by diverse factors (e.g., CPU governor, CNN model size, and image post processing algorithm) in both local and remote executions. We make the following contributions:
\begin{enumerate}
\item Developing two Android benchmark applications that perform real-time object detections: one is running a light CNN model locally on the smartphone and the other is running a large CNN model remotely on an edge server.
\item Measuring and evaluating the energy consumption and latency of each phase in the implemented end-to-end CNN-based object detection processing pipeline. Both local and remote executions are investigated.
\item Comparing the local execution and the remote execution in terms of energy efficiency, latency, detection accuracy, etc.
\item Proposing several insights which can potentially guide the future design of energy-efficient mobile AR systems based on our experimental study.
\end{enumerate}

The rest of this paper is organized as follows. Section \ref{sc:related} discusses related work. Section \ref{sc:methodology} describes our proposed methodology and key performance metrics that we consider in this study. Experimental results of the local execution and remote execution are presented in Section \ref{sc:localmeasurement} and Section \ref{sc:remotemeasurement}, respectively. Finally, Sections \ref{sc:validity} and \ref{sc:conclusion} discusses threats to validity and concludes the paper, respectively.

\section{Related Work}
\label{sc:related}
\textbf{Energy Measurement.} With the popularity of energy constrained mobile devices (e.g., smartphone, AR glass, and smartwatch), a number of research has investigated how the energy is consumed in mobile devices when executing applications through measurement studies \cite{pathak2012energy,de2019recommending,chowdhury2019greenscaler}. \cite{hindle2014greenminer} proposes and implements a measurement framework that can physically measure the energy consumption of mobile devices and automate the reporting of measurement back to researchers. \cite{agolli2017investigating,wan2015detecting} study the energy consumption of GUI colors on OLED displays. In addition, the energy efficiency of network protocols such as HTTP on mobile devices has been discussed in \cite{li2016automated,chowdhury2016client}. However, very few energy measurement studies focus on running deep learning-based applications on mobile devices, especially mobile AR applications. Although \cite{mcintosh2019can} discusses and compares the energy efficiency of different machine learning applications in terms of algorithm, implementation, and operating system (OS), our work focuses on a specific application, object detection, and conducts a comprehensive study on (i) energy efficiency comparison between local and remote executions as well as (ii) how hardware and software configurations impact the energy efficiency of executing object detections on smartphones.

\textbf{Energy Modeling.} Energy modeling has been widely used for investigating the factors that influence the energy consumption of mobile devices. \cite{xiao2013modeling,wang2018rethinking,huang2012close,wang2018smart,wang2017v} propose energy models of WiFi and LTE data transmission with respect to the network performance metrics, such as data and retransmission rates. \cite{shye2009into,walker2016accurate,devogeleer2014modeling,xu2013v,pathak2011fine} propose multiple power consumption models to estimate the energy consumption of mobile CPUs. Tail energy caused by different components, such as disk, Wi-Fi, 3G, and GPS in smartphones has been investigated in \cite{pathak2011fine,pathak2012energy}. However, none of them can be directly applied to estimate the energy consumed by mobile AR applications. This is because mobile AR applications introduce a variety of (i) energy consuming components (e.g., camera sampling and image conversion) that are not considered in the previous models and (ii) configuration variables (e.g., computation model size and camera sample rate) that also significantly influence the energy consumption of mobile devices.

\textbf{CNN.}
In recent years, applying CNNs to object detection has been proven to achieve excellent performance \cite{huynh2017deepmon,yolov3,ren2015faster,wang2018pelee,zhang2018shufflenet,chen2020federated}. In \cite{huang2017speed,liu2018edge}, the speed and accuracy trade-offs of various modern CNN models are compared. However, none of these works considered the performance of running CNNs on smartphones. In addition, although existing papers have extensively investigated how to run CNN models on mobile devices, including model compression of CNNs \cite{howard2017mobilenets}, GPU acceleration \cite{huynh2017deepmon}, and only processing important frames \cite{chen2015glimpse}, none of these works considered the energy consumption of executing CNNs on smartphones. In \cite{ran2017delivering}, a small number of measurements on the battery drain of running a CNN on a powerful smartphone are conducted. However, its battery drain results are reported by the Android OS that can only provide coarse-grained results. For example, it only shows the total battery usage of running a CNN on a smartphone for $30$ minutes. In addition, it only studies running CNNs on smartphones with high computation capabilities and the experimental results are not comparable to smartphones with poor computation capabilities.

\textbf{Computation Offloading.} Most existing research on computation offloading focuses on how to make offloading decisions. \cite{hu2017energy,hu2014energy,luong2017resource,wang2019auto} coordinate the scheduling of offloading requests for multiple applications to further reduce the wireless energy cost caused by the long tail problem. \cite{geng2018energy} proposes an energy-efficient offloading approach for multicore-based mobile devices. \cite{miettinen2010energy} discusses the energy efficiency of computation offloading for mobile clients in cloud computing. However, these solutions cannot be applied to improving the energy efficiency of mobile devices in mobile AR offloading cases. This is because (i) a variety of pre-processing tasks in mobile AR executions, such as camera sampling, screen rendering, and image conversion, are not taken into account and (ii) besides the latency constraint that is considered in most existing computation offloading approaches, detection accuracy is also a key performance metric, which must be considered while designing a mobile AR offloading solution. In addition, although some existing work proposes to study the tradeoffs between the mobile AR service latency and detection accuracy \cite{ran2018deepdecision,liu2018edge,hanhirova2018latency}, none of them considered (i) the energy consumption of the mobile AR device and (ii) the whole processing pipeline of mobile AR (i.e., starting from camera sampling to obtaining detection results).

\textbf{CPU Frequency Scaling.} Our work is also related to CPU frequency scaling. For modern mobile devices, such as smartphones, CPU frequency and the voltage provided to the CPU can be adjusted at run-time, which is called Dynamic Voltage and Frequency Scaling (DVFS). Prior work \cite{chen2007energy,hu2017energy,kwak2014dynamic,lee2009energy} proposes various DVFS strategies to reduce the mobile device energy consumption under various applications, such as video streaming \cite{hu2017energy} and delay-tolerant applications \cite{kwak2014dynamic}. However, to the best of our knowledge, there have been no efforts factoring in the energy efficiency of mobile AR applications in the context of mobile device DVFS.

\section{Proposed Methodology}
\label{sc:methodology}
This section describes the overview of our developed testbed for experimental studies, implemented benchmark applications, our proposed energy measurement process, along with the key performance metrics defined to evaluate the performance of the CNN-based object detection processing pipeline.

\begin{figure}[t]
\centering
\subfigure[Mobile AR client and power monitor]
{\includegraphics[width=0.24\textwidth]{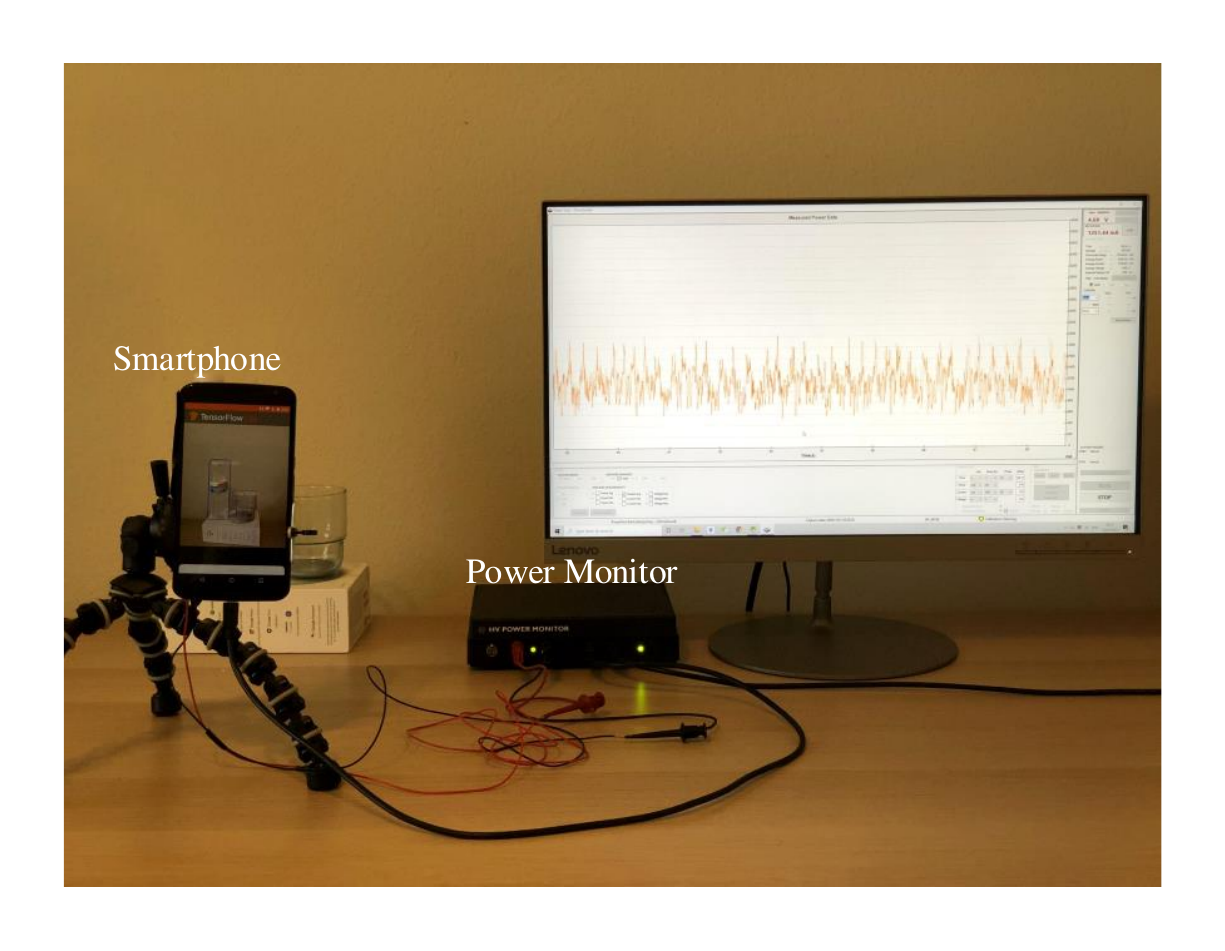}\label{fig:testbed_client}}
\subfigure[Edge server and WiFi AP]
{\includegraphics[width=0.24\textwidth]{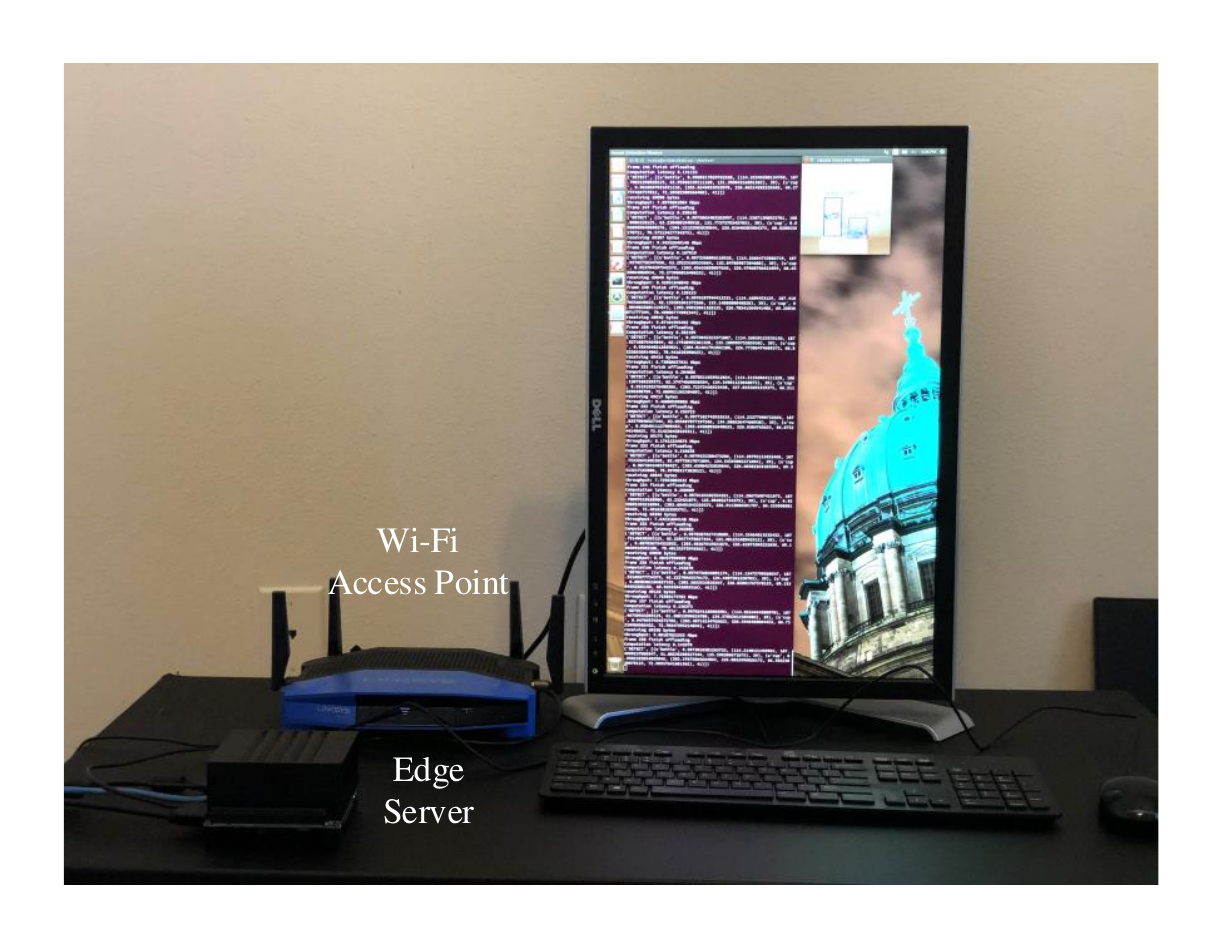}\label{fig:testbed_server}}
\caption{Overview of the developed testbed.}
\label{fig:testbed}   
\end{figure}


\subsection{Overview of the Testbed}
\label{ssc:OverviewTestbed}
As shown in Fig. \ref{fig:testbed}, our testbed consists of three major components: mobile AR client (e.g., smartphone), edge server attached to a WiFi access point (AP), and power monitor.

\textbf{Mobile AR Client.} 
We implement a mobile AR client on a rooted Nexus $6$ smartphone running Android 5.1.1 OS. It is equipped with Qualcomm Snapdragon $805$ SoC (System-on-Chip). The CPU frequency ranges from $0.3$ GHz to $2.649$ GHz. 


\textbf{Edge Server.} The edge server is developed to process received image frames sent from a smartphone and send the detection results back to the smartphone. We implement an edge server on an Nvidia Jetson AGX Xavier which is connected to a WiFi AP through a $1$Gbps Ethernet cable (the length of the cable is less than 1 meter). The transmission latency between the server and AP can be ignored. Two major modules are implemented on the edge server. The first one is the communication service handler module which performs authentication and establishes a TCP socket connection with the mobile AR client. This module is also responsible for dispatching the detection results to the corresponding smartphone. The second one is the object detection module that is designed based on a custom framework called Darknet \cite{darknet13} with GPU acceleration and runs YOLOv3 \cite{yolov3}, a large neural network model with $24$ convolutional layers. The YOLOv3 model used in our experiments is trained on COCO dataset \cite{lin2014microsoft} and can detect $80$ classes.

\begin{figure*}[t]
\centering
\includegraphics[width=0.98\textwidth]{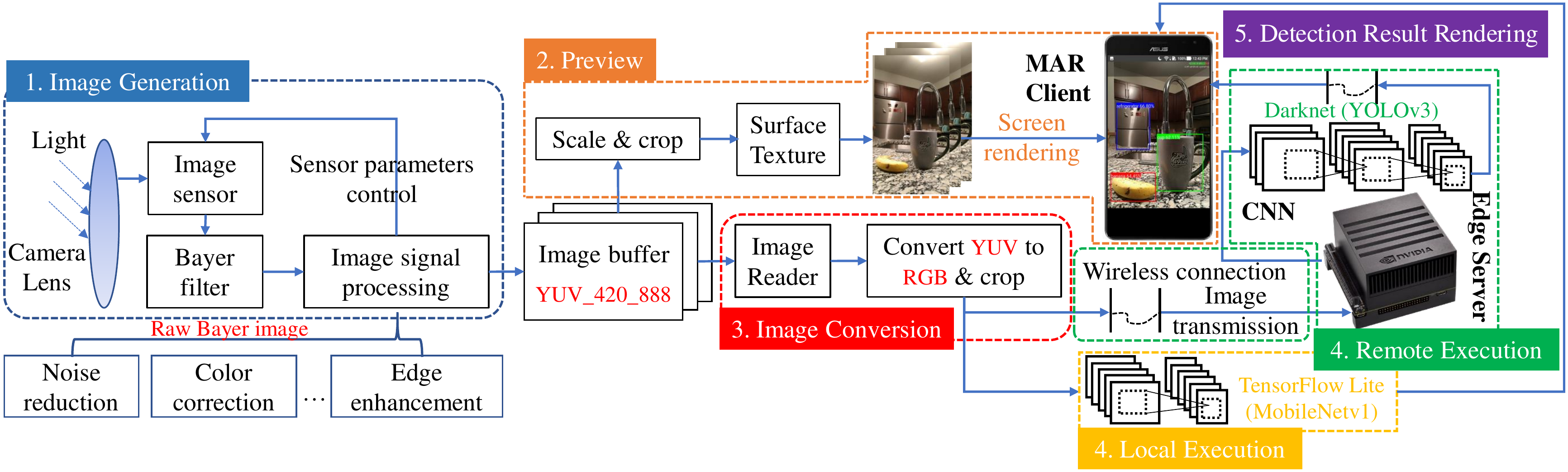}
\caption{Processing pipeline of the CNN-based object detection application implemented in this paper.}
\label{fig:pipeline}
\end{figure*}

\begin{figure*}[t]
\centering
\subfigure[Pipeline for the local execution]
{\includegraphics[width=0.48\textwidth]{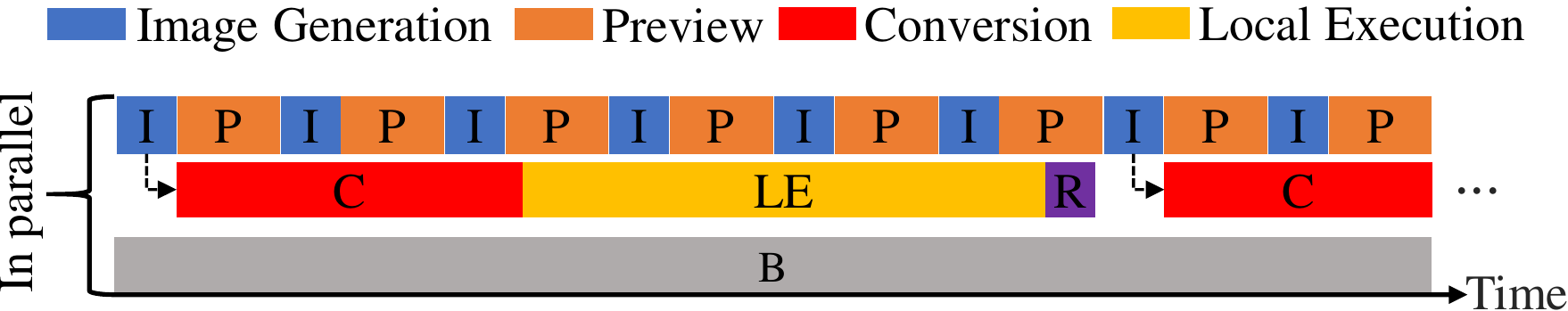}\label{fig:MLocal}}
\subfigure[Pipeline for the remote execution]
{\includegraphics[width=0.48\textwidth]{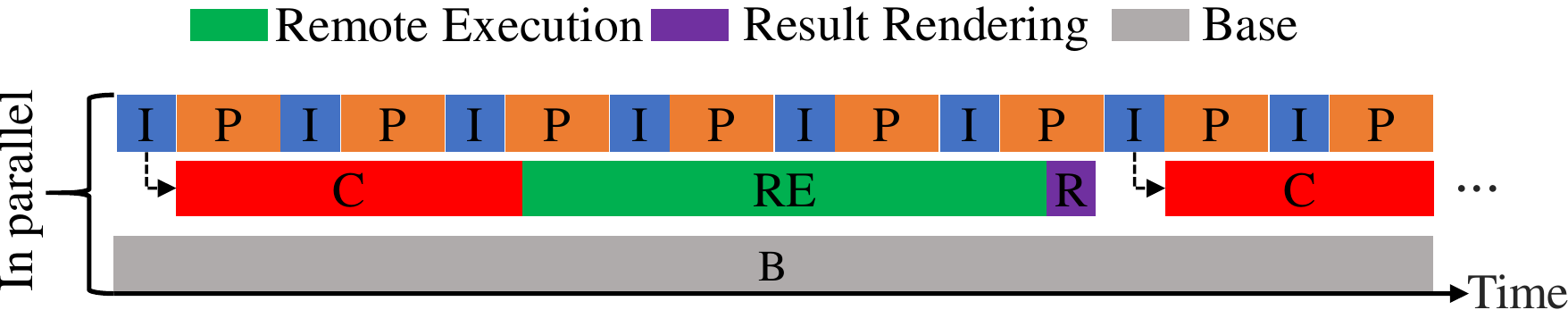}\label{fig:MRemote}}
\caption{The diagrams of the pipelines for benchmark applications.}
\label{fig:Mpipe}   
\end{figure*}

\textbf{Power Monitor.}
To measure the power consumption, we use an external power monitor, a Monsoon Power Monitor \cite{Monsoon}, to provide power supply for the test smartphone. Different from old smartphone models, modern smartphones like Nexus 6 have very tiny battery connectors, making it very challenging to connect the power monitor to them. To solve this problem, we modify the battery connection of Nexus 6 by designing a customized circuit and soldering it to the smartphone's power input interface. In addition, the power measurements are taken with the screen on, with the Bluetooth/LTE radios disabled, and with minimal background application activity, ensuring that the smartphone's \emph{base power} is low and does not vary unpredictably over time. For the measurements of the power consumption in local execution, the base power is defined as the power consumed by the smartphone when its WiFi interface is turned off. For the measurements of the power consumption in remote execution, the base power is defined as the power consumed when the smartphone is connected to the AP without any data transmission activity.

\subsection{Benchmark Applications}

Three benchmark applications are implemented in this paper. The first application is executing CNN-based object detection on tested smartphones, defined as \emph{local execution}. The second application is executing CNN-based object detection on our equipped edge server, defined as \emph{remote execution}. Figs. \ref{fig:pipeline} and \ref{fig:Mpipe} provide an overview of the processing pipeline of these two benchmark applications implemented in this paper, composed of five pipelined operations: image generation, preview, image conversion, local/remote execution, and detection result rendering. These two benchmark applications share the same pipelined operations except Phase 4 (i.e., local execution (yellow box) and remote execution (green box)). The third application only executes the image generation and preview (i.e., Phase 1 and 2).

\textbf{Image Generation (Phase 1).} The input to this phase is continuous light signal and the output is an image frame. In this phase, the image sensor first senses the intensity of light and converts it into an electronic signal. A Bayer filter is responsible for determining the color information. Then, an image signal processor (ISP) takes the raw data from the image senor and converts it into a high-quality image frame. The ISP performs a series of image signal processing operations to deliver a high-quality image, such as noise reduction, color correction, and edge enhancement. In addition, the ISP conducts automated selection of key camera control values according to the environment (e.g., auto-focus (AF), auto-exposure (AE), and auto-white-balance (AWB)). The whole image generation pipeline in our benchmark applications is constructed based on \hytt{android.hardware.camera2} which is a package that provides an interface to individual camera devices connected to an Android device. \hytt{CaptureRequest} is a class in \hytt{android.hardware.camera2} that constructs the configurations for the capture hardware (sensor, lens, and flash), the processing pipeline, and the control algorithms. Therefore, in our implemented benchmark applications, we use \hytt{CaptureRequest} to set up image generation configurations. For example, \hytt{CaptureRequest.CONTROL\_AE\_MODE\_OFF} disables AE and \hytt{CaptureRequest.CONTROL\_AE\_TARGET\_FPS\_RANGE} sets the camera FPS (i.e., the number of frames that the camera samples per second). In this paper, all default image processing operations are enabled and the camera FPS is set to $15$ fps.   

\textbf{Preview (Phase 2).} The input to this phase is a latest generated image frame with \hytt{YUV\_420\_888} format\footnote{For \hytt{android.hardware.camera2}, \hytt{YUV\_420\_888} format is recommended for YUV output \cite{ImageFormat}.} (i.e., the output of Phase 1) and the output is a camera preview rendered on a smartphone's screen with a pre-defined preview resolution. In this phase, the latest generated image frame is first resized to the desired preview resolution and then buffered in a \hytt{SurfaceTexture} which is a class capturing frames from an image stream (e.g., camera preview or video decode) as an OpenGL ES texture. Finally, the camera preview frame in \hytt{SurfaceTexture} is copied and sent to a dedicated drawing surface, \hytt{SurfaceView}, and rendered on the screen. In our benchmark applications, the preview resolution is set via method \hytt{SurfaceTexture.setDefaultBufferSize()}. In this paper, the preview resolution is set to $800\times600$ pixels (different Android devices may have different supported preview resolution sets).

\textbf{Image Conversion (Phase 3).} The input to this phase is a latest generated image frame with \hytt{YUV\_420\_888} format (i.e., the output of Phase 1) and the output is a cropped RGB image frame. In this phase, in order to further process camera captured images (i.e., object detection), an \hytt{ImageReader} class is implemented to acquire the latest generated image frame, where \hytt{ImageReader.OnImageAvailableListener} provides a callback interface for being notified that a new generated image frame is available and method \hytt{ImageReader.acquireLatestImage()} acquires the latest image frame from the \hytt{ImageReader}'s queue while dropping an older image. Additionally, the desired size and format of acquired image frames are configured once an \hytt{ImageReader} is created. In our benchmark applications, the desired size and the preview resolution are the same ($800\times600$ pixels) and the image format in \hytt{ImageReader} is set to \hytt{YUV\_420\_888}. Furthermore, an image converter is implemented to convert the \hytt{YUV\_420\_888} image to an \hytt{RGB} image, because the input to a CNN-based object detection model must be an \hytt{RGB} image. Two image conversion methods are implemented in our benchmark applications: one is Java-based and the other is C-based (we compare these two methods in Section \ref{ssc:impconv_remote}). Finally, the converted \hytt{RGB} image is cropped to the size of the CNN model for object detections.

\textbf{Local/Remote Execution (Phase 4).} The input to this phase is a converted and cropped image frame (i.e., the output of Phase 3) and the output is an object detection result. In our benchmark applications, the object detection result contains one or multiple bounding boxes with labels that identify the locations and classifications of the objects in an image frame. Each bounding box consists of 5 predictions: (x, y, w, h) and a confidence score \cite{yolov3}. The (x, y) coordinates represent the center of the box relative to the bounds of the grid cell. The (h, w) coordinates represent the height and width of the bounding box relative to (x, y). The confidence score reflects how confident the CNN-based object detection model is on the box containing an object and also how accurate it thinks the box is what it predicts. (i) In the local execution, the benchmark application is implemented with a light framework called TensorFlow Lite \cite{Lite} which is TensorFlow's lightweight solution for embedded and mobile devices. It runs a small CNN model, called MobileNetv1 \cite{howard2017mobilenets}. In order to run MobileNetv1 with different frame resolutions in TensorFlow Lite on smartphones, we convert pre-trained MobileNetv1 SSD models to TensorFlow Lite models (i.e., optimized FlatBuffer format identified by the \texttt{.tflite} file extension). (ii) In the remote execution, the benchmark application transmits the converted and cropped image frame to the edge server through a wireless TCP socket connection in real time. To avoid having the server process stale frames, the application always sends the latest generated frame to the server and waits to receive the detection result before sending the next frame for processing.

\textbf{Detection Result Rendering (Phase 5).} The input to this phase is the object detection result of an image frame (i.e., the output of Phase 4) and the output is a view with overlaid augmented objects (specifically, overlaid bounding boxes and labels in this paper) on top of the physical objects (e.g., a cup).

\subsection{Energy Measurement Strategy}
\label{ssc:Measurement Strategy}

In order to measure the energy consumption of running those two benchmark applications on a smartphone and obtain the breakdown of energy consumed by each phase presented in Fig. \ref{fig:pipeline}, we design a measurement strategy. The key idea of the proposed measurement strategy is \textit{synchronizing the recorded time in log files (saved by benchmark applications in the tested Android smartphone) and power measurement data (exported by the Monsoon power monitor)}. However, this is very challenging, because the tested smartphone and the power monitor do not share the same global clock. For example, in Android smartphones, the recorded time of an event can be counted by a system clock, \hytt{uptimeMillis}\footnote{This clock stops when the system enters deep sleep (CPU off, display dark, and device waiting for external input), but is not affected by clock scaling, idle, or other power saving mechanisms. Additionally, it is guaranteed to be monotonic, and is suitable for interval timing when the interval does not span device sleep.}, where the clock is counted in milliseconds since the system is booted (e.g., if an event happens 100 milliseconds after the system is booted, the exported timestamp of the event in the log file is 100). On the other hand, in the power monitor, the timestamp is counted in milliseconds since the power measurement is launched.

\begin{figure*}[t]
\centering
\subfigure[Power consumption recorded by power monitor]
{\includegraphics[width=0.52\textwidth]{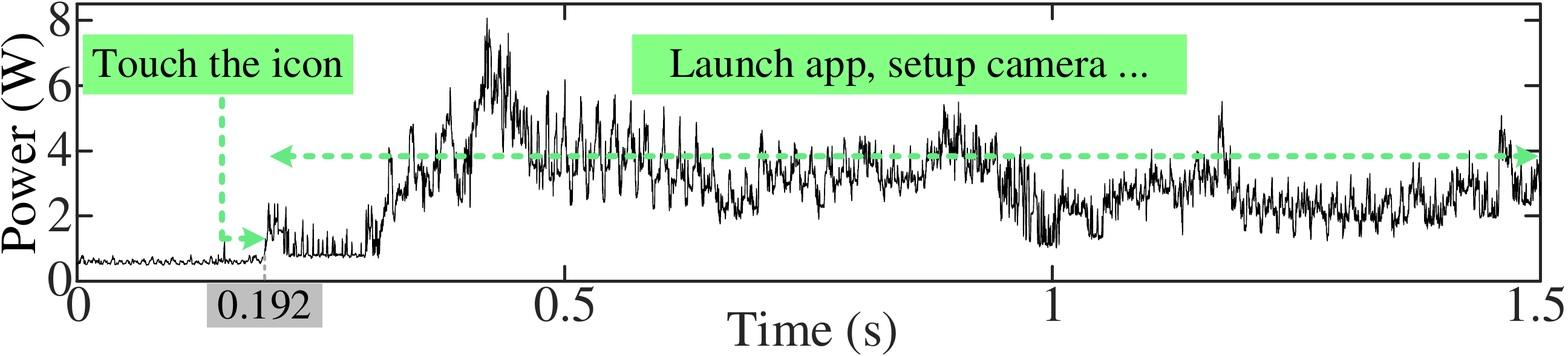}\label{fig:touchPower}}
\subfigure[Timestamp of touch event recorded by smartphone]
{\includegraphics[width=0.42\textwidth]{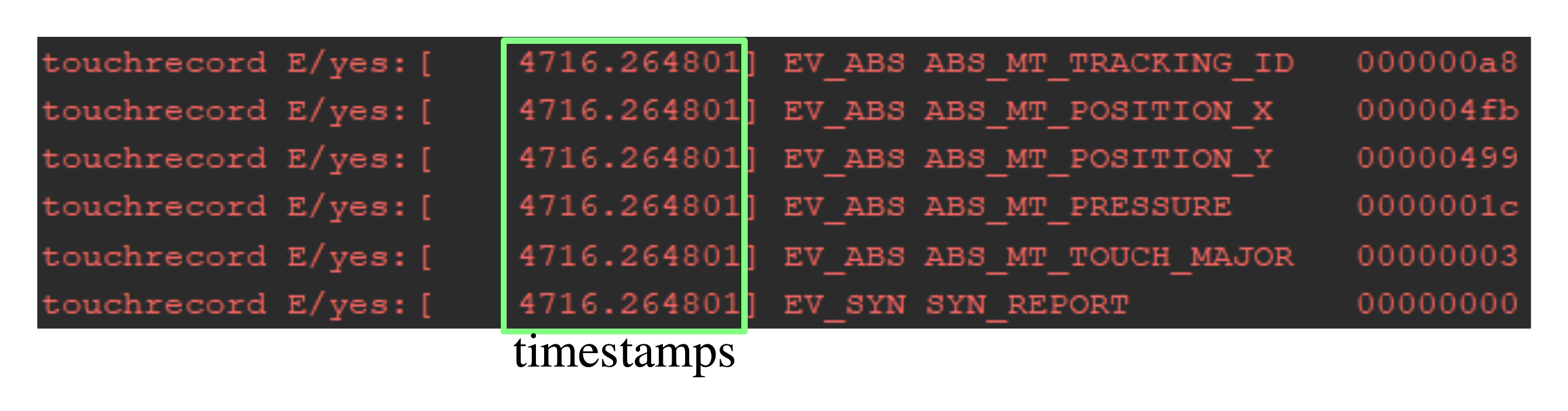}\label{fig:touchLogcat}}
\caption{An example of local clock synchronization.}
\label{fig:touchevent}   
\end{figure*}

\begin{figure*}[t]
\centering
\includegraphics[width=0.98\textwidth]{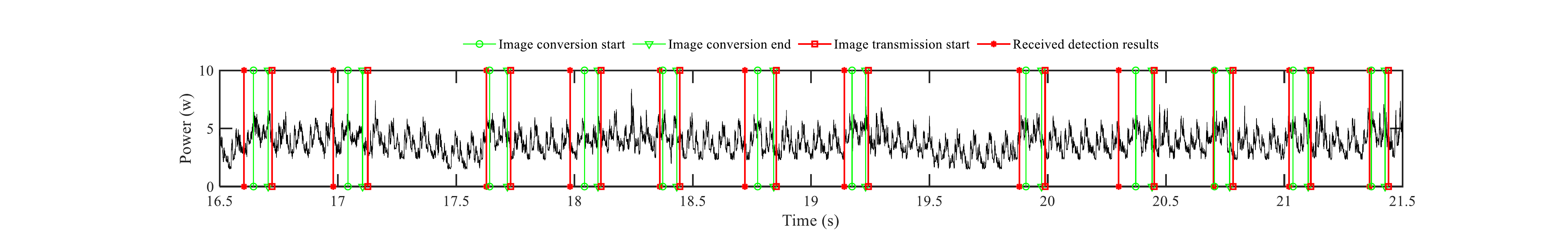}
\caption{An example of event localization in the collected power measurement data (CPU governor: Interactive, remote execution with C-based image conversion method).}
\label{fig:remoteexample}
\end{figure*}

\textbf{Local Clock Synchronization \& Event Localization.} To synchronize the exported timestamps of the Android smartphone and the power monitor, we propose to set up a \textit{flag event} that can be tracked easily and accurately in both of them. The touch event that launches the benchmark application is selected as the flag event to synchronize the timestamps. For example, Fig. \ref{fig:touchPower} illustrates the power consumption of the tested smartphone recorded by a Monsoon power monitor. The power measurement is launched at time 0 and the smartphone only consumes the base power, described in Section \ref{ssc:OverviewTestbed}. Then, we touch the icon of the benchmark application at time $0.192$s (i.e., the moment that the benchmark application is launched). On the other hand, Fig. \ref{fig:touchLogcat} depicts the timestamps recorded by the Android kernel\footnote{These timestamps are also generated by clock \hytt{uptimeMillis} but with microsecond precision.} when the touch event is triggered, which denotes that the touch event happens at time $4716.264801$s. After the timestamp of the flag event is acquired, the local clocks in the tested smartphone and the power monitor can be synchronized easily and accurately. For example, if the start and end time for executing an image conversion are recorded as $4726.136$s and $4726.612$s in the log file generated by the benchmark application, the power consumption of the image conversion can be localized between $10.063$s and $10.539$s in the power measurement data recorded by the power monitor. Fig. \ref{fig:remoteexample} presents an example of event localization in the collected power measurement data through our proposed strategy. 

\begin{figure*}[t]
\centering
\includegraphics[width=0.98\textwidth]{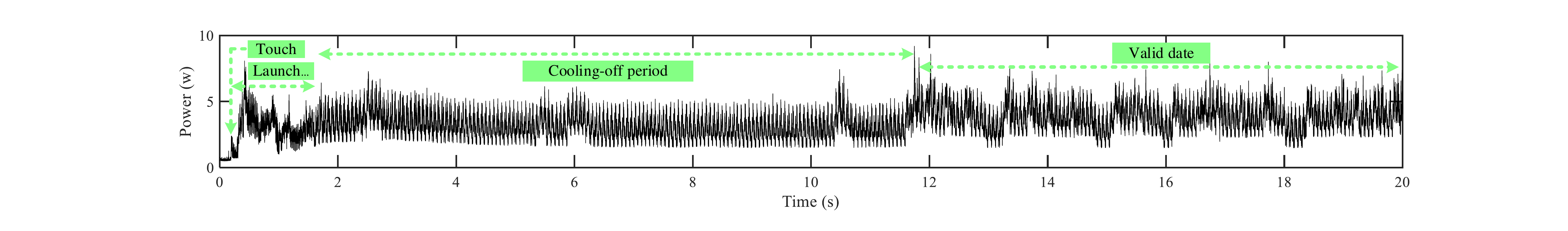}
\caption{Power measurement and valid data collection process.}
\label{fig:measurementoverview}
\end{figure*}

\begin{figure*}[t]
\centering
\includegraphics[width=0.98\textwidth]{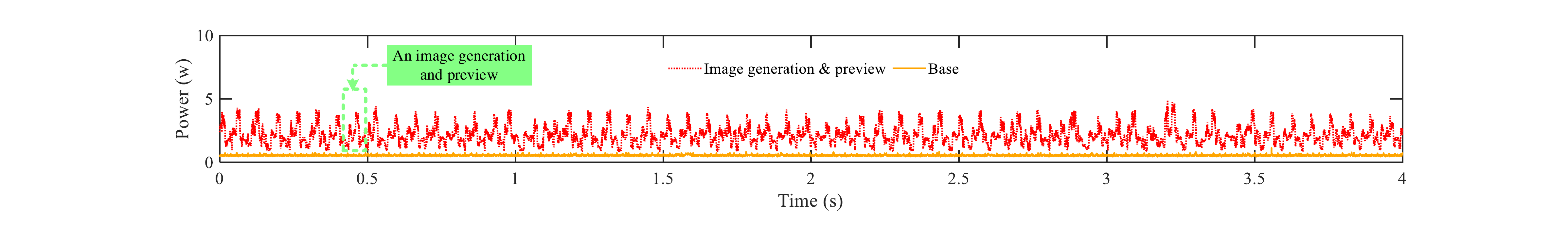}
\caption{A fragment of the power consumption of image generation and preview and base (CPU governor: Interactive).}
\label{fig:cameraexample}
\end{figure*}

\textbf{Script for Capturing the Touch Event.} Since the touch event happens before the benchmark application launching, the function of capturing the touch event cannot be directly added into the benchmark application. In addition, the tested smartphone's USB interface is automatically disabled when the power measurement starts. Thus, the smartphone cannot be instructed to start or terminate a touch event listener through transmitting \hytt{adb shell} commands by a computer. Considering the above mentioned limitations, we design a lightweight application and implement it to record touch events. The function of this application is to run a touch event listener in the background using an \hytt{adb shell} command \hytt{getevent -lt /dev/input/event0}. We evaluate whether running this background touch event listener will impact the power consumption of benchmark applications. The measurement results show that the average power consumption of the smartphone is $3.721$W (running the remote execution benchmark application with the touch event listener) and $3.704$W (running the remote execution benchmark application without the touch event listener), where the two measurements are under the same conditions and each measurement runs for 5 minutes. Therefore, the result demonstrates that our background touch event listener has little impact on the power consumption of benchmark applications. 

\textbf{Cooling-off Period.} Furthermore, in order to mitigate the interference from screen touching, application launching, camera initializing, and CNN model loading in the collected power measurement data, a cooling-off period is set up, as shown in Fig. \ref{fig:measurementoverview}. In the cooling-off period, benchmark applications only executes Phases 1 and 2 for generating a fixed number of image frames (e.g., $150$ frames in this paper). After the cooling-off period, benchmark applications start executing the whole processing pipeline and generating valid power consumption data.

\textbf{Power/Energy Consumption Dissection.} Fig. \ref{fig:Mpipe} illustrate that image generation and preview (Phases 1 and 2), image conversion and local/remote execution (Phases 3 and 4), and base (e.g., OS and screen) are executed in parallel. The workload of running our benchmark applications on the tested smartphone is composed of these three parallel executions. In addition, the power consumption is increased by the workload increment. Therefore, our strategy for dissecting the power consumption of each phase is:

\begin{enumerate}
    \item Measuring the power consumption of the whole processing pipeline, image generation and preview (Phases 1 and 2)\footnote{In order to measure the power consumption of phases 1 and 2, we implement an application that only executes image generation and preview, where it uses the same Android camera package (i.e., \hytt{android.hardware.camera2}) and camera configurations (e.g., preview resolution) with our benchmark applications.}, and base separately with the same configurations (e.g., CPU governor, camera sampling rate, and preview resolution) and conditions (e.g., background activity and screen brightness). Figs. \ref{fig:remoteexample} and \ref{fig:cameraexample} present examples of how the power consumption of these three parallel executions look like.
    \item Isolating the power consumption of (i) image generation and preview $+$ image conversion $+$ base, (ii) image generation and preview $+$ local/remote execution $+$ base, and (iii) image generation and preview $+$ others $+$ base through the proposed clock synchronization and event localization strategy.  
    \item Obtaining the power consumption of image conversion, local/remote execution, and others by subtracting the average power consumption of image generation and preview and base from cases (i), (ii), and (iii), respectively.
    \item Obtaining the energy consumption of each phase via calculating the integral of the power consumption over the latency. For example, the energy consumption of an image conversion is the sum of its power consumption within an image conversion latency.
\end{enumerate}


\textbf{Validation.}
Paper \cite{chowdhury2019greenscaler} observed that a single energy measurement could be misleading due to the variability in energy consumption. Therefore, in this paper, all of our measurement experiments are repeated multiple times, and each energy consumption and latency result shown in Sections \ref{sc:localmeasurement} and \ref{sc:remotemeasurement} is the mean value of completing $200$ object detections. Since we observe that the variance of the mean value of the measured data, such as power consumption, per frame latency, and per frame energy consumption, is negligible after the number of the collected object detections is over $200$ in all of our measurement experiments, collecting measurement results based on $200$ object detection executions is good enough for achieving stable and accurate results. Furthermore, in order to ensure that each measurement is launched with a clean environment, the benchmark application is re-installed on the tested smartphone through Android Studio and the data generated during the execution such as log files are transferred to a workstation and removed from the smartphone after each measurement, even though the configuration of the benchmark application does not require to be changed in the next measurement.

\subsection{Key Performance Metrics}
\label{ssc:metrics}
We define three performance metrics to evaluate the performance of the CNN-based object detection processing pipeline implemented in this paper:

\textbf{Per Frame Latency.} The per frame latency is the total time needed to obtain the detection results on one image frame (i.e., usually shown as one or multiple bounding boxes that identify the locations and classifications of the objects in a frame). In this paper, it is defined as the time period from the moment the \emph{Image Reader} acquires one camera captured image frame to the moment the bounding boxes are drawn on the mobile AR client's screen, as depicted in Fig. \ref{fig:pipeline}. In the local execution, the per frame latency includes the time used for converting the YUV frame to the RGB frame, cropping the frame to the fitted resolution $k\times k$ pixels, and executing CNN, defined as \emph{inference latency}, on the smartphone. In the remote execution, the per frame latency includes, besides the image conversion and crop latency that are both executed locally on the smartphone, the communication latency (i.e., transmitting the frame and receiving the results) and the inference latency on the edge server.

\textbf{Per Frame Energy Consumption.} The per frame energy consumption is the total amount of energy consumed in a mobile AR client by successfully performing the object detection on one image frame. In the local execution, the per frame energy consumption includes the energy consumed by camera sampling (i.e., image generation), screen rendering (i.e., preview), image conversion, inference, and operating system (i.e., base). In the remote execution, it includes the energy consumed by camera sampling, screen rendering, image conversion, communication, and operating system. In a per frame energy consumption, the image generation and preview are usually executed multiple times (depends on the length of the per frame latency), while the image conversion and local/remote execution are executed only once.

\textbf{Detection Accuracy.} The mean average precision (mAP) is a commonly used performance metric in object detection. Better performance is indicated by a higher mAP value. Specifically, the average precision \cite{everingham2010pascal} is computed as the area under the precision/recall curve through numerical integration. The mAP is the mean of the average precision across all classes.


\section{Experimental Results of Local Execution}
\label{sc:localmeasurement}

\begin{figure*}[t]
\centering
\subfigure[Conservative]
{\includegraphics[width=0.16\textwidth]{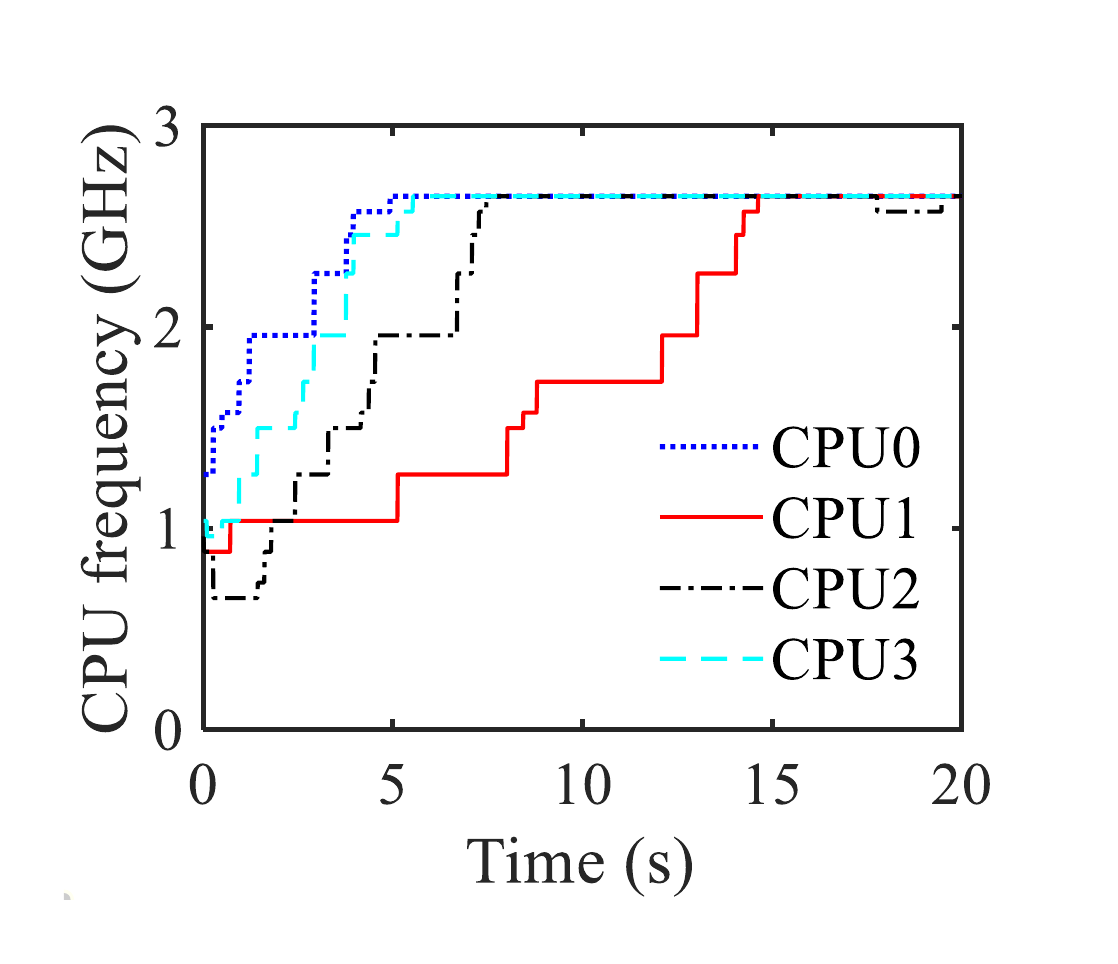}\label{fig:TFLlocal_CPUfreq_Conser}}
\subfigure[Ondemand]
{\includegraphics[width=0.16\textwidth]{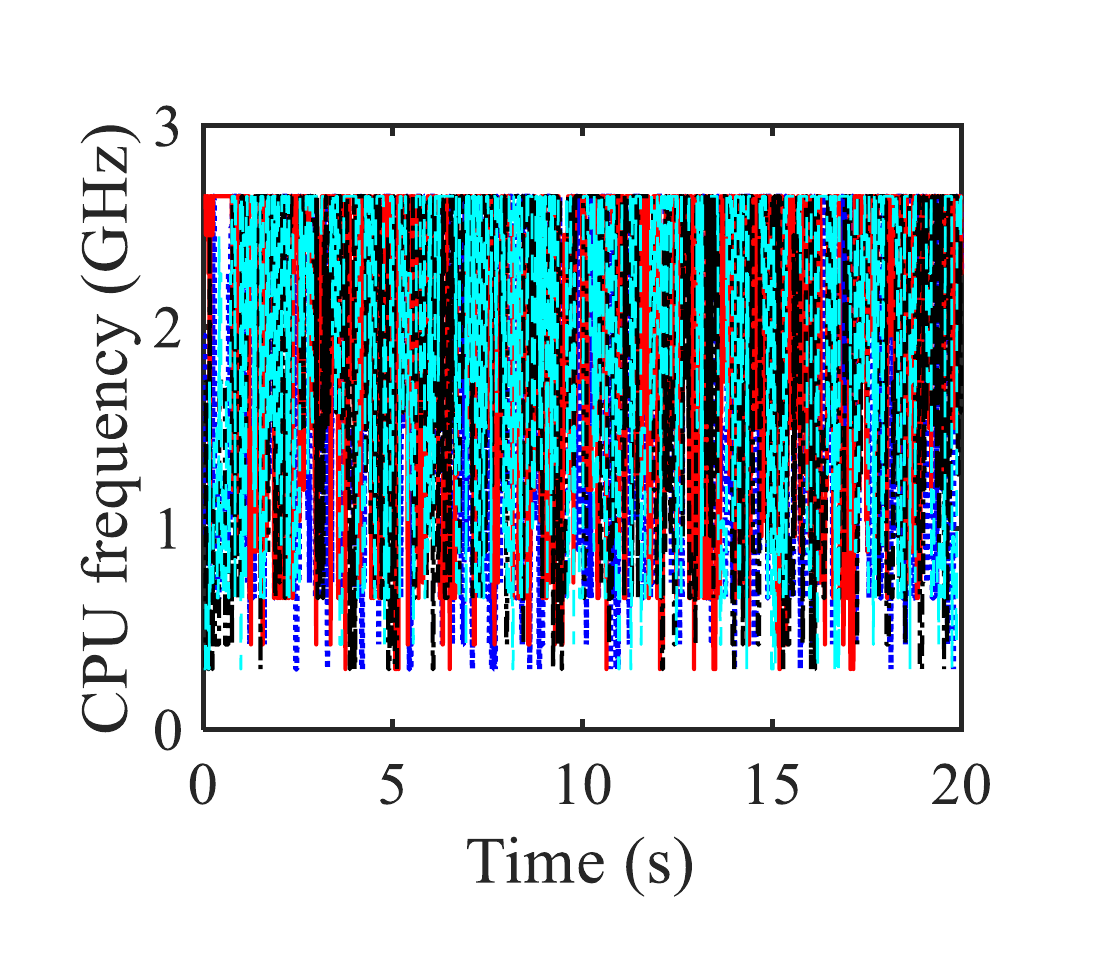}\label{fig:TFLlocal_CPUfreq_Ond}}
\subfigure[Interactive]
{\includegraphics[width=0.16\textwidth]{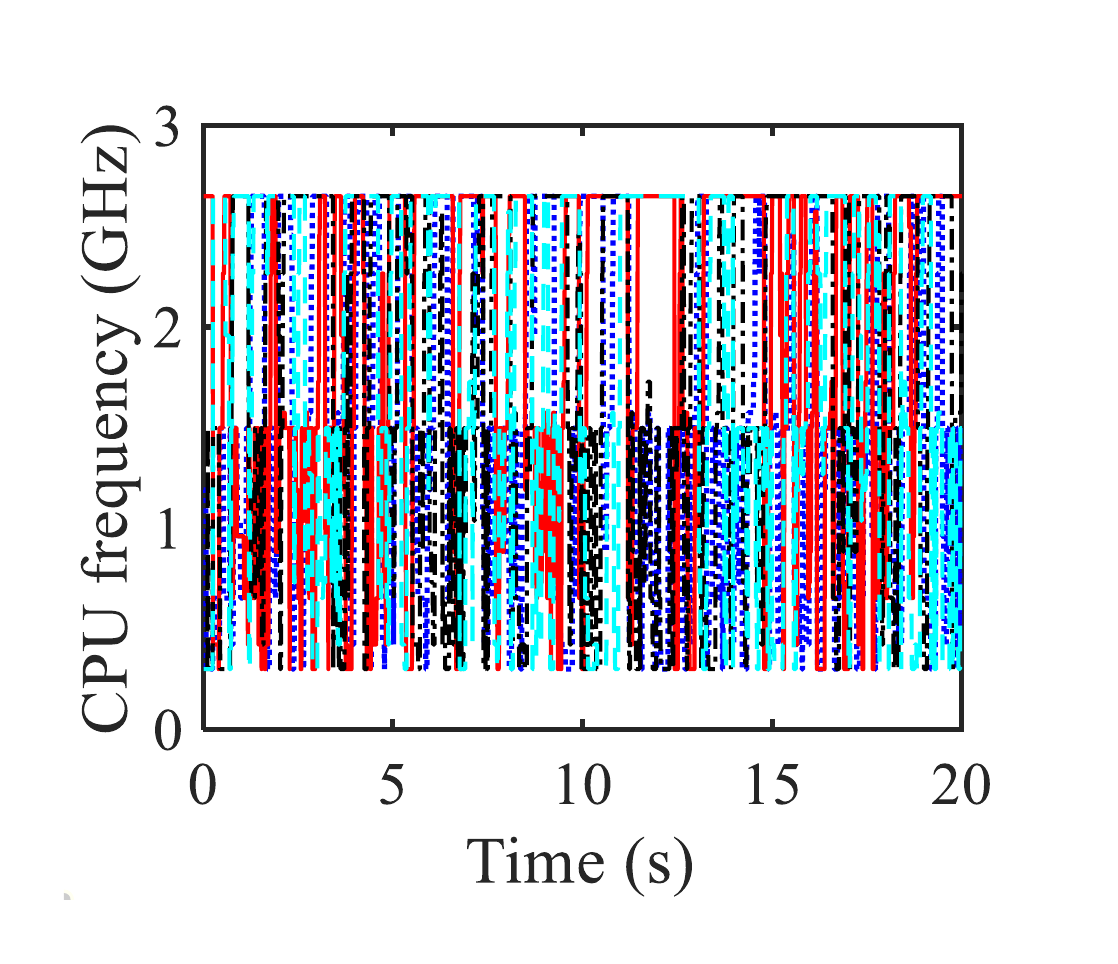}\label{fig:TFLlocal_CPUfreq_Inter}}
\subfigure[Userspace]
{\includegraphics[width=0.16\textwidth]{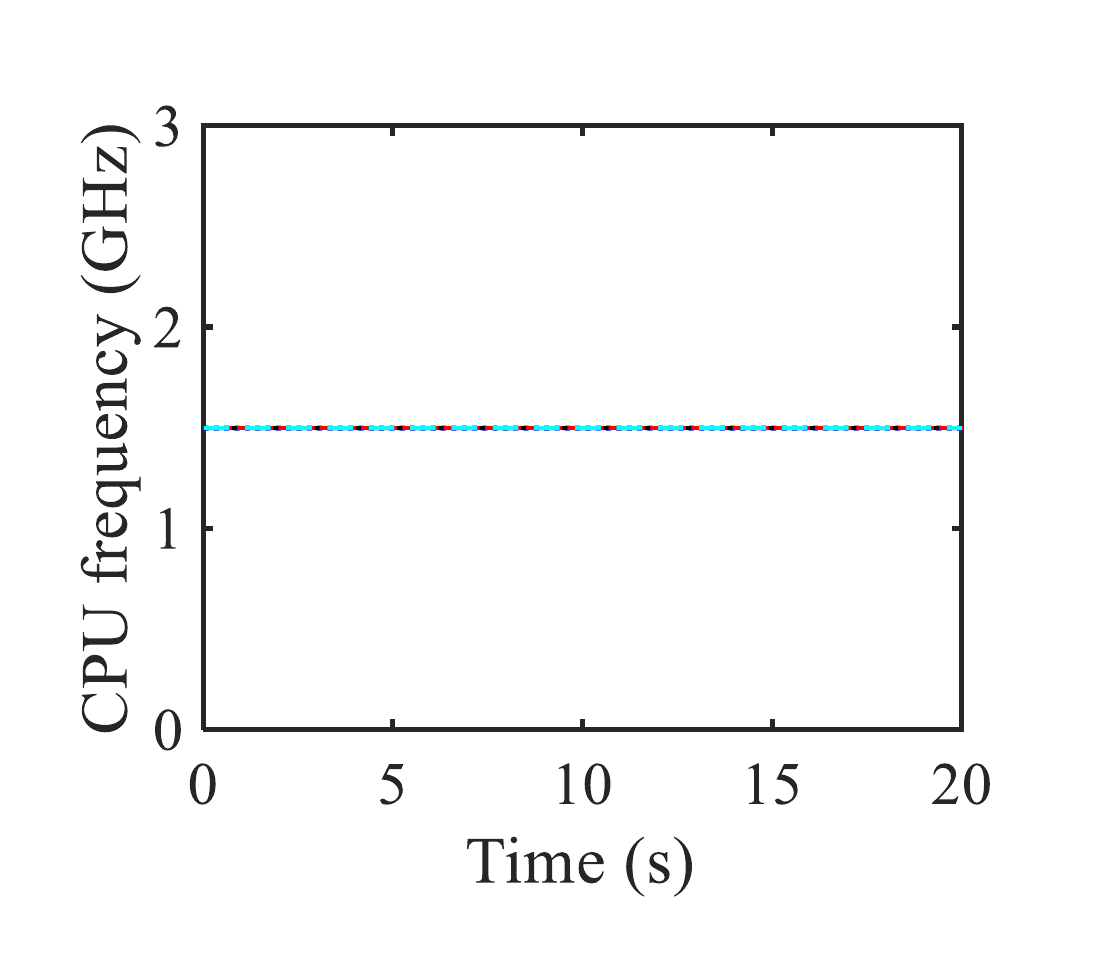}\label{fig:TFLlocal_CPUfreq_Use}}
\subfigure[Powersave]
{\includegraphics[width=0.16\textwidth]{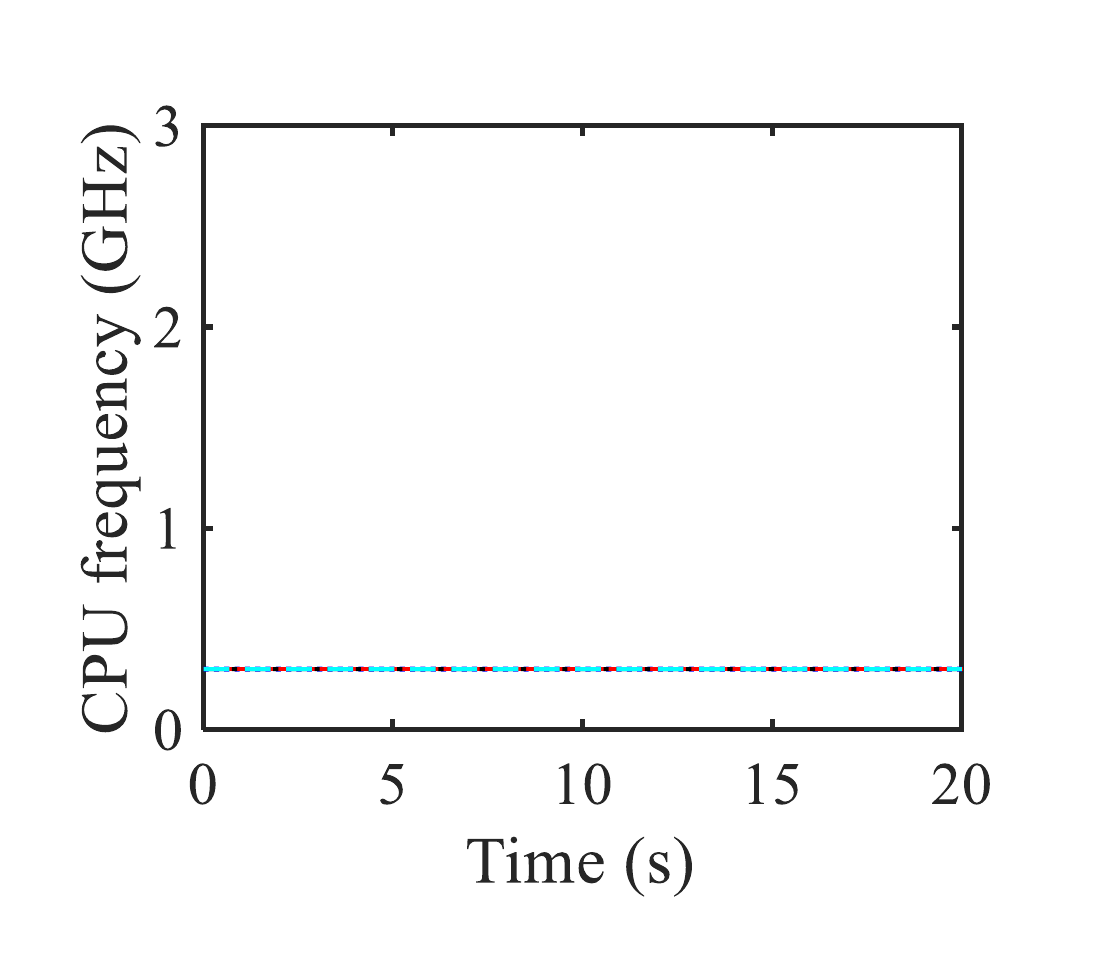}\label{fig:TFLlocal_CPUfreq_Pow}}
\subfigure[Performance]
{\includegraphics[width=0.16\textwidth]{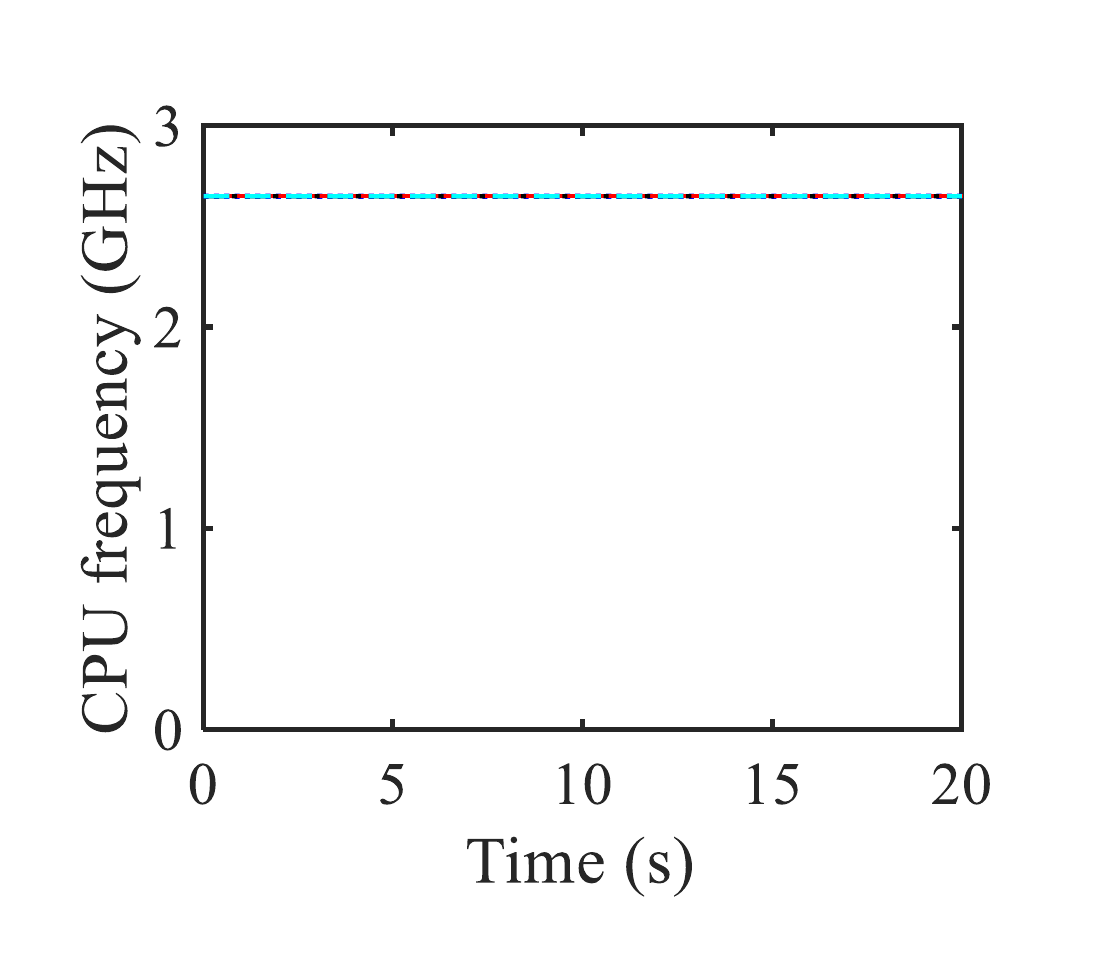}\label{fig:TFLlocal_CPUfreq_Perf}}

\subfigure[Conservative]
{\includegraphics[width=0.16\textwidth]{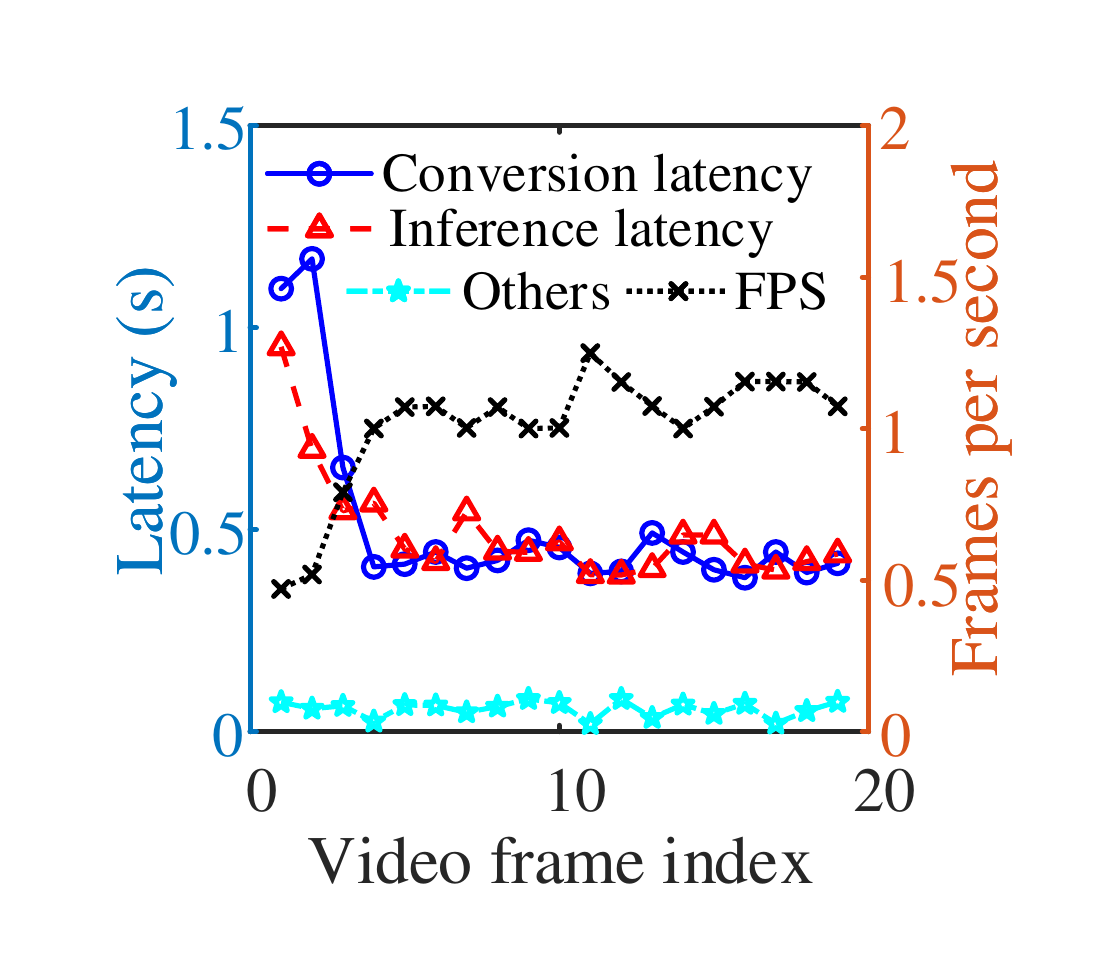}\label{fig:TFLlocal_CPU_Governors_Latency_Conser}}
\subfigure[Ondemand]
{\includegraphics[width=0.16\textwidth]{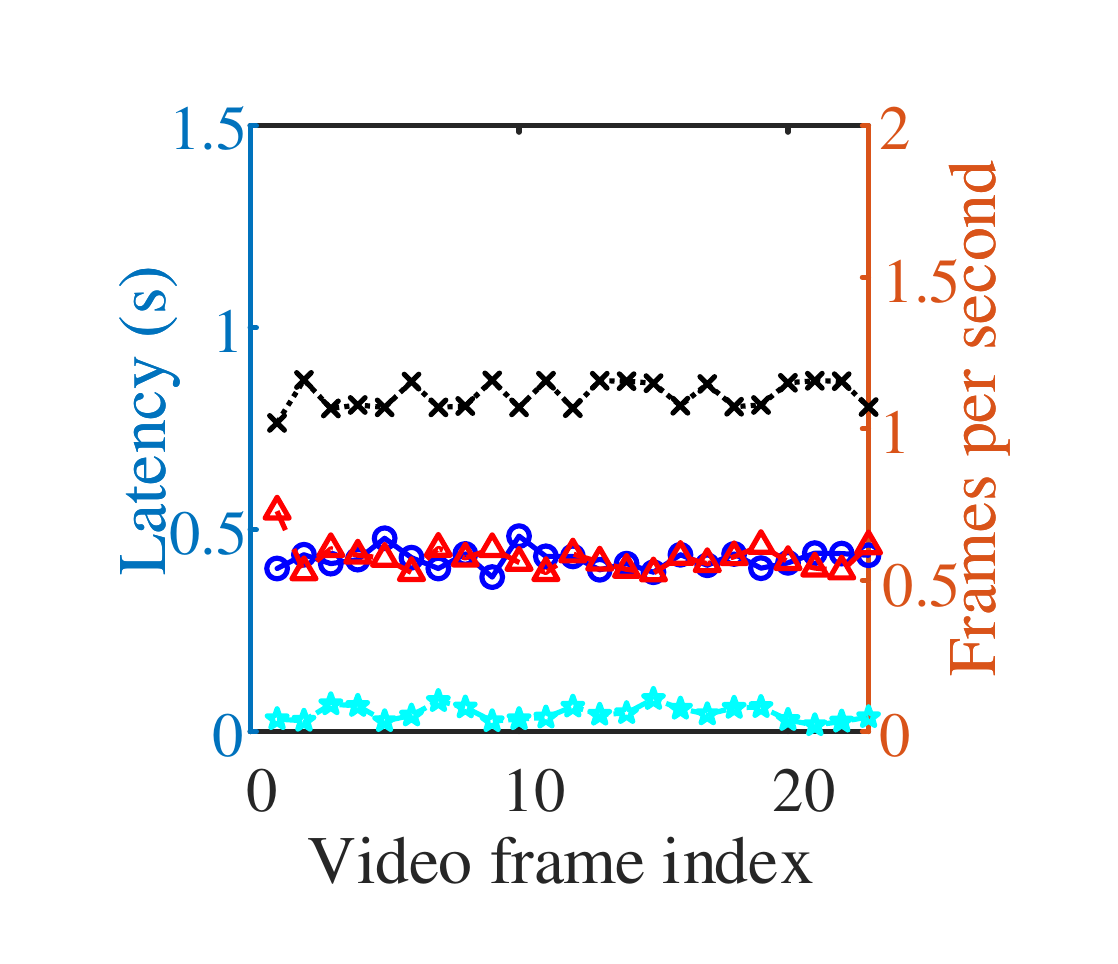}\label{fig:TFLlocal_CPU_Governors_Latency_Ond}}
\subfigure[Interactive]
{\includegraphics[width=0.16\textwidth]{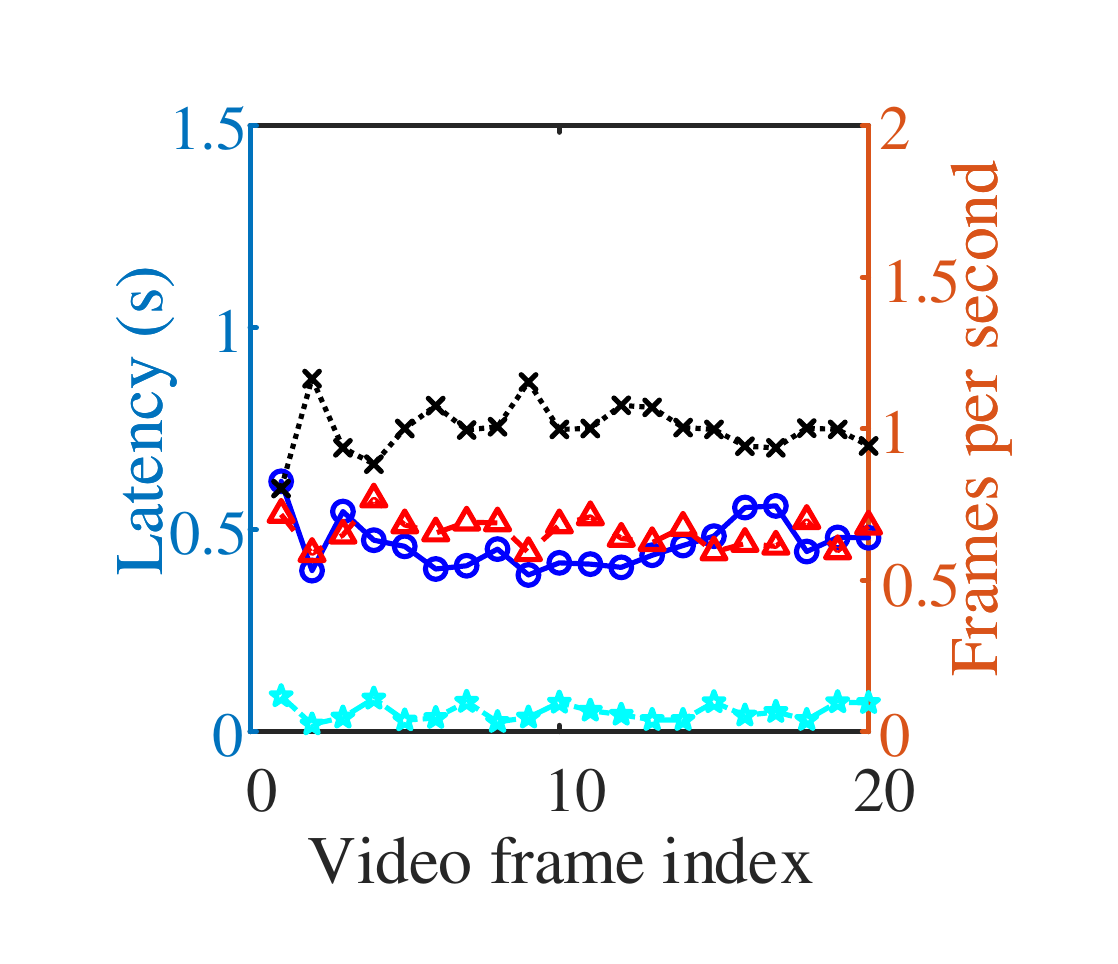}\label{fig:TFLlocal_CPU_Governors_Latency_Inter}}
\subfigure[Userspace]
{\includegraphics[width=0.16\textwidth]{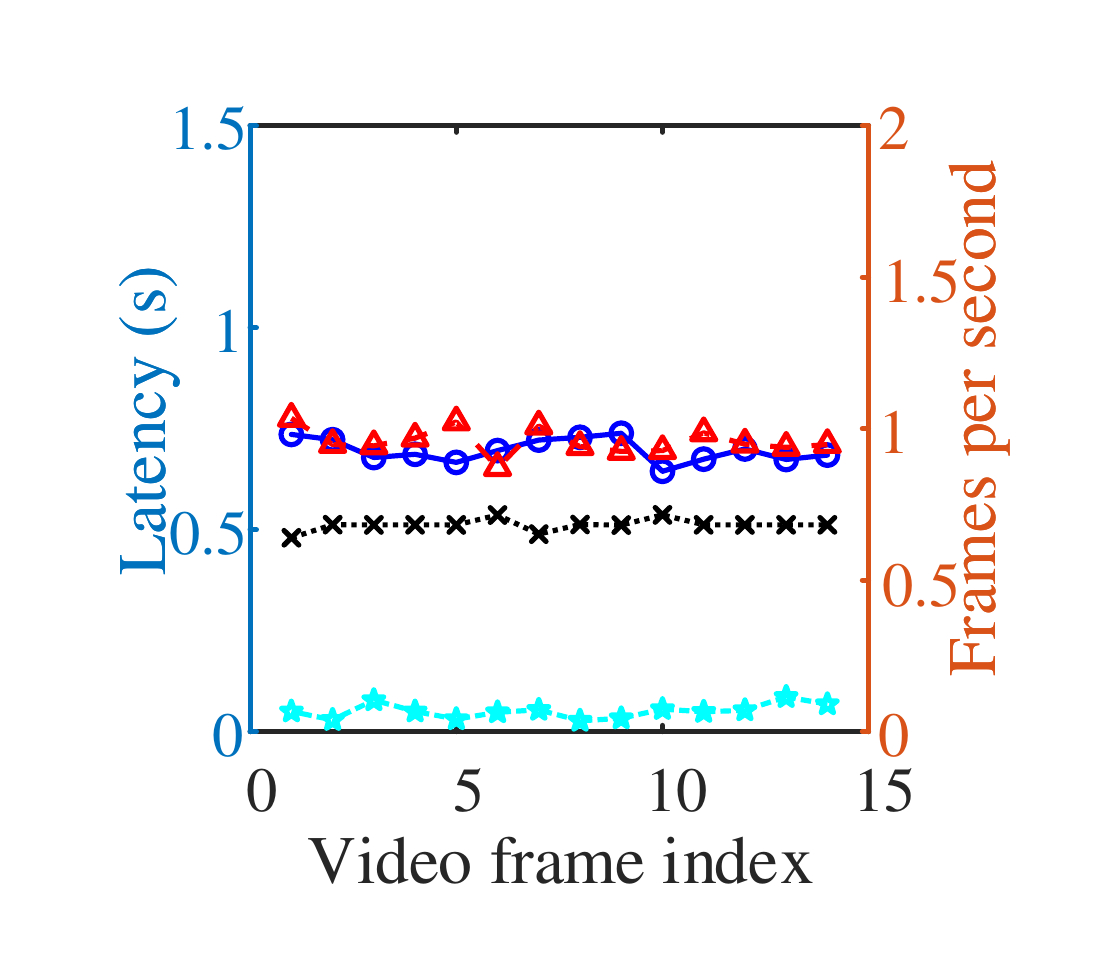}\label{fig:TFLlocal_CPU_Governors_Latency_Use}}
\subfigure[Powersave]
{\includegraphics[width=0.16\textwidth]{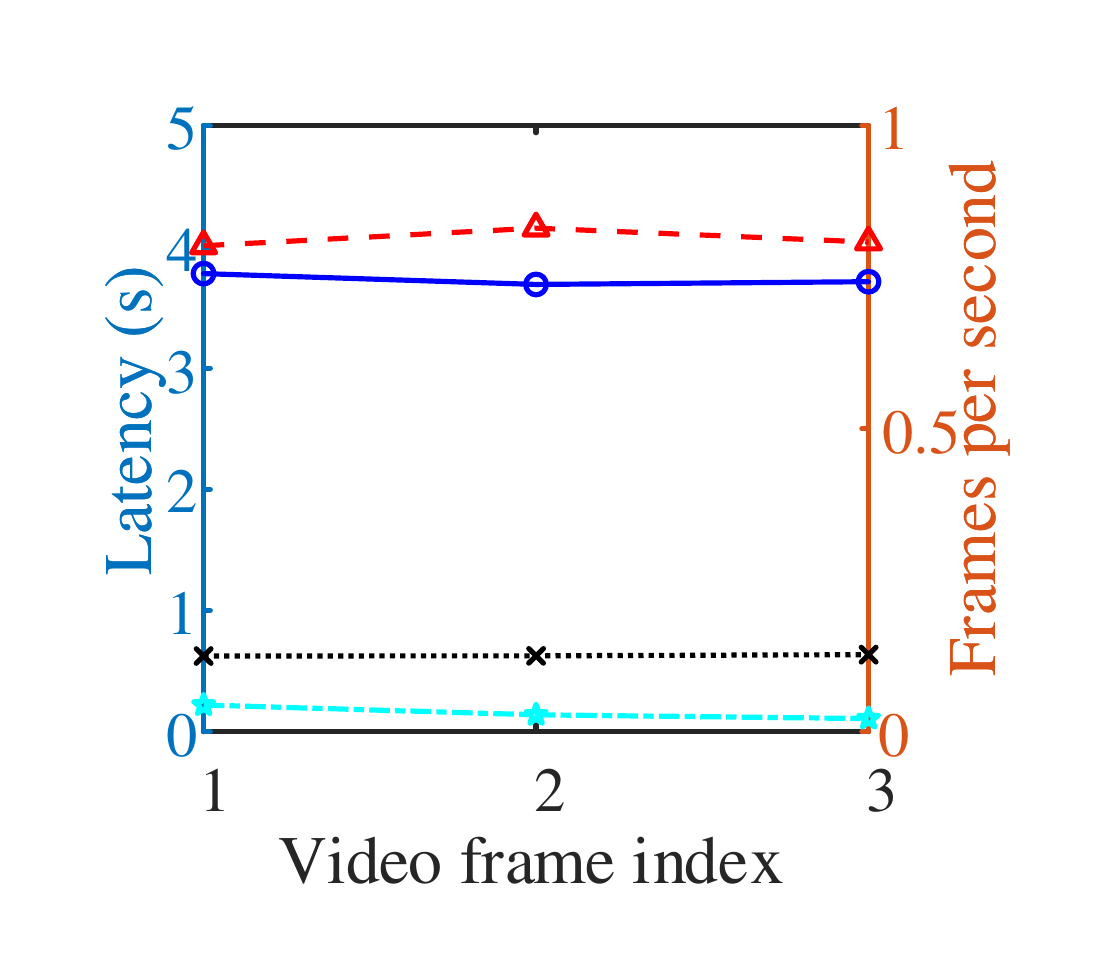}\label{fig:TFLlocal_CPU_Governors_Latency_Pow}}
\subfigure[Performance]
{\includegraphics[width=0.16\textwidth]{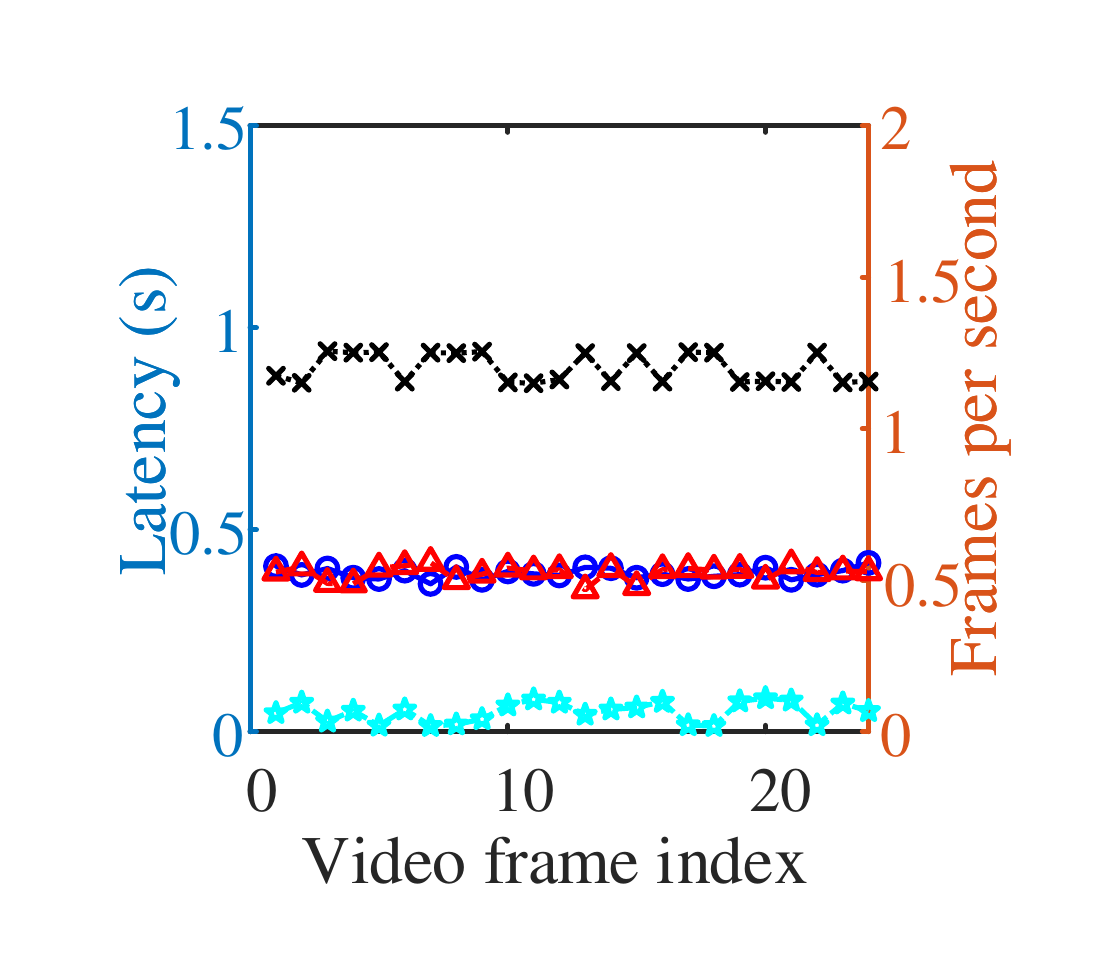}\label{fig:TFLlocal_CPU_Governors_Latency_Perf}}
\caption{CPU governor vs. per frame latency (CNN model size: $300\times300$ pixels).}
\label{fig:cpugovernors_latencylocal}   
\end{figure*}

\begin{figure*}[t]
\centering
\subfigure[Conservative]
{\includegraphics[width=0.16\textwidth]{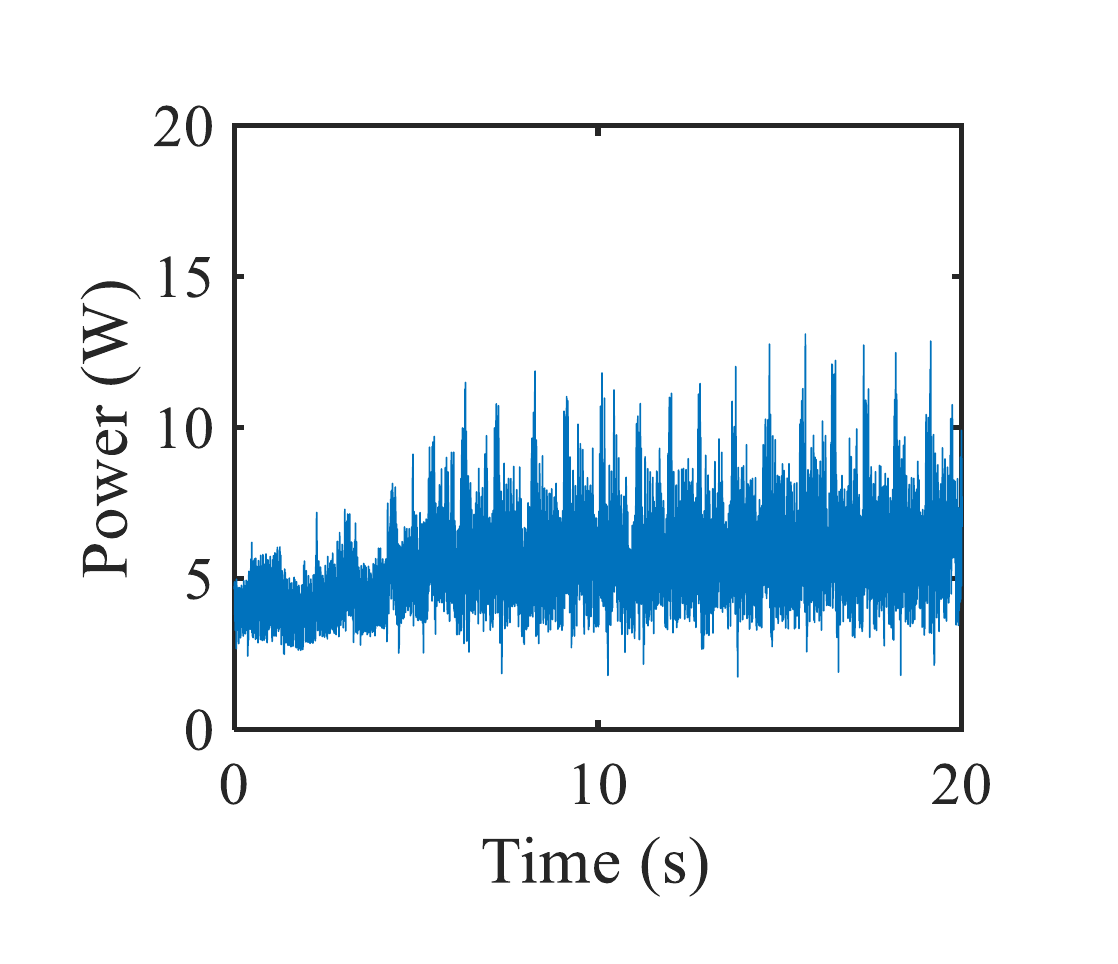}\label{fig:TFLlocal_CPU_Governors_Power_Conser}}
\subfigure[Ondemand]
{\includegraphics[width=0.16\textwidth]{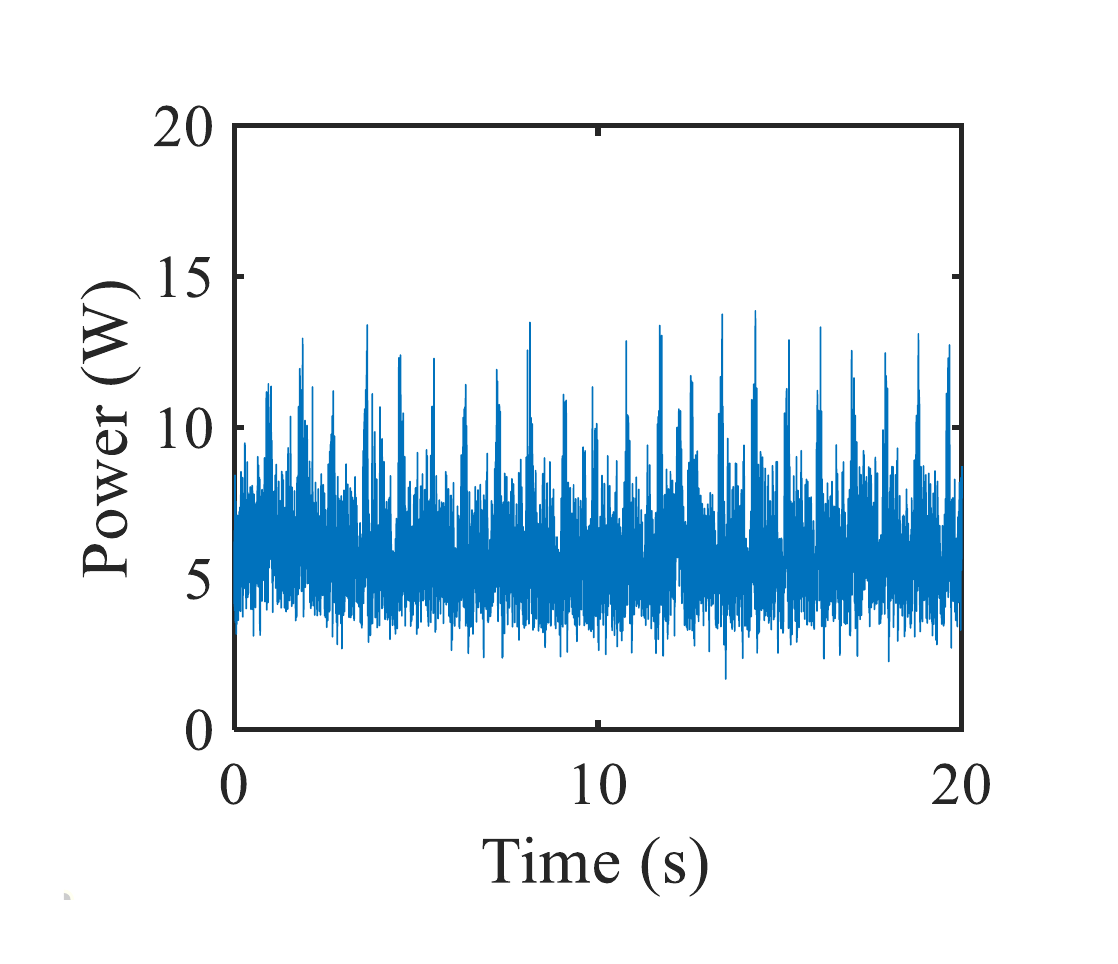}\label{fig:TFLlocal_CPU_Governors_Power_Ond}}
\subfigure[Interactive]
{\includegraphics[width=0.16\textwidth]{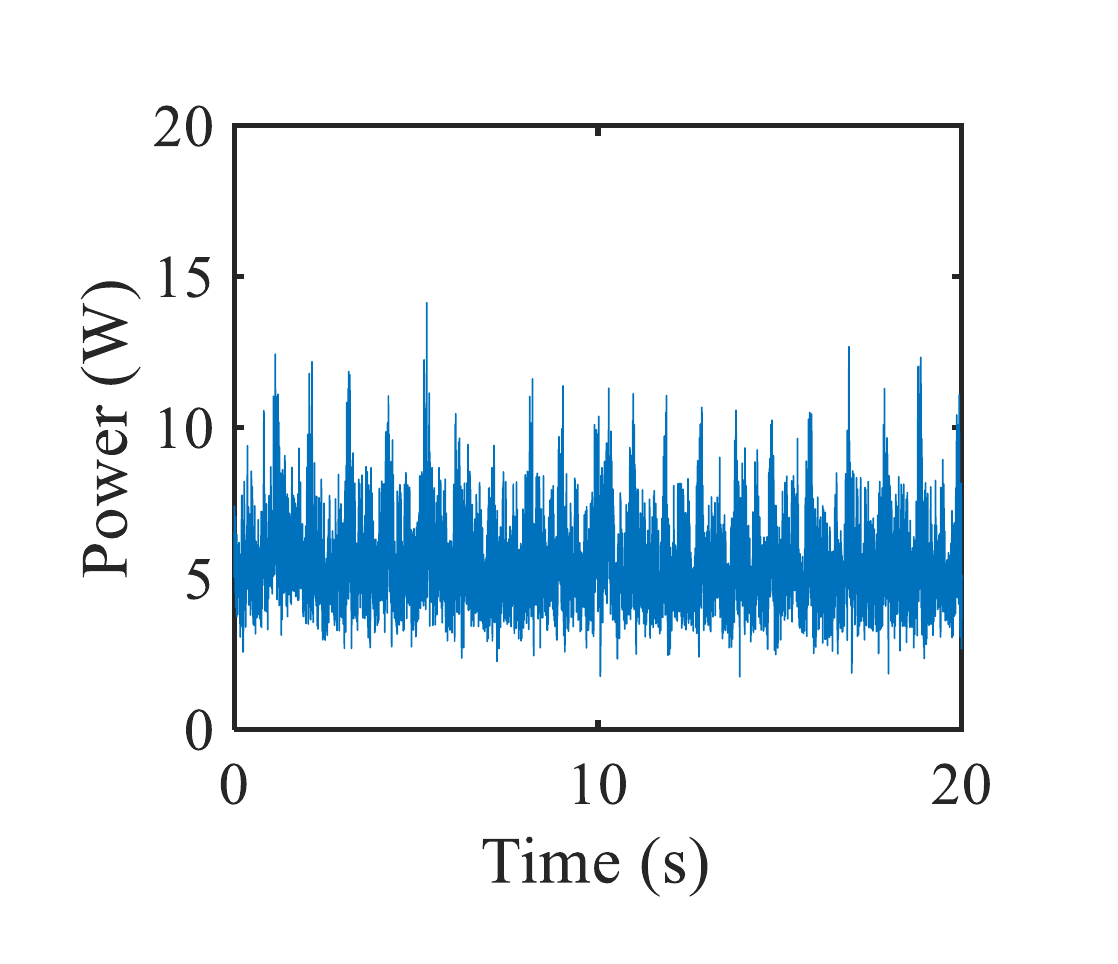}\label{fig:TFLlocal_CPU_Governors_Power_Inter}}
\subfigure[Userspace]
{\includegraphics[width=0.16\textwidth]{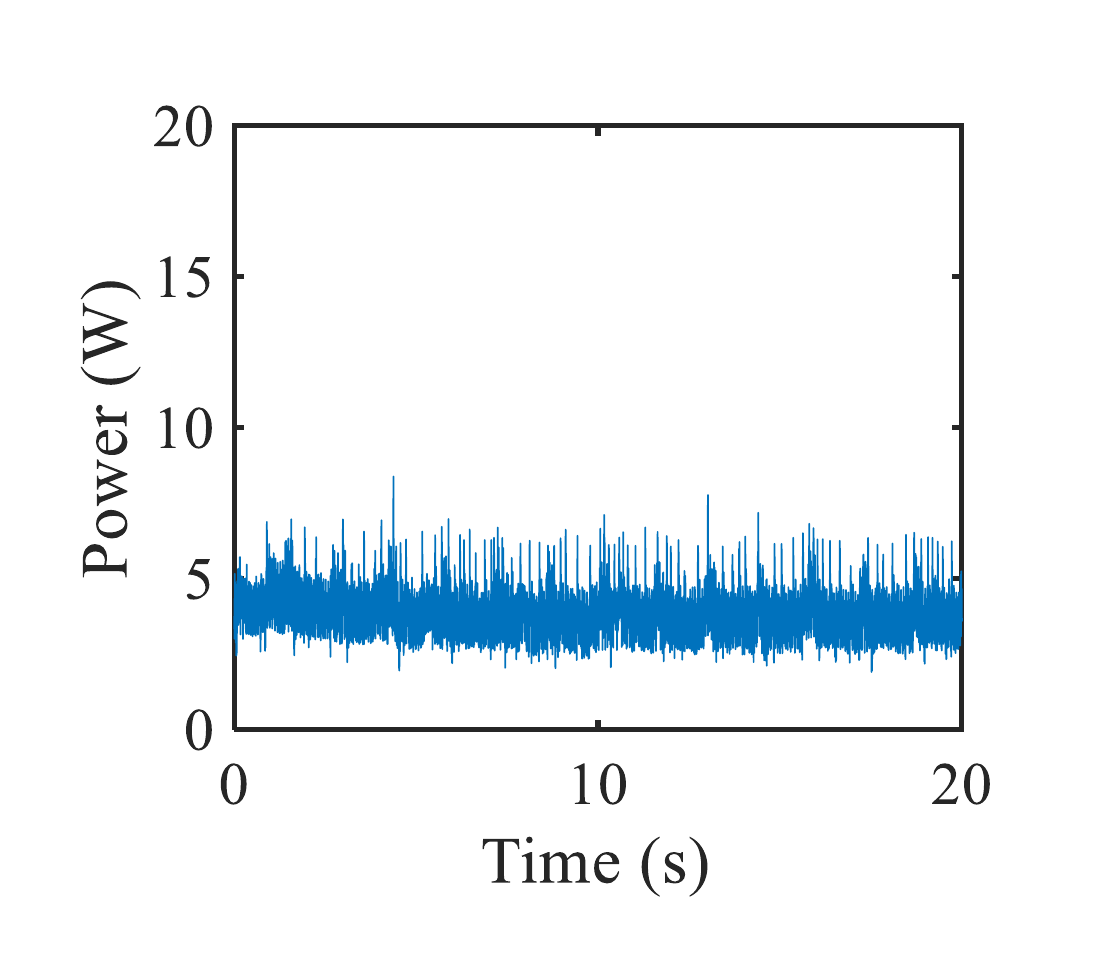}\label{fig:TFLlocal_CPU_Governors_Power_Use}}
\subfigure[Powersave]
{\includegraphics[width=0.16\textwidth]{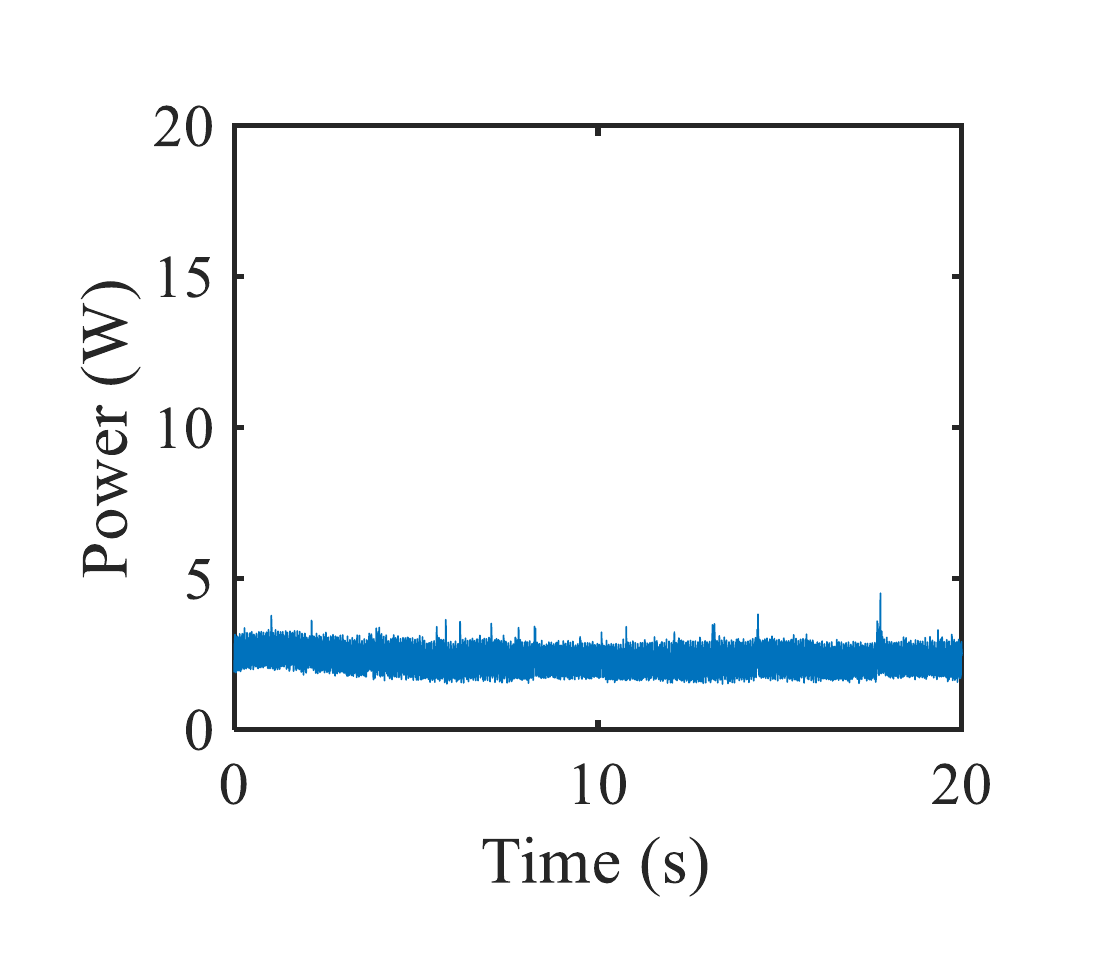}\label{fig:TFLlocal_CPU_Governors_Power_Pow}}
\subfigure[Performance]
{\includegraphics[width=0.16\textwidth]{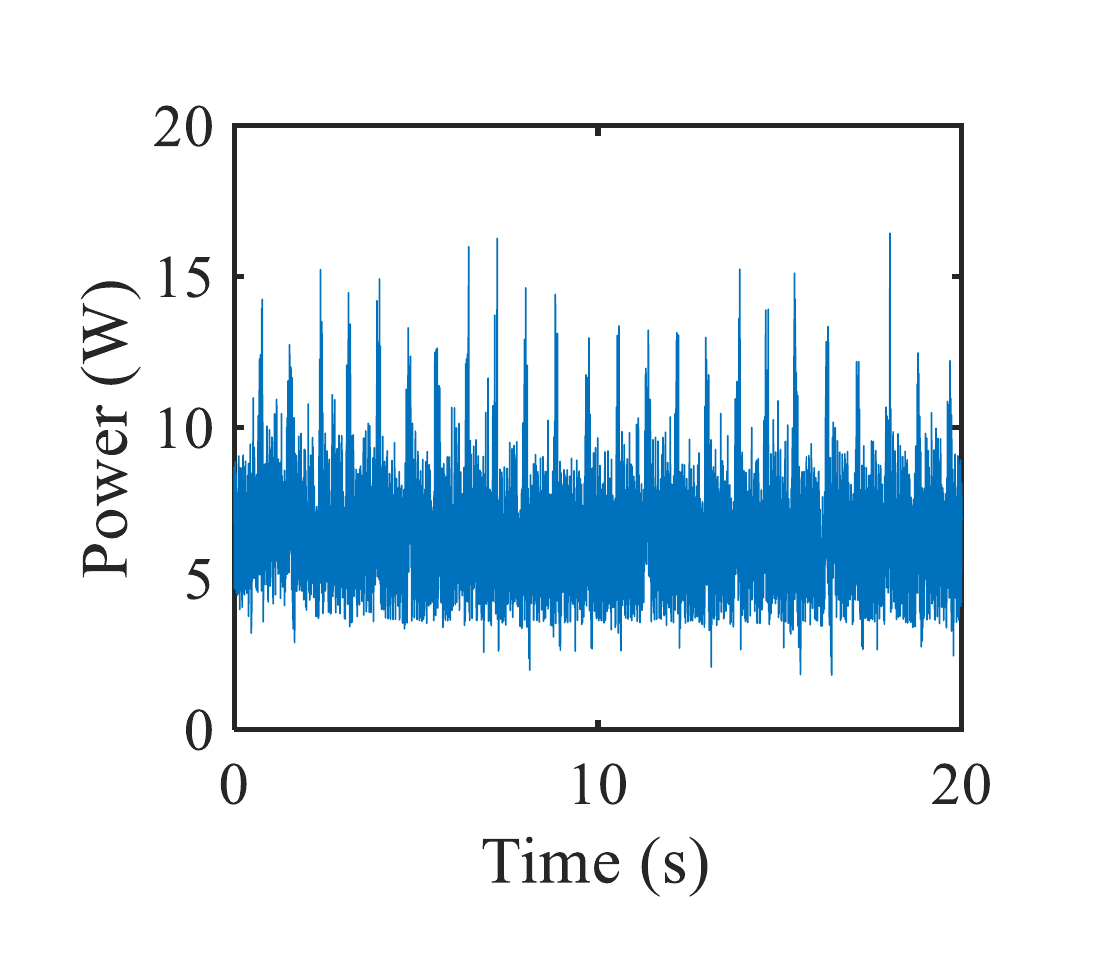}\label{fig:TFLlocal_CPU_Governors_Power_Perf}}

\subfigure[Conservative]
{\includegraphics[width=0.16\textwidth]{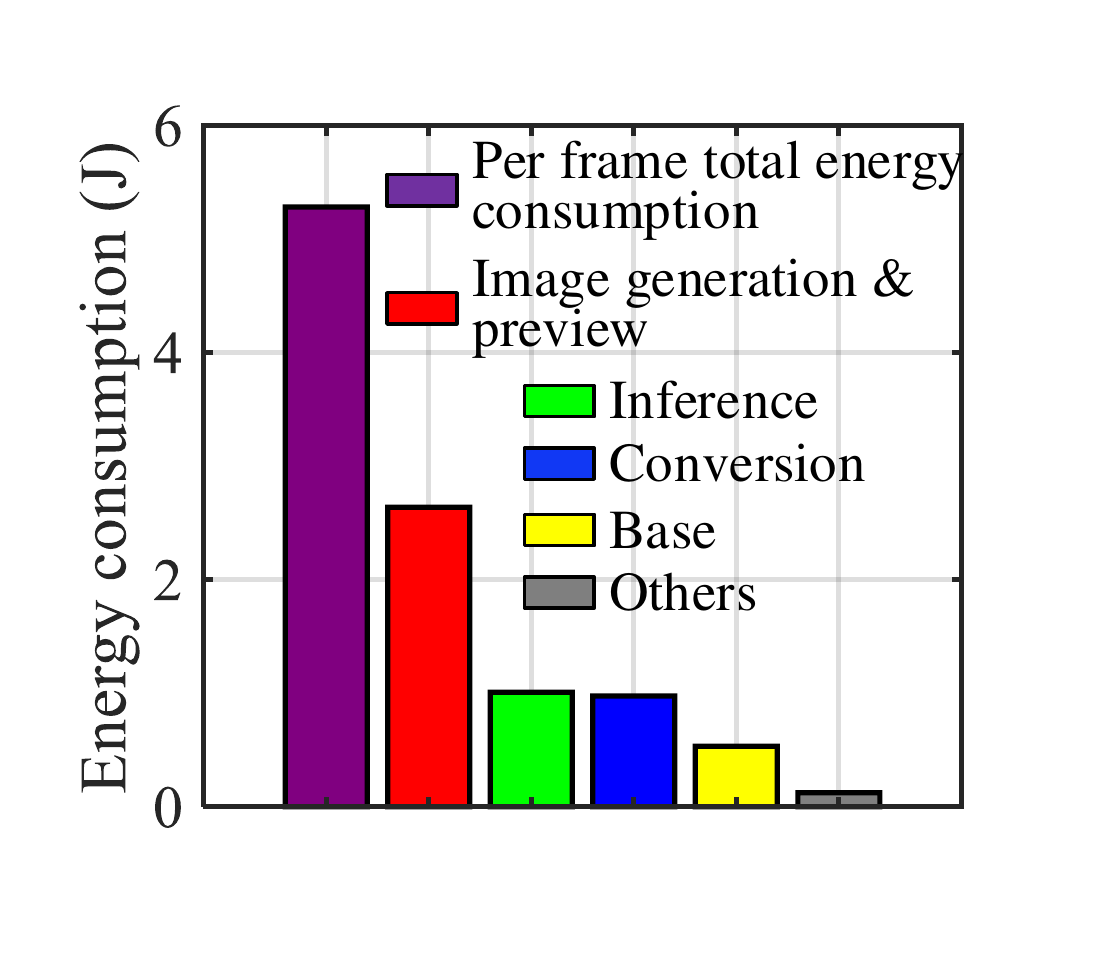}\label{fig:TFLlocal_CPU_Governors_Energy_Conser}}
\subfigure[Ondemand]
{\includegraphics[width=0.16\textwidth]{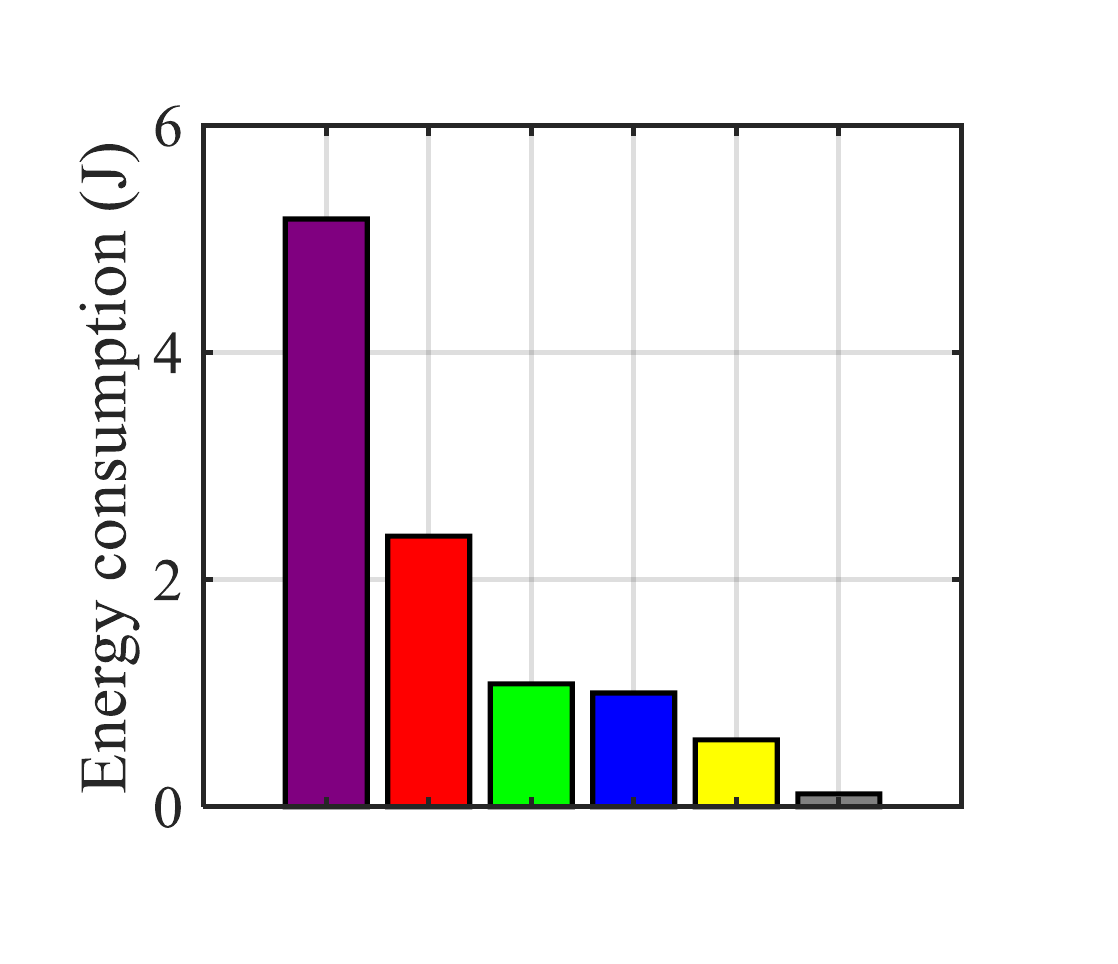}\label{fig:TFLlocal_CPU_Governors_Energy_Ond}}
\subfigure[Interactive]
{\includegraphics[width=0.16\textwidth]{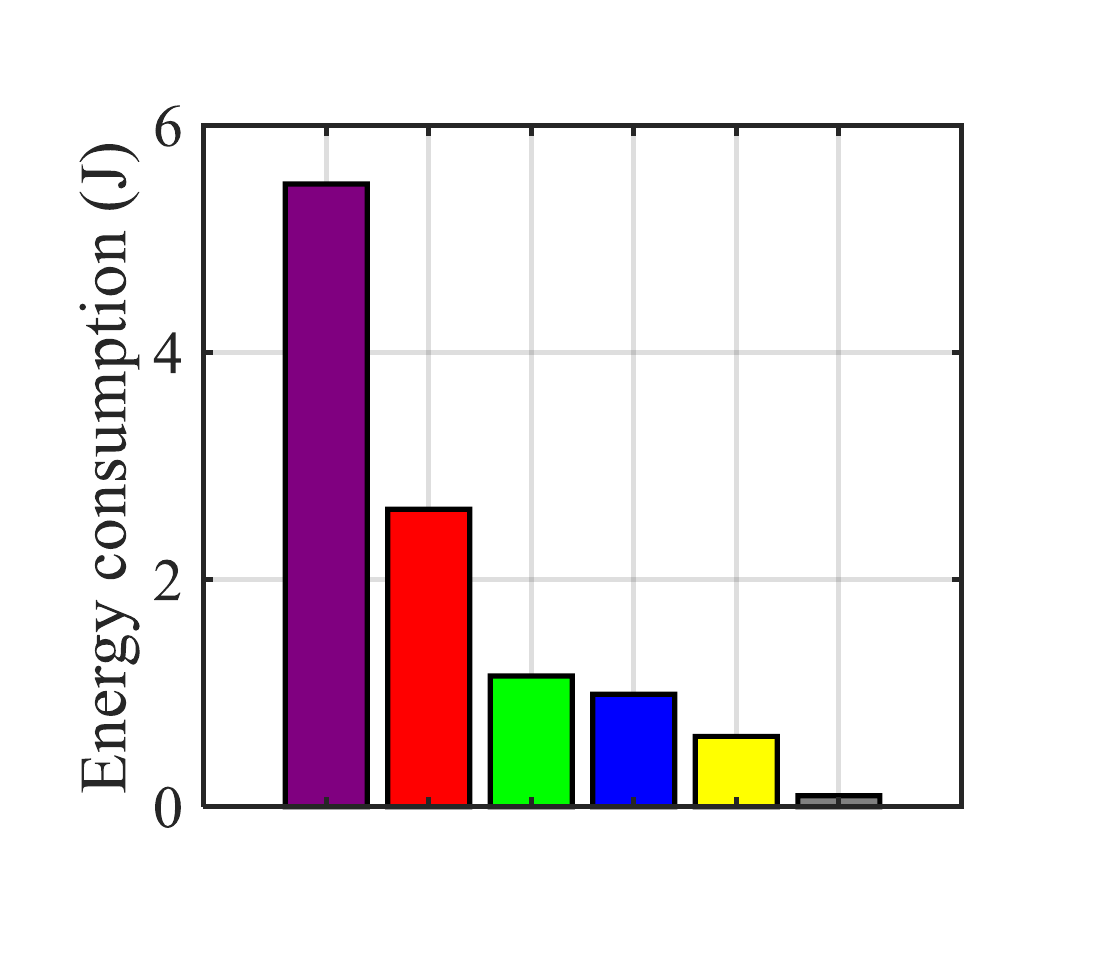}\label{fig:TFLlocal_CPU_Governors_Energy_Inter}}
\subfigure[Userspace]
{\includegraphics[width=0.16\textwidth]{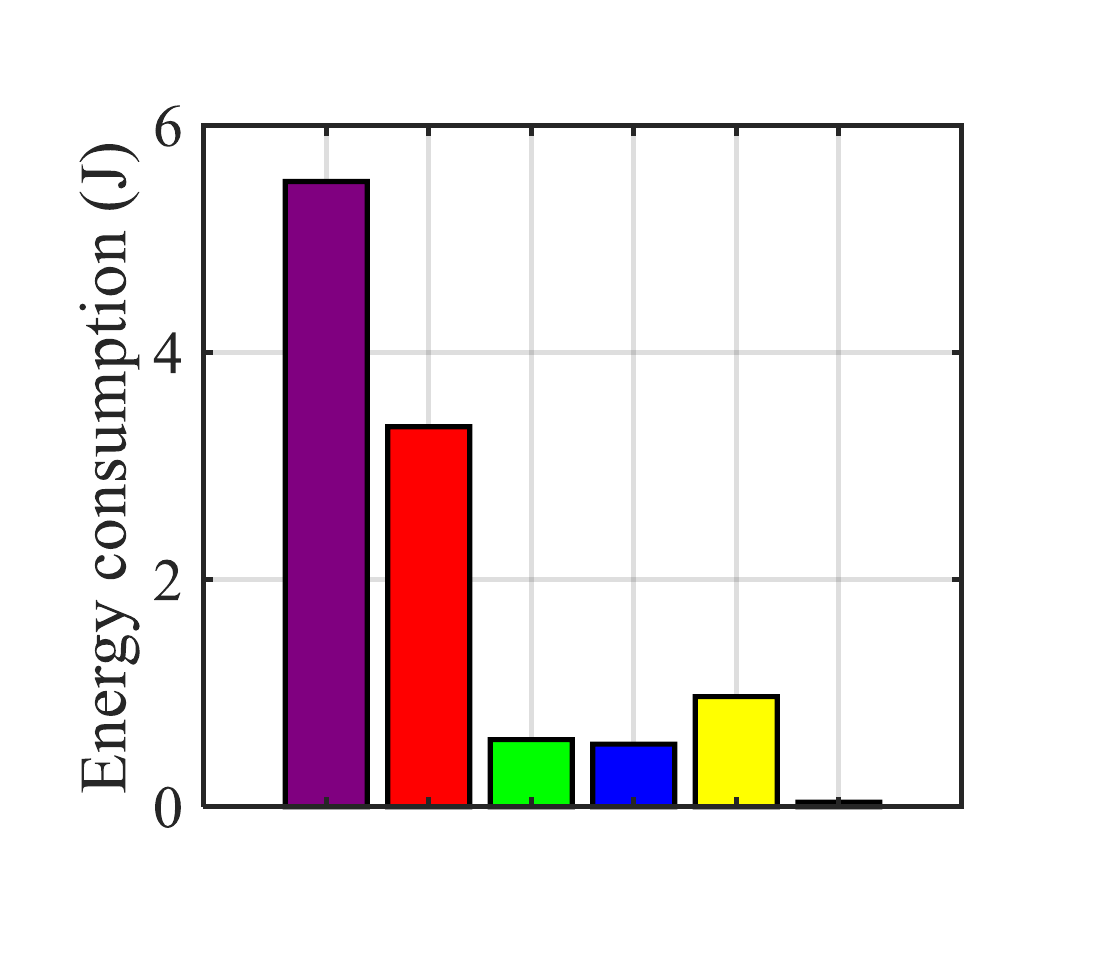}\label{fig:TFLlocal_CPU_Governors_Energy_Use}}
\subfigure[Powersave]
{\includegraphics[width=0.16\textwidth]{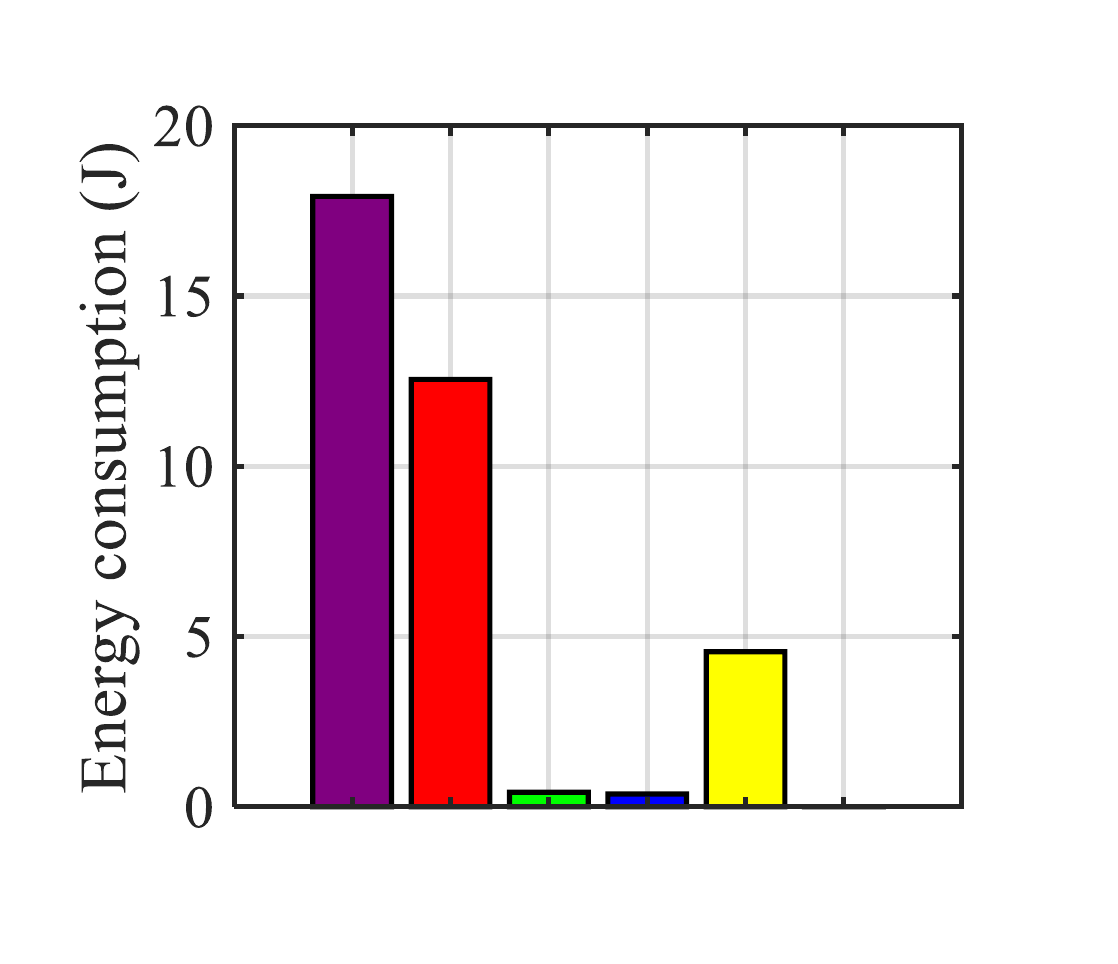}\label{fig:TFLlocal_CPU_Governors_Energy_Pow}}
\subfigure[Performance]
{\includegraphics[width=0.16\textwidth]{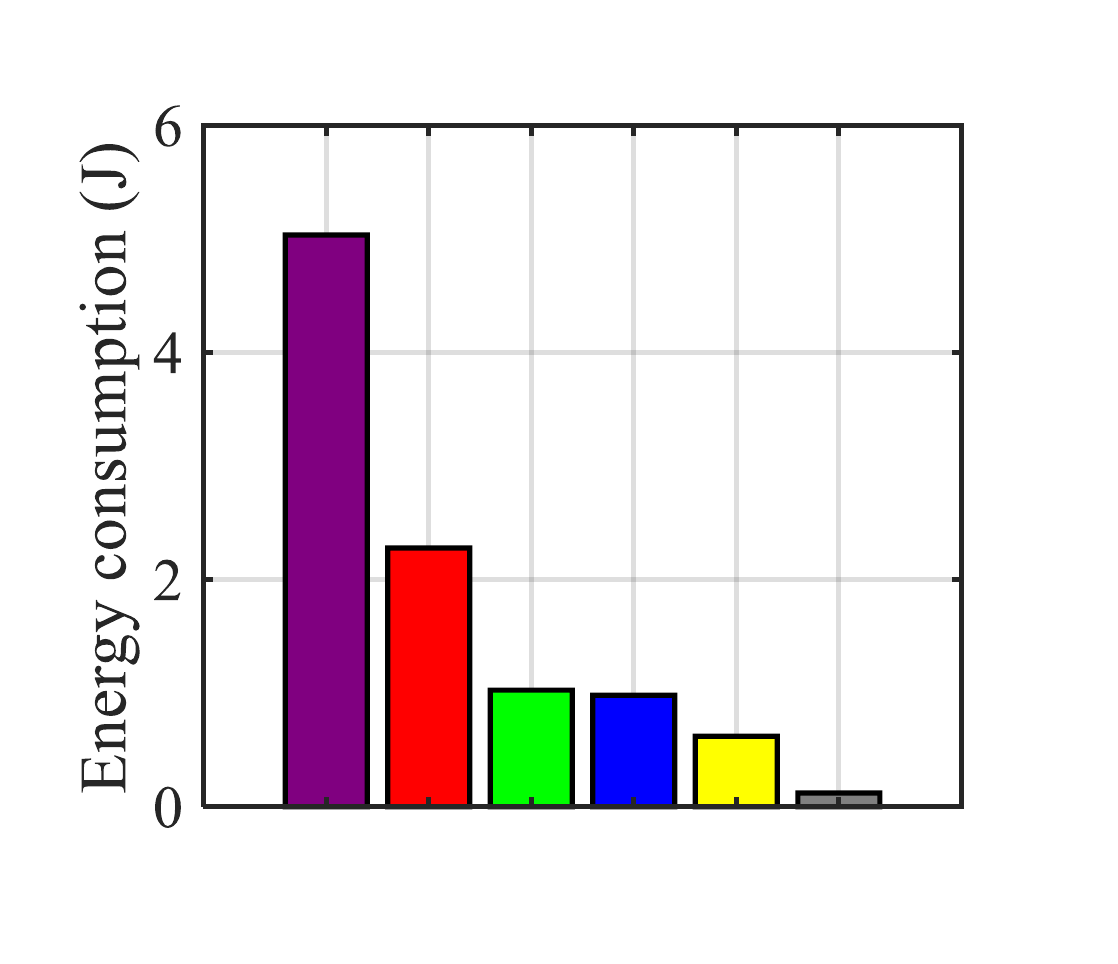}\label{fig:TFLlocal_CPU_Governors_Energy_Perf}}

\subfigure[Conservative]
{\includegraphics[width=0.16\textwidth]{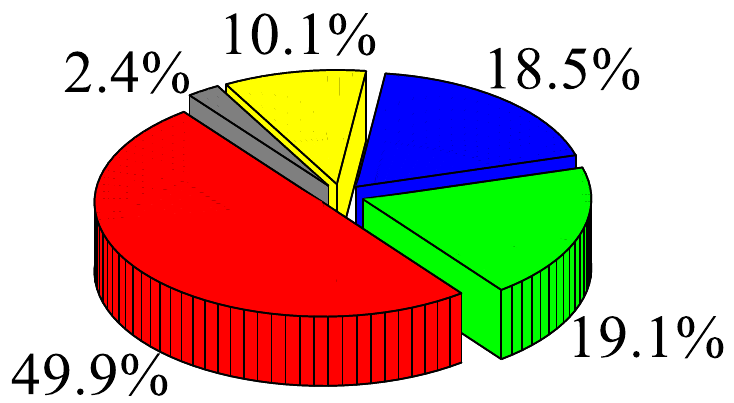}\label{fig:TFLlocal_CPU_Governors_Epercent_Conser}}
\subfigure[Ondemand]
{\includegraphics[width=0.16\textwidth]{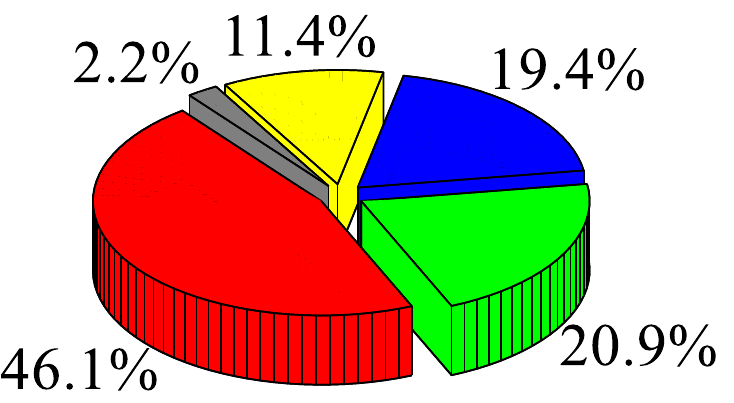}\label{fig:TFLlocal_CPU_Governors_Epercent_Ond}}
\subfigure[Interactive]
{\includegraphics[width=0.16\textwidth]{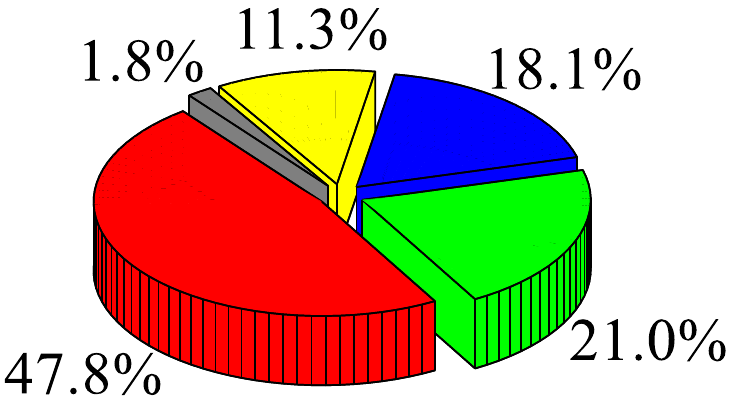}\label{fig:TFLlocal_CPU_Governors_Epercent_Inter}}
\subfigure[Userspace]
{\includegraphics[width=0.16\textwidth]{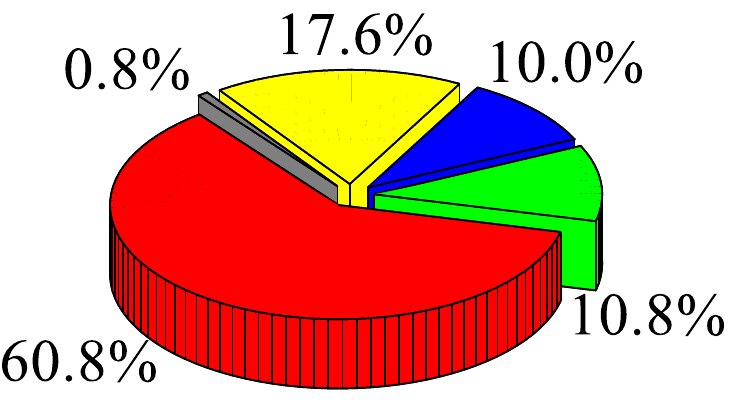}\label{fig:TFLlocal_CPU_Governors_Epercent_Use}}
\subfigure[Powersave]
{\includegraphics[width=0.16\textwidth]{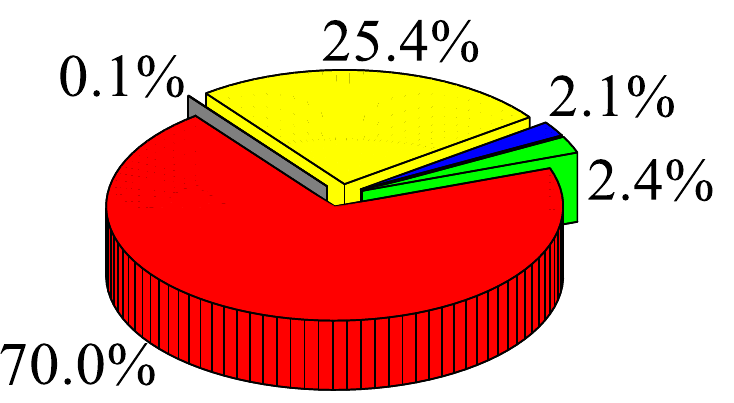}\label{fig:TFLlocal_CPU_Governors_Epercent_Pow}}
\subfigure[Performance]
{\includegraphics[width=0.16\textwidth]{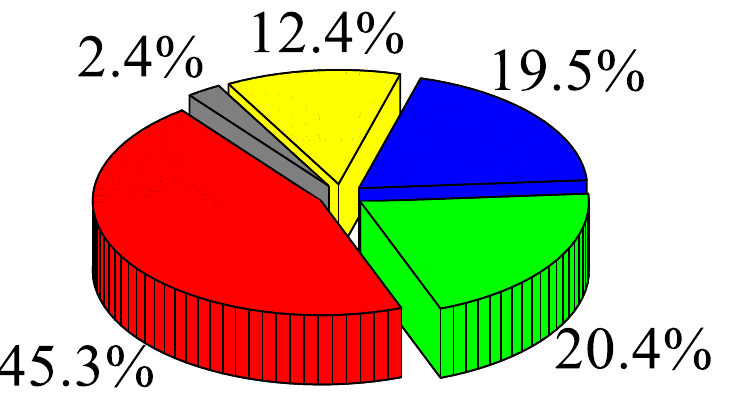}\label{fig:TFLlocal_CPU_Governors_Epercent_Perf}}
\caption{CPU governor vs. power and average per frame energy consumption (CNN model size: $300\times300$ pixels).}
\label{fig:cpugovernors_energylocal}   
\end{figure*}

\textbf{RQ 1.} How is energy consumed when a CNN-based object detection application is executed locally on a mobile AR client? To answer this question, in this section, we describe our efforts towards measuring and understanding the energy consumption and the performance of running CNN models on smartphones locally. We begin by measuring the per frame latency and the per frame energy consumption of executing CNN-based object detection under different smartphone's CPU governors in Section \ref{ssc:cpugovernor_local}. In addition, we explore the impact of the CNN model size on the per frame  latency and the per frame energy consumption in Section \ref{ssc:cnnsize_local}. Lastly, in Section \ref{ssc:sumdis_local}, we summarize the insights from our measurement studies and discuss potential research opportunities for improving the energy efficiency of locally executing CNN-based object detection on smartphones. 

\subsection{The Impact of CPU Governor}
\label{ssc:cpugovernor_local}
\textbf{CPU Governor\footnote{We change the Android smartphone's CPU governor manually by writing files in \hytt{/sys/devices/system/cpu/[cpu\#]/cpufreq/} \hytt{scaling\_governor} virtual file system with root privilege.}} Dynamic voltage and frequency scaling (DVFS) is a technique commonly used for dynamically adjusting the voltage and frequency of a mobile device's CPU in order to balance the trade-off between the power consumption of the device and the required performance. In order to offer DVFS, the CPU provides a set of valid voltages and frequencies that can be dynamically selected by a power management policy which is usually called a CPU governor. Different CPU governors adjust the CPU voltage and frequency based on variant criteria such as CPU usage. The six most popular CPU governors are described as follows:

\begin{itemize}
\item \textit{Conservative governor:} It adjusts the CPU frequency based on the current usage and it biases the mobile device to prefer the lowest possible CPU frequency as often as possible. In other words, a large and persistent load can be placed on the CPU only before the CPU frequency is raised. Thus, the conservative governor is good for the mobile device's battery life.
\item \textit{Ondemand governor:} It adjusts the CPU frequency based on the current usage, which is similar to the conservative governor. However, the ondemand governor immediately boosts the CPU to the highest possible frequency when there is a load on the CPU and switches back to the lowest possible frequency when the CPU is idle rather than gradually increases and decreases the CPU frequency. Thus, it offers excellent interface fluidity due to its high-frequency bias.
\item \textit{Interactive governor:} It is the default CPU governor for most android mobile devices. Similar to conservative and ondemand governors, it sets the CPU frequency based on the current usage. However, the interactive governor is designed for latency-sensitive and interactive workloads, so it is more aggressive about scaling the CPU speed up in response to CPU-intensive activities. 
\item \textit{Userspace governor:} It allows the user or any userspace program to set the CPU to a specific frequency (i.e., the CPU frequency is set to 1.497 GHz in this work), whereas it only allows the CPU frequency to be set to predefined fixed values.
\item \textit{Powersave governor:} It sets the CPU statically to the lowest possible frequency to minimize the energy consumption of the mobile device's CPU.
\item \textit{Performance governor:} It sets the CPU statically to the highest possible frequency to maximize the performance of the mobile device's CPU.
\end{itemize}

\begin{table*}[t]
 \begin{center}
    \caption{Latency results of the local execution with different CPU governors.}
    \label{tb:local_cpu_latency}
  \begin{tabular}{|l||c|c|c|c|c|c|}
    \hline
    CPU Governor & Conservative & Ondemand & Interactive & Userspace & Powersave & Performance \\ \hline
    Per Frame Latency (second) & 0.915 & 0.904 & 1.013 & 1.444 & 7.766 & \textbf{0.823}\\ \hline
    Image Conversion Latency (second) & 0.436 & 0.425 & 0.466 & 0.693 & 3.713 & \textbf{0.391}\\ \hline
    Inference Latency (Second) & 0.424 & 0.431 & 0.501 & 0.699 & 3.971 & \textbf{0.386}\\ \hline
    Others (Second) & 0.055 & 0.047 & 0.047 & 0.052 & 0.082 & \textbf{0.046}\\ \hline
  \end{tabular}
  \end{center}
\end{table*}

\begin{table*}[t]
 \begin{center}
    \caption{Per frame energy consumption results of the local execution with different CPU governors.}
    \label{tb:local_cpu_energy}
  \begin{tabular}{|l||c|c|c|c|c|c|}
    \hline
    CPU Governor & Conservative & Ondemand & Interactive & Userspace & Powersave & Performance \\ \hline
    Power Consumption (watt)   & 5.357 & 5.725 & 5.415 & 3.814 & \textbf{2.308}  & 6.115        \\ \hline
    Per Frame Energy Consumption (Joule) & 5.284 & 5.179 & 5.487 & 5.508 & 17.926 & \textbf{5.037}\\ \hline
    Image Generation \& Preview Energy Consumption (Joule) & 2.639 & 2.385 & 2.622 & 3.349 & 12.552 & \textbf{2.281}\\ \hline
    Inference Energy Consumption (Joule) & 1.009 & 1.084 & 1.153 & 0.593 & \textbf{0.432} & 1.028\\ \hline
    Image Conversion Energy Consumption (Joule) & 0.977 & 1.004 & 0.992 & 0.553 & \textbf{0.380} & 0.983\\ \hline
    Base Energy Consumption (Joule) & \textbf{0.534} & 0.591 & 0.621 & 0.971 & 4.555 & 0.622\\ \hline
    Others (Joule) & 0.125 & 0.115 & 0.099 & 0.042 & \textbf{0.007} & 0.123\\ \hline
  \end{tabular}
  \end{center}
\end{table*}

\textbf{Per Frame Latency.} We first seek to investigate how the CPU governor impacts the per frame latency of object detection in the local execution scenario, where a CNN model is executed on a smartphone and the model size is $300\times 300$ pixels. The experimental results are shown in Fig. \ref{fig:cpugovernors_latencylocal}, where Figs. \ref{fig:TFLlocal_CPUfreq_Conser}-\ref{fig:TFLlocal_CPUfreq_Perf} depict the frequency variations of the tested smartphone's CPUs and Figs. \ref{fig:TFLlocal_CPU_Governors_Latency_Conser}-\ref{fig:TFLlocal_CPU_Governors_Latency_Perf} illustrate the latency of each phase in the object detection processing pipeline. We show the latency of the two highest time-consuming phases, image conversion and inference latency, which comes up to $95$\% of the per frame latency. We observe that (1) as described above, the conservative governor provides a graceful CPU frequency increase, which causes temporarily high per frame latency and low frame per second (FPS) when the object detection application is launched, as shown in Figs. \ref{fig:TFLlocal_CPUfreq_Conser} and \ref{fig:TFLlocal_CPU_Governors_Latency_Conser}. This observation demonstrates that the conservative governor may not be suitable for CNN-based object detection applications because object detection requires a high fluidity to interact with the user. (2) Although both ondemand and interactive governors provide aggressive responses to the execution of object detection, as depicted in Figs. \ref{fig:TFLlocal_CPUfreq_Ond} and \ref{fig:TFLlocal_CPUfreq_Inter}, the ondemand governor offers a relatively steadier 
latency performance than the interactive governor due to its high-frequency bias. In Figs. \ref{fig:TFLlocal_CPUfreq_Use}, \ref{fig:TFLlocal_CPUfreq_Pow}, and \ref{fig:TFLlocal_CPUfreq_Perf}, the CPU is set to a user-defined, the lowest, and the highest possible frequencies, respectively. (3) It is not surprising to find that the powersave governor is the worst-performing governor in terms of the latency, where its per frame latency is almost eight times higher than that of the performance governor. (4) The performance governor outperforms other presented CPU governors in terms of latency, and the measured average per frame latency is shown in Table \ref{tb:local_cpu_latency}.

\begin{figure*}[t]
\centering
\subfigure[$100^2$ pixels]
{\includegraphics[width=0.16\textwidth]{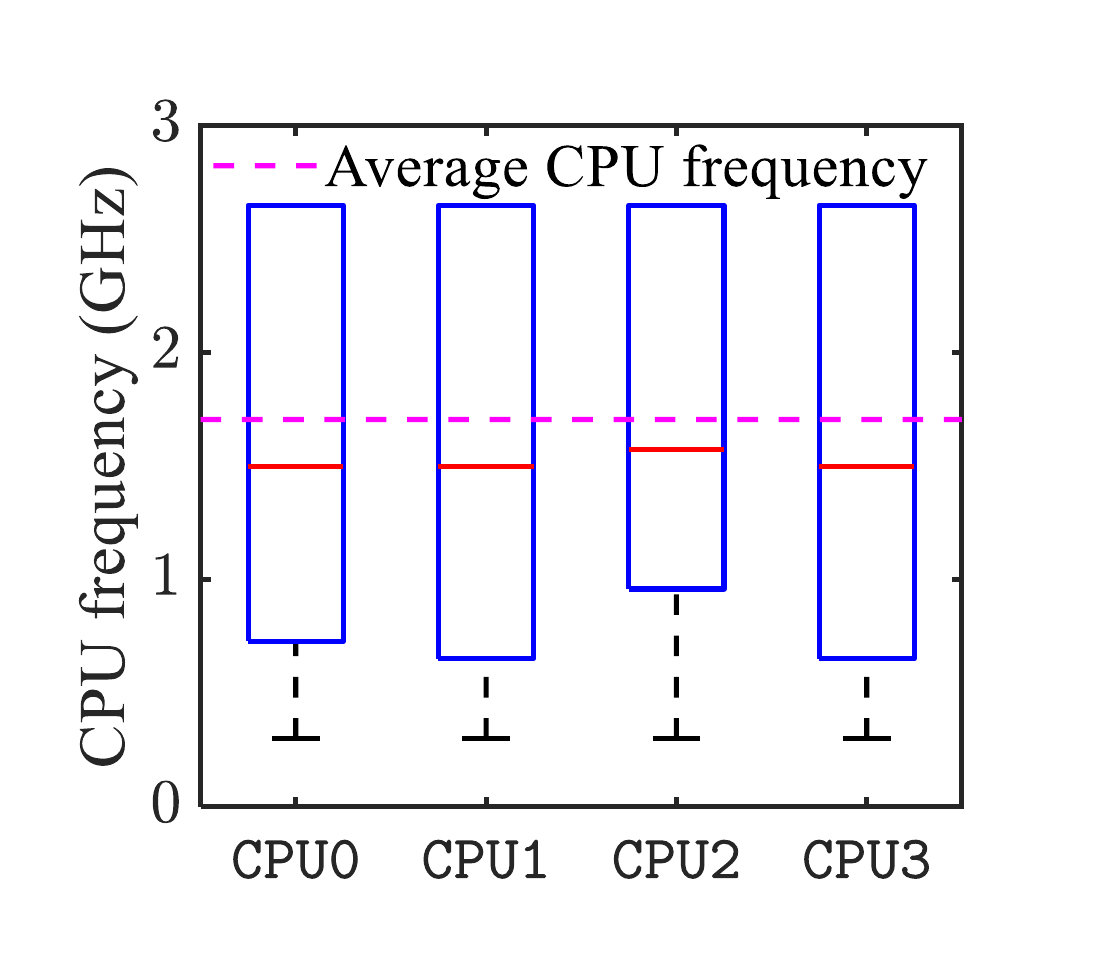}\label{fig:100InterCPUlocal}}
\subfigure[$200^2$ pixels]
{\includegraphics[width=0.16\textwidth]{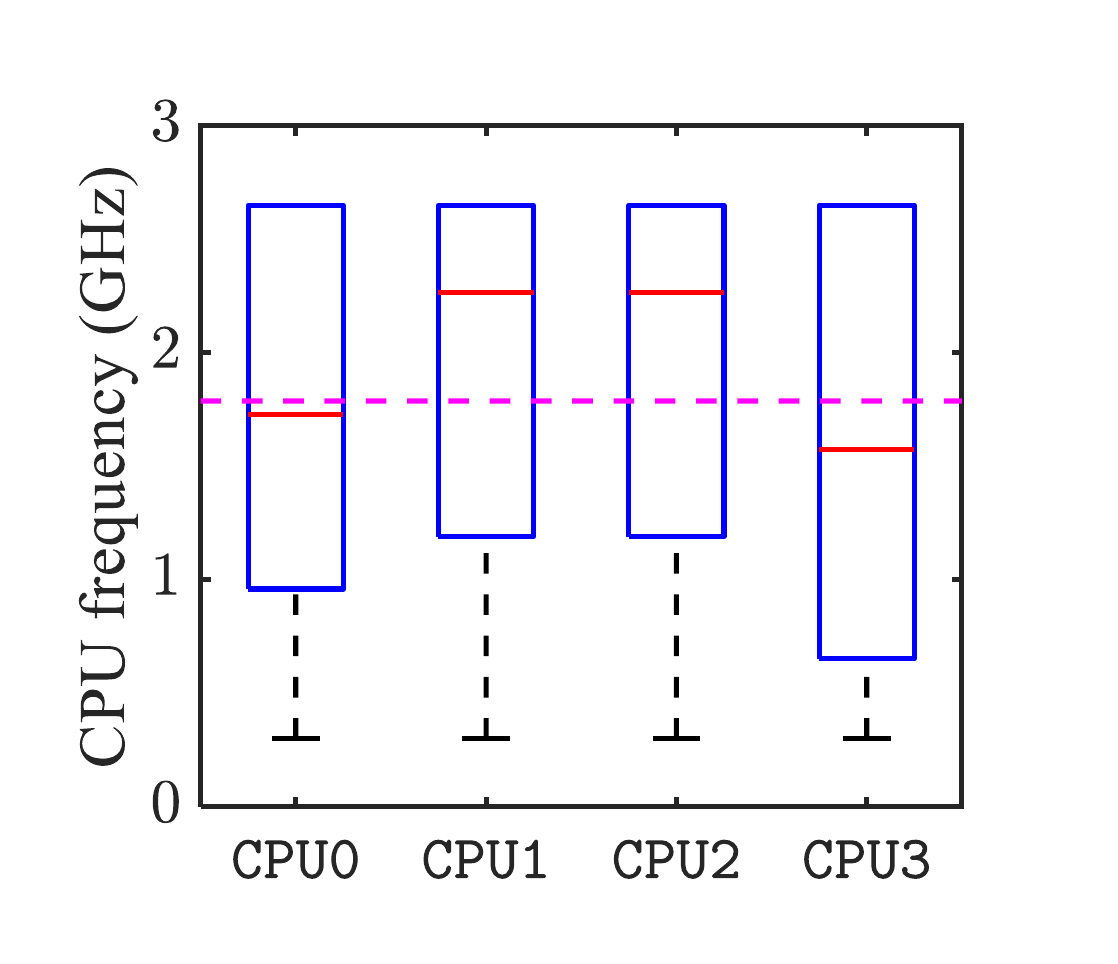}\label{fig:200InterCPUlocal}}
\subfigure[$300^2$ pixels]
{\includegraphics[width=0.16\textwidth]{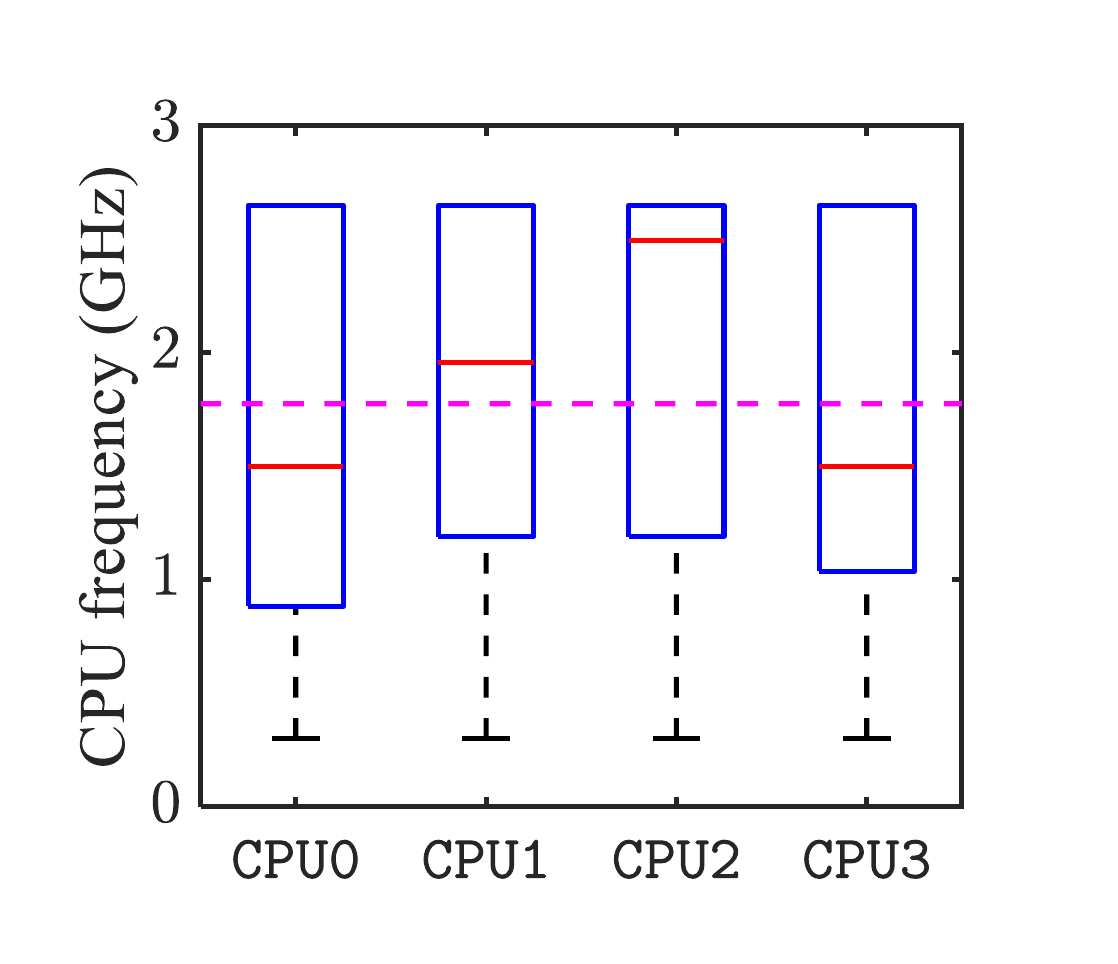}\label{fig:300InterCPUlocal}}
\subfigure[$400^2$ pixels]
{\includegraphics[width=0.16\textwidth]{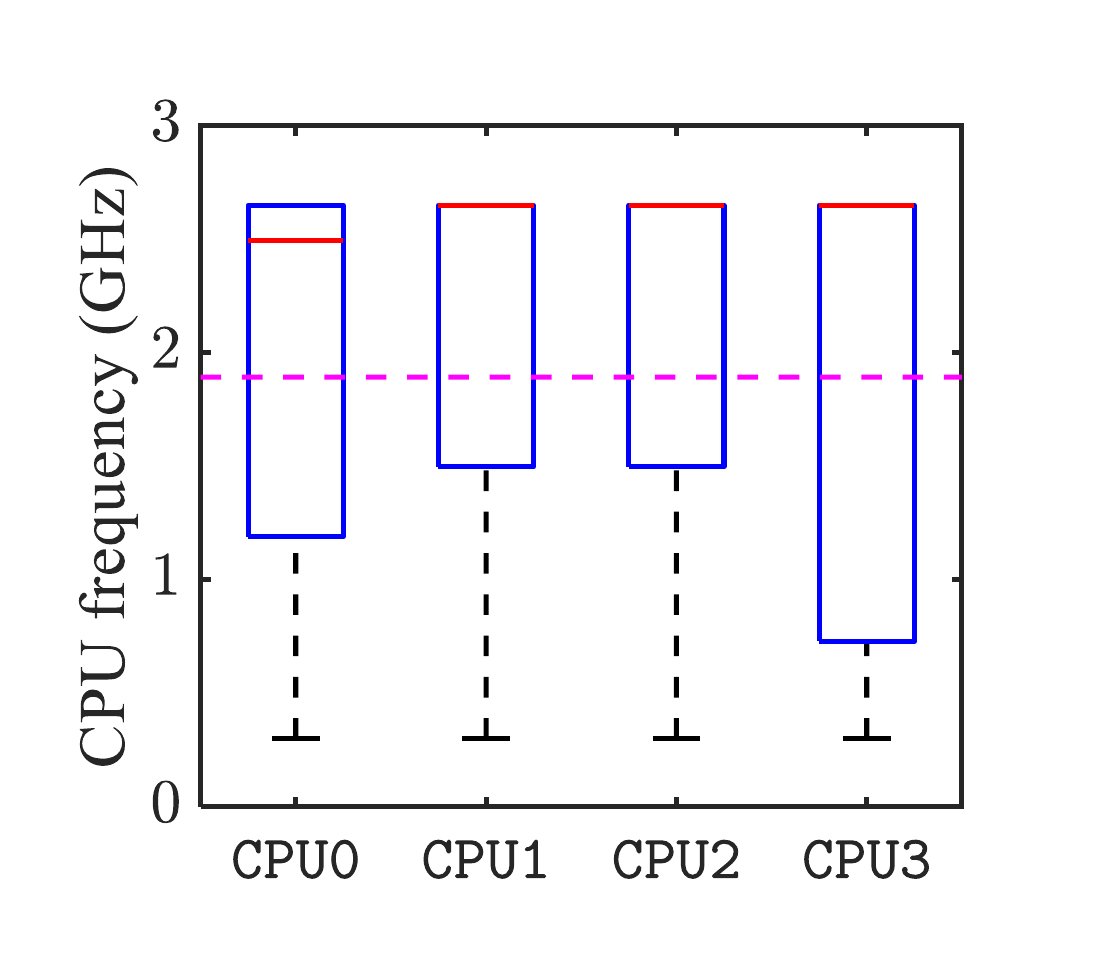}\label{fig:400InterCPUlocal}}
\subfigure[$500^2$ pixels]
{\includegraphics[width=0.16\textwidth]{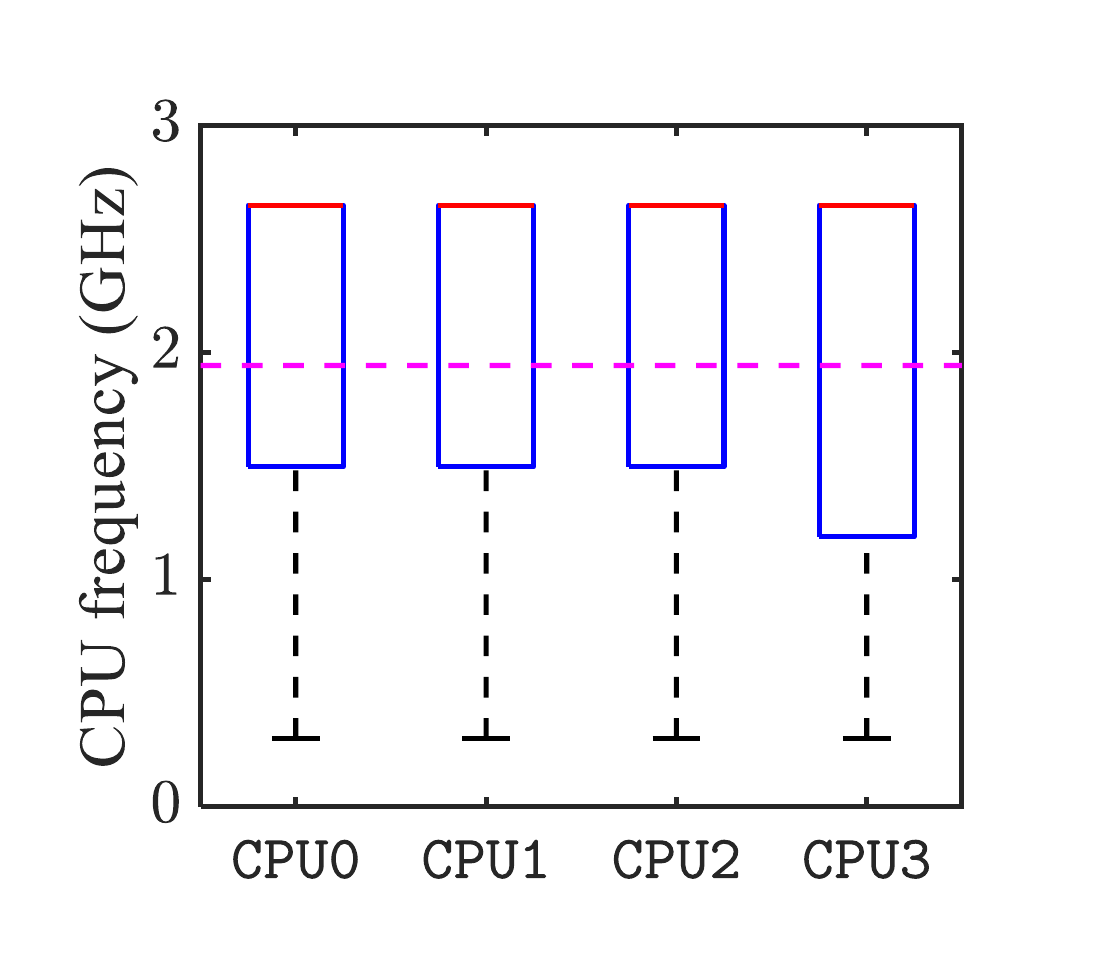}\label{fig:500InterCPUlocal}}
\subfigure[$600^2$ pixels]
{\includegraphics[width=0.16\textwidth]{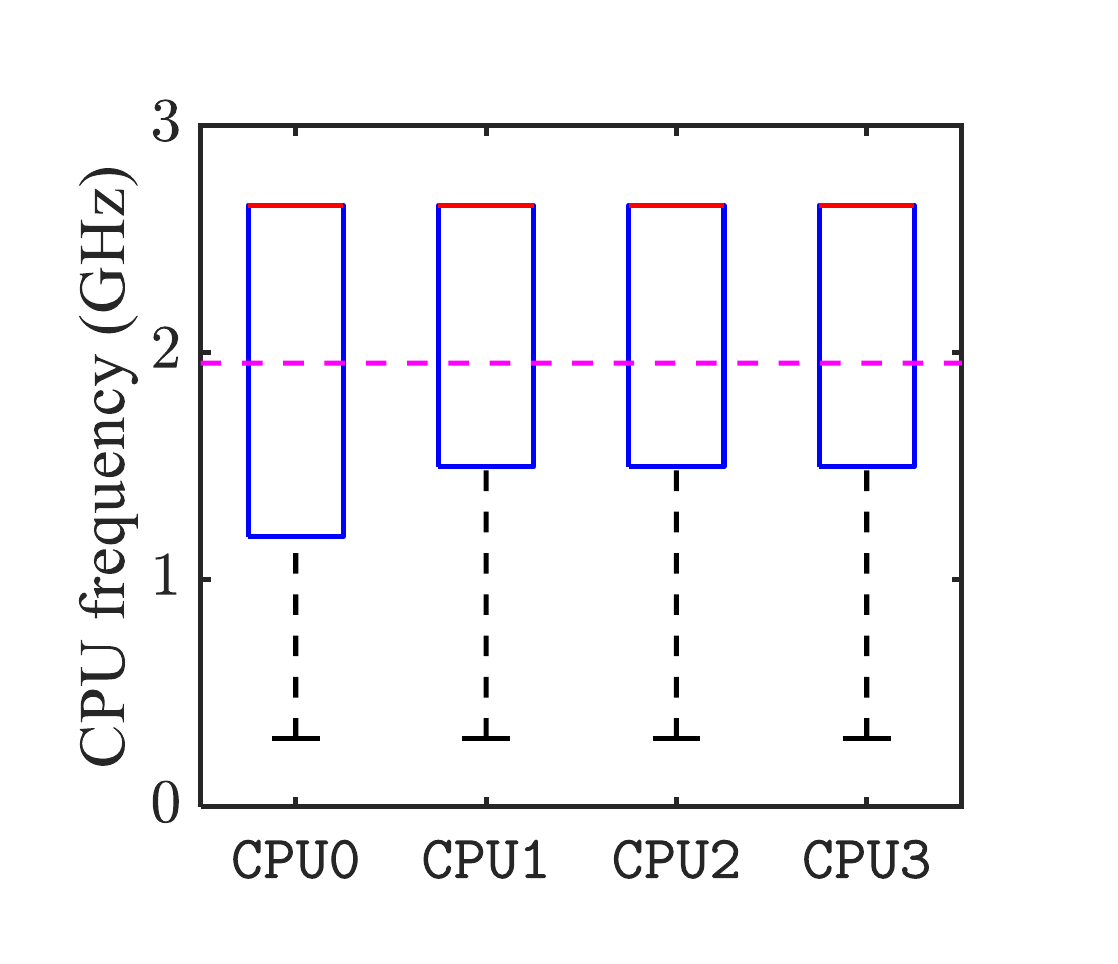}\label{fig:600InterCPUlocal}}

\subfigure[Per frame latency]
{\includegraphics[width=0.325\textwidth]{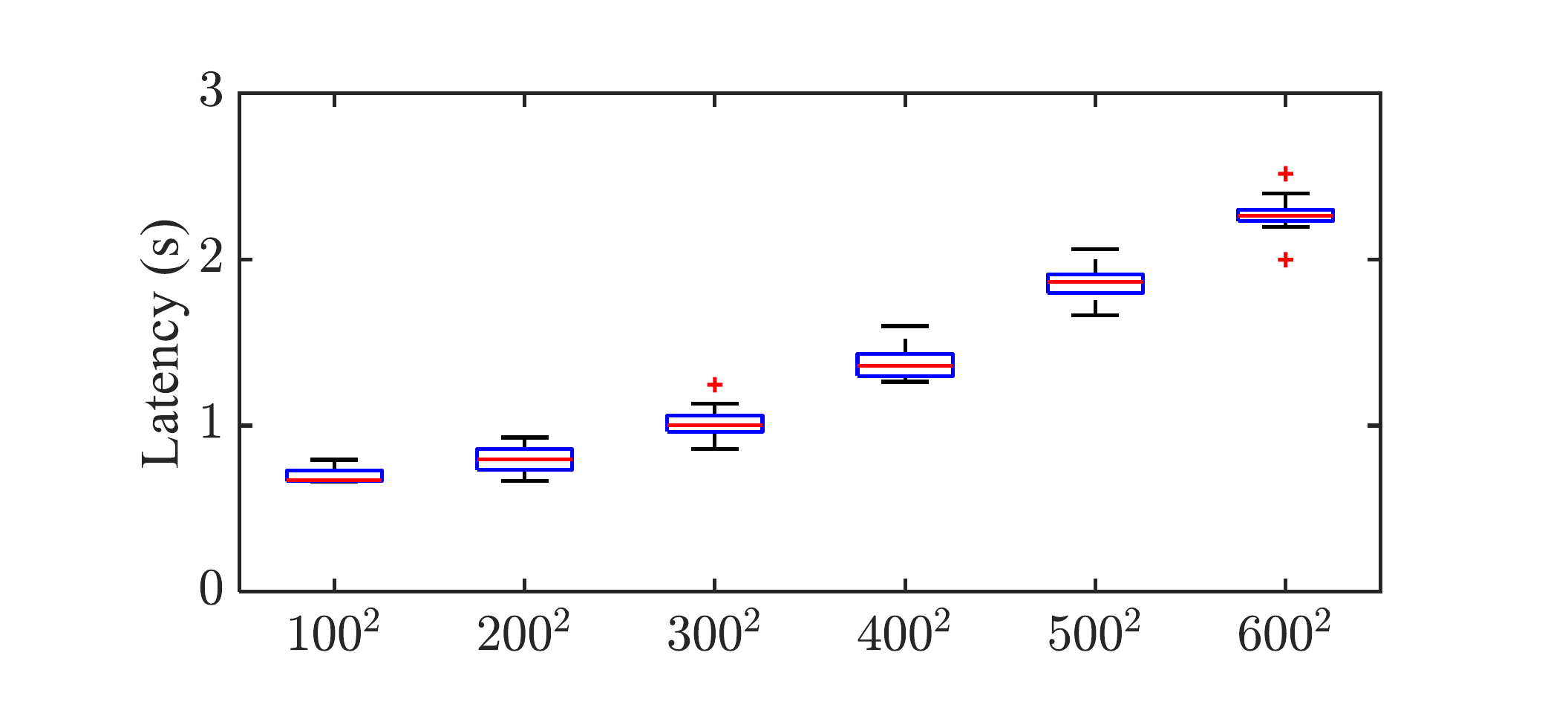}\label{fig:Inter_alllatency_local}}
\subfigure[Image conversion latency]
{\includegraphics[width=0.33\textwidth]{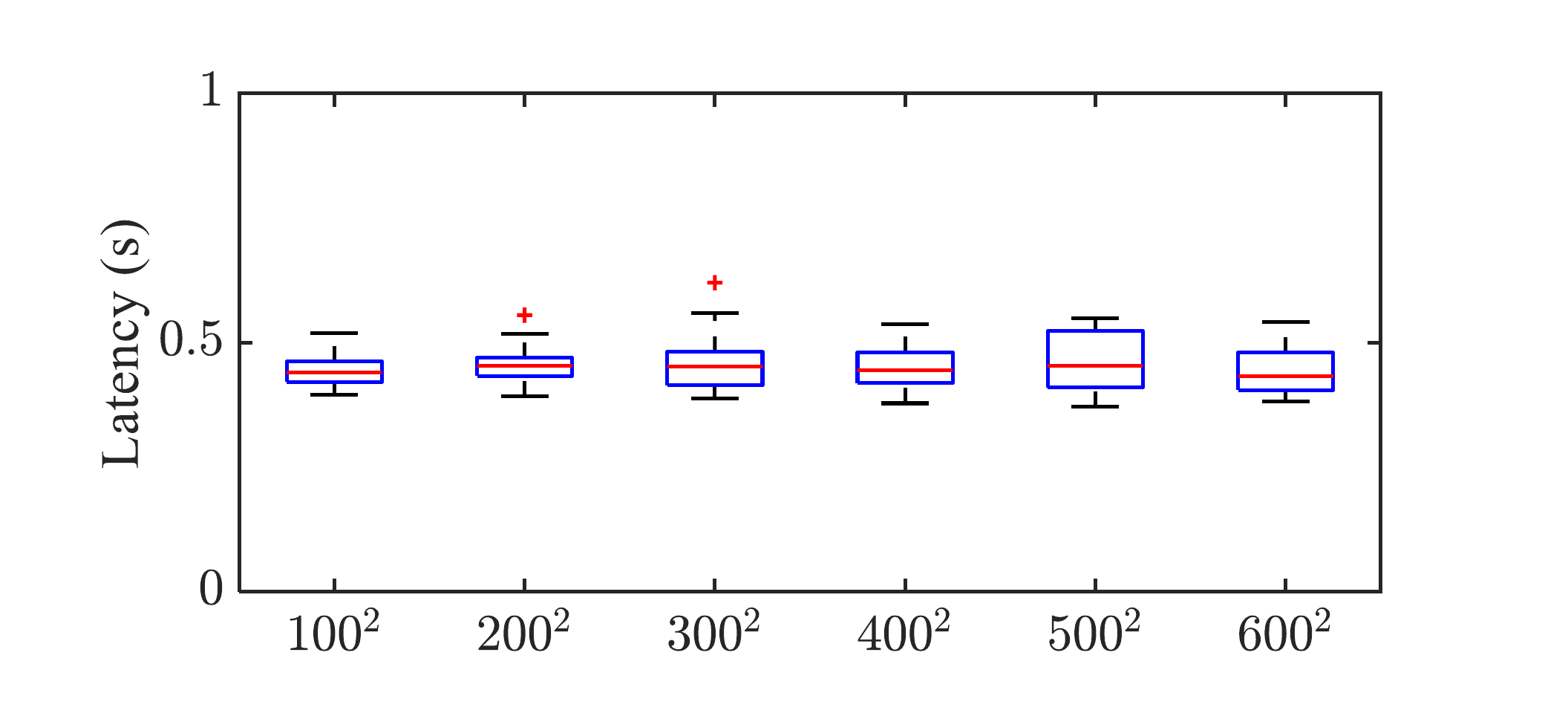}\label{fig:Inter_convlatency_local}}
\subfigure[Inference latency]
{\includegraphics[width=0.325\textwidth]{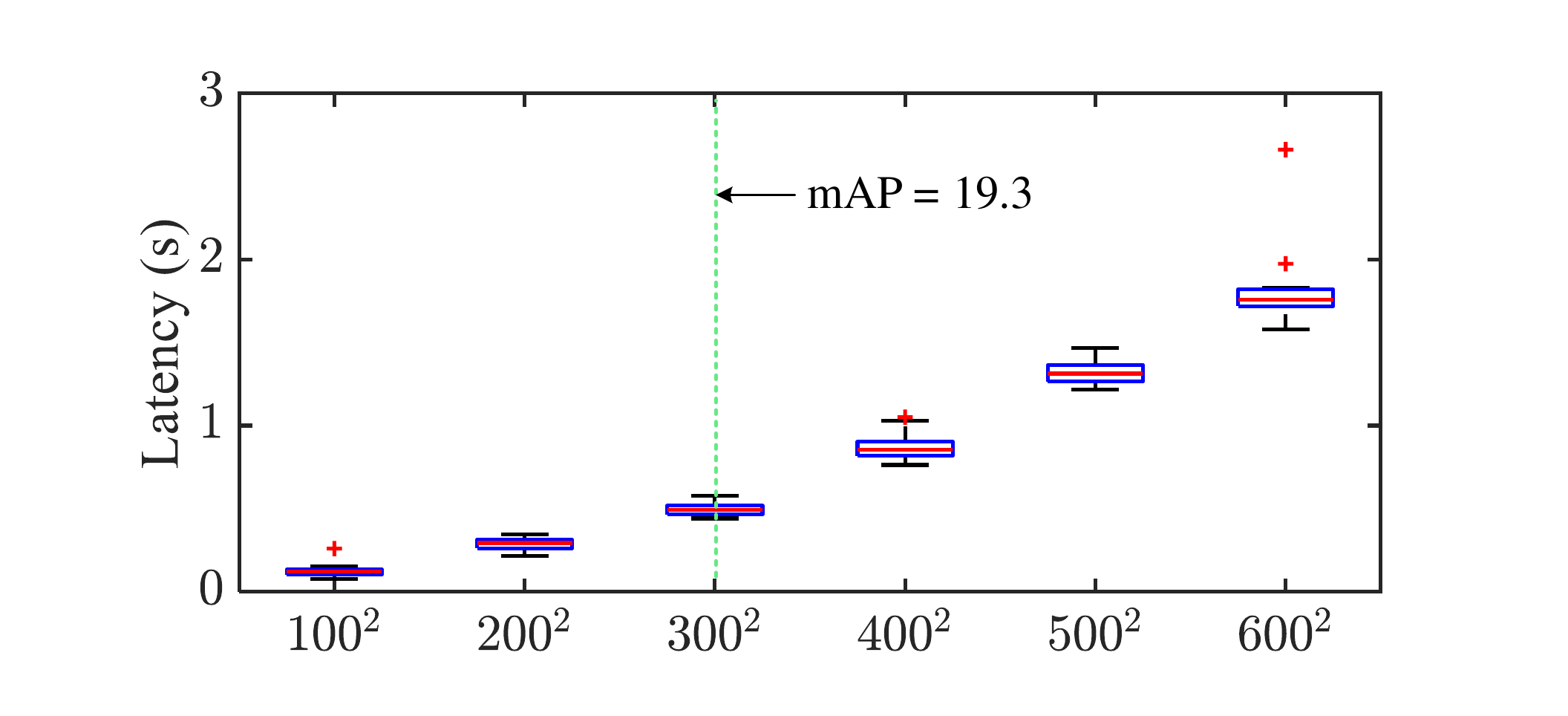}\label{fig:Inter_inflatency_local}}
\caption{CNN model size vs. CPU frequency \& latency (CPU governor: interactive).}
\label{fig:cnn_latency_local_inter}   
\end{figure*}

\begin{figure*}[t]
\centering
\subfigure[Per frame latency]
{\includegraphics[width=0.325\textwidth]{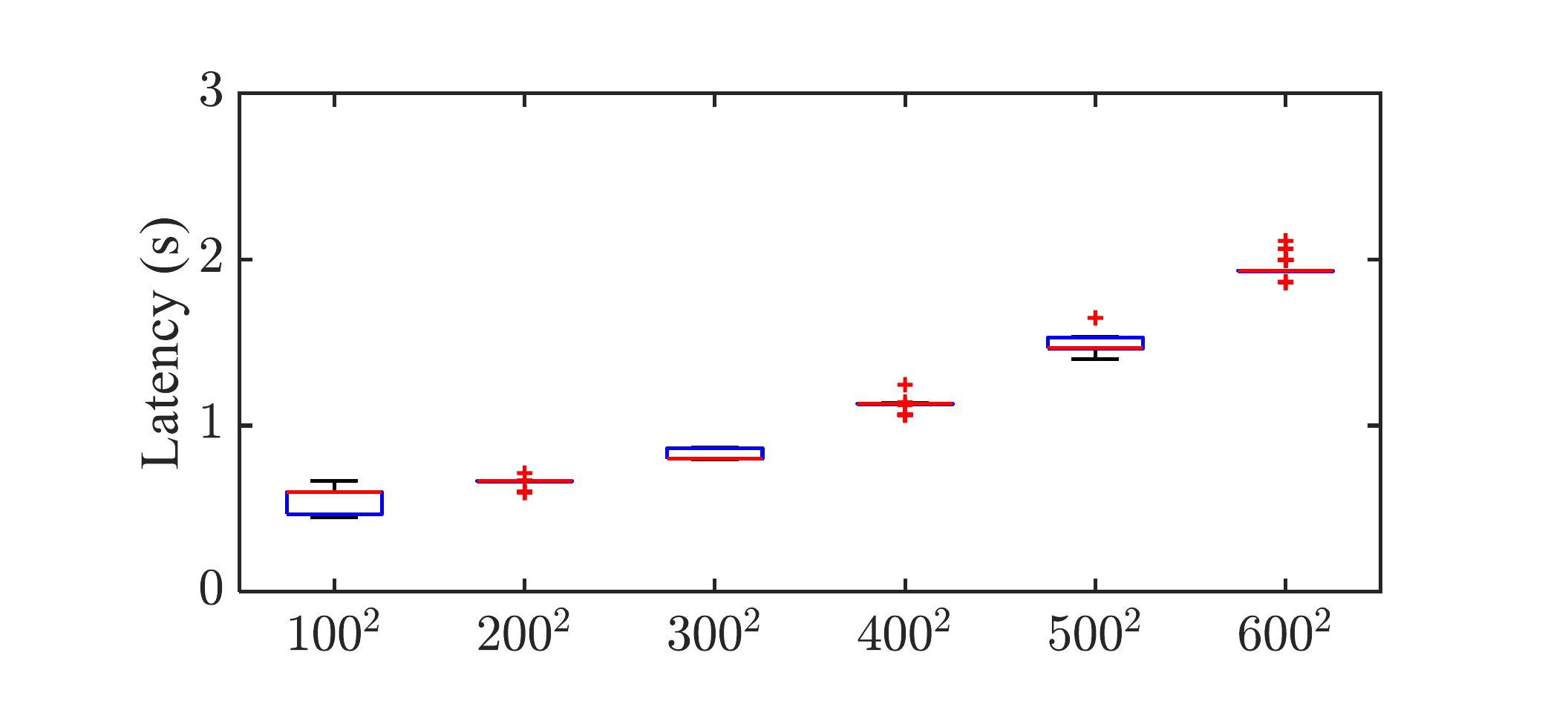}\label{fig:Perf_alllatency_local}}
\subfigure[Image conversion latency]
{\includegraphics[width=0.33\textwidth]{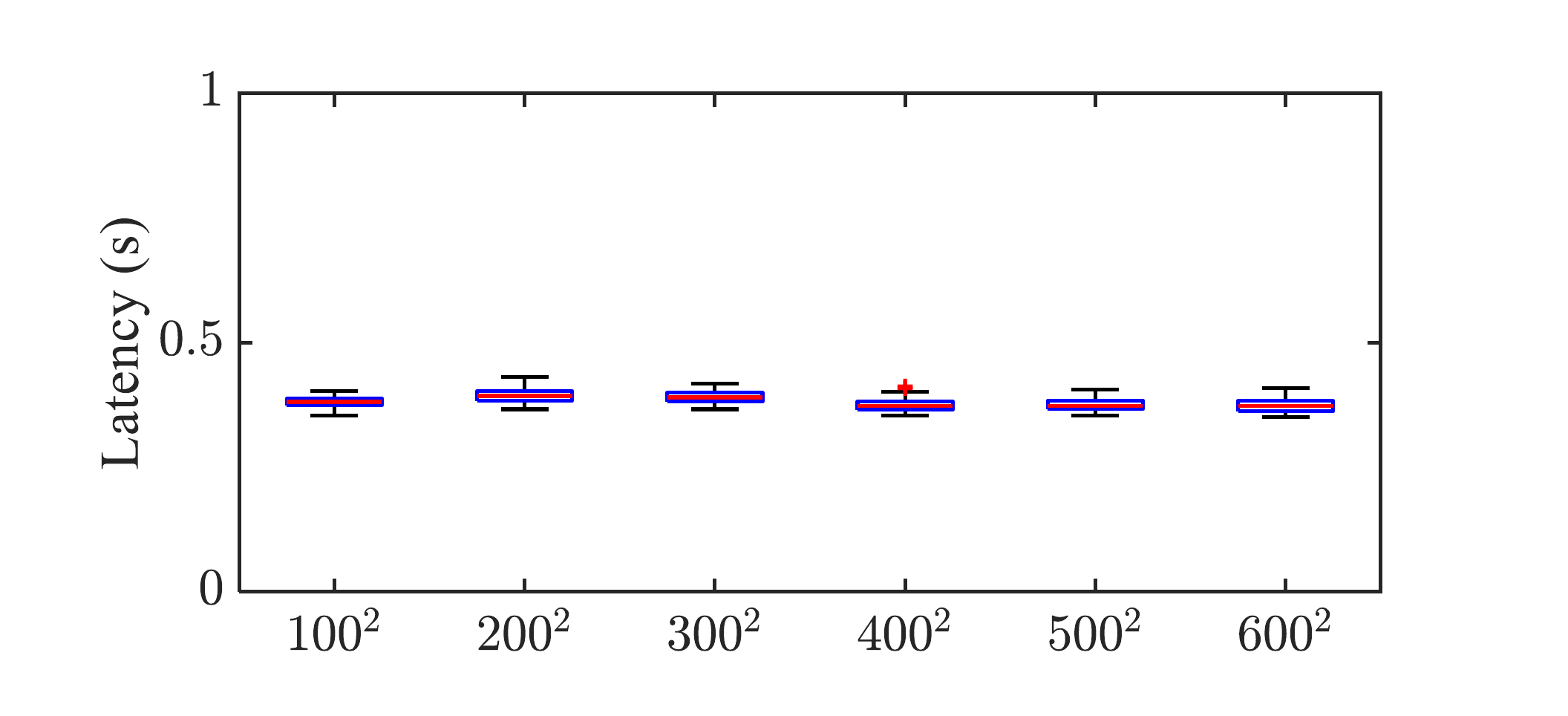}\label{fig:Perf_convlatency_local}}
\subfigure[Inference latency]
{\includegraphics[width=0.325\textwidth]{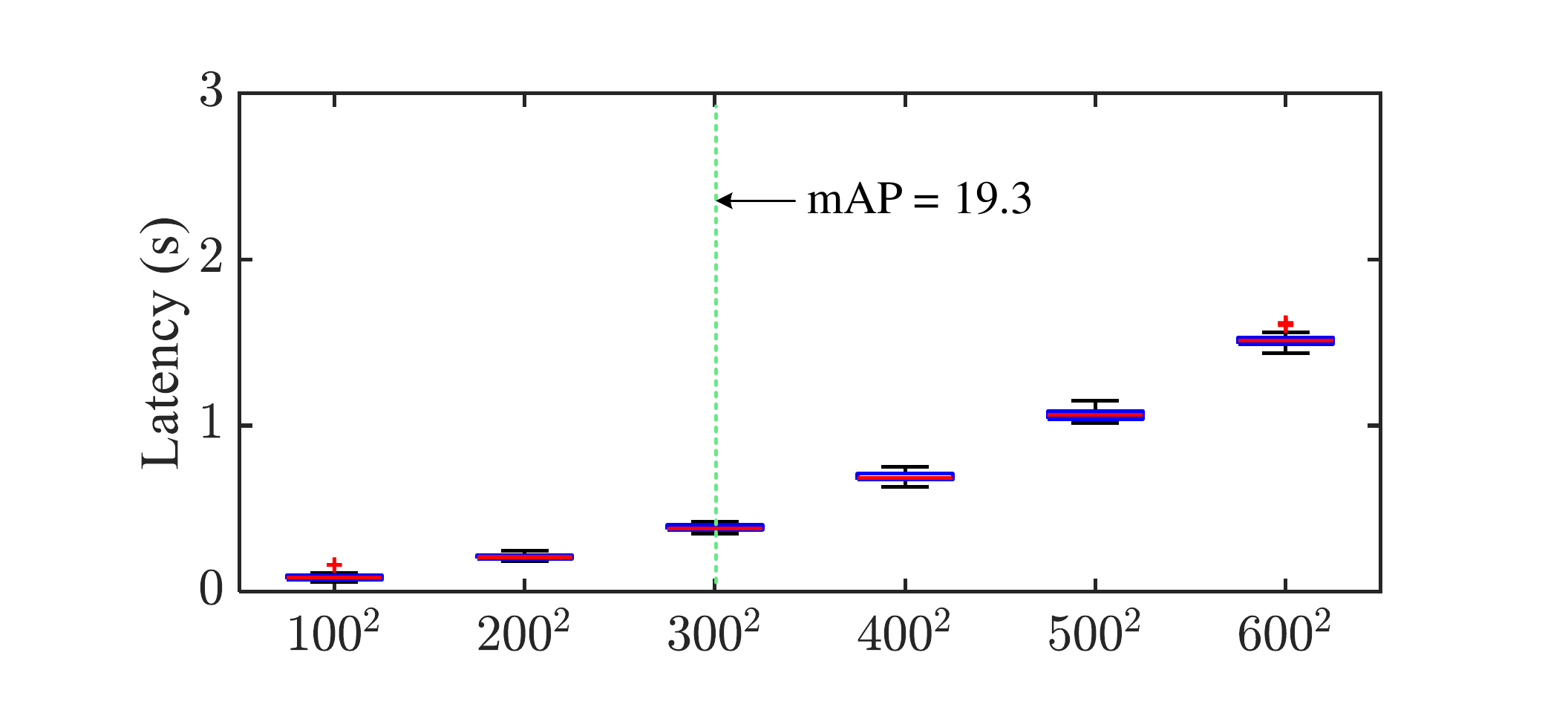}\label{fig:Perf_inflatency_local}}
\caption{CNN model size vs. latency (CPU governor: performance).}
\label{fig:cnn_latency_local_perf}   
\end{figure*}

\textbf{Per Frame Energy Consumption.} We next examine how the CPU governor impacts the per frame energy consumption of executing object detection on the smartphone. The experimental results are shown in Fig. \ref{fig:cpugovernors_energylocal}, where Figs. \ref{fig:TFLlocal_CPU_Governors_Power_Conser}-\ref{fig:TFLlocal_CPU_Governors_Power_Perf} depict the power consumption; Figs. \ref{fig:TFLlocal_CPU_Governors_Energy_Conser}-\ref{fig:TFLlocal_CPU_Governors_Energy_Perf} illustrate the average per frame energy consumption; and Figs. \ref{fig:TFLlocal_CPU_Governors_Epercent_Conser}-\ref{fig:TFLlocal_CPU_Governors_Epercent_Perf} depict the average percentage breakdown of energy consumed by each phase in the processing pipeline. We make the following observations. (5) The performance governor consumes the highest power consumption, as shown in Fig. \ref{fig:TFLlocal_CPU_Governors_Power_Perf}, because the processors always run with the highest possible CPU frequency. Although it is capable of providing the best latency performance, continuously run with the highest CPU frequency may cause the smartphone overheating and trigger CPU throttling mechanisms to avoid thermal emergencies by sacrificing the performance. (6) Interestingly, the performance governor provides the lowest per frame energy consumption, while the powersave governor offers the highest per frame energy consumption, as shown in Table \ref{tb:local_cpu_energy}. \textit{This observation indicates a critical trade-off between the battery life (i.e., power consumption) and per frame energy consumption in CNN-based object detection applications.} 

In order to dissect the energy drain through different processing pipeline phases, we break down the per frame energy consumption as follows: image generation and preview, inference, image conversion, base, and others. We find that (7) the image generation and preview phase always contributes the highest energy consumption (i.e., approximately $45.3\%$ - $70.0\%$). The reason it consumes considerably high energy is executing the $3$A (i.e., AF, AE, and AWB) and multiple fine-grained image post processing algorithms (e.g., noise reduction (NR), color correction (CC), and edge enhancement (EE)) on ISP. These sophisticated algorithms are designed to make an image that is captured by the smartphone camera look perfect. \emph{However, is it always necessary for the camera captured frame to be processed by all of those energy-hungry image processing algorithms in order to achieve a successful object detection result?} In addition, the number of frames captured by the camera per second is a fixed value (e.g., $24$ or $30$ frames/second) or in a range (e.g., $[7,30]$ frames/second), which is controlled by the AE algorithm. However, due to the limited computation capacity of smartphones, usually the detection FPS is far slower than the camera capture frame rate. On the other hand, the CNN always extracts the latest camera captured frame, which indicates that, from the perspective of the energy efficiency of the object detection pipeline, capturing frames with a fast rate is unnecessary and energy-inefficient. \textit{Therefore, both raising CPU frequency and decreasing camera capture frame rate are efficient approaches to reduce the energy consumption of image generation and preview.}

Besides the energy consumption of image generation and preview, inference and image conversion phases consume a large amount of energy, as depicted in Fig. \ref{fig:cpugovernors_energylocal} and Table \ref{tb:local_cpu_energy}. (8) Although a low CPU frequency incurs high per frame energy consumption, it decreases the energy consumption of both inference and image conversion phases. This observation indicates that \textit{there is a trade-off between the energy consumption reduction of image generation and preview phases and inference and image conversion phases.} For example, raising the CPU frequency can decrease the energy consumption of image generation and preview phases but concurrently increases the energy consumption of inference and image conversion phases. Furthermore, (9) the conservative governor provides the lowest base energy consumption, which demonstrates that existing CPU governors are capable of scaling CPU frequency for the workload of the smartphone's operating system.

\subsection{The Impact of CNN Model Size}
\label{ssc:cnnsize_local}

\begin{figure*}[t]
\centering
\subfigure[$100^2$ pixels]
{\includegraphics[width=0.16\textwidth]{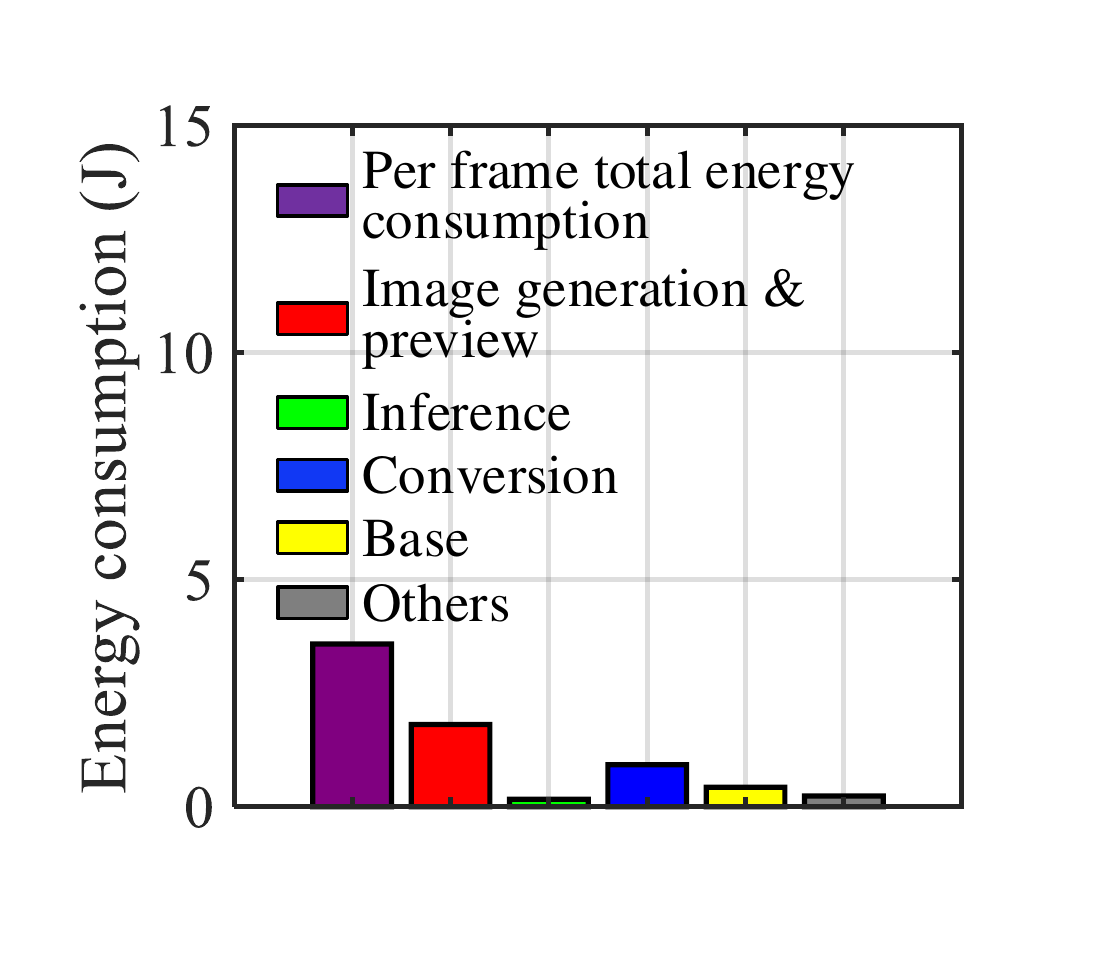}\label{fig:100InterCPU_Energy}}
\subfigure[$200^2$ pixels]
{\includegraphics[width=0.16\textwidth]{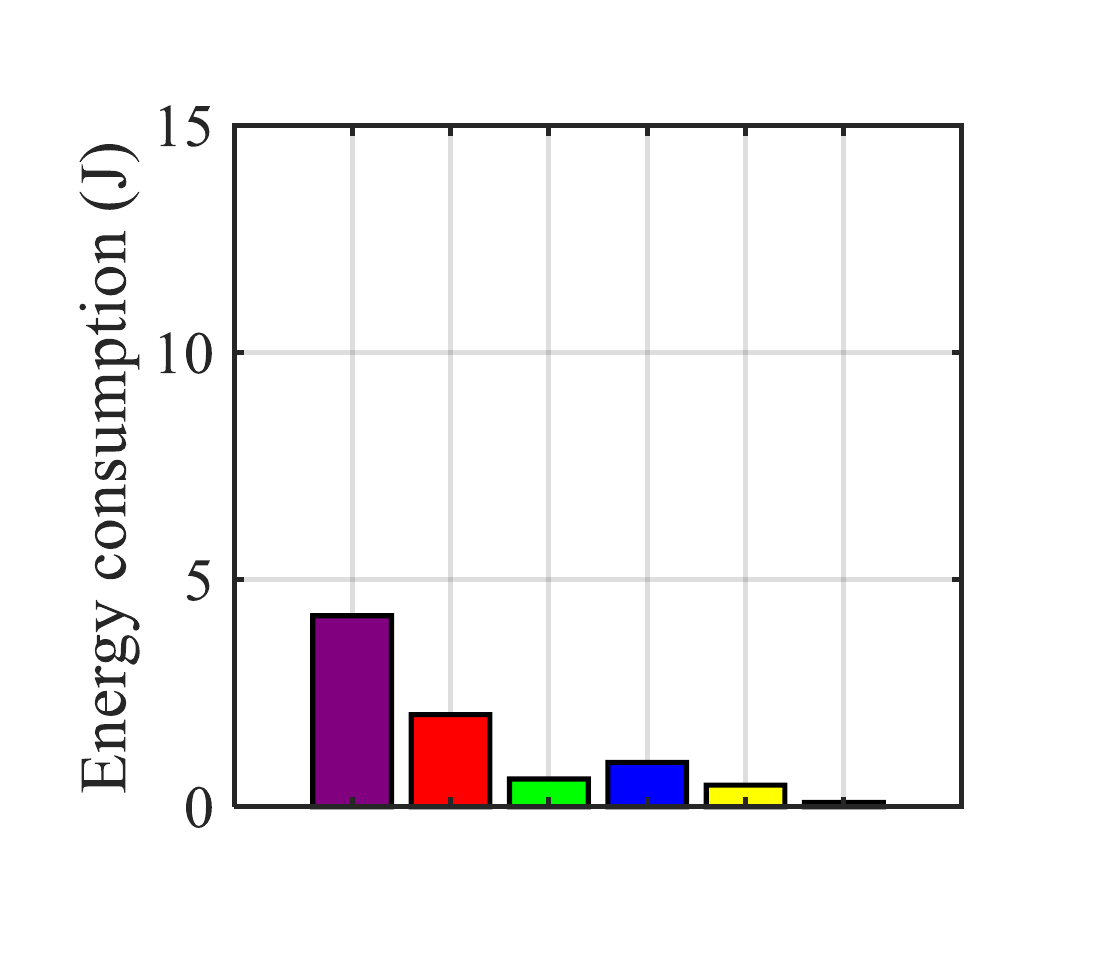}\label{fig:200InterCPU_Energy}}
\subfigure[$300^2$ pixels]
{\includegraphics[width=0.16\textwidth]{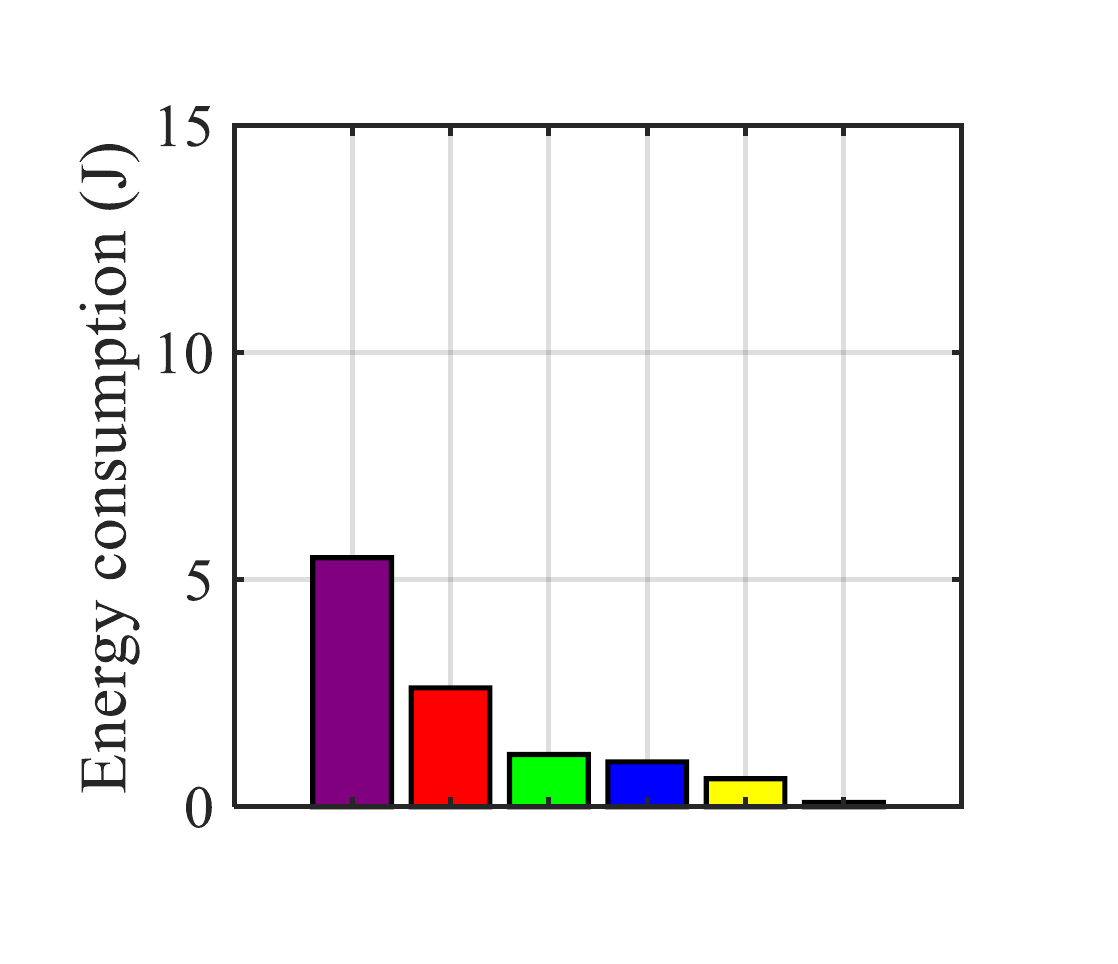}\label{fig:300InterCPU_Energy}}
\subfigure[$400^2$ pixels]
{\includegraphics[width=0.16\textwidth]{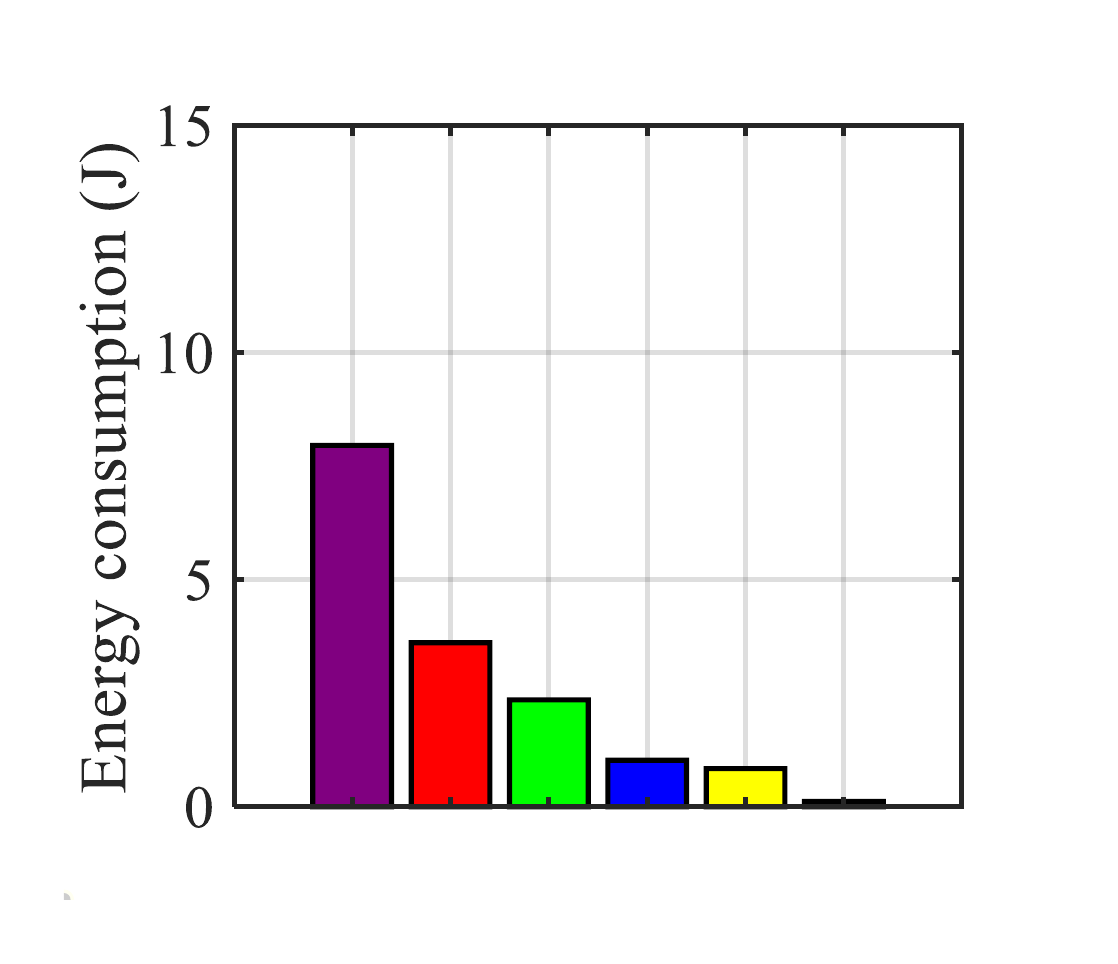}\label{fig:400InterCPU_Energy}}
\subfigure[$500^2$ pixels]
{\includegraphics[width=0.16\textwidth]{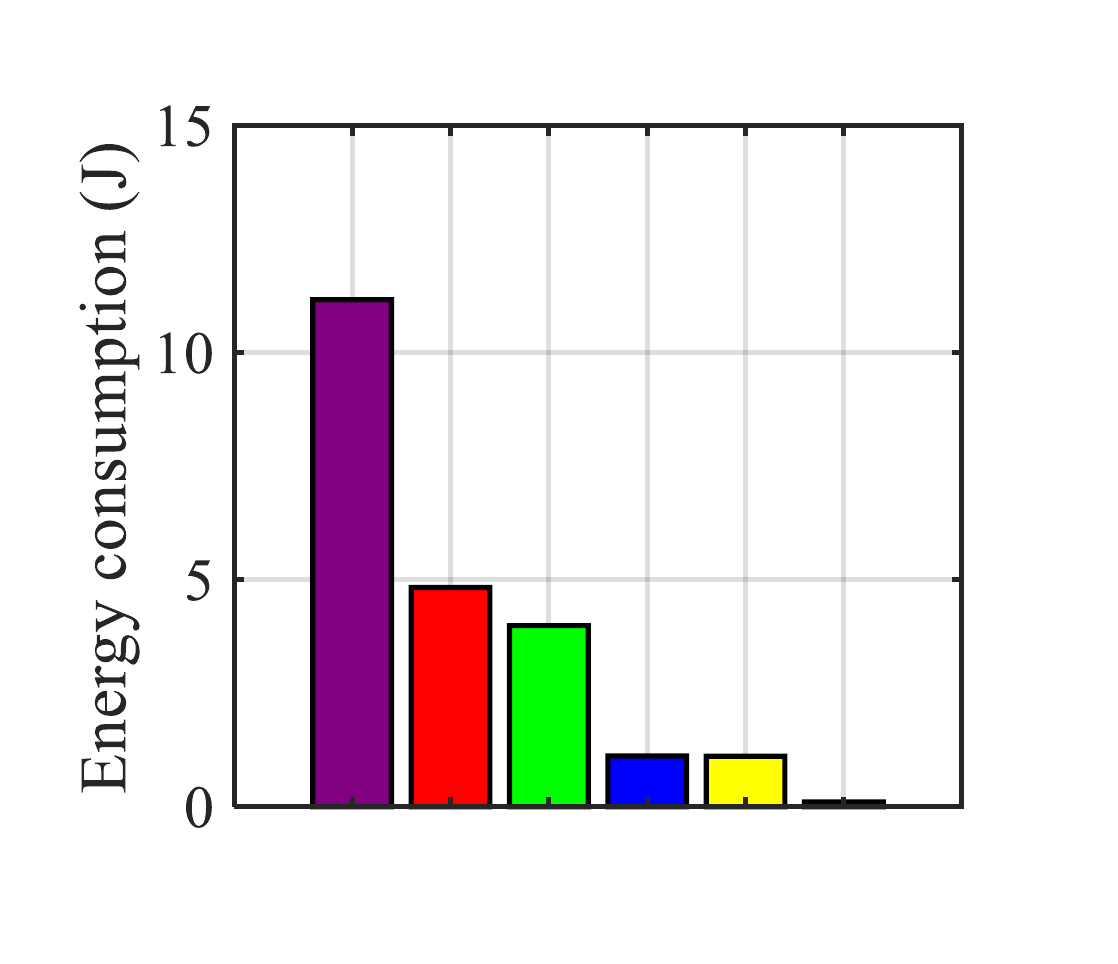}\label{fig:500InterCPU_Energy}}
\subfigure[$600^2$ pixels]
{\includegraphics[width=0.16\textwidth]{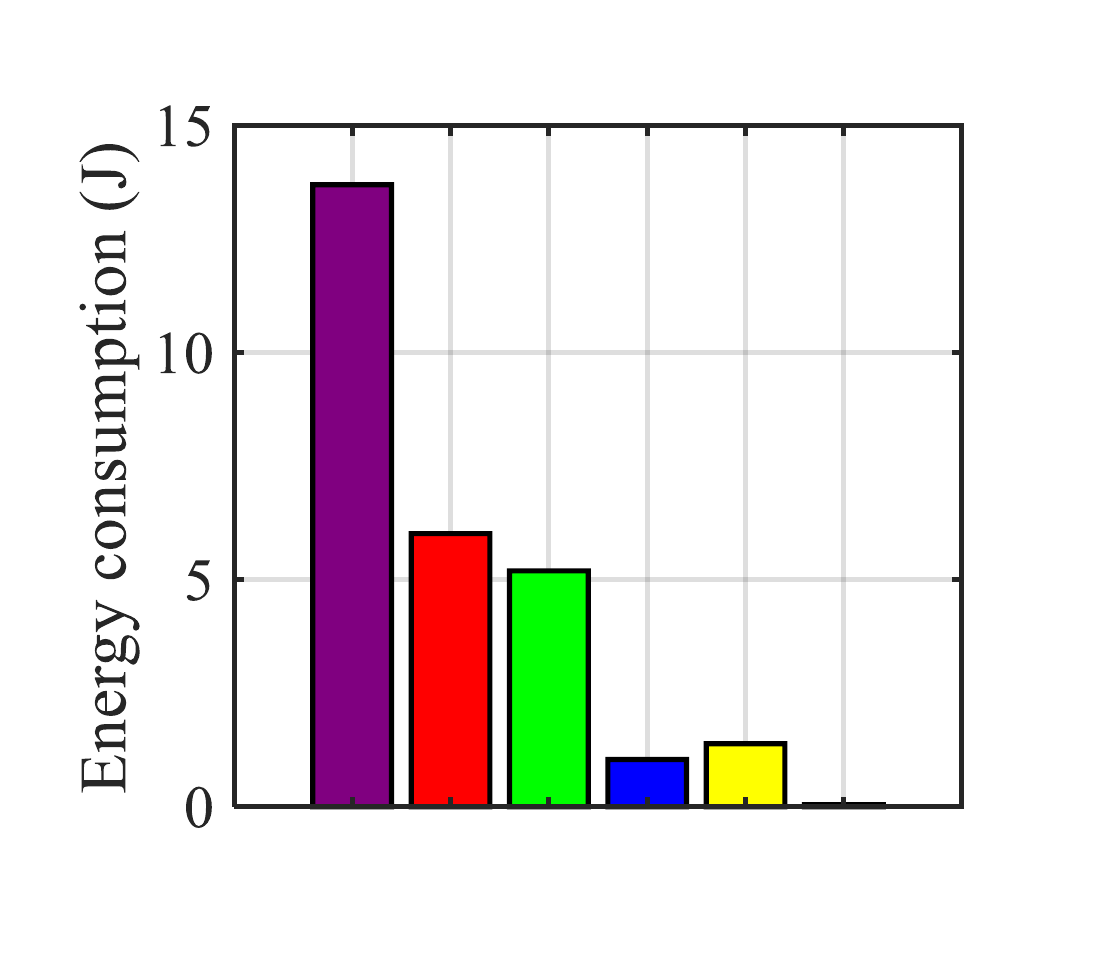}\label{fig:600InterCPU_Energy}}

\subfigure[$100^2$ pixels]
{\includegraphics[width=0.16\textwidth]{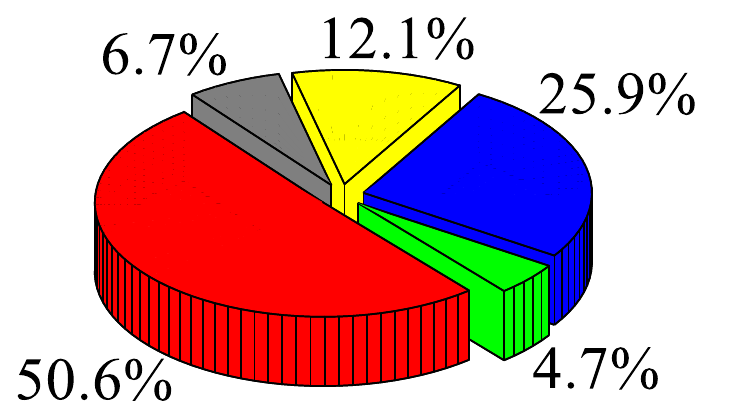}\label{fig:100InterCPU_Percentage}}
\subfigure[$200^2$ pixels]
{\includegraphics[width=0.16\textwidth]{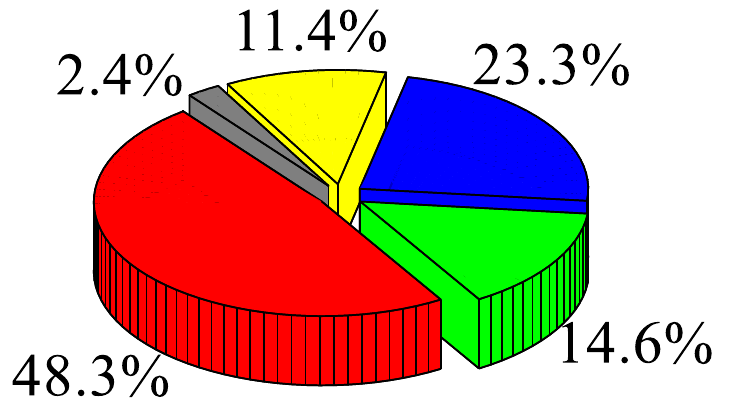}\label{fig:200InterCPU_Percentage}}
\subfigure[$300^2$ pixels]
{\includegraphics[width=0.16\textwidth]{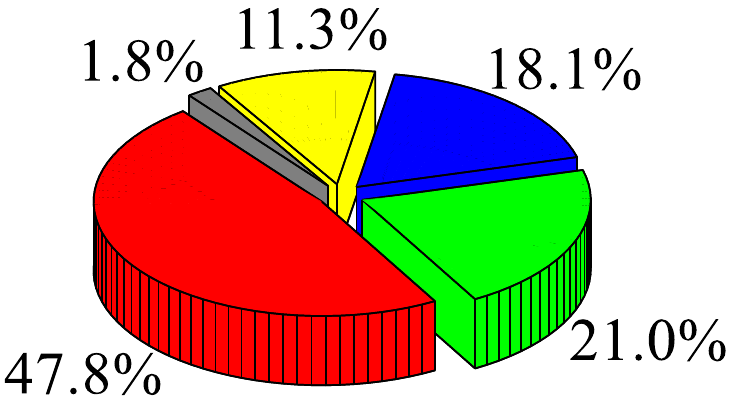}\label{fig:300InterCPU_Percentage}}
\subfigure[$400^2$ pixels]
{\includegraphics[width=0.16\textwidth]{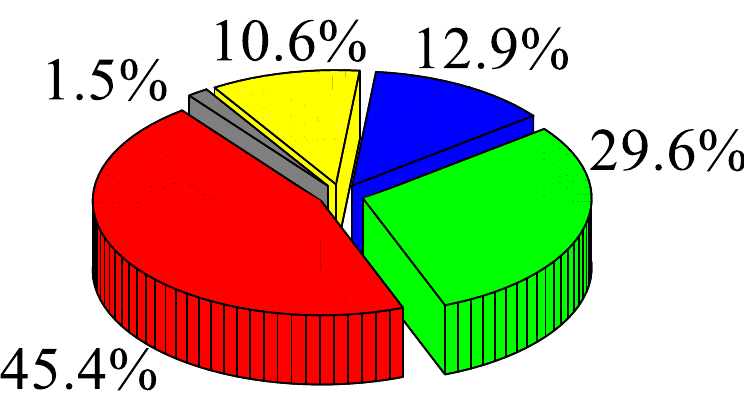}\label{fig:400InterCPU_Percentage}}
\subfigure[$500^2$ pixels]
{\includegraphics[width=0.16\textwidth]{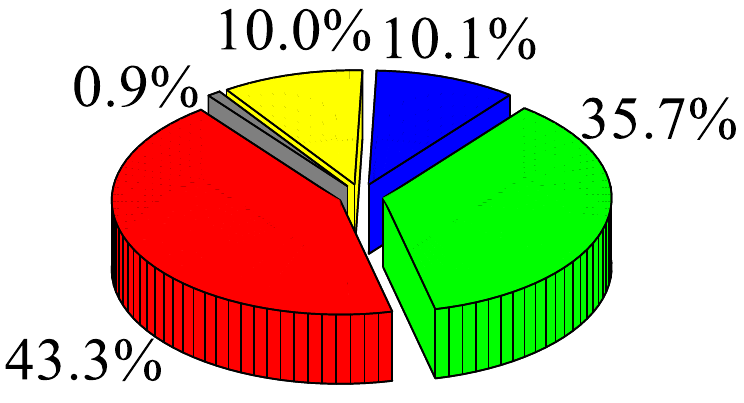}\label{fig:500InterCPU_Percentage}}
\subfigure[$600^2$ pixels]
{\includegraphics[width=0.16\textwidth]{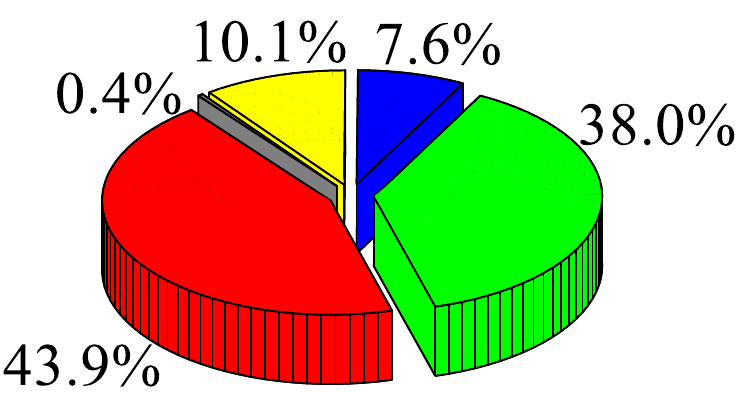}\label{fig:600InterCPU_Percentage}}
\caption{CNN model size vs. per frame energy consumption (CPU governor: interactive).}
\label{fig:cnn_energy_local_inter}   
\end{figure*}

\begin{figure*}[t]
\centering
\subfigure[$100^2$ pixels]
{\includegraphics[width=0.16\textwidth]{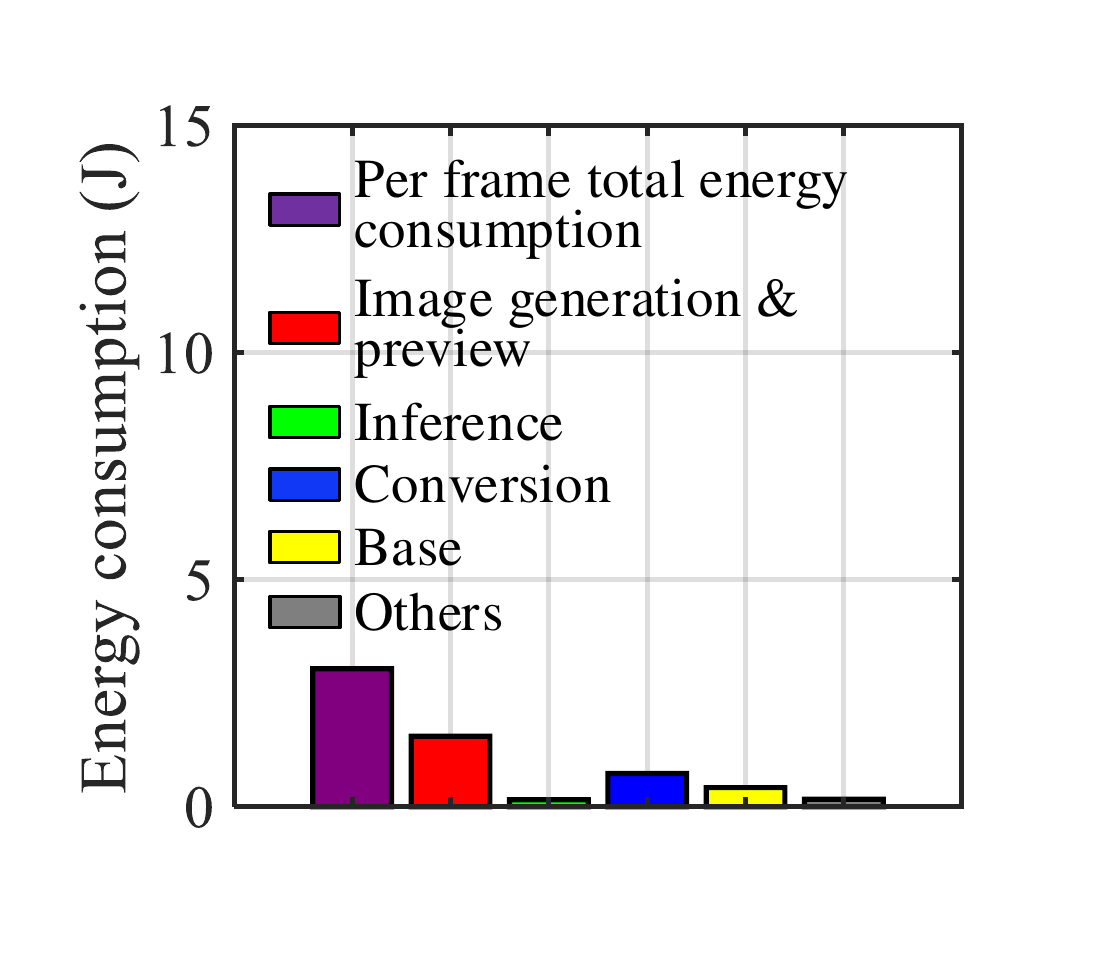}\label{fig:100PerfCPU_Energy}}
\subfigure[$200^2$ pixels]
{\includegraphics[width=0.16\textwidth]{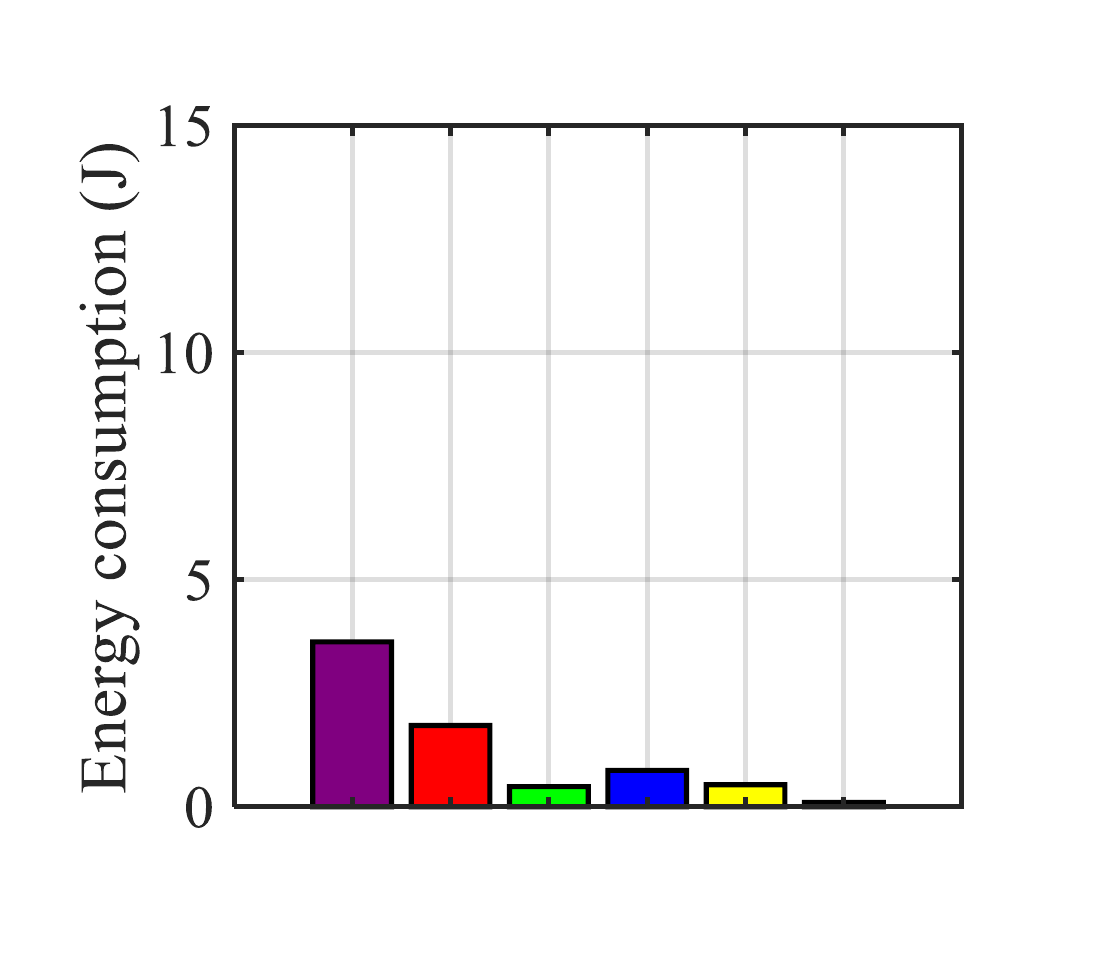}\label{fig:200PerfCPU_Energy}}
\subfigure[$300^2$ pixels]
{\includegraphics[width=0.16\textwidth]{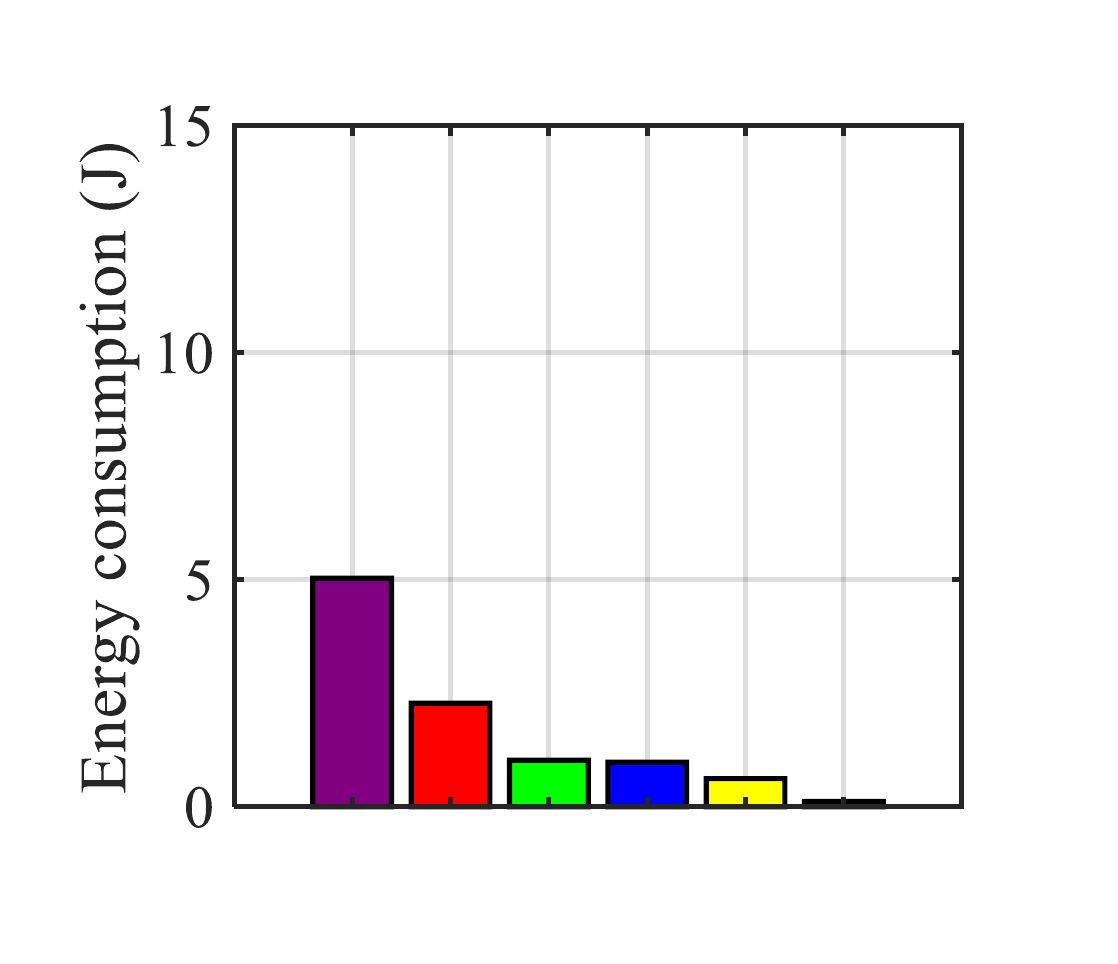}\label{fig:300PerfCPU_Energy}}
\subfigure[$400^2$ pixels]
{\includegraphics[width=0.16\textwidth]{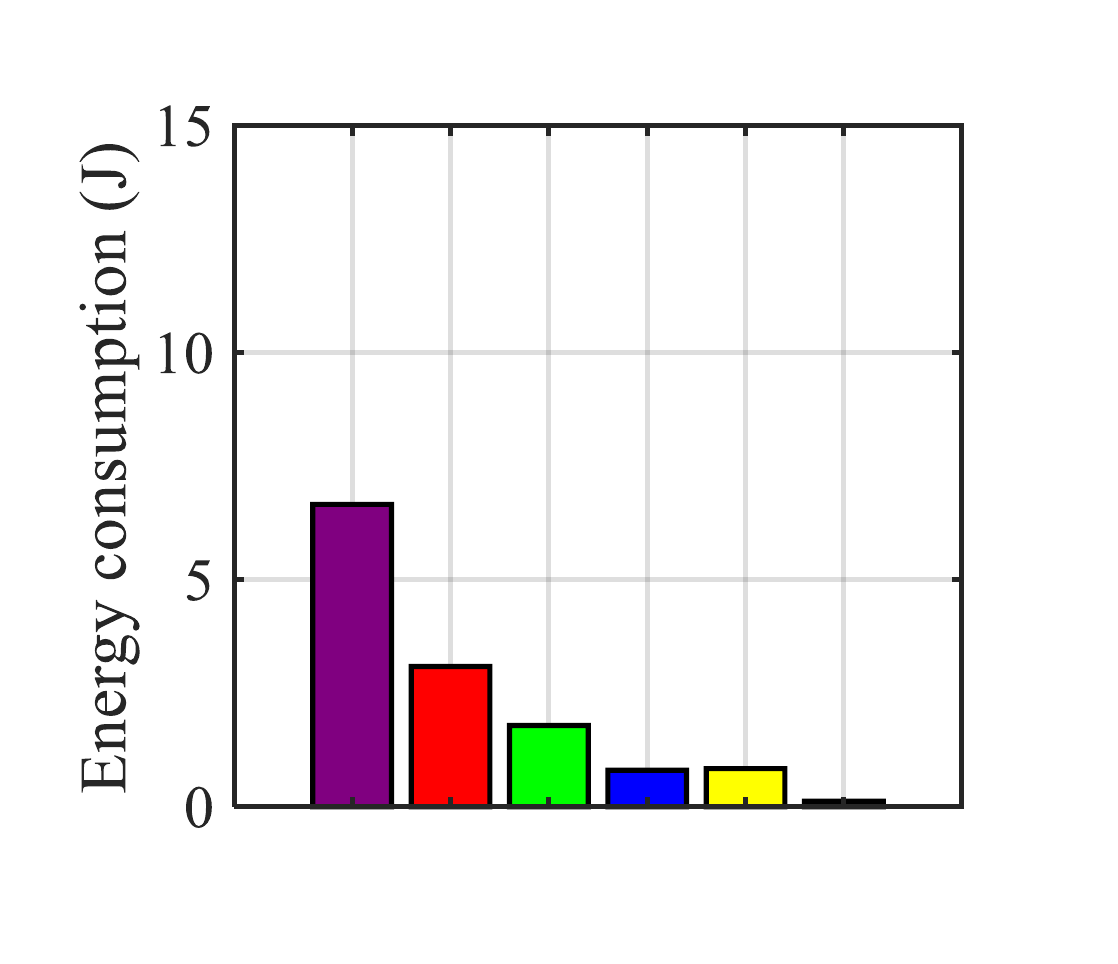}\label{fig:400PerfCPU_Energy}}
\subfigure[$500^2$ pixels]
{\includegraphics[width=0.16\textwidth]{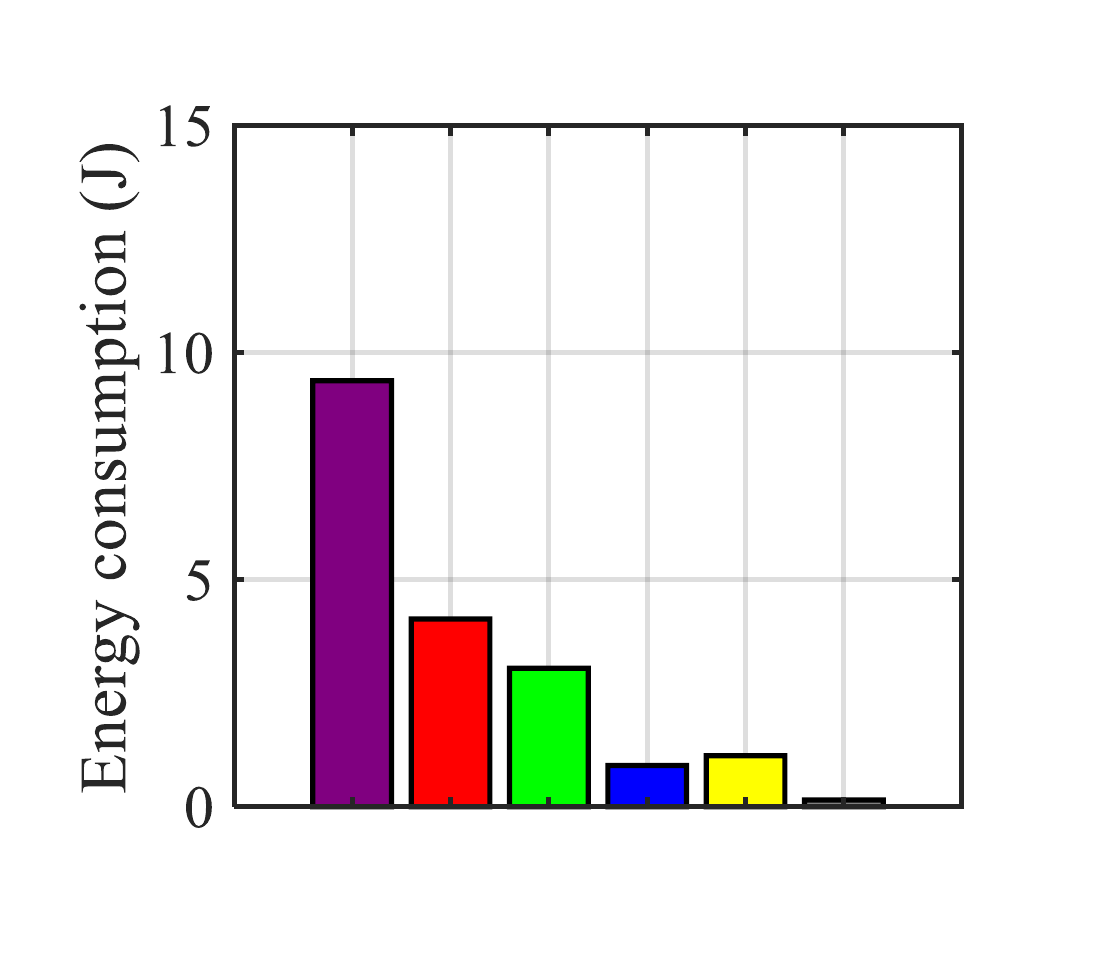}\label{fig:500PerfCPU_Energy}}
\subfigure[$600^2$ pixels]
{\includegraphics[width=0.16\textwidth]{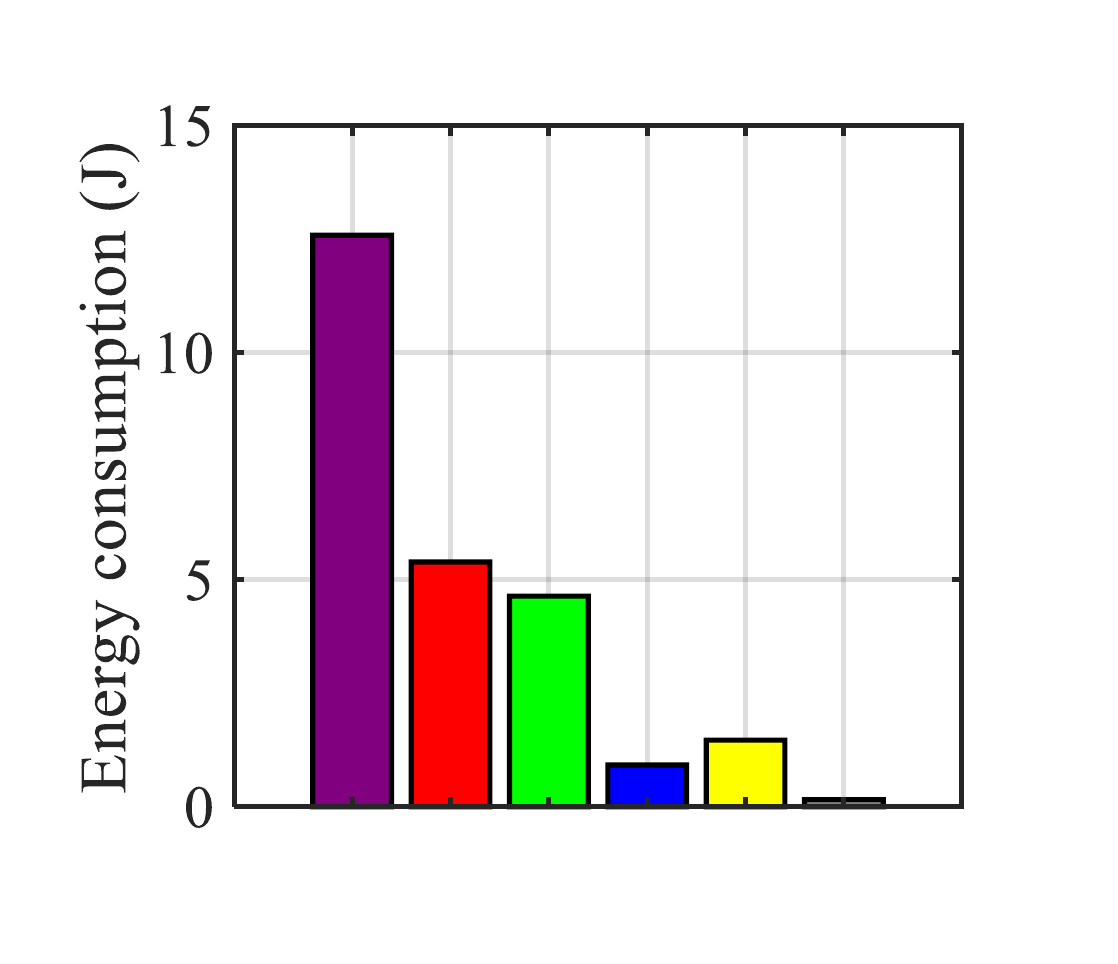}\label{fig:600PerfCPU_Energy}}

\subfigure[$100^2$ pixels]
{\includegraphics[width=0.16\textwidth]{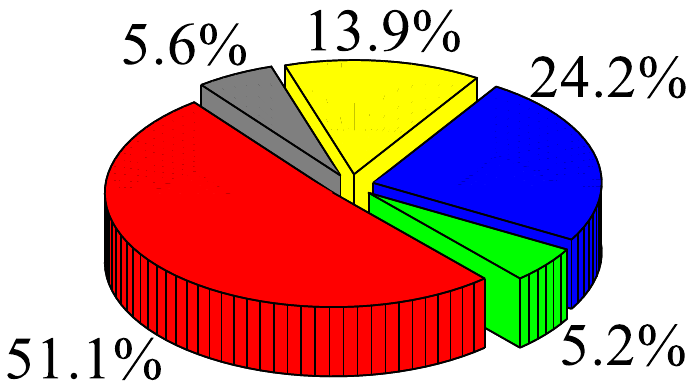}\label{fig:100PerfCPU_Percentage}}
\subfigure[$200^2$ pixels]
{\includegraphics[width=0.16\textwidth]{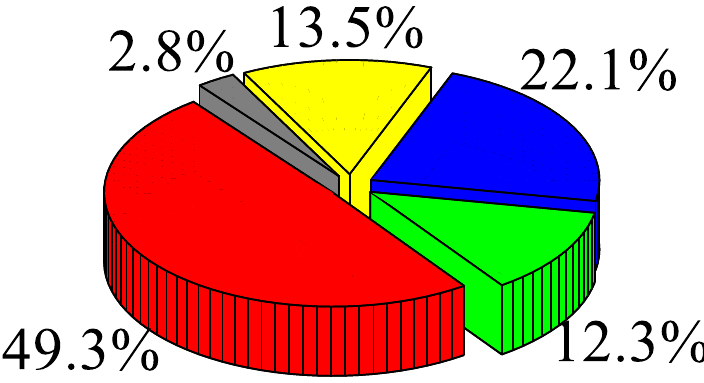}\label{fig:200PerfCPU_Percentage}}
\subfigure[$300^2$ pixels]
{\includegraphics[width=0.16\textwidth]{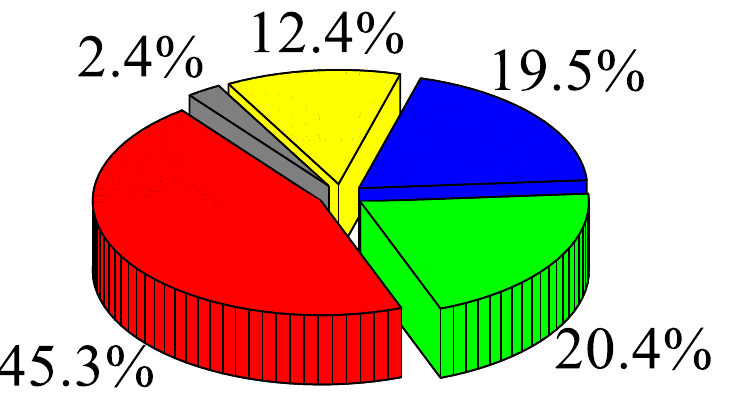}\label{fig:300PerfCPU_Percentage}}
\subfigure[$400^2$ pixels]
{\includegraphics[width=0.16\textwidth]{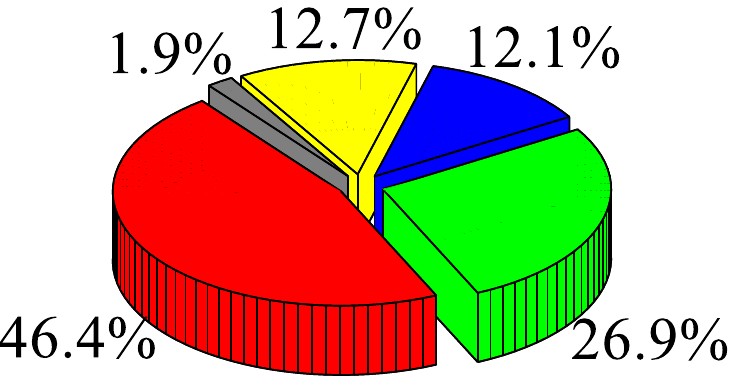}\label{fig:400PerfCPU_Percentage}}
\subfigure[$500^2$ pixels]
{\includegraphics[width=0.16\textwidth]{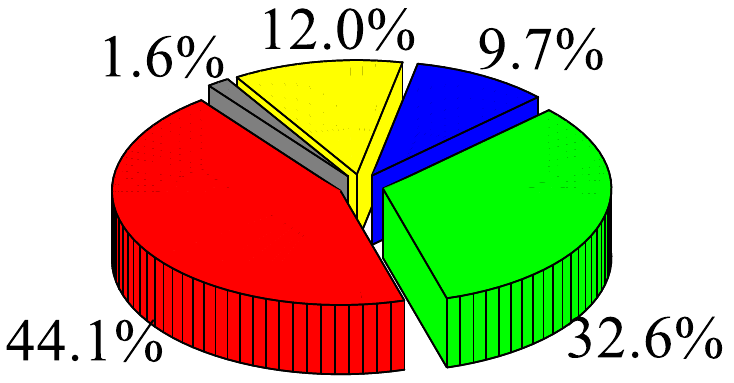}\label{fig:500PerfCPU_Percentage}}
\subfigure[$600^2$ pixels]
{\includegraphics[width=0.16\textwidth]{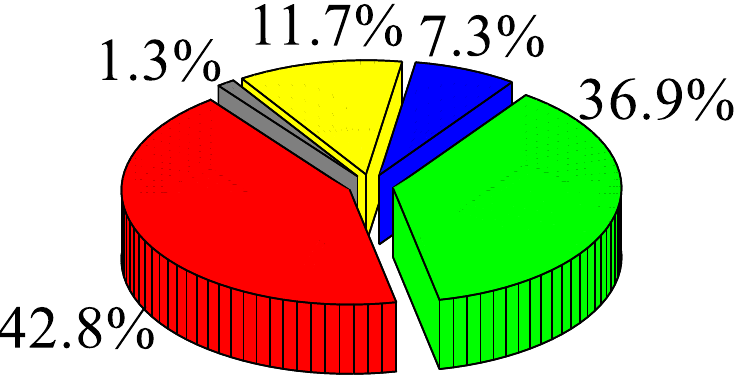}\label{fig:600PerfCPU_Percentage}}
\caption{CNN model size vs. per frame energy consumption (CPU governor: performance).}
\label{fig:cnn_energy_local_perf}   
\end{figure*}

\textbf{CNN Model Size.} Recently, CNN-based methods have become the leading approach for achieving high quality object detection. The CNN model size determines the detection accuracy (i.e., mAP). Increasing the CNN model size always results in a gain of mAP \cite{wang2020user,wang2019globe}. In this section, we seek to investigate how the CNN model size impacts the per frame latency and energy consumption of executing the object detection on the smartphone.

\textbf{Per Frame Latency.} In this experiment, we implement the MobileNets \cite{howard2017mobilenets} with six different CNN model sizes (i.e., from $100\times 100$ to $600\times 600$ pixels). Figs. \ref{fig:cnn_latency_local_inter} and \ref{fig:cnn_latency_local_perf} depict the latency results of running CNN-based object detection with different model sizes, where the smartphone works on interactive and performance governors, respectively. We make the following observations. (10) Running a large CNN model increases the average CPU frequency under the interactive governor, as shown in Figs. \ref{fig:100InterCPUlocal}-\ref{fig:600InterCPUlocal}. This observation demonstrates that a larger CNN model will generate more workload on the smartphone's CPU. (11) A larger CNN model always results in a higher per frame latency for both interactive and performance governors, as depicted in Figs. \ref{fig:Inter_alllatency_local} and \ref{fig:Perf_alllatency_local}, where the per frame latency of the interactive and performance boosts $220\%$ and $247\%$, respectively, when the CNN model size increases from $100\times 100$ to $600\times 600$ pixels. (12) The per frame latency increment is mainly from the raise of the inference latency, while the image conversion latency does not vary much when the CNN model size increases, as illustrated in Figs. \ref{fig:Inter_convlatency_local}, \ref{fig:Inter_inflatency_local}, \ref{fig:Perf_convlatency_local}, and \ref{fig:Perf_inflatency_local}. This is because no matter what the CNN model size $k\times k$ is configured, every YUV frame is converted to an RGB frame with the preview resolution $k_1\times k_2$ first. After the image conversion is completed, the RGB frame is resized to $k\times k$ pixels. (13) For each CNN model size, the performance governor provides a lower per frame latency (i.e., $14\%$-$21\%$) than the interactive governor, which demonstrates that our observation (4) can be applied to diverse CNN model sizes.

\textbf{Per Frame Energy Consumption.} We next examine how the CNN model size impacts the per frame energy consumption of executing object detection on the smartphone. Figs. \ref{fig:cnn_energy_local_inter} and \ref{fig:cnn_energy_local_perf} depict the measured per frame energy consumption results of running CNN-based object detection with different model sizes, where the smartphone works on interactive and performance governors, respectively. Figs. \ref{fig:100InterCPU_Energy}-\ref{fig:600InterCPU_Energy} and \ref{fig:100PerfCPU_Energy}-\ref{fig:600PerfCPU_Energy} illustrate the average per frame energy consumption; and Figs. \ref{fig:100InterCPU_Percentage}-\ref{fig:600InterCPU_Percentage} and Figs. \ref{fig:100PerfCPU_Percentage}-\ref{fig:600PerfCPU_Percentage} depict the average percentage breakdown of energy consumed by each phase in the processing pipeline. We observe that (14) the per frame energy consumption grows dramatically as the CNN model size increases, which is mainly contributed by the inference energy consumption increment. For example, the inference energy consumption accounts for $4.7$\% and $38.0$\% of the per frame energy consumption when the CNN model size is $100\times 100$ and $600\times 600$ pixels, respectively, as shown in Fig. \ref{fig:cnn_energy_local_inter}. In addition, although increasing the CNN model size always results in a gain of mAP, the gain of mAP becomes smaller as the increase of the model size \cite{yolov3}. \textit{This observation inspires us to trade mAP for the per frame energy consumption reduction when the CNN model size is large.} (15) There is a reduction in the proportion of the energy consumption of both the image generation and preview phase and base phase when the CNN model size grows. As we discussed in Section \ref{ssc:cpugovernor_local}, a large proportion of the image generation and preview energy consumption to the per frame energy consumption may result in the smartphone expending significant reactive energy for sampling non-detectable image frames. \textit{These two observations indicate that there is a trade-off between reducing the per frame energy consumption and decreasing the proportion of the reactive energy.} Therefore, \textit{a comprehensive approach for improving the energy efficiency of executing CNN-based object detection on smartphones must take into account the reduction of both the per frame energy consumption and the proportion of the reactive energy.}

\subsection{Insights and Research Opportunities}
\label{ssc:sumdis_local}

\textbf{Insights.}

\begin{itemize}

\item Ondemand and performance CPU governors achieve lower per frame energy consumption and latency than the other four popular CPU governors when the smartphone locally executes the CNN-based object detection application. However, as the smartphone's CPUs keep running at the highest frequency in the performance governor, it may cause the smartphone overheating and trigger CPU throttling mechanism to avoid thermal emergencies by sacrificing the performance. Therefore, the ondemand governor is recommended as the default CPU governor of the local execution, which supports sustainable and low per frame energy consumption and latency object detections.

\item Both the CPU governor (i.e., CPU frequency) and the CNN model size significantly impact the per frame latency and energy consumption. However, simply increasing the CPU frequency or decreasing the CNN model size is inadequate to minimize the per frame energy consumption because different phases may have opposite reactions.

\item Increasing the CNN model size always results in a gain of mAP and an increment of the per frame energy consumption. However, the amount of the increment of mAP becomes smaller as the increase of the model size, while the increment of the per frame energy consumption becomes larger as the increase of the model size. Therefore, this observation inspires us to trade mAP for the per frame energy consumption reduction when the CNN model size is large. 

\item In order to improve the energy efficiency of smartphones that locally execute the CNN-based object detection, we must jointly consider the per frame energy consumption, the proportion of the reactive energy, and the battery life.

\end{itemize}

\textbf{Research Opportunities.} 
\begin{itemize}
\item Current CPU governors cannot achieve energy-efficient object detection on smartphones (i.e., jointly considering the per frame energy consumption, the proportion of the reactive energy, and the battery life). A CPU governor specifically designed for CNN-based object detection applications is critical and desirable.

\item An intelligent configuration adaption algorithm that is capable of selecting the best combination of the CPU governor, CNN model size, and camera sample rate according to the smartphone's battery life, processor's temperature, and detection accuracy requirement might be a potential solution for achieving energy-efficient and high-performance object detection.

\end{itemize}

\begin{figure*}[t!]
\centering
\subfigure[Conservative]
{\includegraphics[width=0.16\textwidth]{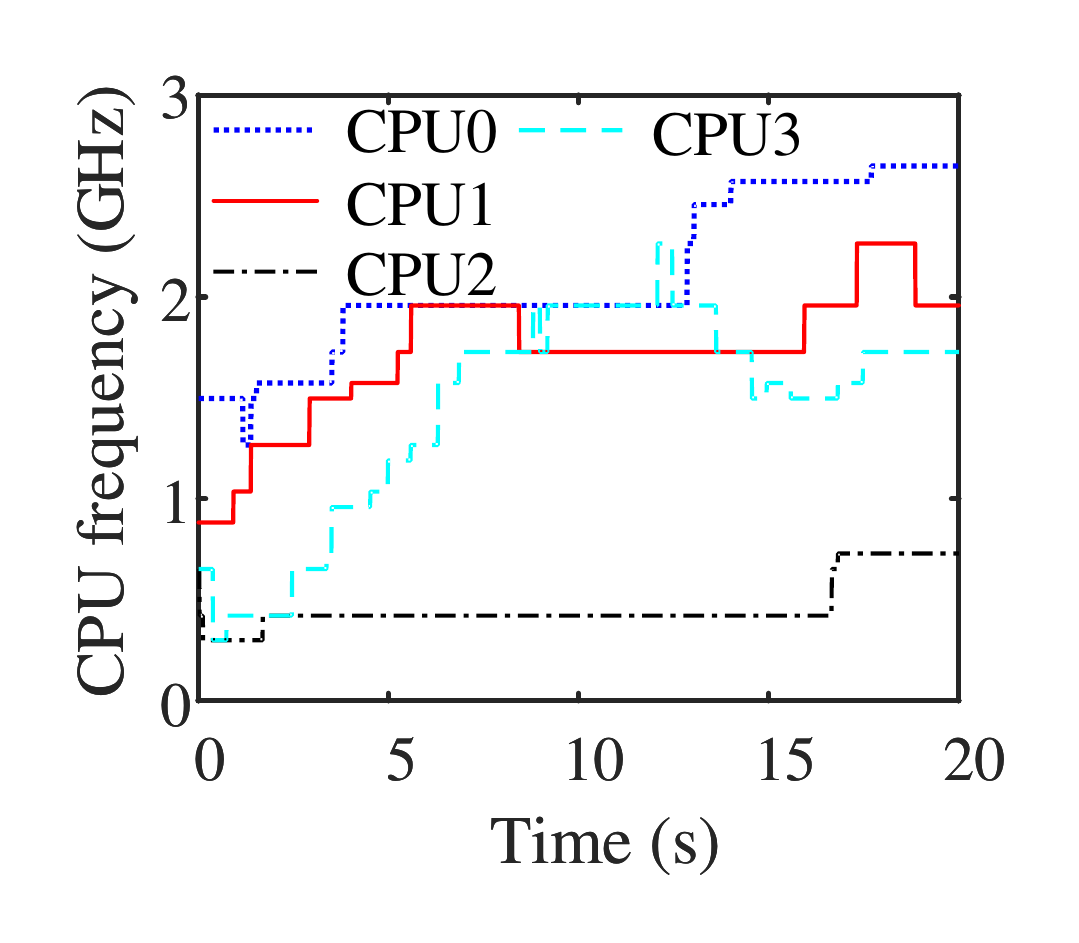}\label{fig:TFLremote_CPUfreq_Conser}}
\subfigure[Ondemand]
{\includegraphics[width=0.16\textwidth]{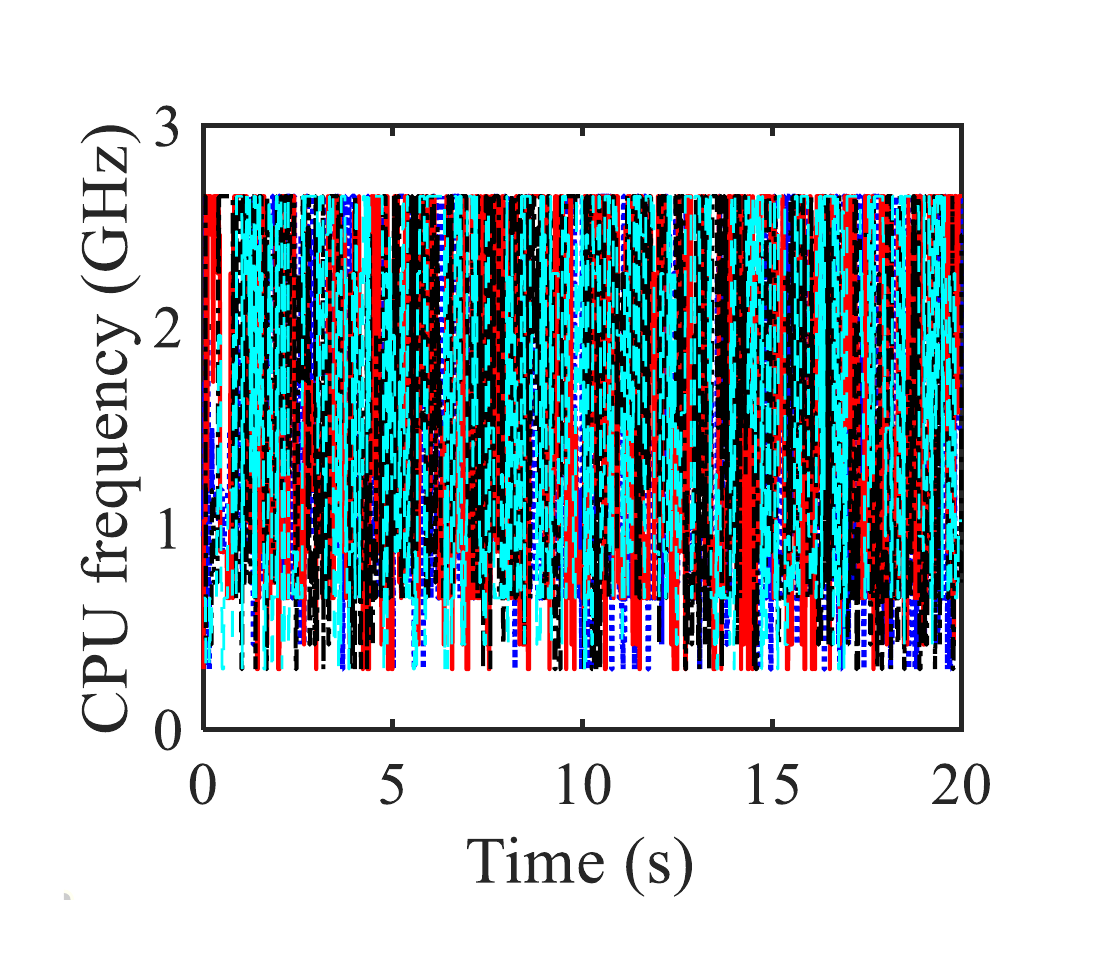}\label{fig:TFLremote_CPUfreq_Ond}}
\subfigure[Interactive]
{\includegraphics[width=0.16\textwidth]{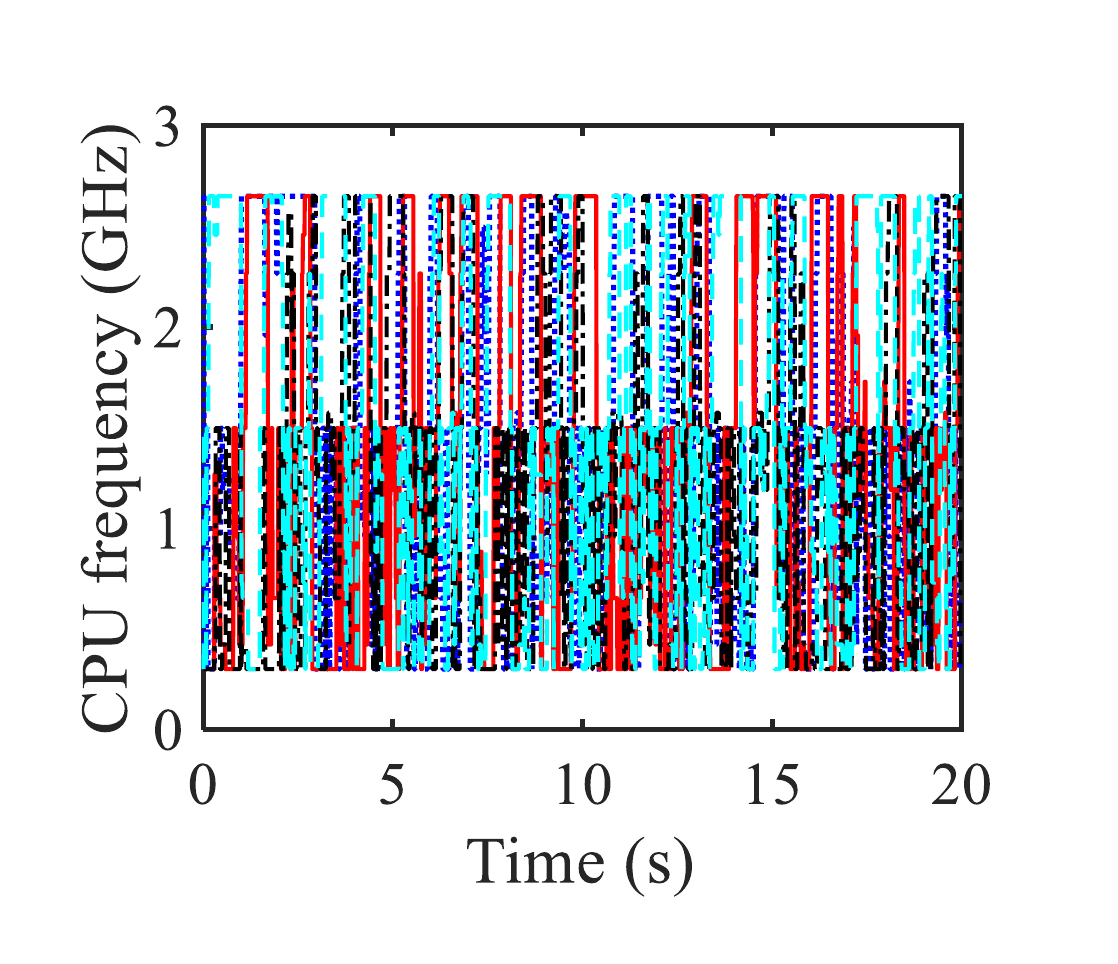}\label{fig:TFLremote_CPUfreq_Inter}}
\subfigure[Userspace]
{\includegraphics[width=0.16\textwidth]{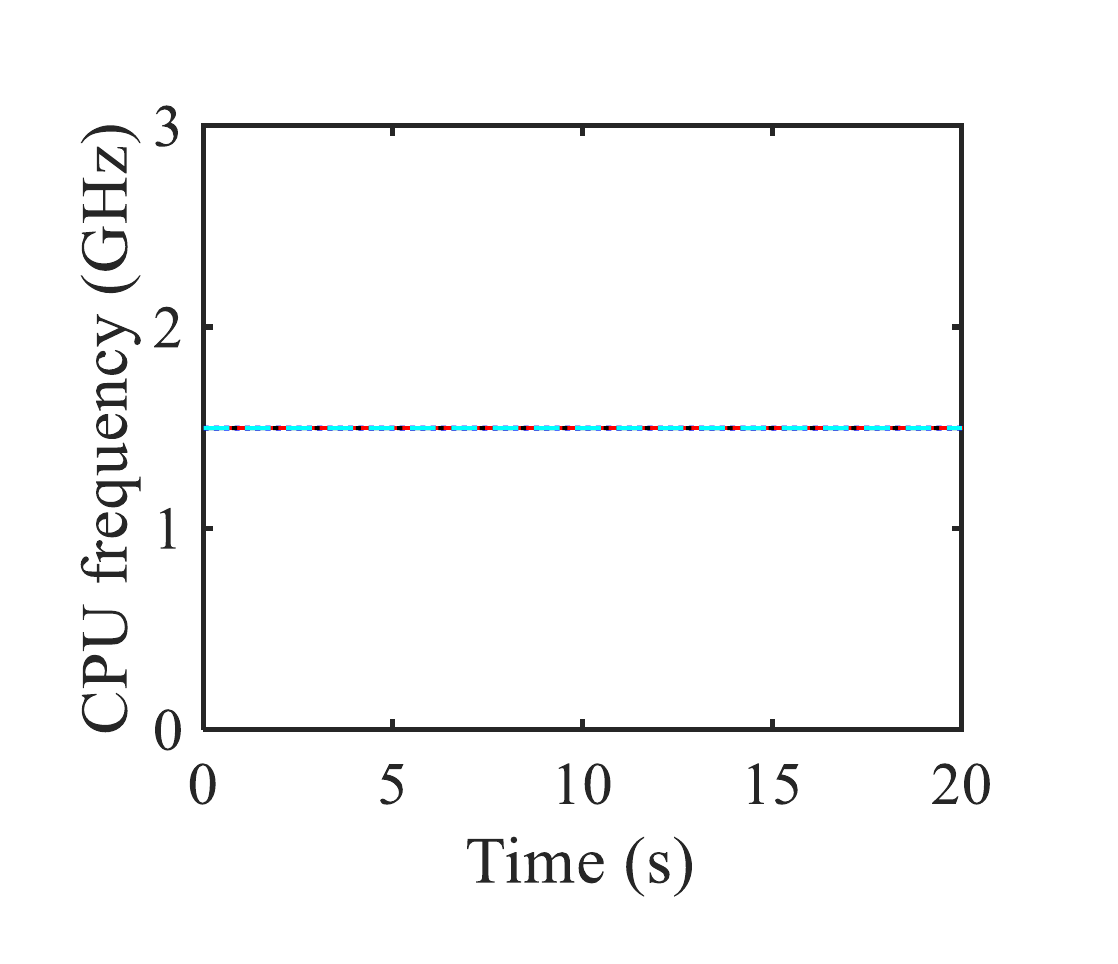}\label{fig:TFLremote_CPUfreq_Use}}
\subfigure[Powersave]
{\includegraphics[width=0.16\textwidth]{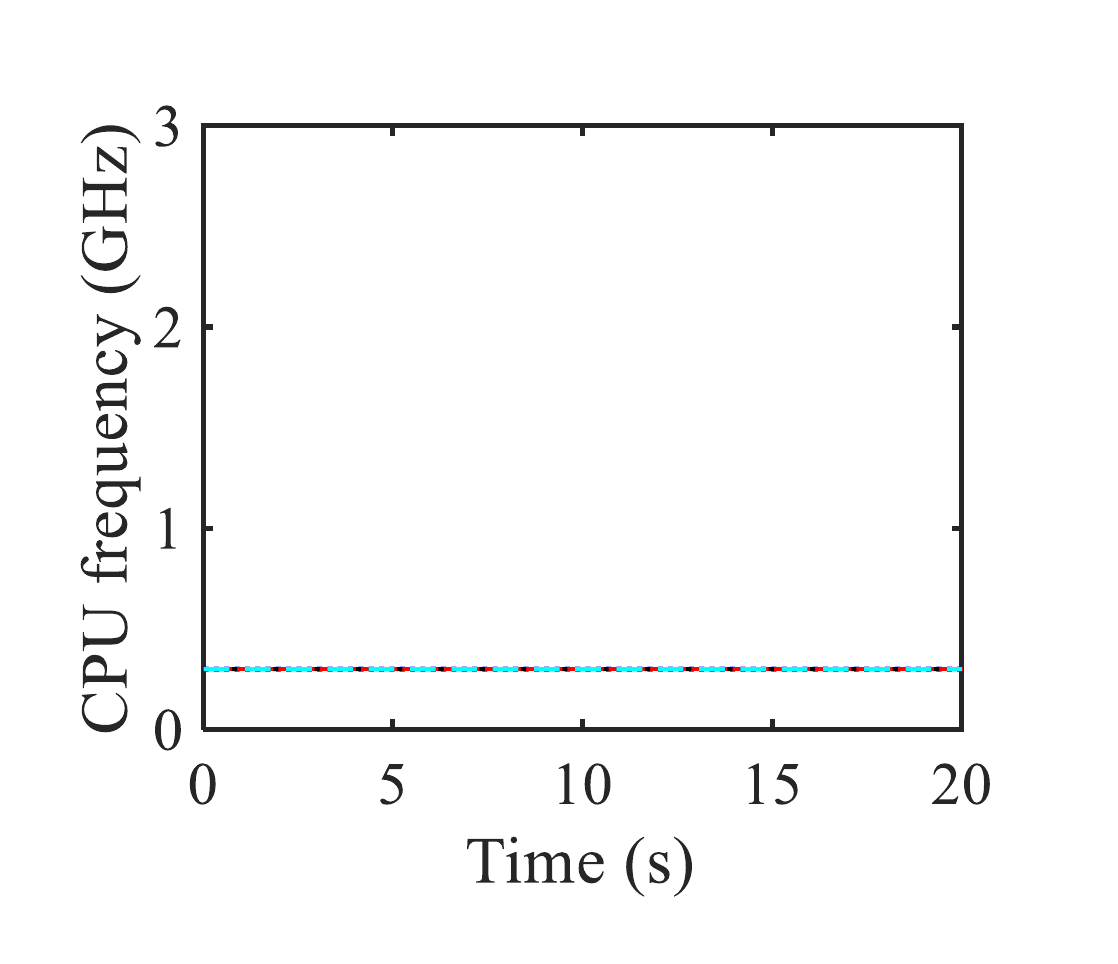}\label{fig:TFLremote_CPUfreq_Pow}}
\subfigure[Performance]
{\includegraphics[width=0.16\textwidth]{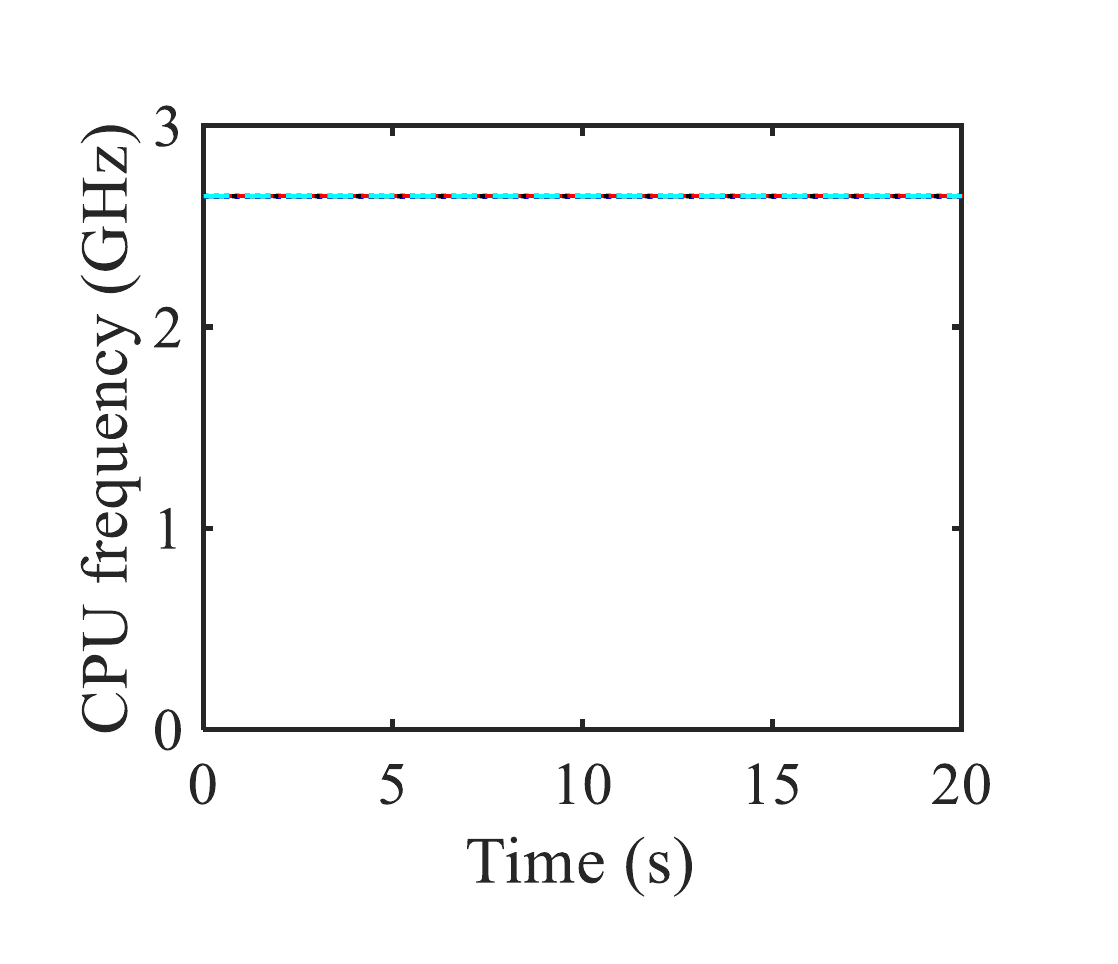}\label{fig:TFLremote_CPUfreq_Perf}}

\subfigure[Conservative]
{\includegraphics[width=0.16\textwidth]{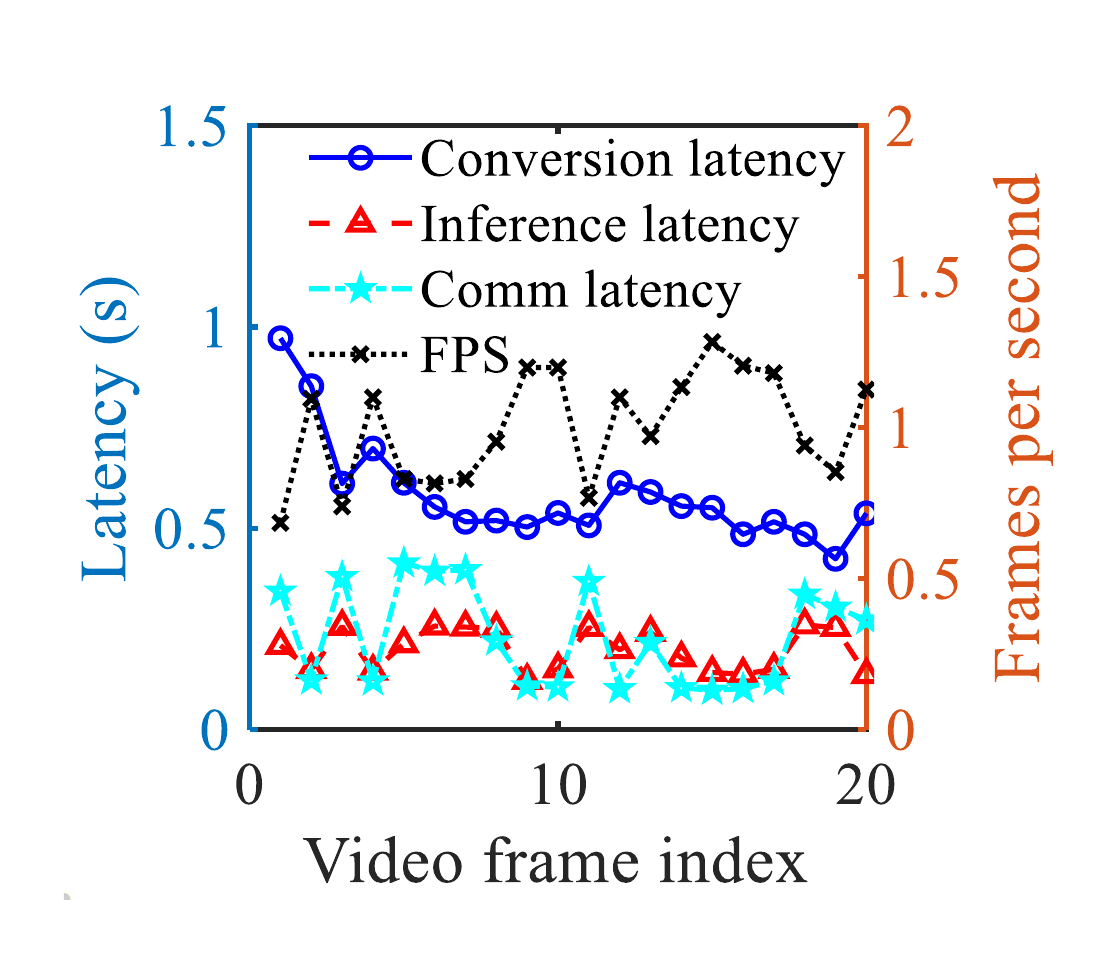}\label{fig:TFLremote_CPU_Governors_Latency_Conser}}
\subfigure[Ondemand]
{\includegraphics[width=0.16\textwidth]{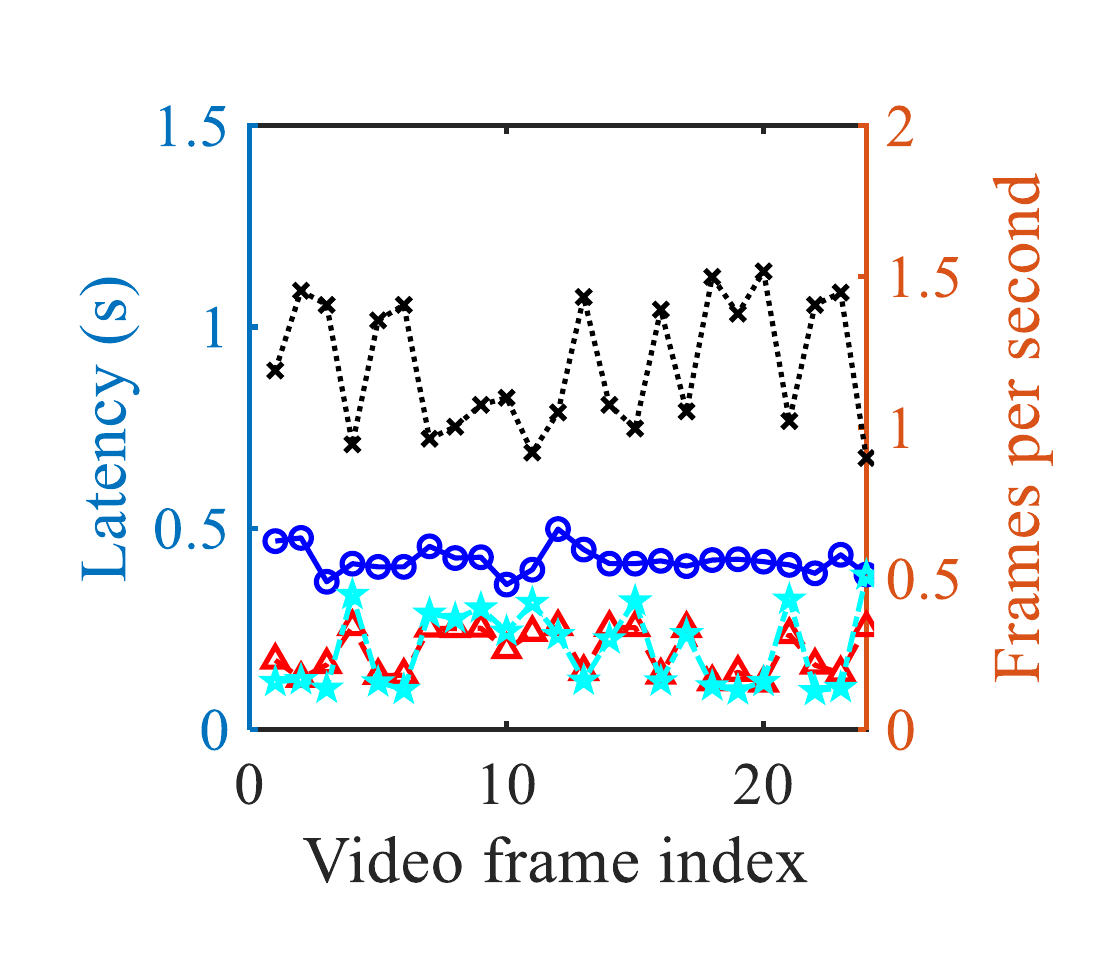}\label{fig:TFLremote_CPU_Governors_Latency_Ond}}
\subfigure[Interactive]
{\includegraphics[width=0.16\textwidth]{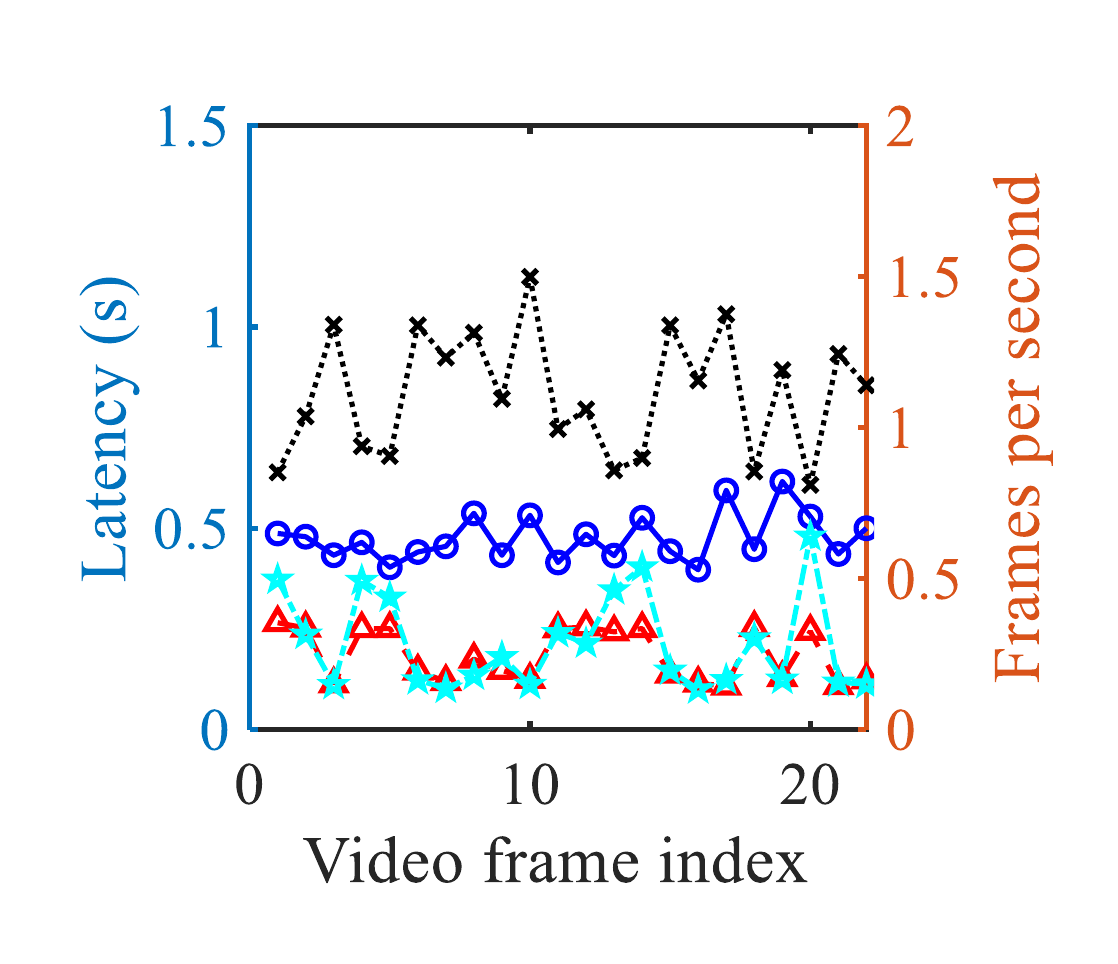}\label{fig:TFLremote_CPU_Governors_Latency_Inter}}
\subfigure[Userspace]
{\includegraphics[width=0.16\textwidth]{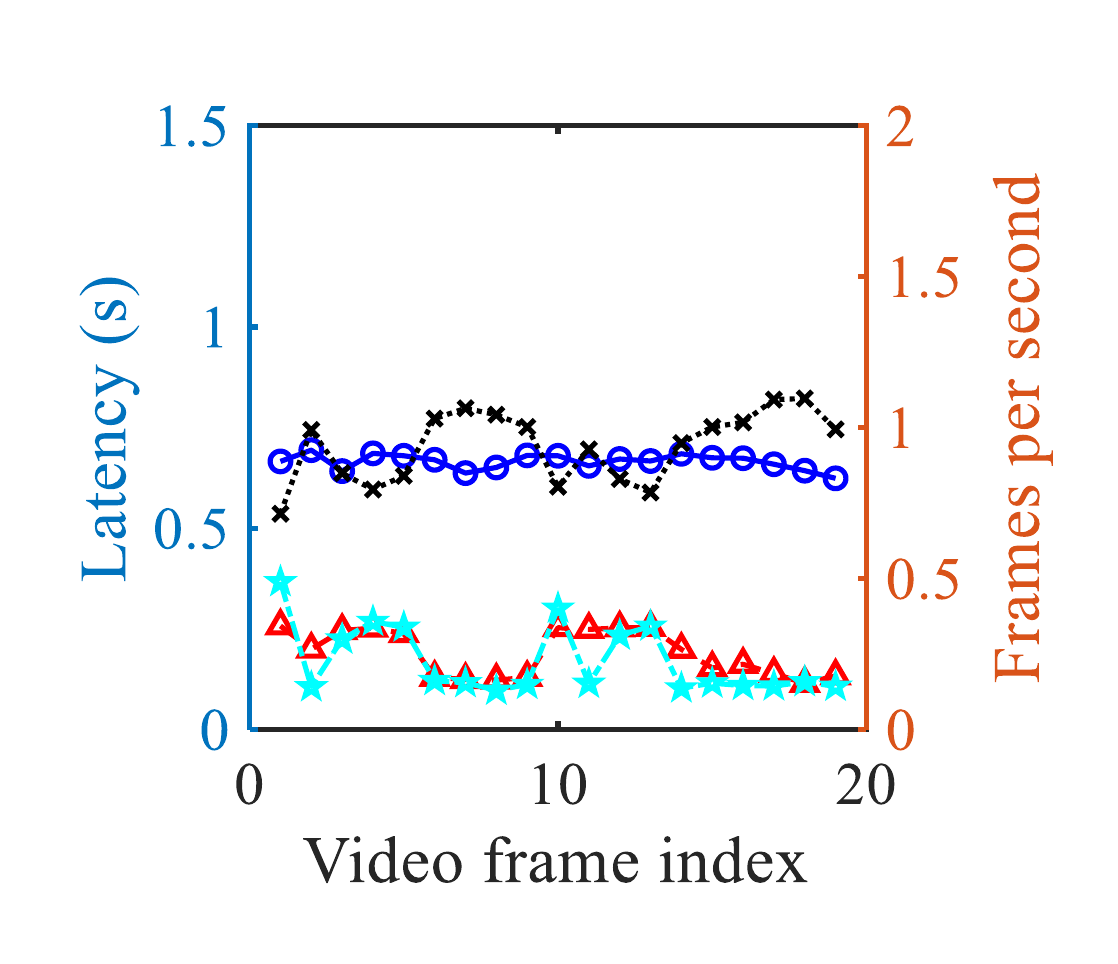}\label{fig:TFLremote_CPU_Governors_Latency_Use}}
\subfigure[Powersave]
{\includegraphics[width=0.16\textwidth]{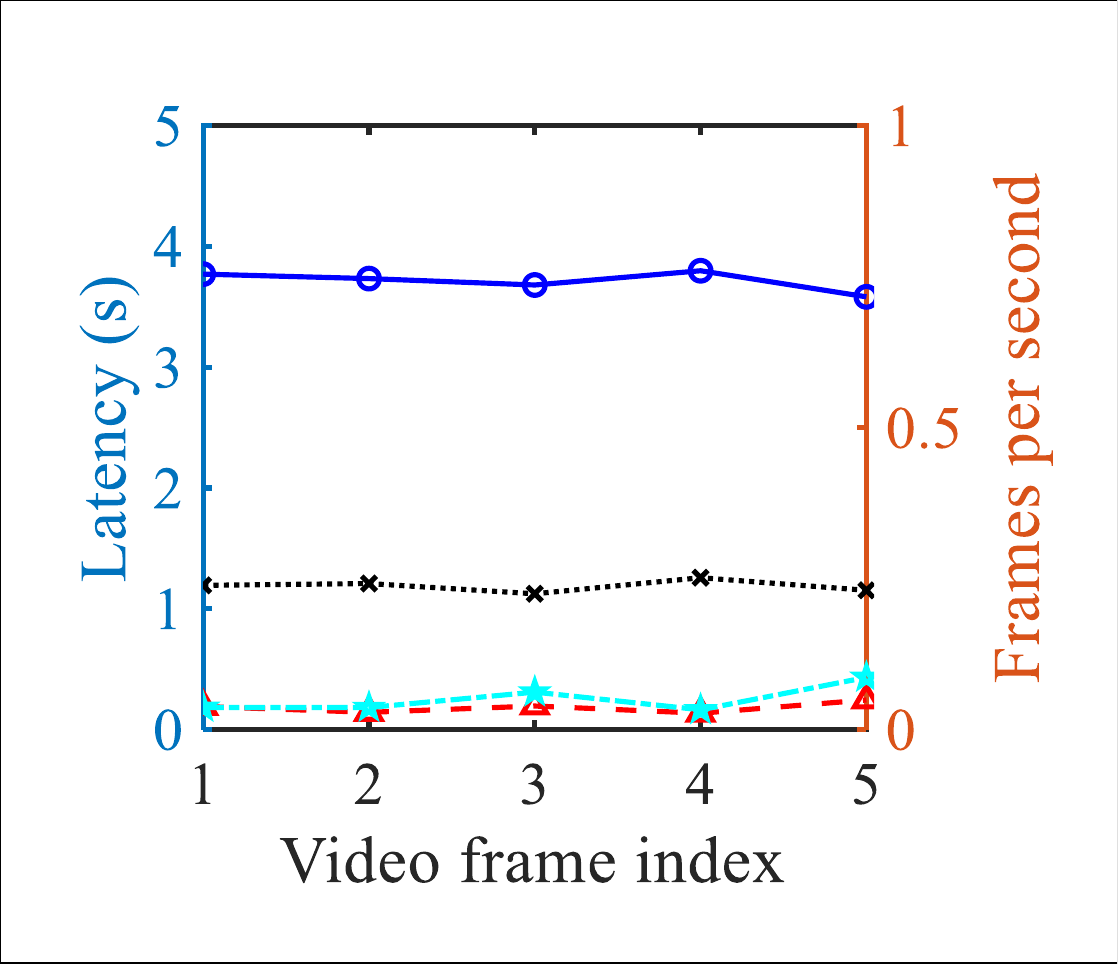}\label{fig:TFLremote_CPU_Governors_Latency_Pow}}
\subfigure[Performance]
{\includegraphics[width=0.16\textwidth]{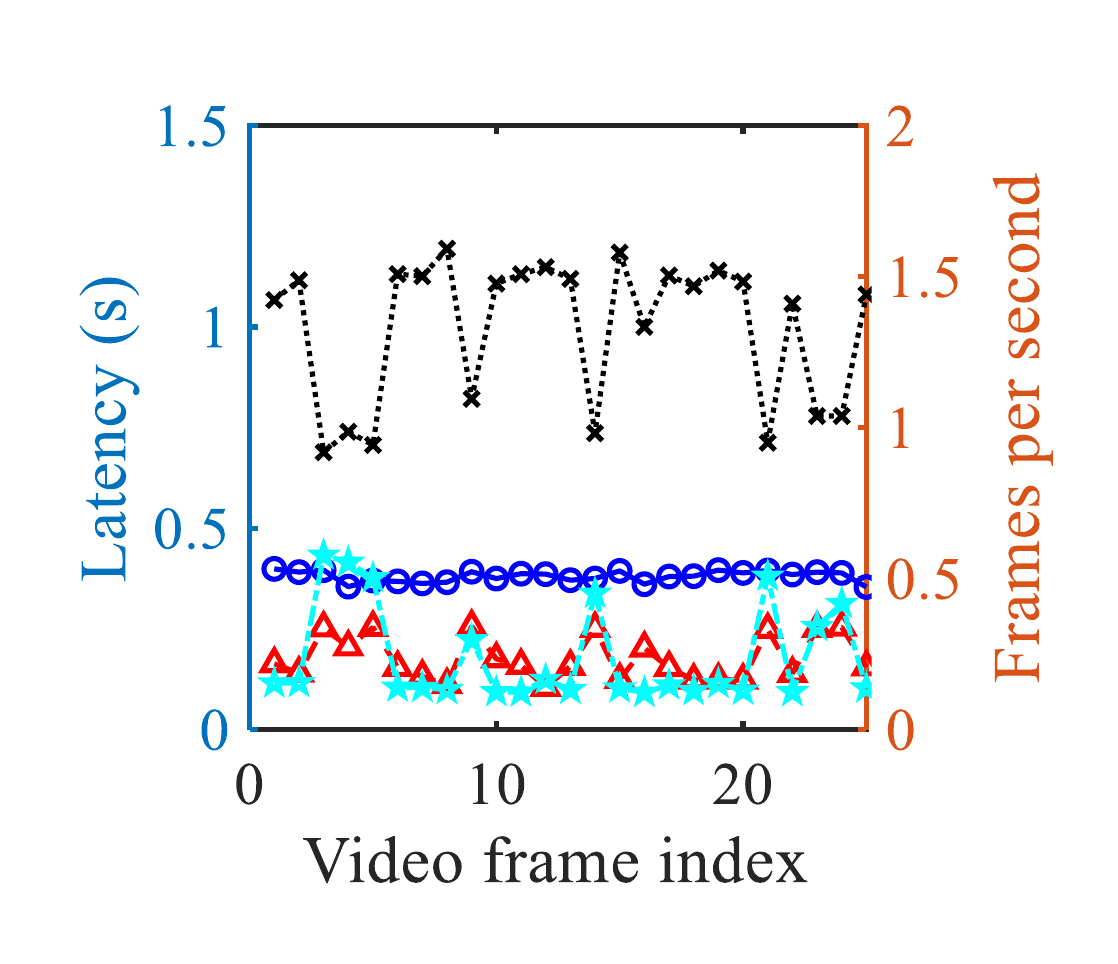}\label{fig:TFLremote_CPU_Governors_Latency_Perf}}
\caption{CPU governor vs. per frame latency (CNN model size: $320\times320$ pixels).}
\label{fig:cpugovernors_latencyremote}   
\end{figure*}

\section{Experimental Results of Remote Execution}
\label{sc:remotemeasurement}
\textbf{RQ 2.} Does offloading the object detection tasks to a powerful infrastructure significantly decrease both the energy consumption and latency? To answer this question, in this section, we describe the experimental results on evaluating the impact of various factors on the energy consumption of a mobile AR client, latency, and detection accuracy of remotely executing CNN-based object detection on smartphones. We begin by measuring the per frame latency and the per frame energy consumption of executing CNN-based object detection under different smartphone's CPU governors in Section \ref{ssc:cpugovernor_remote}. In addition, we explore the impact of the CNN model size on the per frame latency and the per frame energy consumption in Section \ref{ssc:cnnsize_remote}. Furthermore, the image generation and preview phase and image conversion phase are discussed in Sections \ref{ssc:impimage_remote} and \ref{ssc:impconv_remote}, respectively. Lastly, in Section \ref{ssc:sumdis_remote}, we summarize the insights from our measurement studies and discuss potential research opportunities for improving the energy efficiency of remotely executing CNN-based object detection on smartphones.

\begin{table*}[t!]
 \begin{center}
    \caption{Latency results of the remote execution with different CPU governors.}
    \label{tb:remote_cpu_latency}
  \begin{tabular}{|l||c|c|c|c|c|c|}
    \hline
    CPU Governor                      & Conservative & Ondemand & Interactive & Userspace & Powersave & Performance \\ \hline
    Per Frame Latency (second)        & 1.009 & 0.853 & 0.905 & 1.075 & 4.214 & \textbf{0.819}\\ \hline
    Image Conversion Latency (second) & 0.571 & 0.409 & 0.477 & 0.666 & 3.622 & \textbf{0.376}\\ \hline
    Inference Latency (Second)        & 0.189 & 0.189 & \textbf{0.179} & 0.180 & 0.185 & 0.192\\ \hline
    Communication Latency (Second)    & 0.205 & 0.210 & 0.200 & \textbf{0.174} & 0.293 & 0.207\\ \hline
    Others (Second)                   & 0.044 & 0.045 & 0.049 & 0.055 & 0.114 & \textbf{0.044}\\ \hline
    Per Frame Latency Reduction ($\%$)          & -10.3 & 5.6 & 10.7 & 25.6 & \textbf{45.7} & 0.5\\ \hline
  \end{tabular}
  \end{center}
\end{table*}

\subsection{The Impact of CPU Governor}
\label{ssc:cpugovernor_remote}
\textbf{Per Frame Latency.} We first seek to investigate how the CPU governor impacts the per frame latency of object detection in the remote execution scenario, where a CNN model is executed on the implemented edge server with a 5GHz WiFi link to the smartphone. The executed CNN model size is $320\times 320$ pixels. The experimental results are shown in Fig. \ref{fig:cpugovernors_latencyremote}, where Figs. \ref{fig:TFLremote_CPUfreq_Conser}-\ref{fig:TFLremote_CPUfreq_Perf} depict the frequency variations of the tested smartphone's CPUs and Figs. \ref{fig:TFLremote_CPU_Governors_Latency_Conser}-\ref{fig:TFLremote_CPU_Governors_Latency_Perf} illustrate the latency of each phase in the object detection processing pipeline. Compared to the local execution, a new time-consuming phase named communication is introduced into the processing pipeline of the remote execution besides image conversion and inference phases. We obtain similar observations to (3) and (4). In addition, (16) Fig. \ref{fig:TFLremote_CPUfreq_Conser} shows that only one core of the smartphone's processor reaches to the highest possible frequency under the conservative governor, which indicates that running CNN models on the edge server is capable of reducing the workload on the smartphone's CPUs. (17) However, because of the workload reduction and the conservative governor's low-frequency bias, the per frame latency of the remote execution under the conservative governor is approximately $10.3\%$ larger than that of the local execution, as shown in Table \ref{tb:remote_cpu_latency}. This observation demonstrates that the conservative governor is not suitable for the remote execution either and performs worse in the remote execution. 

\begin{table*}[t]
  \begin{center}
    \caption{Smartphones and the edge server used in this experiment.}
    \label{tb:charoftest}
    \begin{tabular}{l||c|c|c|c} \toprule
     Manufacturer & Samsung & Google & Asus & Nvidia\\\midrule
      Model & Galaxy S5 & Nexus 6 & ZenFone AR & Jetson AGX Xavier\\
      OS & Android 6.0.1 & Android 5.1.1 & Android 7.0 & Ubuntu 18.04 LTS aarch64\\
      SoC & Snapdragon 801 (28 nm) & Snapdragon 805 (28 nm) & Snapdragon 821 (14 nm) & Xavier \\
      CPU & 32-bit 4-core 2.5GHz Krait 400 & 32-bit 4-core 2.7GHz Krait 450 & 64-bit 4-core 2.4GHz Kryo & 64-bit 8-core 2.26GHz Carmel \\
      \multirow{2}*{GPU} & \multirow{2}*{578MHz Adreno 330} & \multirow{2}*{600MHz Adreno 420} & \multirow{2}*{653MHz Adreno 530}  & 512-core 1377MHz Volta\\
      & & & & with 64-TensorCores \\
      RAM & 2GB & 3GB  & 6GB  & 16GB \\
      WiFi & 802.11n/ac, MIMO $2\times2$ & 802.11n/ac, MIMO $2\times2$ & 802.11n/ac/ad, MIMO $2\times2$ & --- \\
      Release date & April 2014 & November 2014 & July 2017 & September 2018\\
      \bottomrule
    \end{tabular}
  \end{center}
\end{table*}

Interestingly, we find that (18) the remote execution achieves significantly distinct per frame latency reduction when the smartphone works on different CPU governors. For example, as shown in Table \ref{tb:remote_cpu_latency}, the remote execution achieves a per frame latency reduction of $45.7\%$ in the powersave governor compared to the local execution, while it only obtains a per frame latency reduction of $0.5\%$ in the performance governor. This observation may infer that \textit{locally executing CNN-based object detection on the smartphone with advanced processors and working on a high CPU frequency is capable of achieving a comparable latency performance as the remote execution.} This inference is important for guiding whether a smartphone has to offload its object detection tasks to the edge server for reducing the service latency. 

\begin{table}[t!]
\newcommand{\tabincell}[2]{\begin{tabular}{@{}#1@{}}#2\end{tabular}}
 \begin{center}
    \caption{Classification \& latency results of different smartphones.}
    \label{tb:Antutuperformance}
  \begin{tabular}{|l|l||c|c|c|}
    \hline
    \multicolumn{2}{|l||}{Smartphone}& S5 & Nexus 6 & ZenFone AR\\ \hline
    \multicolumn{2}{|l||}{CPU Score} & 36871 & 37521 & \textbf{58531}\\ \hline
    \multicolumn{2}{|l||}{GPU Score} & 6678 & 18063 & \textbf{67286} \\\hline
    \multicolumn{2}{|l||}{Image Processing Score} & 3103 & 6862 & \textbf{11321}\\\hline
    \multicolumn{2}{|l||}{Total Score} & 66414 & 80047 & \textbf{173472} \\\hline
    \multicolumn{2}{|l||}{Class} & Low-end & Low-end & High-end \\\hline
    \multirow{2}{*}{\tabincell{l}{Per Frame\\ Latency (s)}} & Local & 1.098 & 1.013 & \textbf{0.225} \\\cline{2-5}
      & Remote & 0.956 & 0.905 & \textbf{0.312} \\\hline
    \multicolumn{2}{|l||}{Latency Reduction ($\%$)} & \textbf{12.9} & 10.7 & -38.7 \\\hline
  \end{tabular}
  \end{center}
\end{table}

In order to verify this inference, we conduct a measurement study using three smartphones with different computation capacities, where their characteristics are summarized in Table \ref{tb:charoftest}. We classify them into two classes, \emph{low-end} and \emph{high-end} smartphones, according to their general hardware performance tested by using an Antutu benchmark \cite{Antutu}. The testing results are shown in Table \ref{tb:Antutuperformance}. All these three smartphones work on the interactive governor. The results verify our inference above, where the per frame latency of the low-end smartphones is decreased around $12\%$, whereas the per frame latency of the high-end smartphone is increased approximately $38.7\%$ when offloading the object detection tasks to the edge server (note that the value of the latency reduction may differ depending on how powerful the edge server's GPU is). This observation supports the fact that lots of recently released smartphones with high computation power possess the capability of running a light CNN model with low latency. However, the detection accuracy of the large CNN model on the edge server is better than that of the light CNN model on the smartphone (e.g., $mAP = 51.5$ on the server and $mAP = 19.3$ on the smartphone when the frame resolution is around $300\times 300$ pixels). Furthermore, in general, different use cases may have variant latency/accuracy requirements. For example, the AR cognitive assistance case where a high-end wearable device helps visually impaired people to navigate on a street may need a low latency but can tolerate a relatively high number of false positives (i.e., false alarms are fine but missing any potential threats on the street is costly) \cite{ran2017delivering}. In contrast, an AR used for recommending products in shopping malls or supermarkets may tolerate a relatively long latency but require high detection accuracy. \textit{Therefore, both the smartphone's computation capacity and the use case should be considered when determining the appropriate execution approach (i.e., local or remote).}

\begin{figure*}[t]
\centering
\subfigure[Conservative]
{\includegraphics[width=0.16\textwidth]{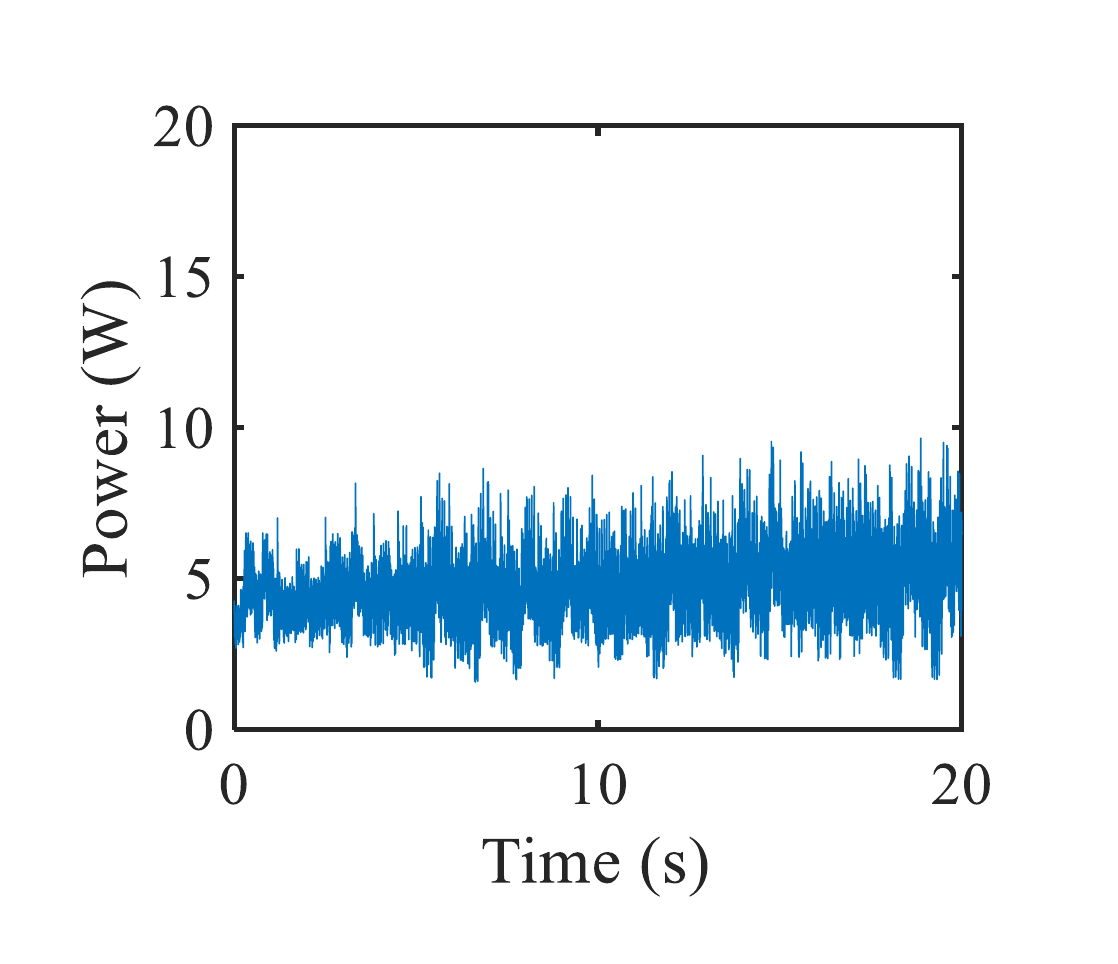}\label{fig:TFLremote_CPU_Governors_Power_Conser}}
\subfigure[Ondemand]
{\includegraphics[width=0.16\textwidth]{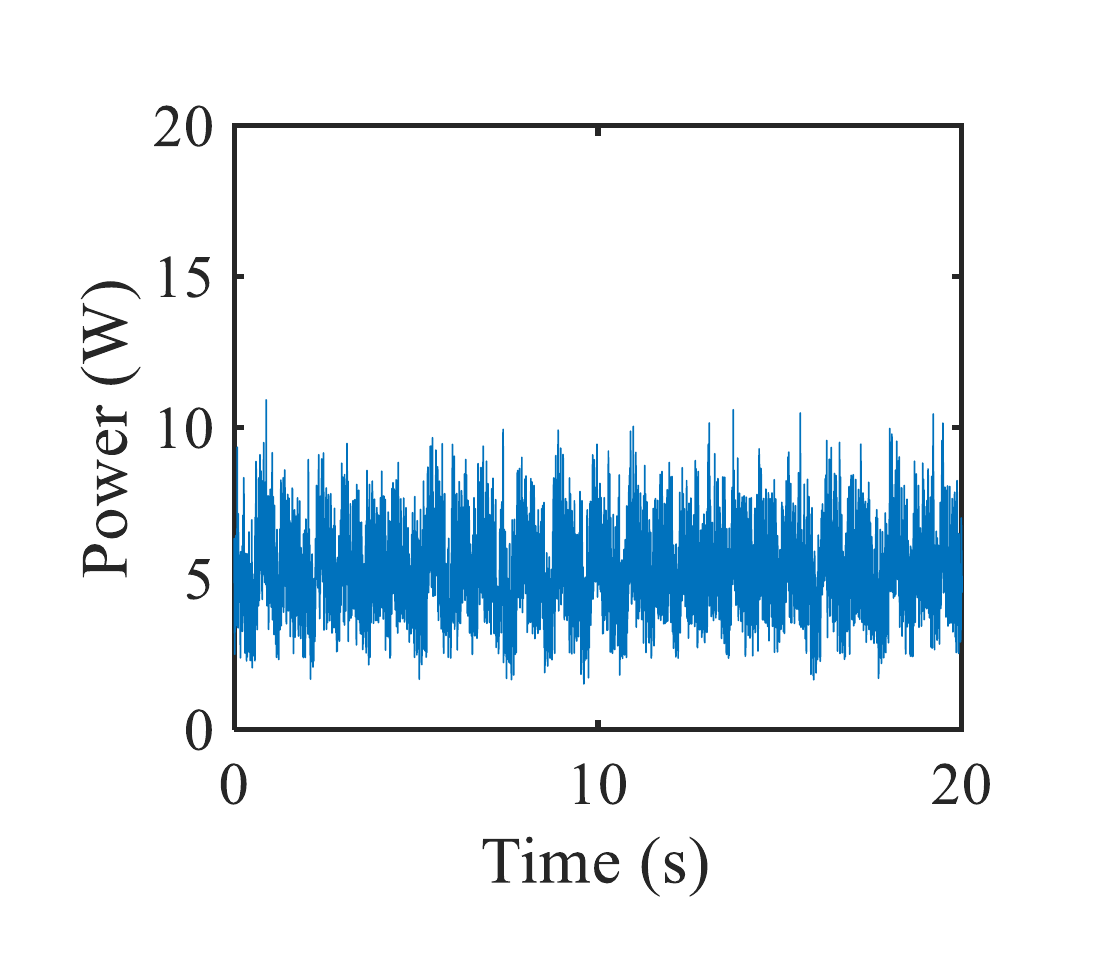}\label{fig:TFLremote_CPU_Governors_Power_Ond}}
\subfigure[Interactive]
{\includegraphics[width=0.16\textwidth]{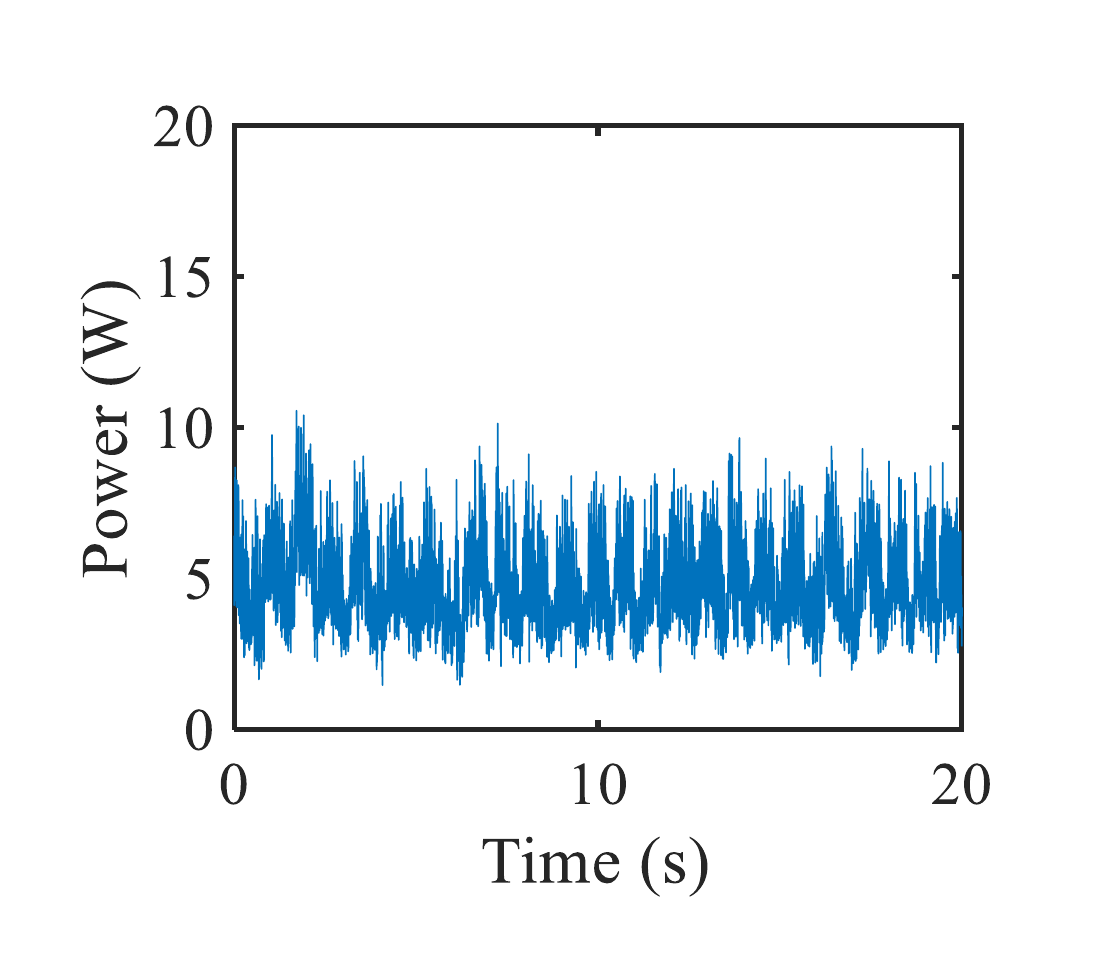}\label{fig:TFLremote_CPU_Governors_Power_Inter}}
\subfigure[Userspace]
{\includegraphics[width=0.16\textwidth]{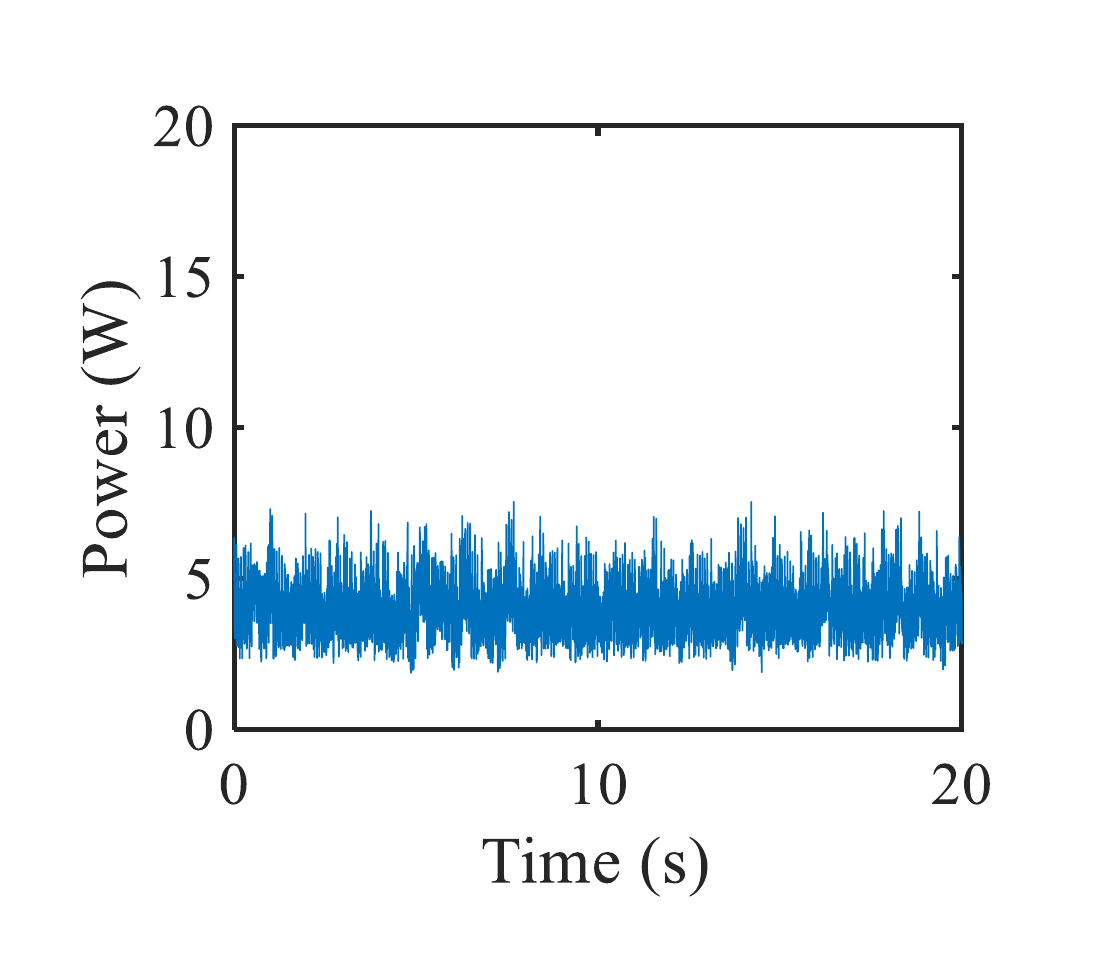}\label{fig:TFLremote_CPU_Governors_Power_Use}}
\subfigure[Powersave]
{\includegraphics[width=0.16\textwidth]{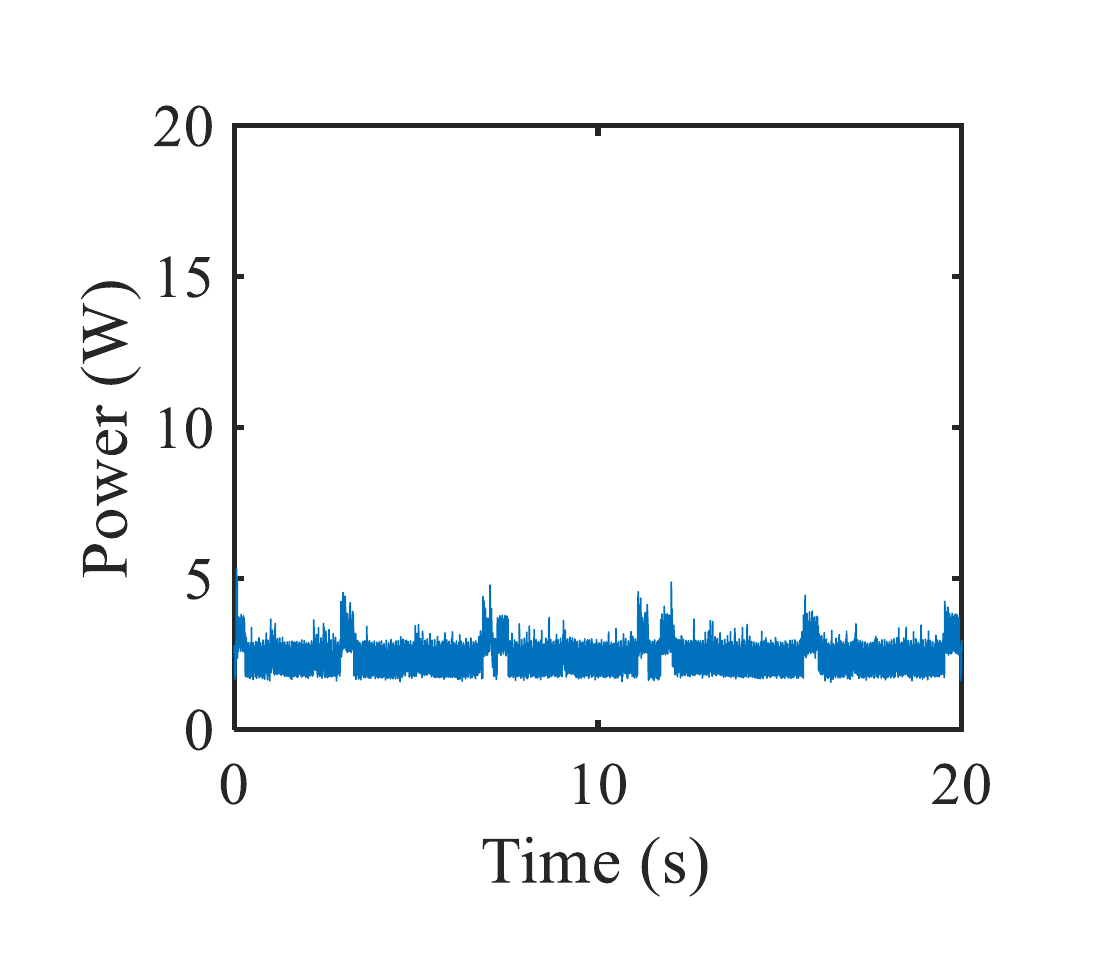}\label{fig:TFLremote_CPU_Governors_Power_Pow}}
\subfigure[Performance]
{\includegraphics[width=0.16\textwidth]{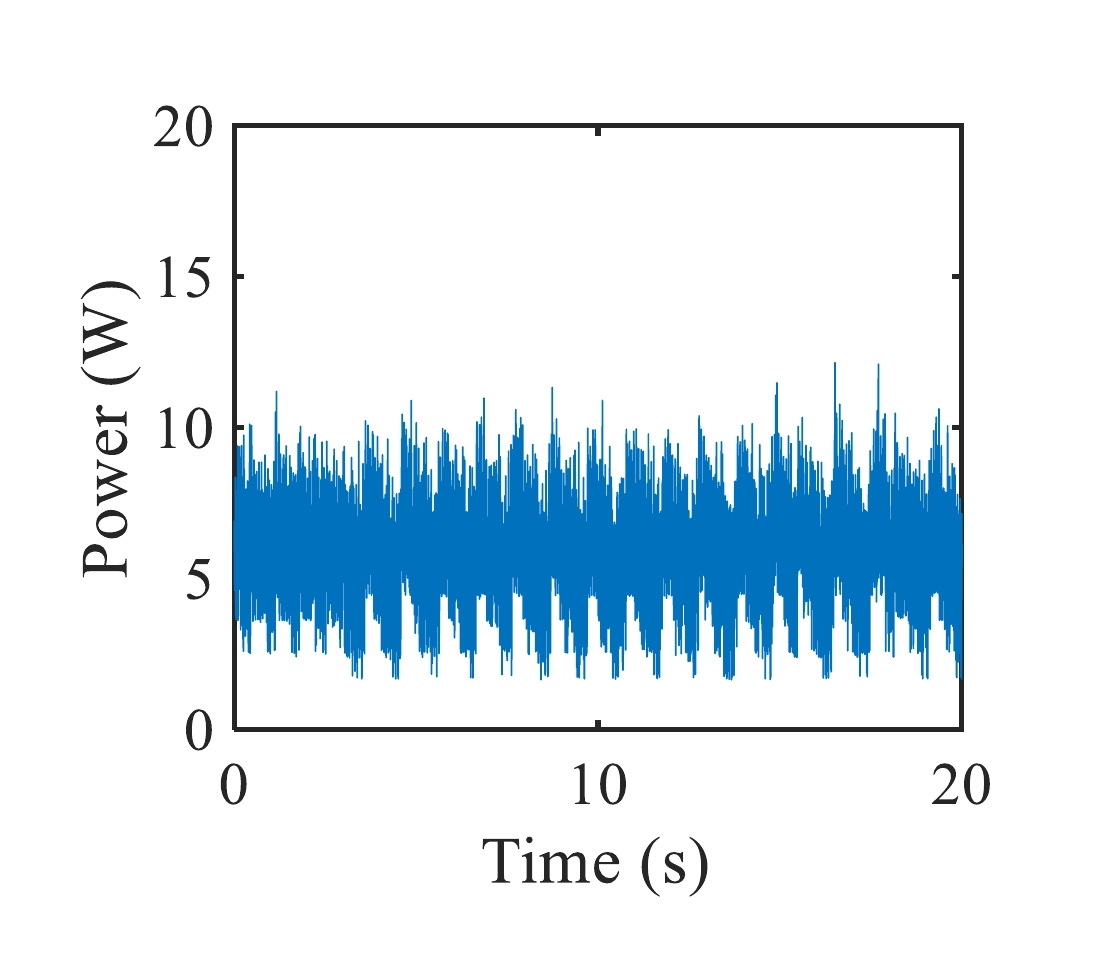}\label{fig:TFLremote_CPU_Governors_Power_Perf}}

\subfigure[Conservative]
{\includegraphics[width=0.16\textwidth]{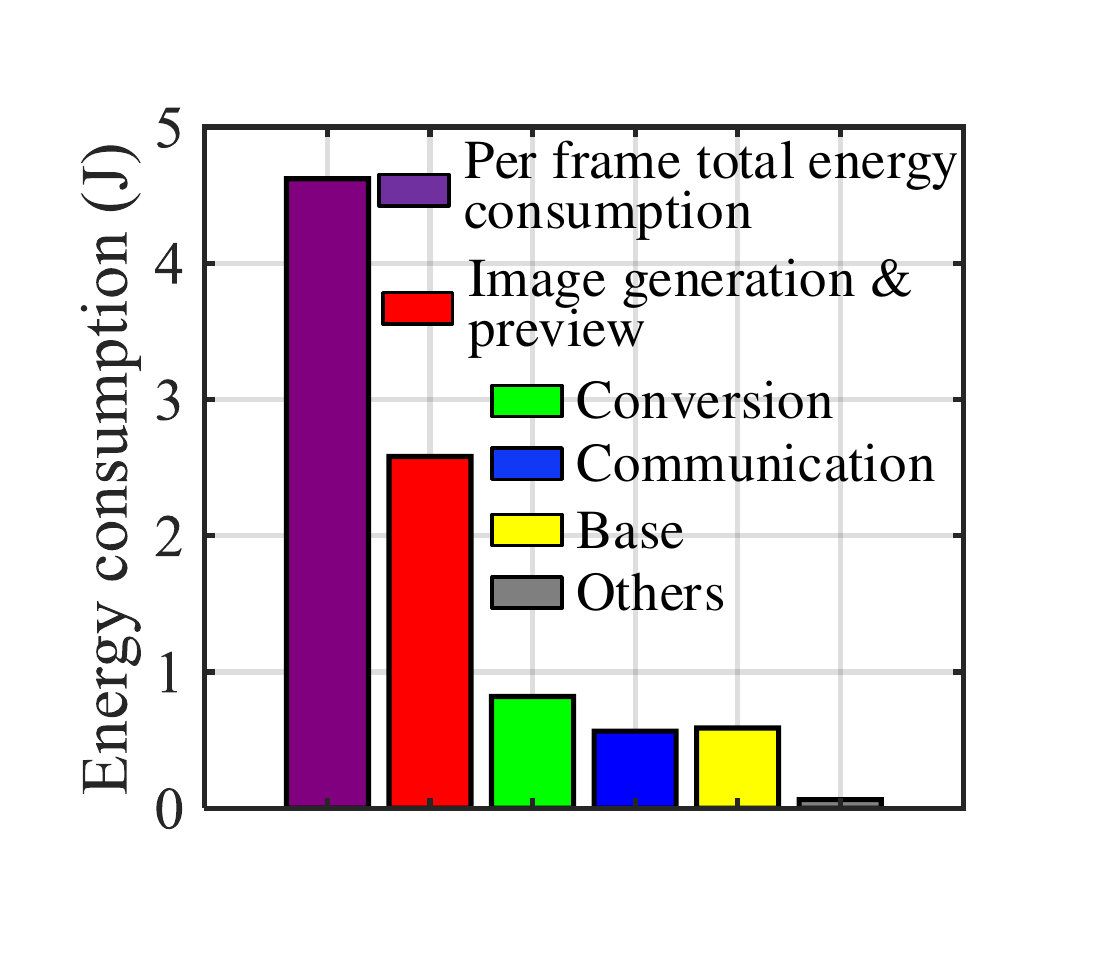}\label{fig:TFLremote_CPU_Governors_Energy_Conser}}
\subfigure[Ondemand]
{\includegraphics[width=0.16\textwidth]{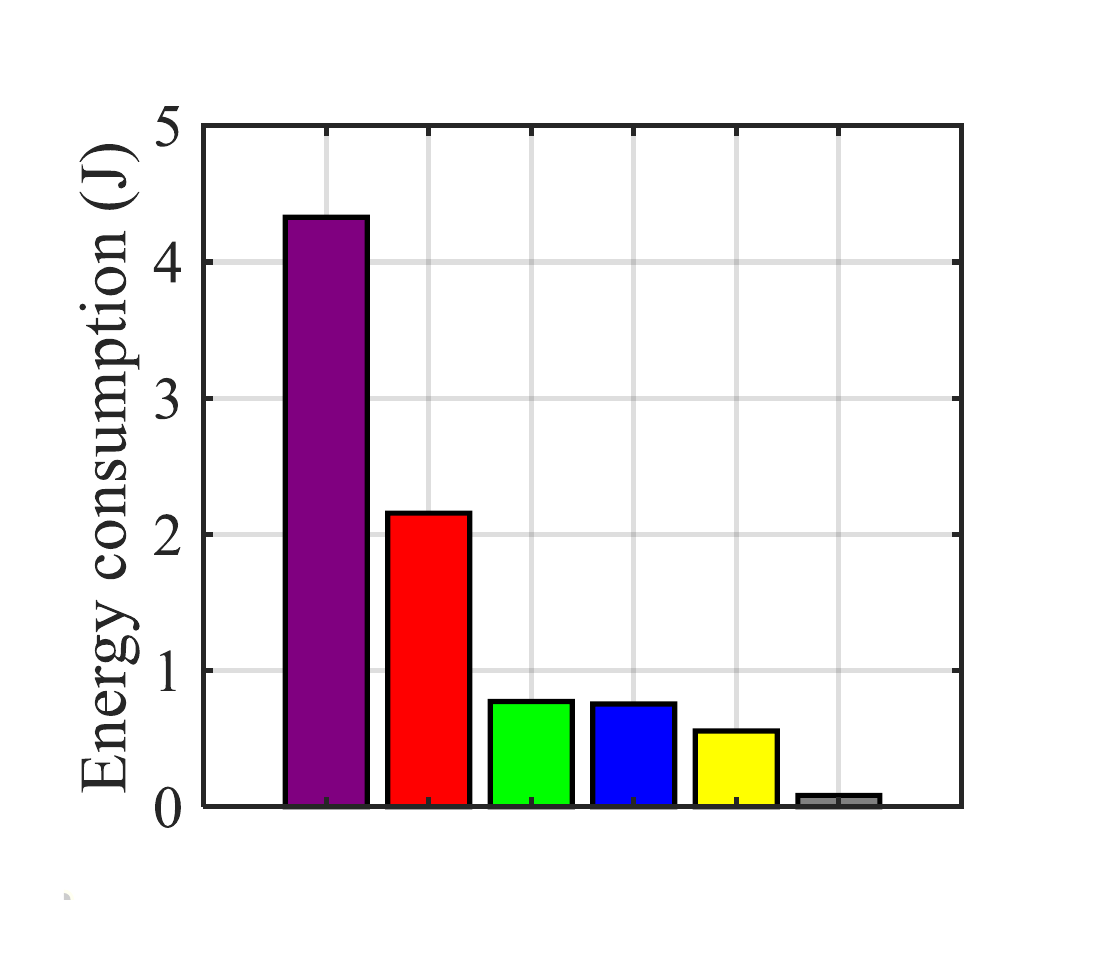}\label{fig:TFLremote_CPU_Governors_Energy_Ond}}
\subfigure[Interactive]
{\includegraphics[width=0.16\textwidth]{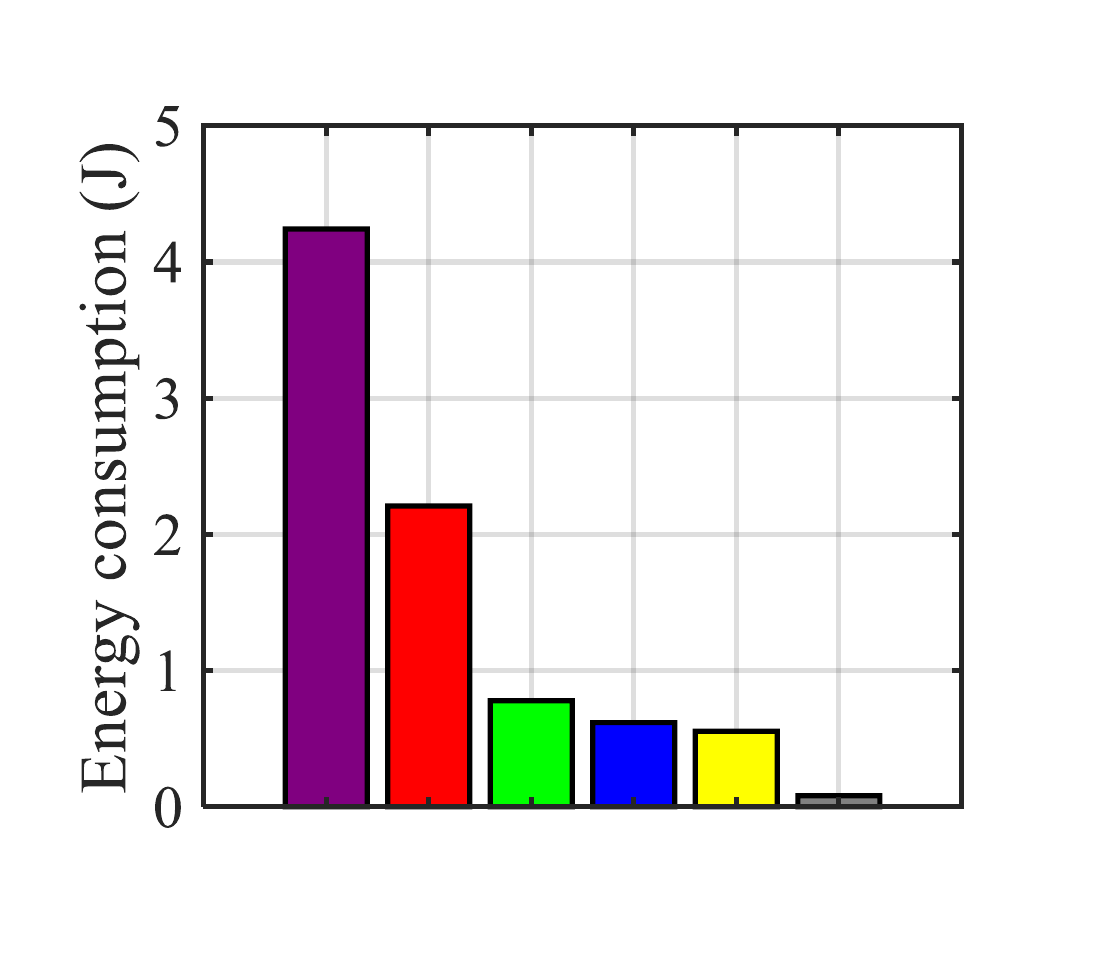}\label{fig:TFLremote_CPU_Governors_Energy_Inter}}
\subfigure[Userspace]
{\includegraphics[width=0.16\textwidth]{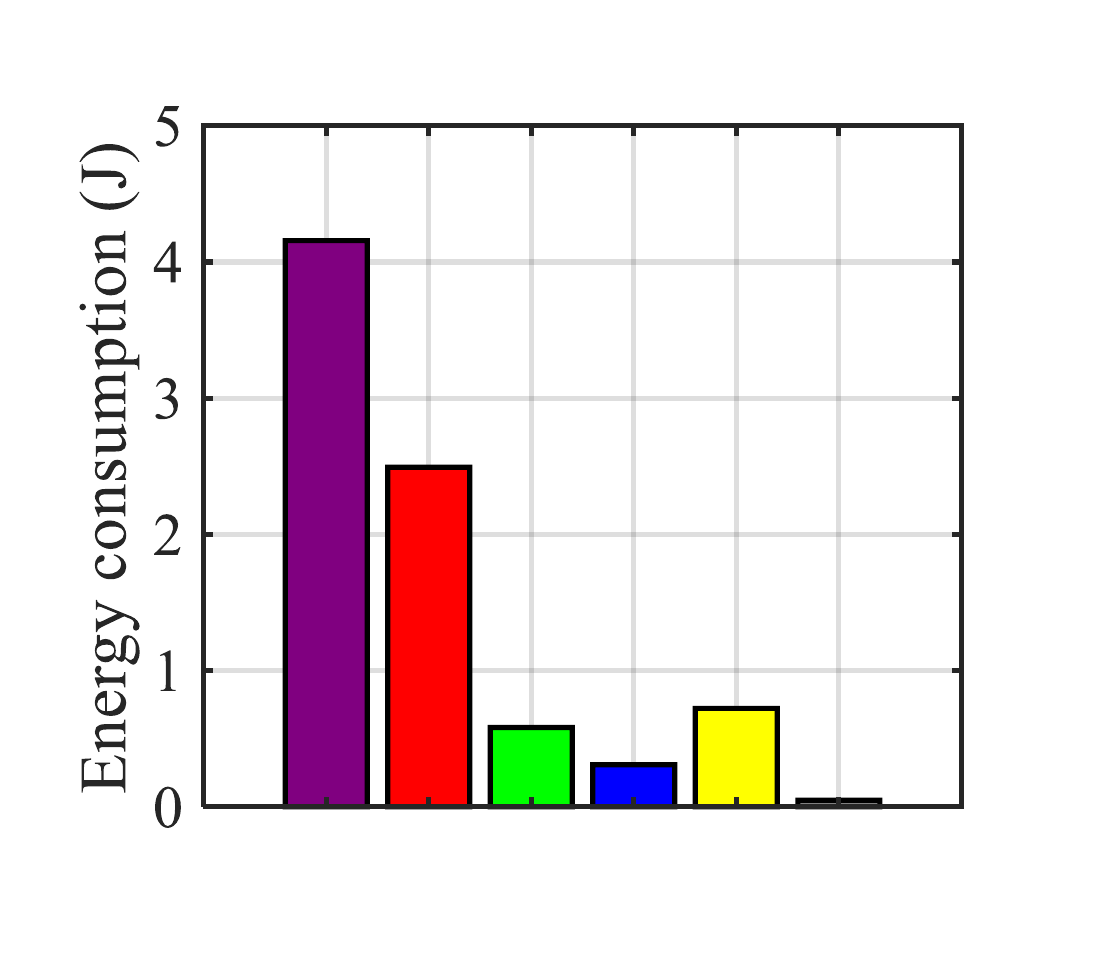}\label{fig:TFLremote_CPU_Governors_Energy_Use}}
\subfigure[Powersave]
{\includegraphics[width=0.16\textwidth]{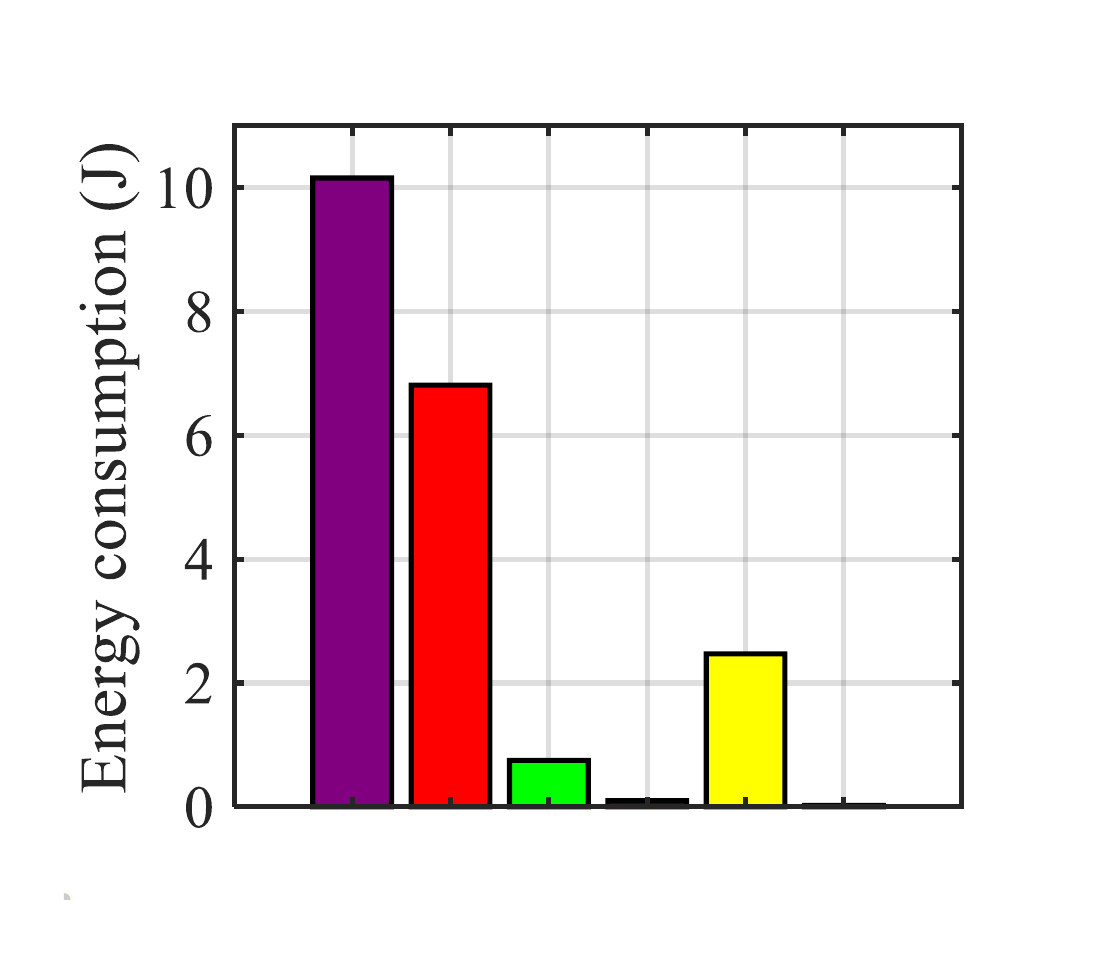}\label{fig:TFLremote_CPU_Governors_Energy_Pow}}
\subfigure[Performance]
{\includegraphics[width=0.16\textwidth]{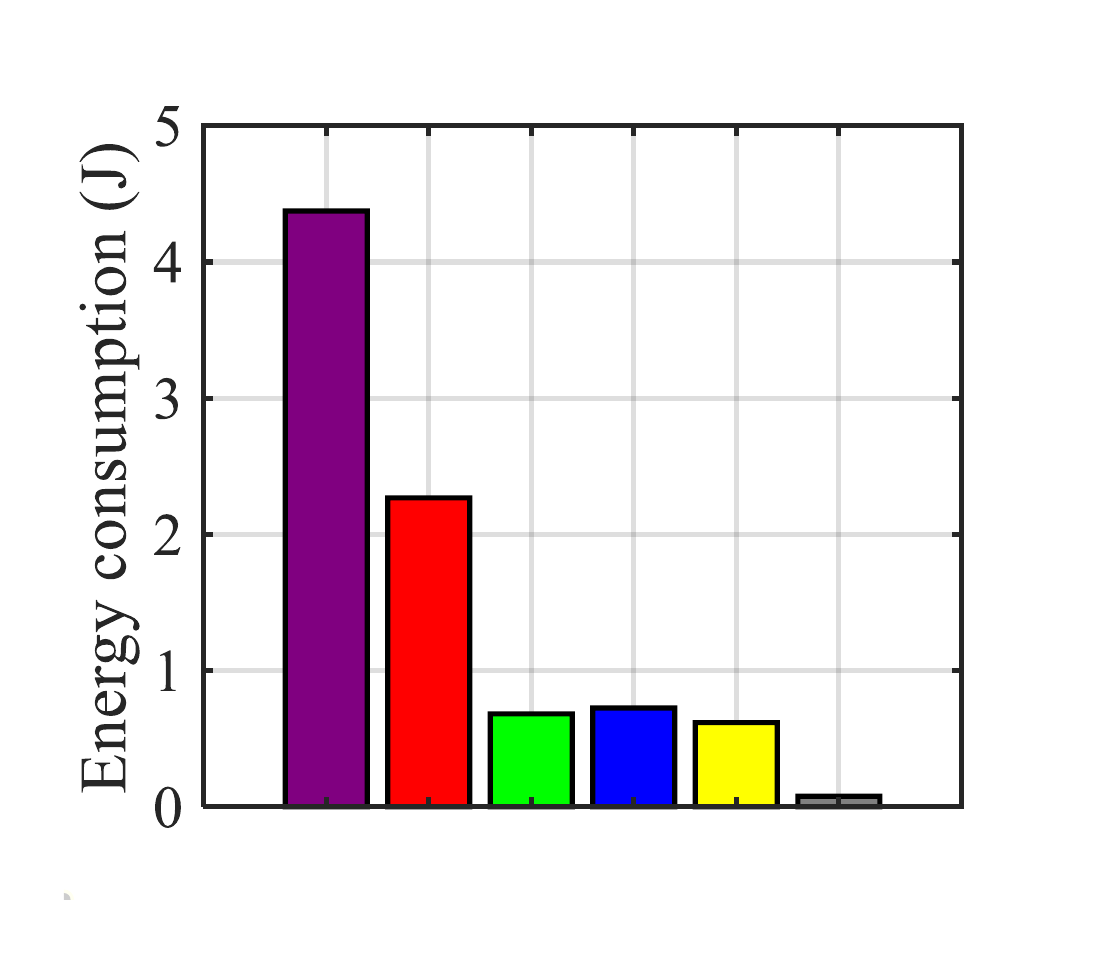}\label{fig:TFLremote_CPU_Governors_Energy_Perf}}

\subfigure[Conservative]
{\includegraphics[width=0.16\textwidth]{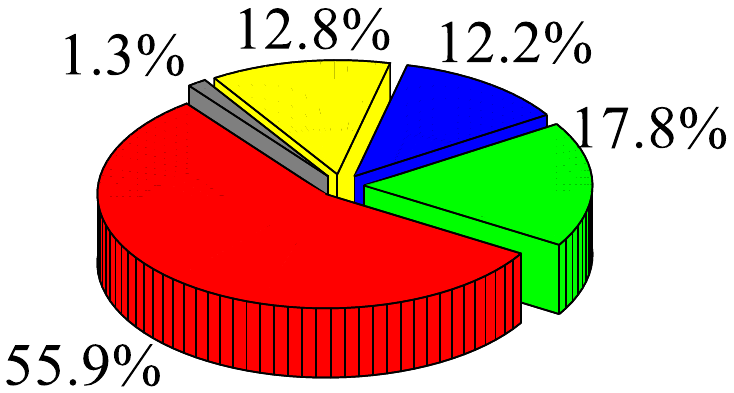}\label{fig:TFLremote_CPU_Governors_Epercent_Conser}}
\subfigure[Ondemand]
{\includegraphics[width=0.16\textwidth]{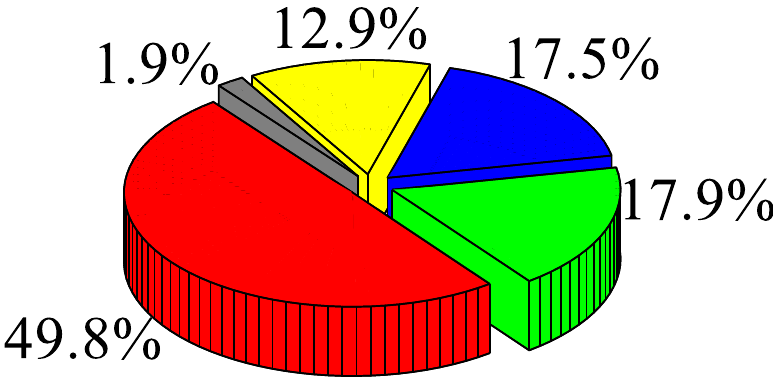}\label{fig:TFLremote_CPU_Governors_Epercent_Ond}}
\subfigure[Interactive]
{\includegraphics[width=0.16\textwidth]{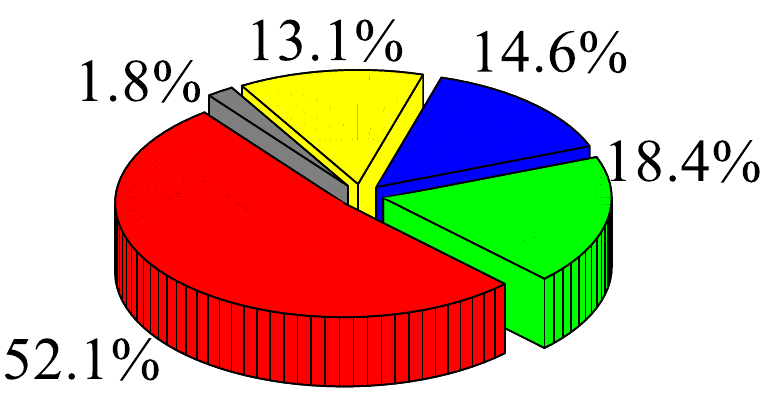}\label{fig:TFLremote_CPU_Governors_Epercent_Inter}}
\subfigure[Userspace]
{\includegraphics[width=0.16\textwidth]{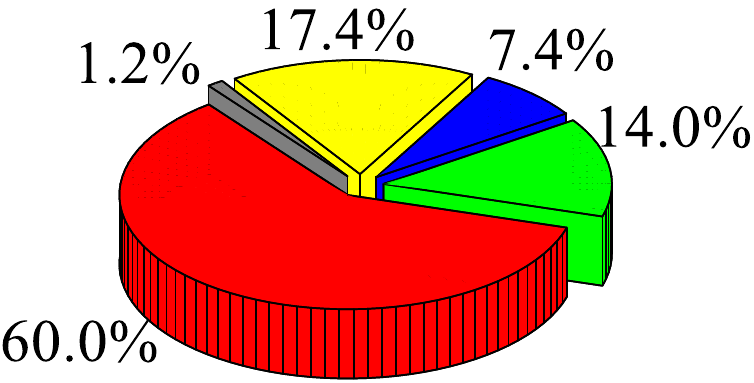}\label{fig:TFLremote_CPU_Governors_Epercent_Use}}
\subfigure[Powersave]
{\includegraphics[width=0.16\textwidth]{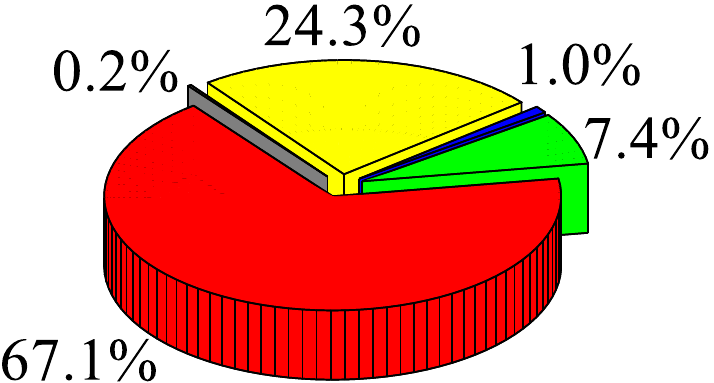}\label{fig:TFLremote_CPU_Governors_Epercent_Pow}}
\subfigure[Performance]
{\includegraphics[width=0.16\textwidth]{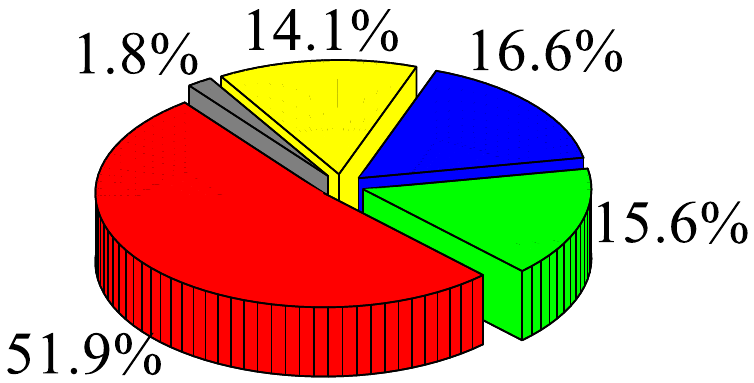}\label{fig:TFLremote_CPU_Governors_Epercent_Perf}}
\caption{CPU governor vs. power and average per frame energy consumption (CNN model size: $320\times320$ pixels).}
\label{fig:cpugovernors_energyremote}   
\end{figure*}

\textbf{Per Frame Energy Consumption.} We next explore how the CPU governor impacts the per frame energy consumption of object detection in the remote execution scenario, where the smartphone works on the interactive CPU governor. The experimental results are shown in Fig. \ref{fig:cpugovernors_energyremote}, where Figs. \ref{fig:TFLremote_CPU_Governors_Power_Conser}-\ref{fig:TFLremote_CPU_Governors_Power_Perf} depict the power consumption; Figs. \ref{fig:TFLremote_CPU_Governors_Energy_Conser}-\ref{fig:TFLremote_CPU_Governors_Energy_Perf} illustrate the average per frame energy consumption; and Figs. \ref{fig:TFLremote_CPU_Governors_Epercent_Conser}-\ref{fig:TFLremote_CPU_Governors_Epercent_Perf} show the average percentage breakdown of energy consumed by each phase in the processing pipeline. We find that (19) the remote execution decreases the power consumption compared to the local execution when the smartphone works on conservative, ondemand, interactive, and performance CPU governors. However, when the smartphone works on userspace and powersave CPU governors, the remote execution consumes more power than the local execution, as shown in Table \ref{tb:remote_cpu_energy}. This observation is a supplement to observation (16), which indicates that \textit{(i) offloading the object detection tasks to the edge server may not be able to reduce the workload on the smartphone when the smartphone's CPUs run at a low frequency; (ii) the communication phase (i.e., remote execution) is more power-consuming than the inference phase (i.e., local execution) when the CPU frequency is low.} (20) As depicted in Figs. \ref{fig:TFLremote_CPU_Governors_Energy_Conser}-\ref{fig:TFLremote_CPU_Governors_Energy_Perf} and Table \ref{tb:remote_cpu_energy}, the remote execution is capable of reducing the per frame energy consumption compared to the local execution when the smartphone works on these six tested CPU governors. In addition, observation (18) and its corresponding inference are also applicable for the per frame energy consumption.

\begin{table*}[t]
 \begin{center}
    \caption{Per frame energy consumption results of the remote execution with different CPU governors.}
    \label{tb:remote_cpu_energy}
  \begin{tabular}{|l||c|c|c|c|c|c|}
    \hline
    CPU Governor                                  & Conservative & Ondemand & Interactive & Userspace & Powersave & Performance \\ \hline
    Power Consumption (watt)  & 4.579 & 5.074 & 4.728 & 3.867 & \textbf{2.410} & 5.341\\ \hline
    Per Frame Energy Consumption (Joule)          & 4.620 & 4.328 & 4.278 & \textbf{4.157} & 10.156 & 4.375\\ \hline
    Power Consumption Reduction ($\%$)            & \textbf{14.5}  & 11.4  & 12.7  &  -1.4 &  -4.4  &  12.7    \\ \hline
    Per Frame Energy Consumption Reduction ($\%$) &  12.6 &  16.4   &  22.0  &  24.5  &  \textbf{43.4}   &  13.1   \\ \hline
  \end{tabular}
  \end{center}
\end{table*}

\begin{figure*}[t!]
\centering
\subfigure[CPU frequency vs. per frame latency]
{\includegraphics[width=0.3\textwidth]{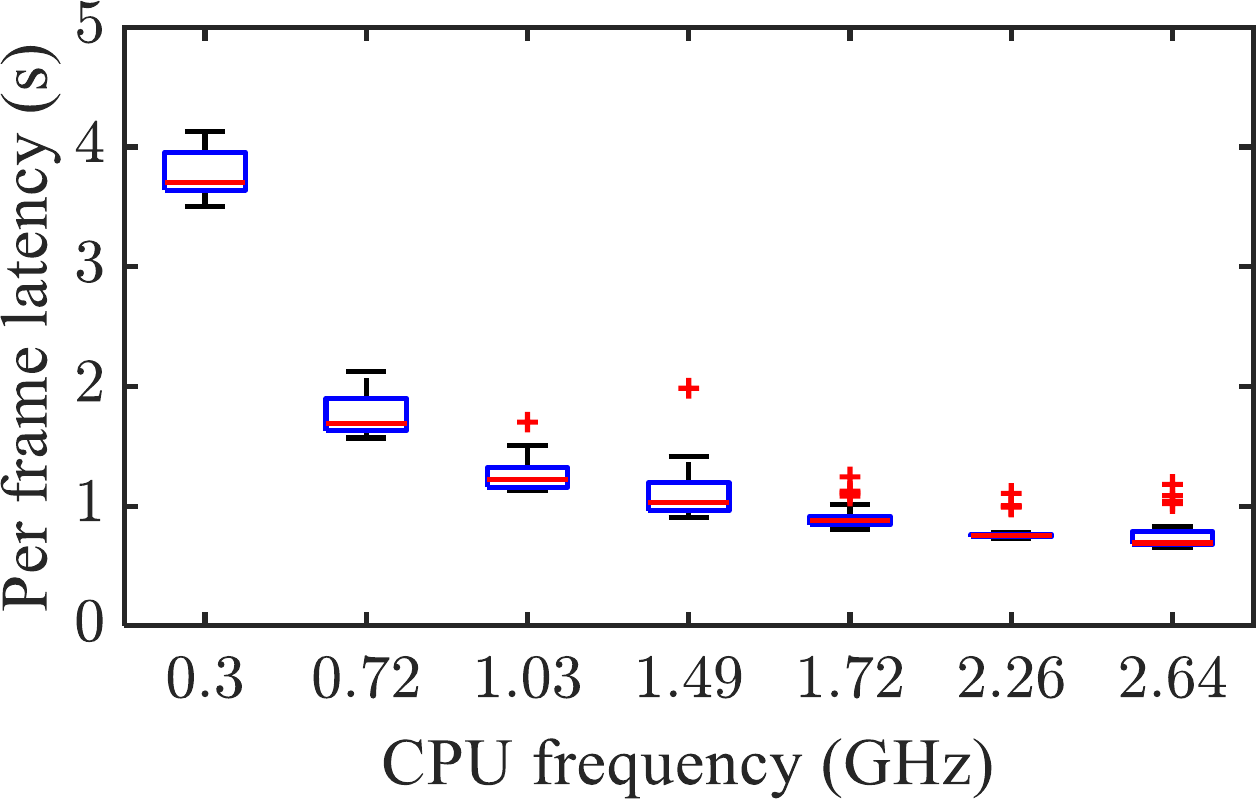}\label{fig:TFLremote_CPU_Frequency_latency}}
\subfigure[CPU frequency vs. power consumption]
{\includegraphics[width=0.3\textwidth]{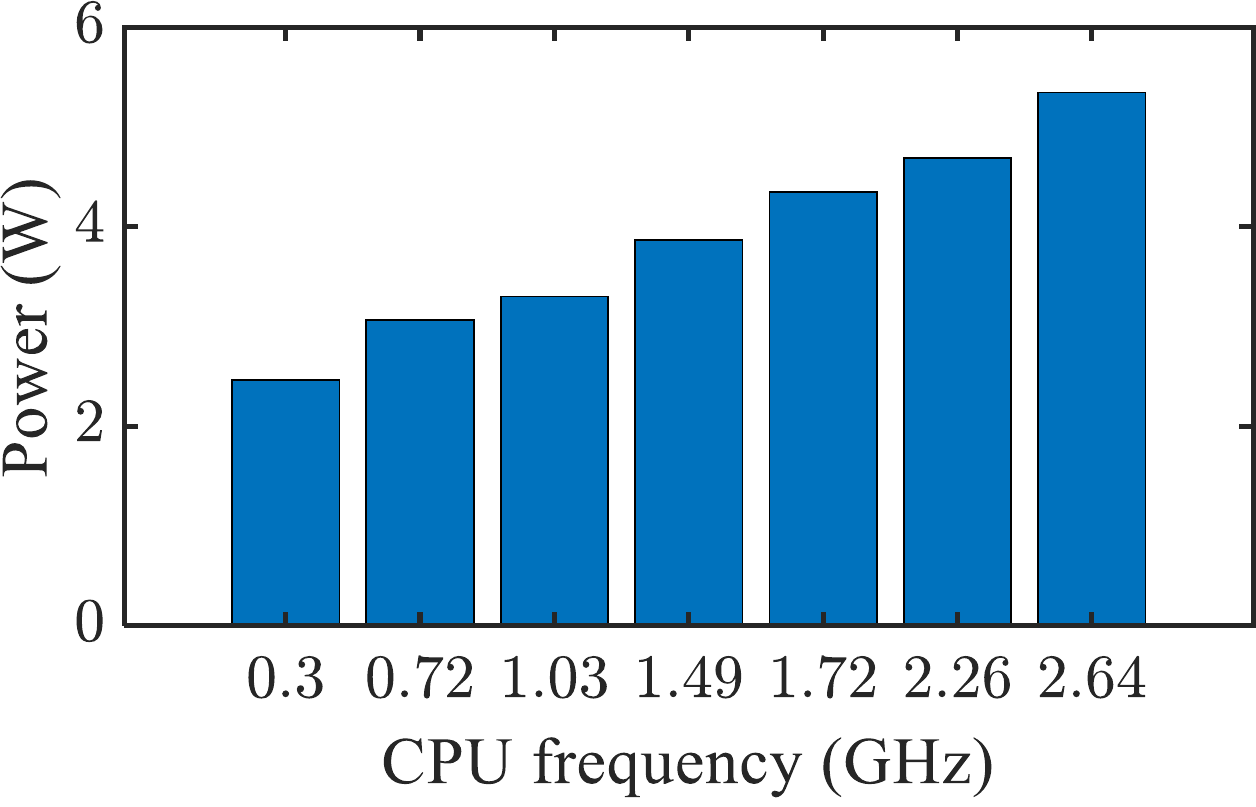}\label{fig:TFLremote_CPU_Frequency_power}}
\subfigure[CPU frequency vs. per frame energy consumption]
{\includegraphics[width=0.3\textwidth]{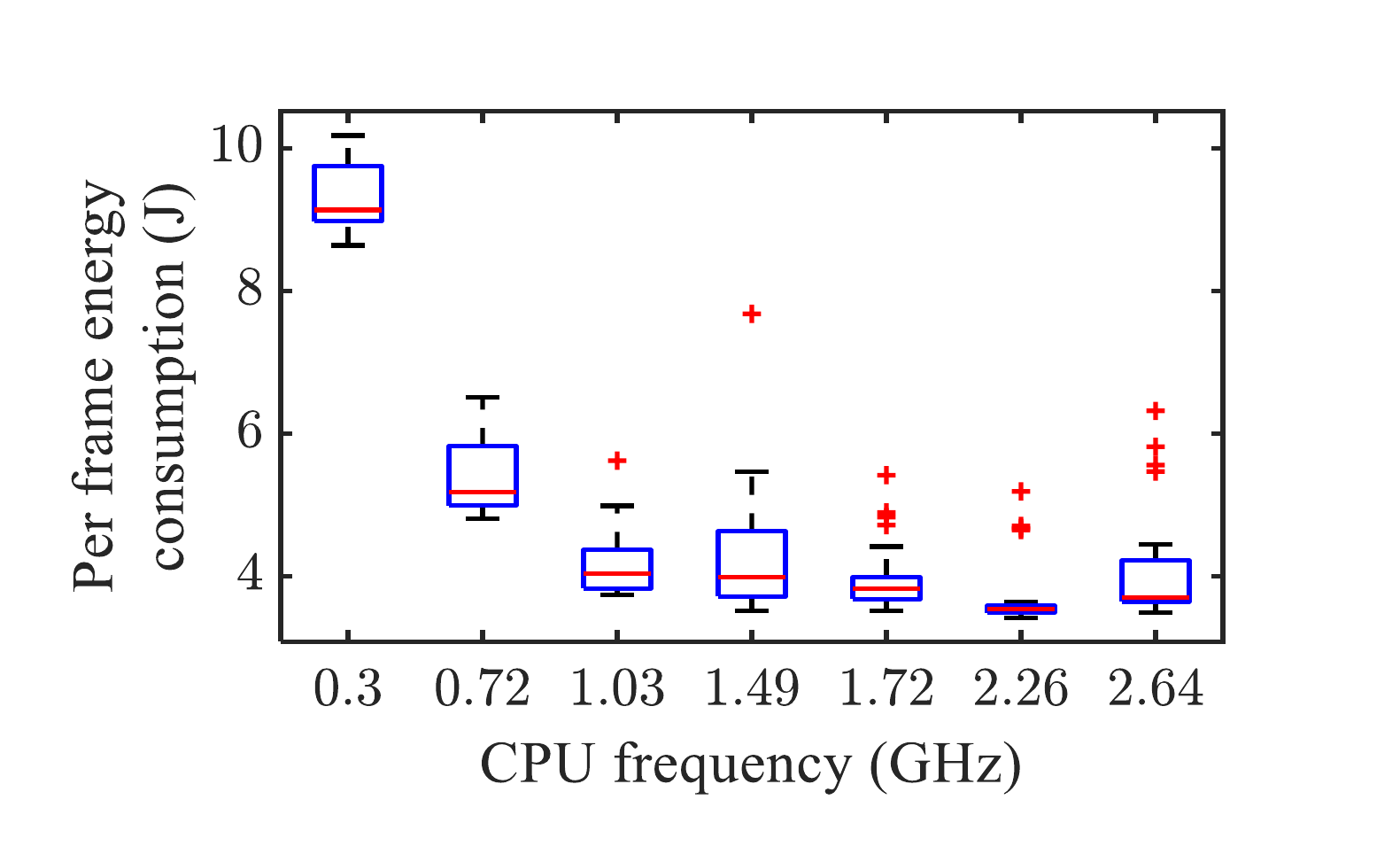}\label{fig:TFLremote_CPU_Frequency_energy}}
\caption{Performance variations with increasing the CPU frequency in remote execution (CNN model size: $320\times320$ pixels).}
\label{fig:cpufrequency_remote}   
\end{figure*}

Interestingly, (21) the userspace governor (i.e., the CPU frequency is set to $1.49$GHz) achieves the lowest per frame energy consumption in the remote execution, as illustrated in Figs. \ref{fig:TFLremote_CPU_Governors_Energy_Conser}-\ref{fig:TFLremote_CPU_Governors_Energy_Perf} and Table \ref{tb:remote_cpu_energy}. This observation is different from the local execution, where the CPU with the highest frequency achieves the lowest per frame energy consumption. We conduct an experiment study to explore the reason, where we set the test smartphone to the userspace governor and gradually raise its CPU frequency from the lowest to the highest. The experimental results are shown in Fig. \ref{fig:cpufrequency_remote}. We find that (22) the higher the CPU frequency, the lower per frame latency the smartphone derives and the higher power it consumes. However, the reduction of the per frame latency and the increase of the power consumption are disproportional, as depicted in Figs. \ref{fig:TFLremote_CPU_Frequency_latency} and \ref{fig:TFLremote_CPU_Frequency_power}. For example, as compared to $2.26$GHz, $2.64$GHz only reduces about $5\%$ latency but increases about $14\%$ power consumption. As compared to $0.3$GHz, $0.72$GHz reduces about $55\%$ latency but only increases about $24\%$ power consumption. \textit{This observation advocates adapting the smartphone’s CPU frequency for the per frame latency reduction by trading as little increase of the per frame energy consumption as possible.} For example, Fig. \ref{fig:TFLremote_CPU_Frequency_energy} illustrates that selecting the CPU frequency around $2.26$GHz achieves the lowest per frame energy consumption, a comparable per frame latency, and a lower power consumption compared to $2.64$GHz.


\subsection{The Impact of CNN Model Size}
\label{ssc:cnnsize_remote}

\begin{figure*}[t]
\centering
\subfigure[$128^2$ pixels]
{\includegraphics[width=0.16\textwidth]{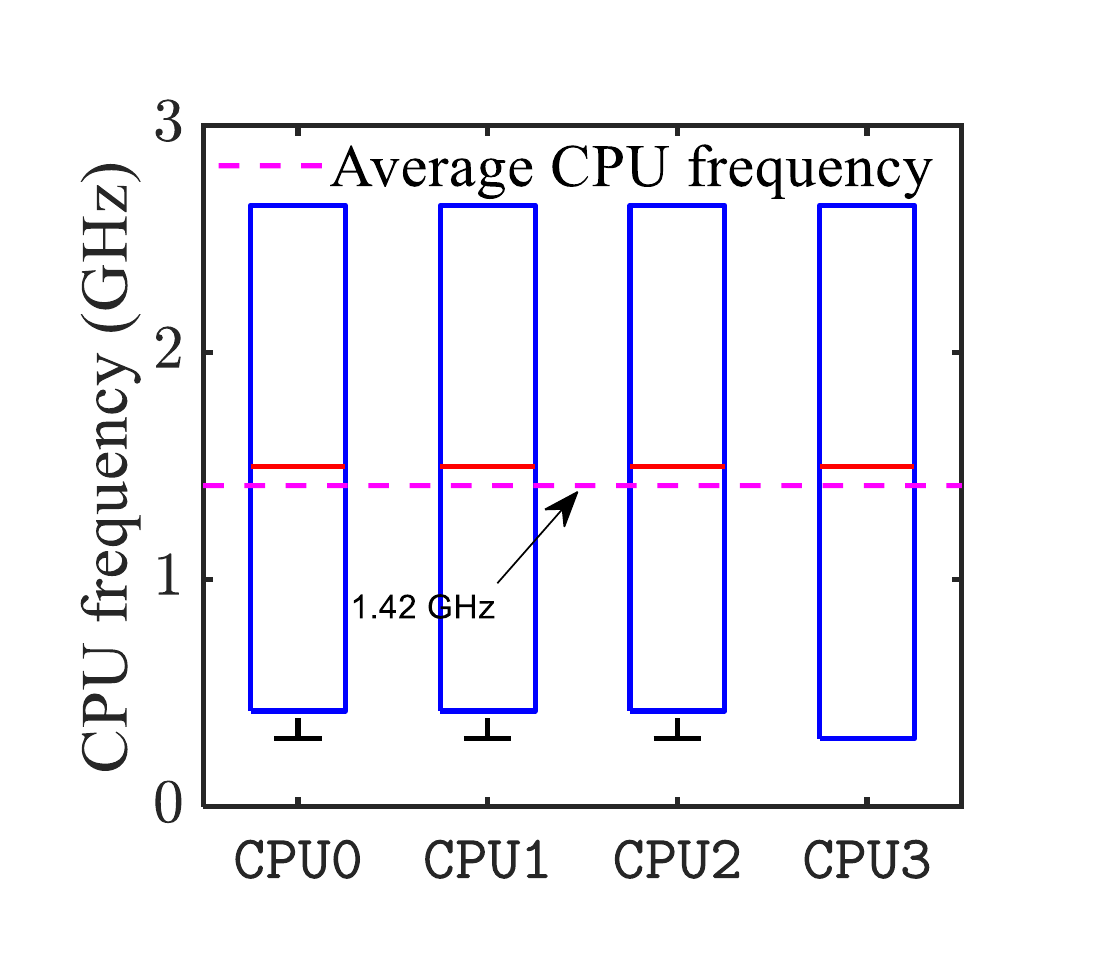}\label{fig:128InterCPUremote}}
\subfigure[$224^2$ pixels]
{\includegraphics[width=0.16\textwidth]{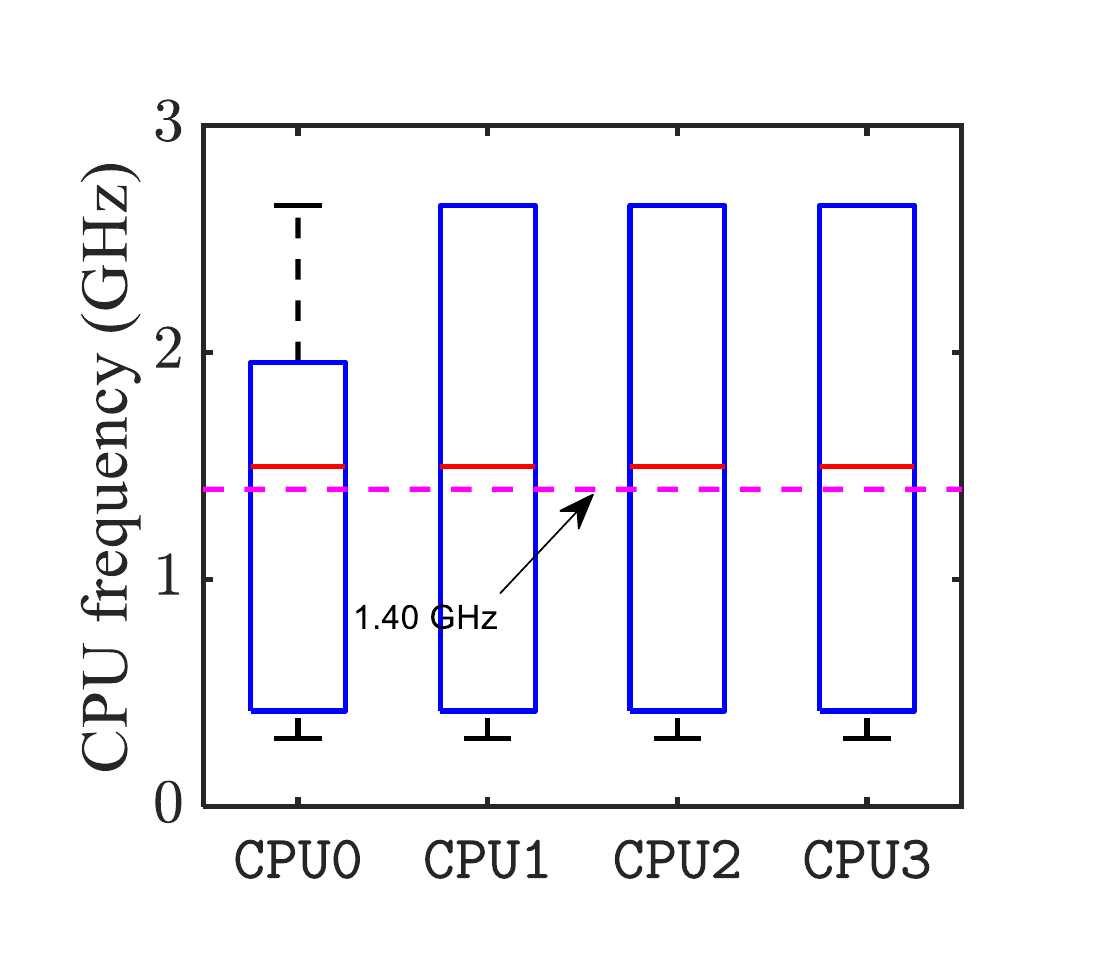}\label{fig:224InterCPUremote}}
\subfigure[$320^2$ pixels]
{\includegraphics[width=0.16\textwidth]{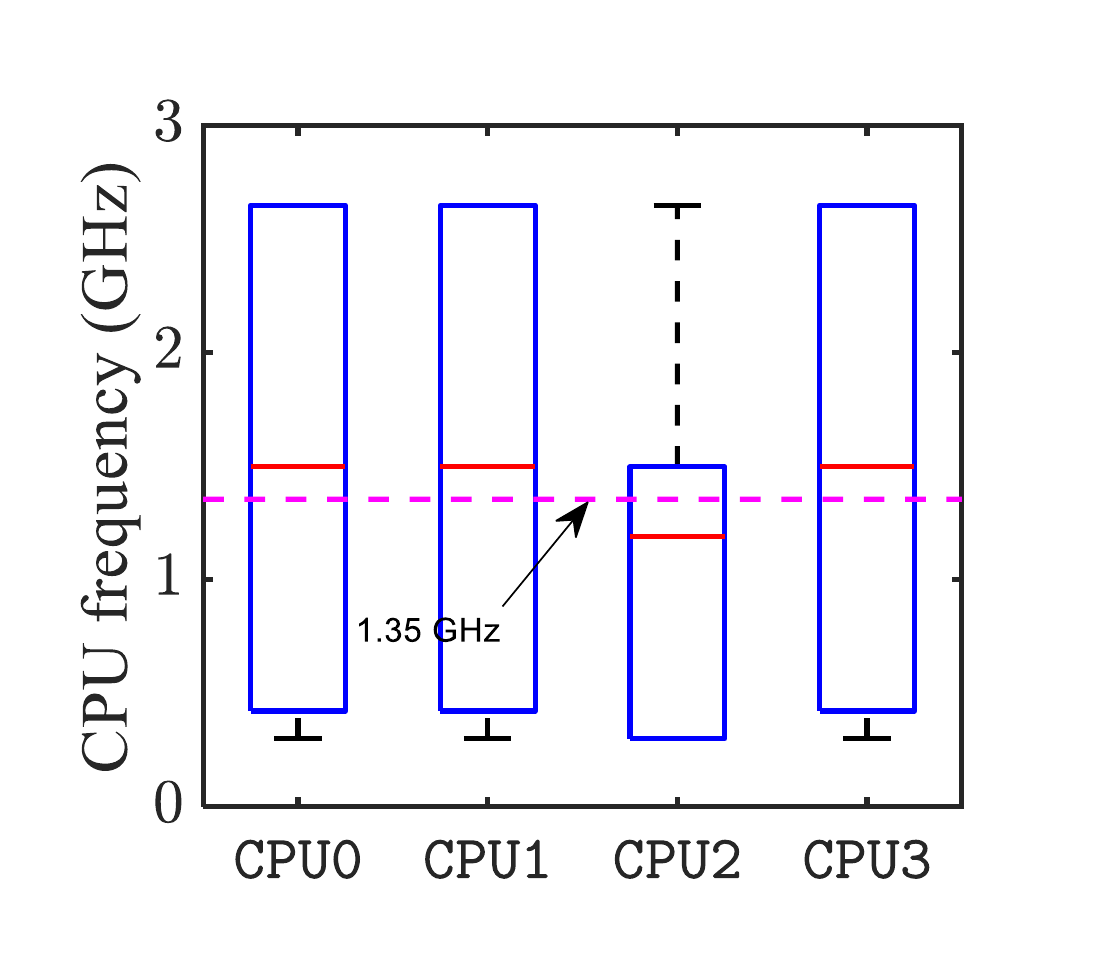}\label{fig:320InterCPUremote}}
\subfigure[$416^2$ pixels]
{\includegraphics[width=0.16\textwidth]{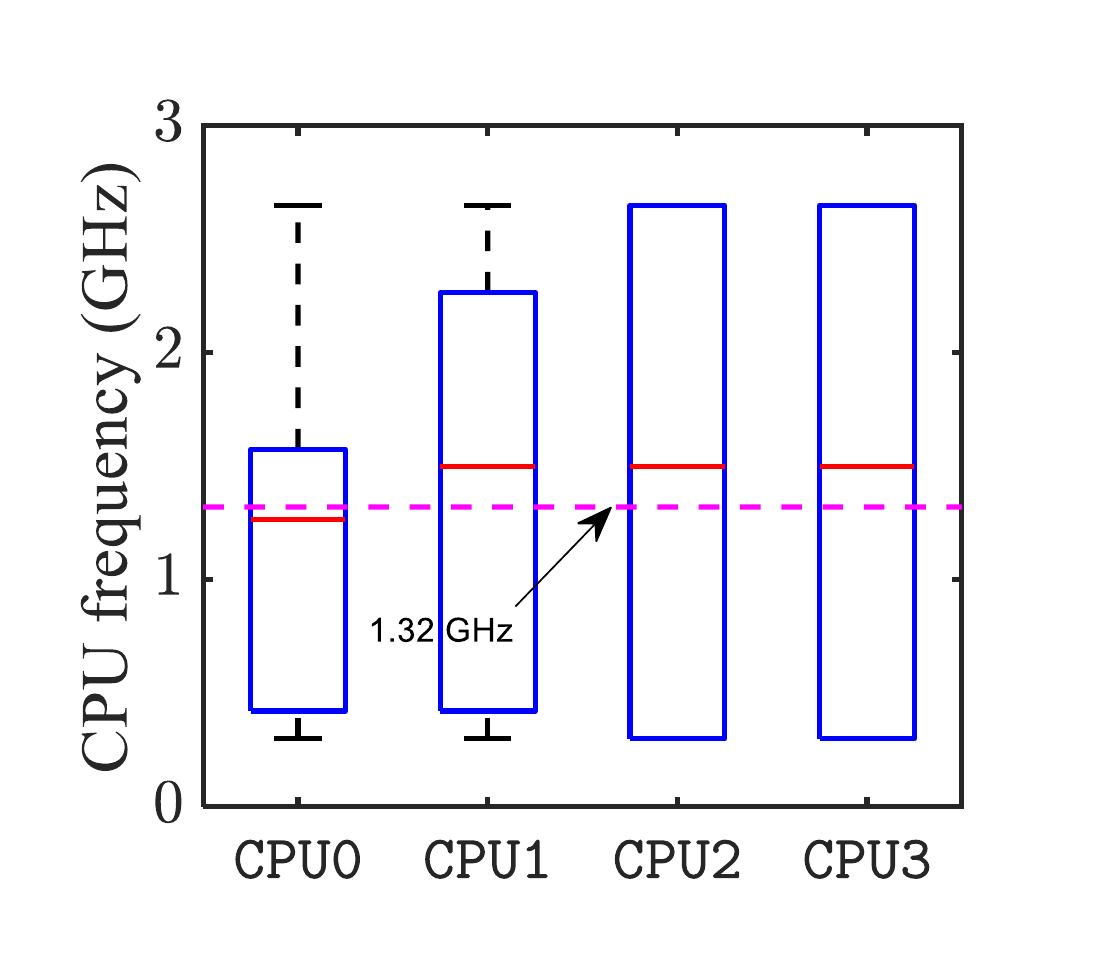}\label{fig:416InterCPUremote}}
\subfigure[$512^2$ pixels]
{\includegraphics[width=0.16\textwidth]{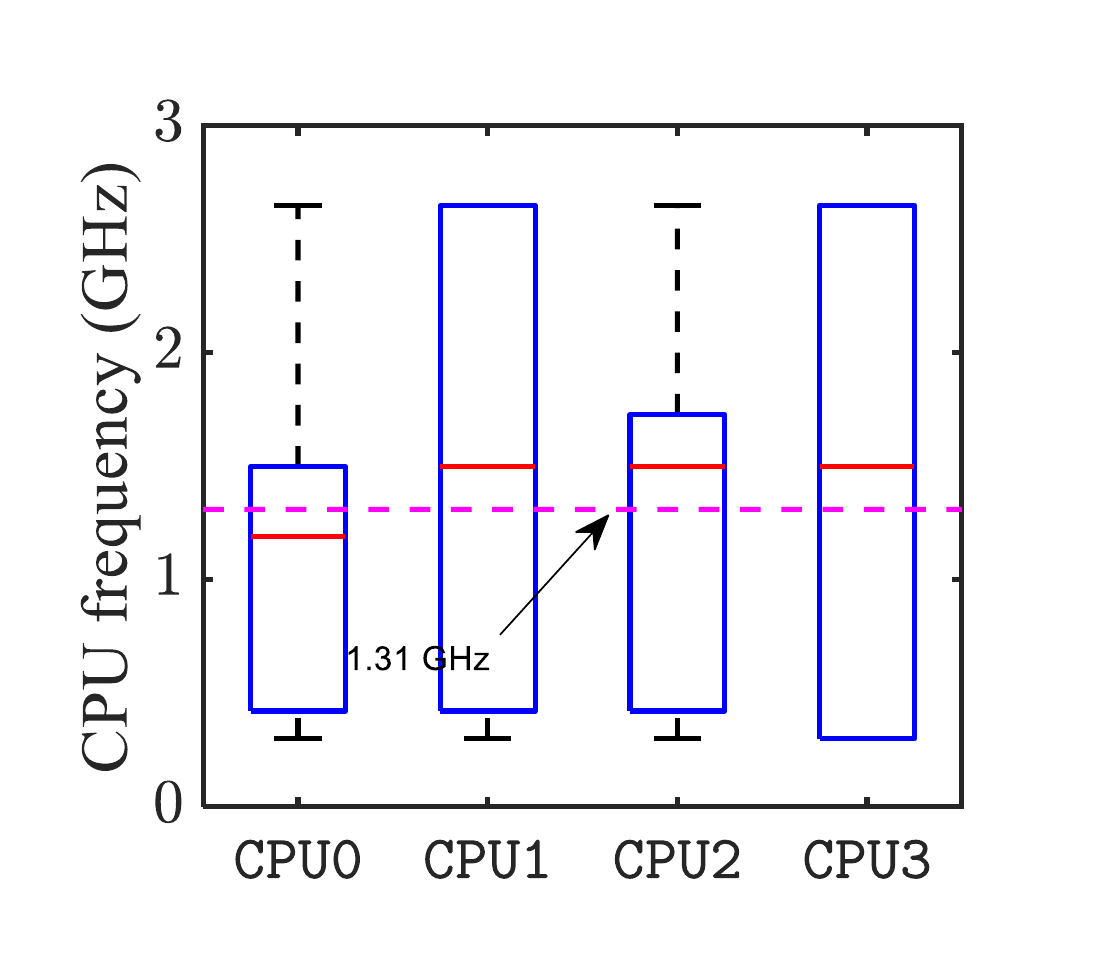}\label{fig:512InterCPUremote}}
\subfigure[$608^2$ pixels]
{\includegraphics[width=0.16\textwidth]{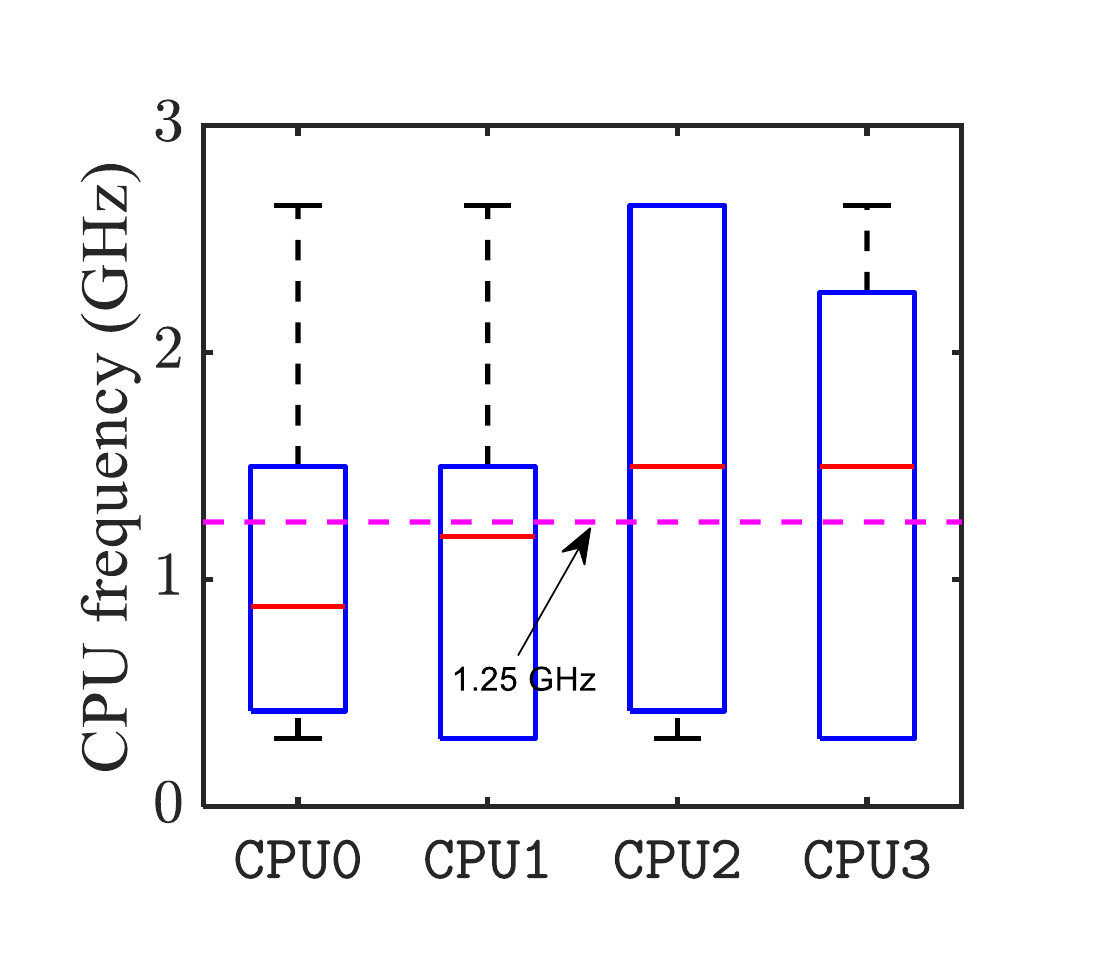}\label{fig:608InterCPUremote}}

\subfigure[Per frame latency]
{\includegraphics[width=0.195\textwidth]{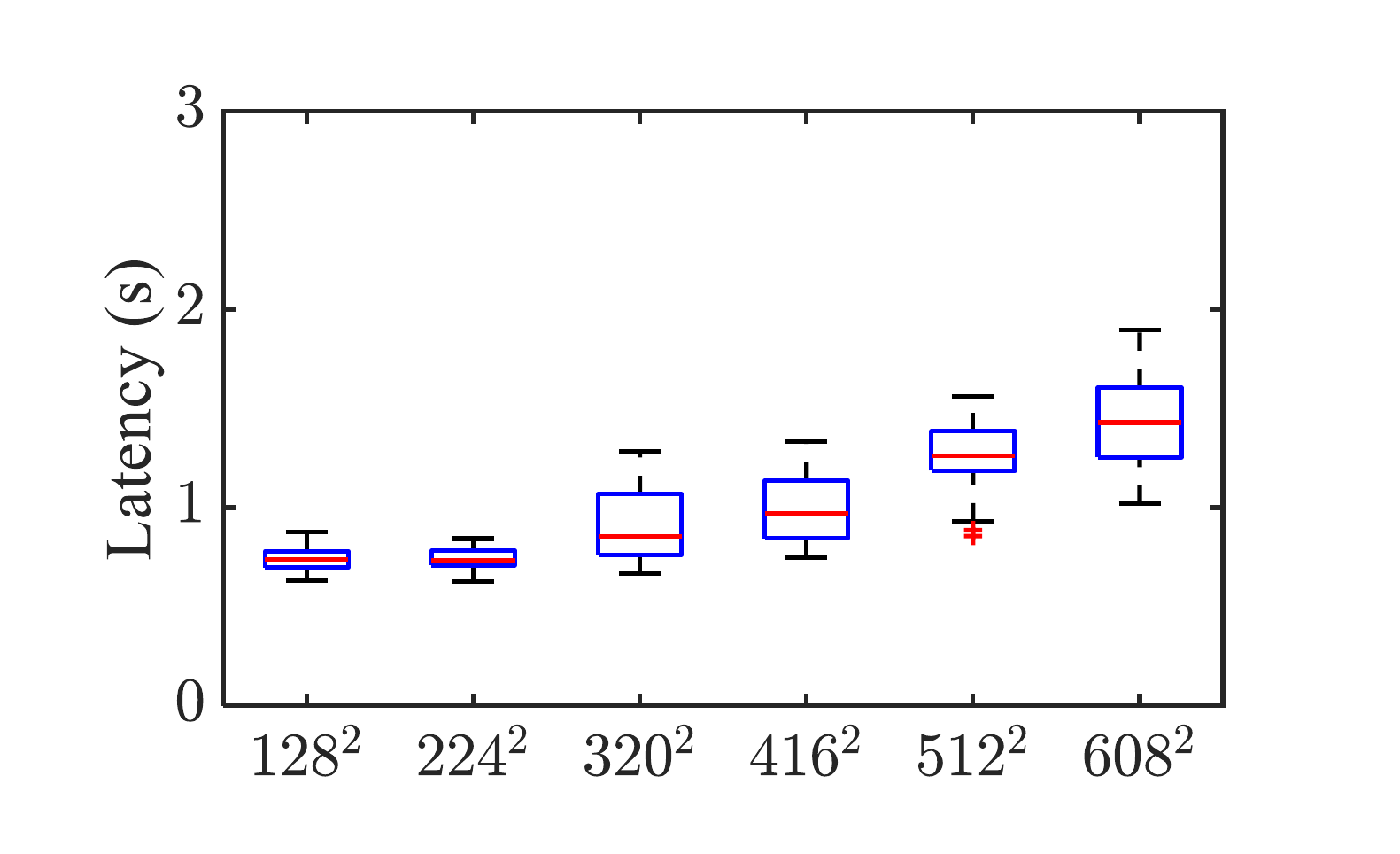}\label{fig:Inter_alllatency_remote}}
\subfigure[Image conversion latency]
{\includegraphics[width=0.195\textwidth]{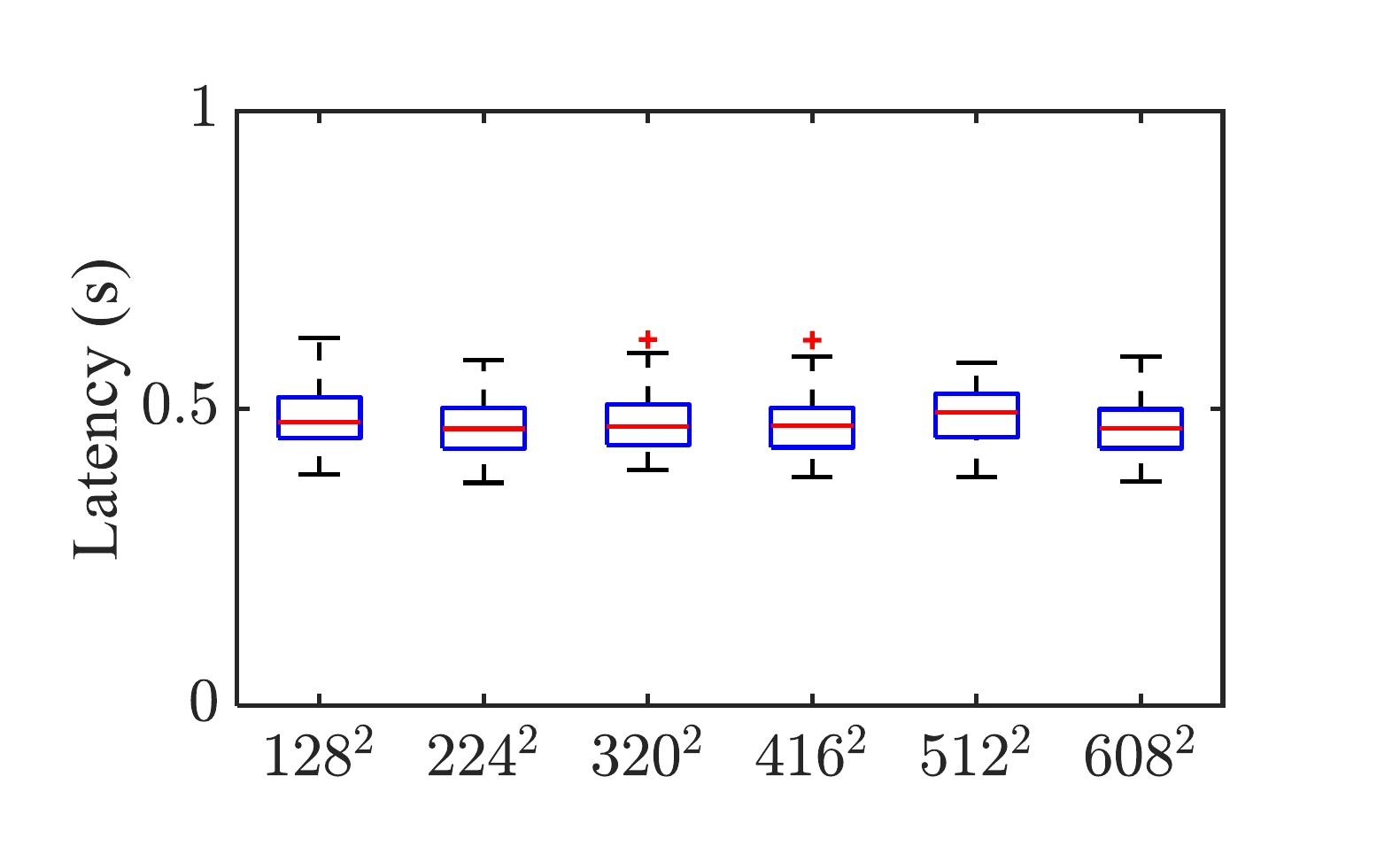}\label{fig:Inter_convlatency_remote}}
\subfigure[Communication latency]
{\includegraphics[width=0.195\textwidth]{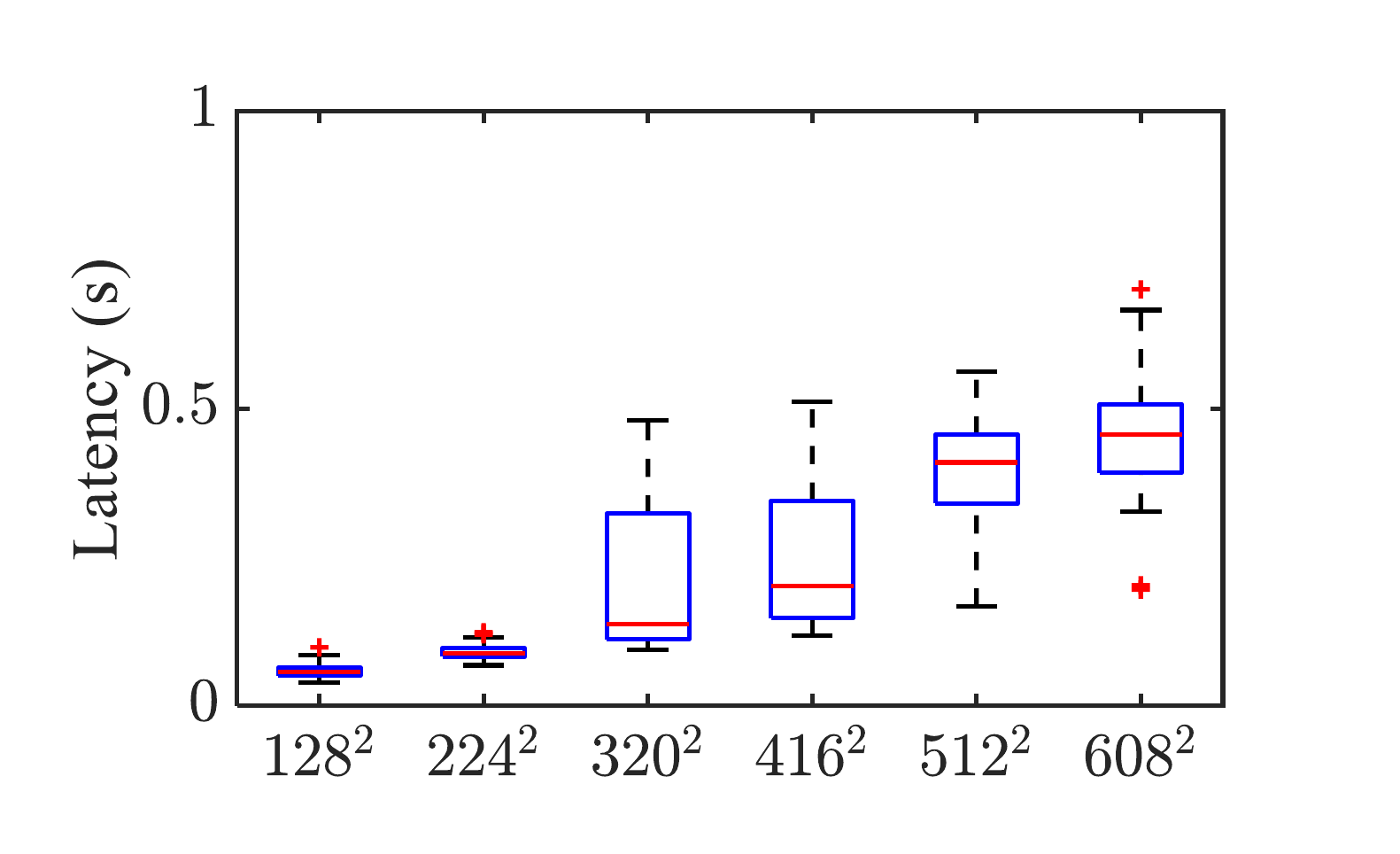}\label{fig:Inter_commlatency_remote}}
\subfigure[Inference latency]
{\includegraphics[width=0.195\textwidth]{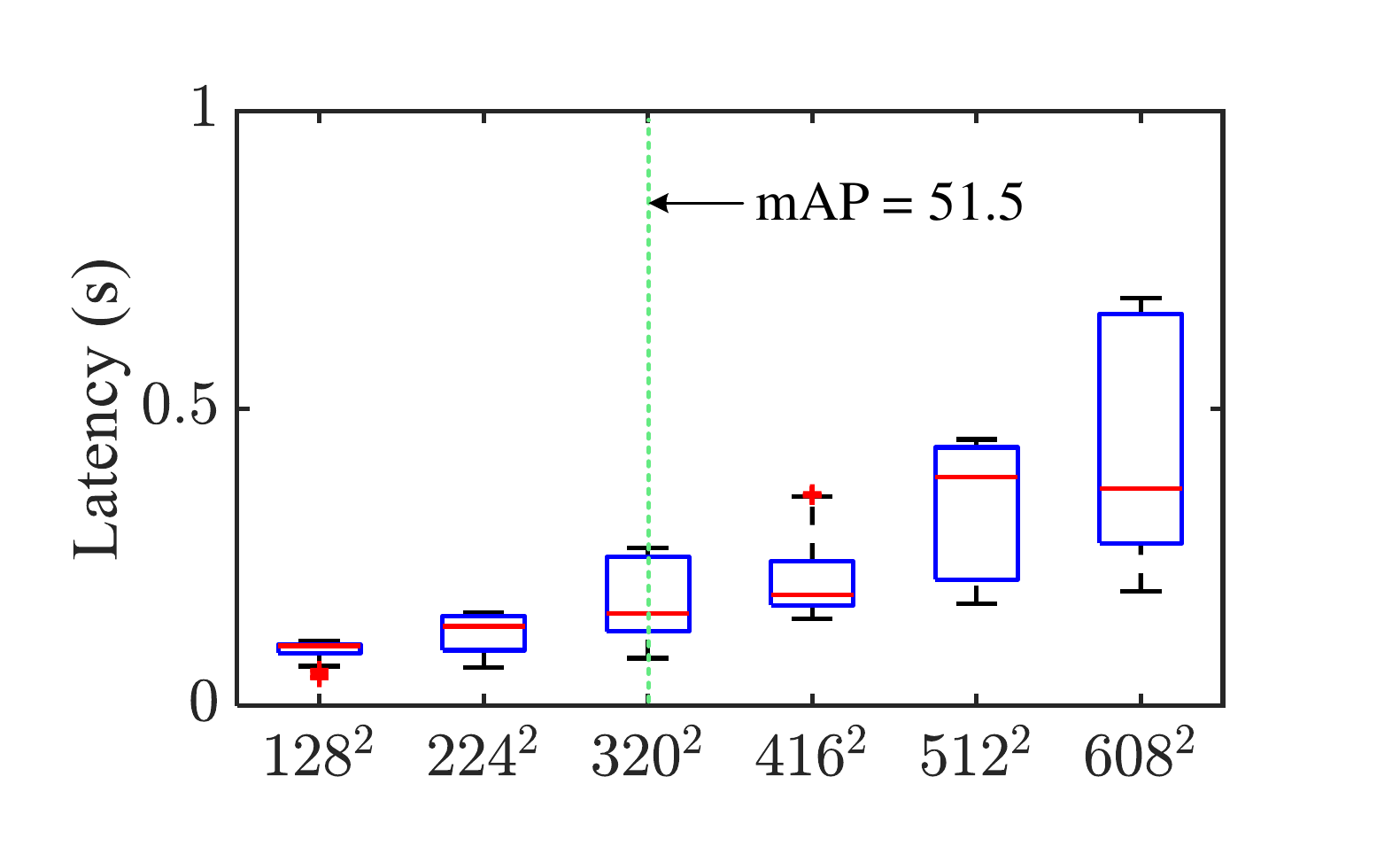}\label{fig:Inter_inflatency_remote}}
\subfigure[Detection accuracy]
{\includegraphics[width=0.195\textwidth]{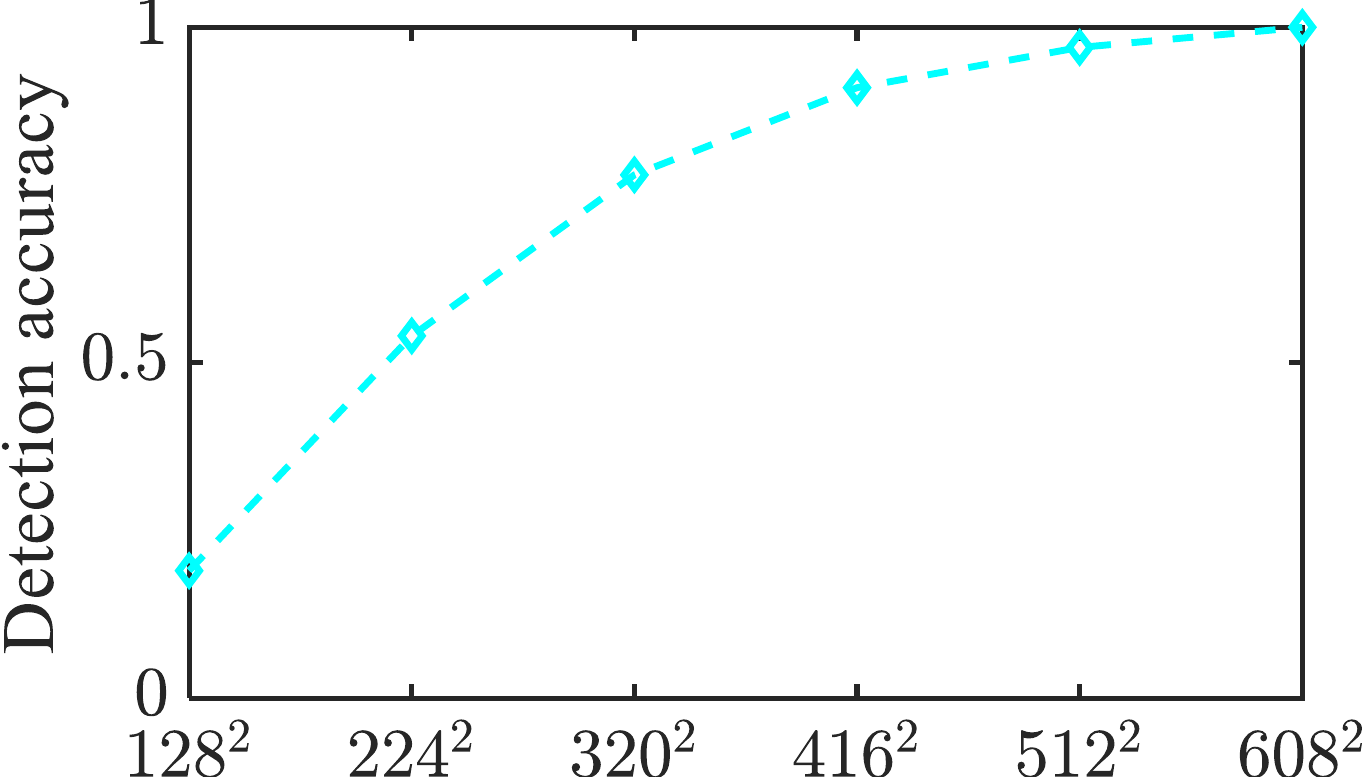}\label{fig:Inter_accuracy_remote}}
\caption{CNN model size vs. CPU frequency, latency, and detection accuracy (CPU governor: interactive).}
\label{fig:cnn_latency_remote_inter}   
\end{figure*}

\textbf{Per Frame Latency.} In this experiment, we implement six object detection models based on the YOLOv3 framework \cite{yolov3} with different CNN model sizes (i.e., from $128\times 128$ to $608 \times 608$ pixels). The test smartphone works on the default CPU governor, interactive. Fig. \ref{fig:cnn_latency_remote_inter} depicts the per frame latency of running CNN-based object detection with different model sizes in the remote execution. We make the following observations. (23) In contrary to the local execution, raising the CNN model size in the remote execution decreases the average CPU frequency, as shown in Figs. \ref{fig:128InterCPUremote}-\ref{fig:608InterCPUremote}. This is because the smartphone experiences a relatively long idle period (i.e., waiting for the detection results from the edge server) when the CNN model size is large (i.e., a long inference latency at the edge server side). In WiFi networks, when transmitting a single image frame, the smartphone’s wireless interface experiences four phases: promotion, data transmission, tail, and idle. When an image transmission request comes, the wireless interface enters the promotion phase. Then, it enters the data transmission phase to send the image frame to the edge server. After completing the transmission, the wireless interface is forced to stay in the tail phase for a fixed duration and waits for other data transmission requests and the detection results. If the smartphone does not receive the detection result in the tail phase, it enters the idle phase and waits for the feedback from its associated edge server. \textit{Therefore, in contrary to the local execution, using a large CNN model size in the remote execution can extend the battery life and improve the detection accuracy.} (24) Similar to the local execution, a larger CNN model size always results in a higher per frame latency in the remote execution, where the per frame latency increment is mainly from the raise of the communication and the inference latency, as shown in Figs. \ref{fig:Inter_alllatency_remote}-\ref{fig:Inter_inflatency_remote}. In addition, Fig. \ref{fig:Inter_accuracy_remote} depicts the detection accuracy of the YOLO under different CNN model sizes, where the detection accuracy is defined as the ratio of the number of correctly recognized objects to that of the total objects in an image frame (on calculating the accuracy, we assume that the YOLO is capable of detecting all objects in an image frame when the CNN model size is $608 \times 608$ pixels). We find that (25) although a higher CNN model size enables a better detection accuracy, the accuracy gain narrows down at a high CNN model size. However, (26) the speed of the per frame latency and the inference latency increases becomes faster at a higher CNN model size, as illustrated in Figs. \ref{fig:Inter_alllatency_remote} and \ref{fig:Inter_inflatency_remote}. \textit{These two observations inspire us to trade detection accuracy (i.e., mAP) for the per frame latency reduction when the CNN model size is large.}

\begin{figure*}[t]
\centering
\subfigure[$128^2$ pixels]
{\includegraphics[width=0.16\textwidth]{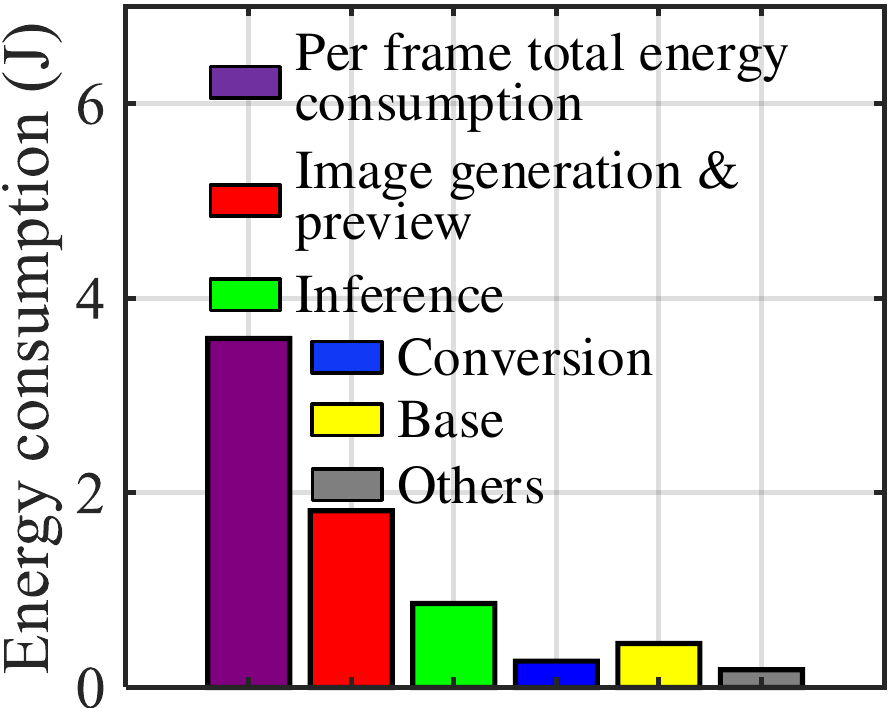}\label{fig:128InterCPU_Energy}}
\subfigure[$224^2$ pixels]
{\includegraphics[width=0.16\textwidth]{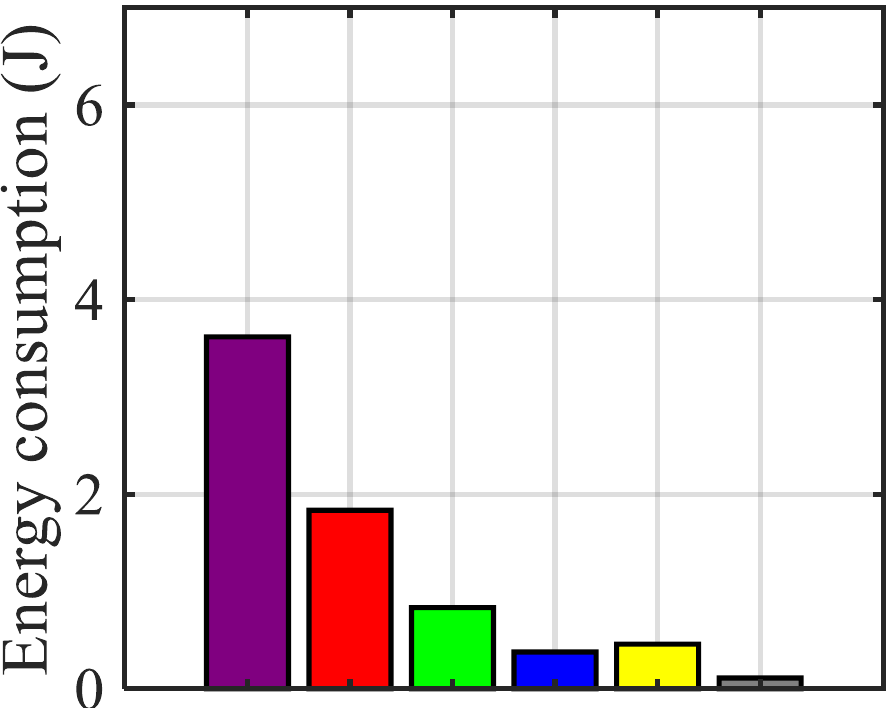}\label{fig:224InterCPU_Energy}}
\subfigure[$320^2$ pixels]
{\includegraphics[width=0.16\textwidth]{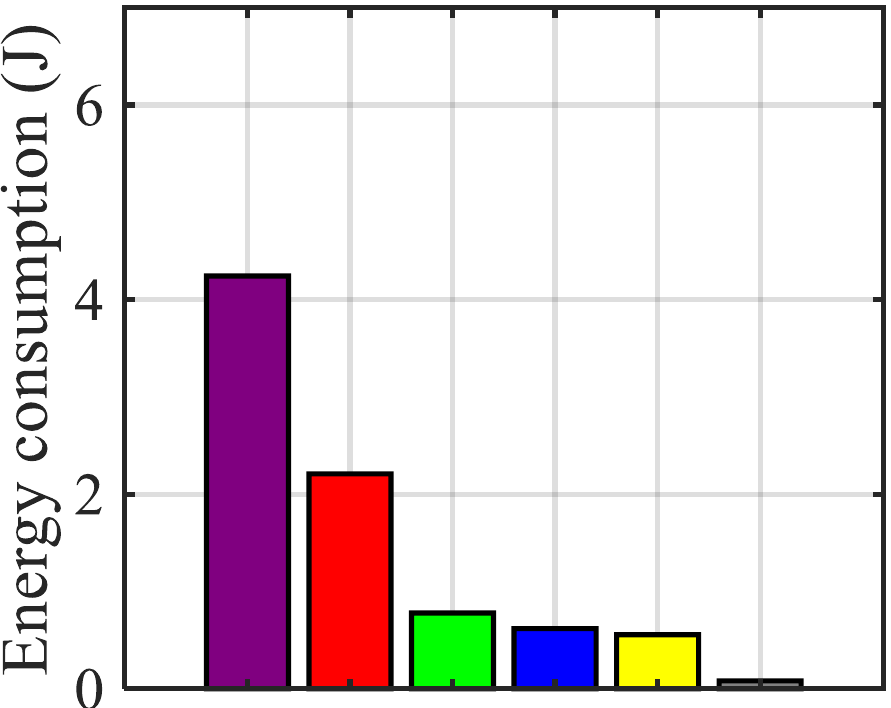}\label{fig:320InterCPU_Energy}}
\subfigure[$416^2$ pixels]
{\includegraphics[width=0.16\textwidth]{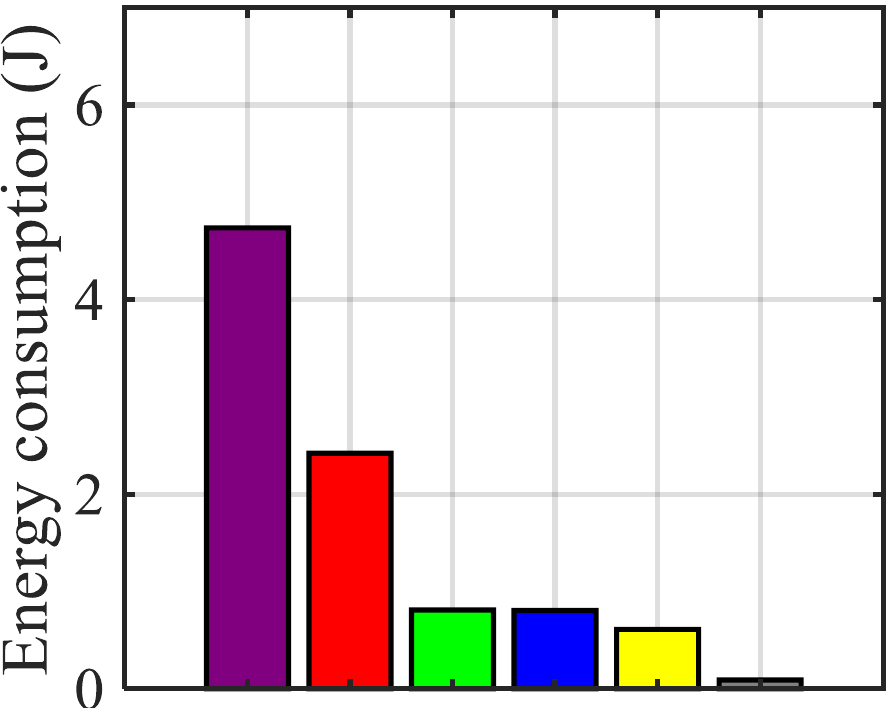}\label{fig:416InterCPU_Energy}}
\subfigure[$512^2$ pixels]
{\includegraphics[width=0.16\textwidth]{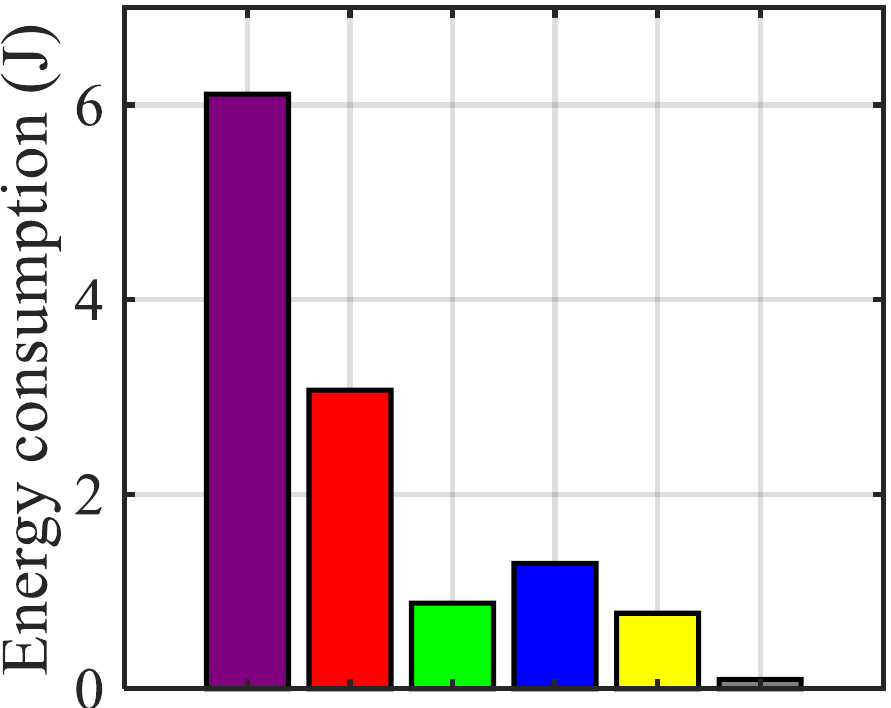}\label{fig:512InterCPU_Energy}}
\subfigure[$608^2$ pixels]
{\includegraphics[width=0.16\textwidth]{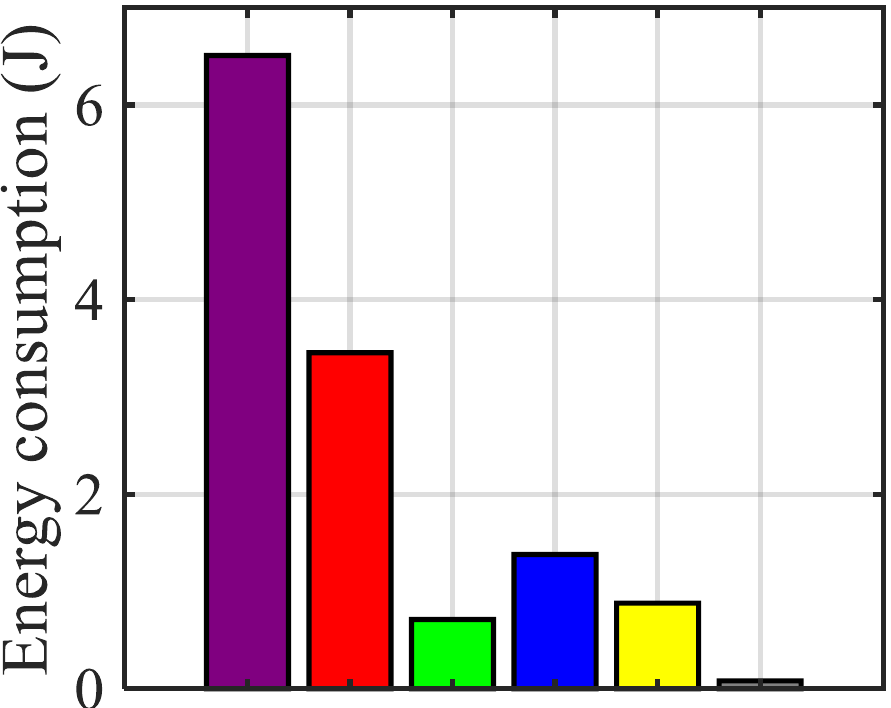}\label{fig:608InterCPU_Energy}}

\subfigure[$128^2$ pixels]
{\includegraphics[width=0.16\textwidth]{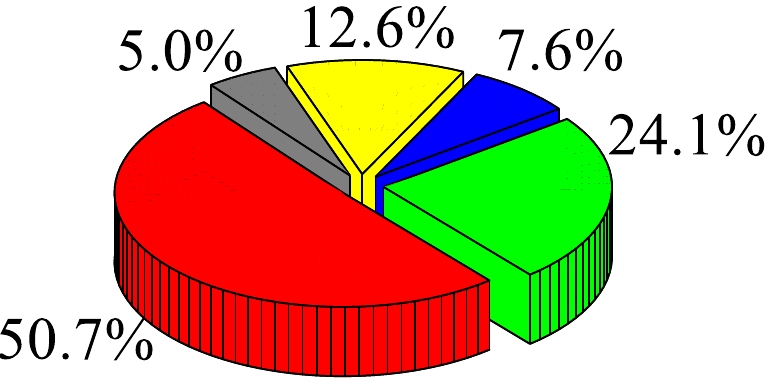}\label{fig:128InterCPU_Percentage}}
\subfigure[$224^2$ pixels]
{\includegraphics[width=0.16\textwidth]{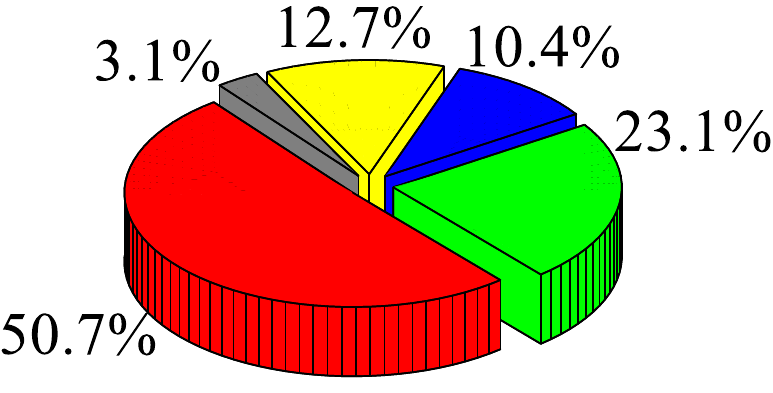}\label{fig:224InterCPU_Percentage}}
\subfigure[$320^2$ pixels]
{\includegraphics[width=0.16\textwidth]{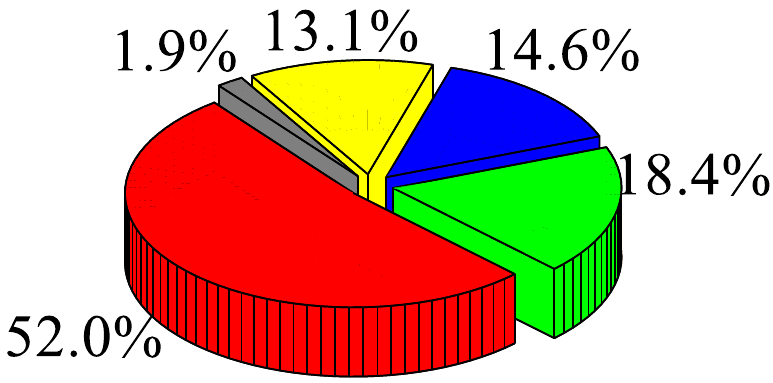}\label{fig:320InterCPU_Percentage}}
\subfigure[$416^2$ pixels]
{\includegraphics[width=0.16\textwidth]{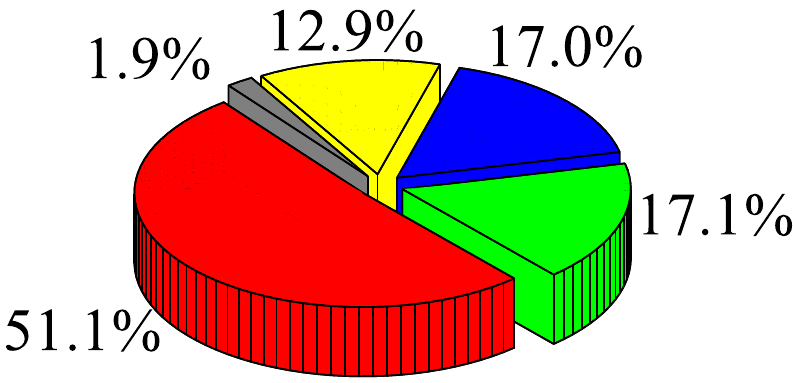}\label{fig:416InterCPU_Percentage}}
\subfigure[$512^2$ pixels]
{\includegraphics[width=0.16\textwidth]{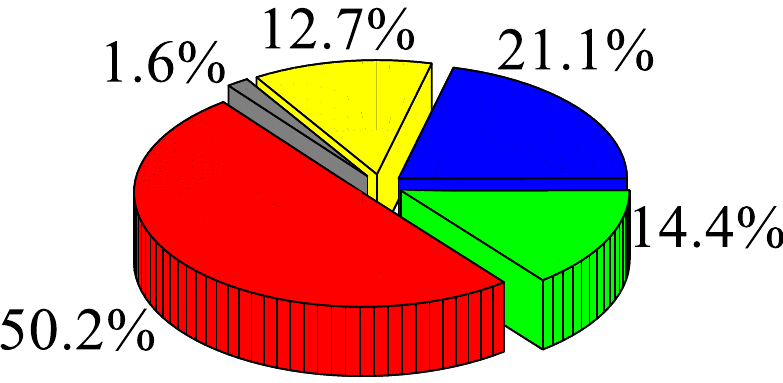}\label{fig:512InterCPU_Percentage}}
\subfigure[$608^2$ pixels]
{\includegraphics[width=0.16\textwidth]{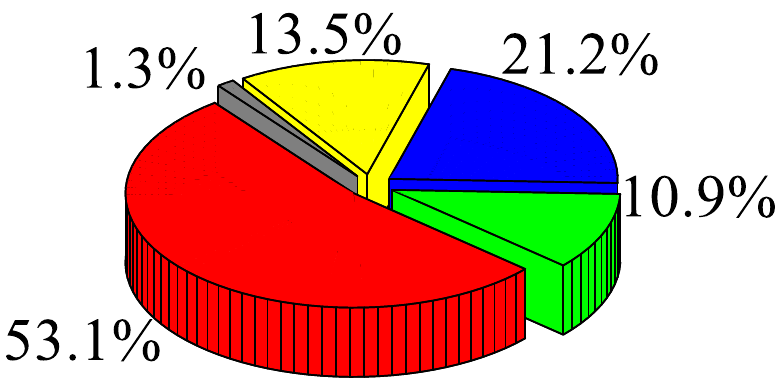}\label{fig:608InterCPU_Percentage}}
\caption{CNN model size vs. per frame energy consumption (CPU governor: interactive).}
\label{fig:cnn_energy_remote_inter}   
\end{figure*}

\textbf{Per Frame Energy Consumption.} We next investigate how the CNN model size impacts the per frame energy consumption in the remote execution. Fig. \ref{fig:cnn_energy_remote_inter} shows the measured energy consumption results, where the smartphone works on the interactive CPU governor. We observe that (27) the remote execution saves approximately $52.5$\% per frame energy on average when the frame resolution is $608\times 608$ pixels, as shown in Table \ref{tb:remote_cpu_energy_cnn}. However, it consumes slightly more per frame energy than the local execution when the frame resolution is $128\times 128$ pixels. This observation is rather significant, which demonstrates that \emph{running CNN-based object detection remotely does not always consume less energy than the local execution.} In addition, (28) the larger the CNN model size, the more per frame energy reduction the remote execution derives compared to the local execution. For example, running a $224\times 224$ pixels model only reduces about $14.1\%$ per frame energy, while executing a $608\times 608$ pixels model decreases about $52.5\%$ per frame energy consumption. \textit{Therefore, in order to take the best advantage of the remote execution, executing a CNN with a larger model size is recommended.}

\begin{table*}[t!]
\newcommand{\tabincell}[2]{\begin{tabular}{@{}#1@{}}#2\end{tabular}}
 \begin{center}
    \caption{Per frame energy consumption results of the remote execution with different CNN model sizes.}
    \label{tb:remote_cpu_energy_cnn}
  \begin{tabular}{|l|l||c|c|c|c|c|c|}
    \hline
    \multicolumn{2}{|l||}{CNN Model Size (pixels)} & $128\times 128$ & $224\times 224$ & $320\times 320$ & $416\times416$ & $512\times512$ & $608\times608$ \\\hline
    \multirow{2}{*}{\tabincell{l}{Per Frame Energy Consumption (J)}} & Local  & \textbf{3.584}  &  4.210  &  5.923  & 7.961 & 11.169 &  13.699  \\\cline{2-8}
                                                                     & Remote & \textbf{3.586}  &  3.616  &  4.242  & 4.736  & 6.111  &  6.508  \\\hline
    \multicolumn{2}{|l||}{Per Frame Energy Consumption Reduction ($\%$)} & 0 & 14.1 & 28.4  & 40.5 & 45.3 & \textbf{52.5} \\\hline
  \end{tabular}
  \end{center}
\end{table*}

\subsection{The Impact of Image Generation and Preview}
\label{ssc:impimage_remote}
\textbf{RQ 3.} Besides the network condition, what else impacts the energy consumption and latency when executed remotely, and how?
As we presented in the aforementioned observations, the image generation and preview is the most energy-consuming phase in both local and remote execution scenarios. Thus, to improve the energy efficiency of the object detection processing pipeline, we must reduce the energy consumption of image generation and preview phases. We seek to understand the interactions between the power consumption and various factors (e.g., the preview resolution, 3A, and several image post processing algorithms) as follows.

\begin{figure*}[t]
\centering
\subfigure[Preview resolution vs. power consumption.]
{\includegraphics[width=0.32\textwidth]{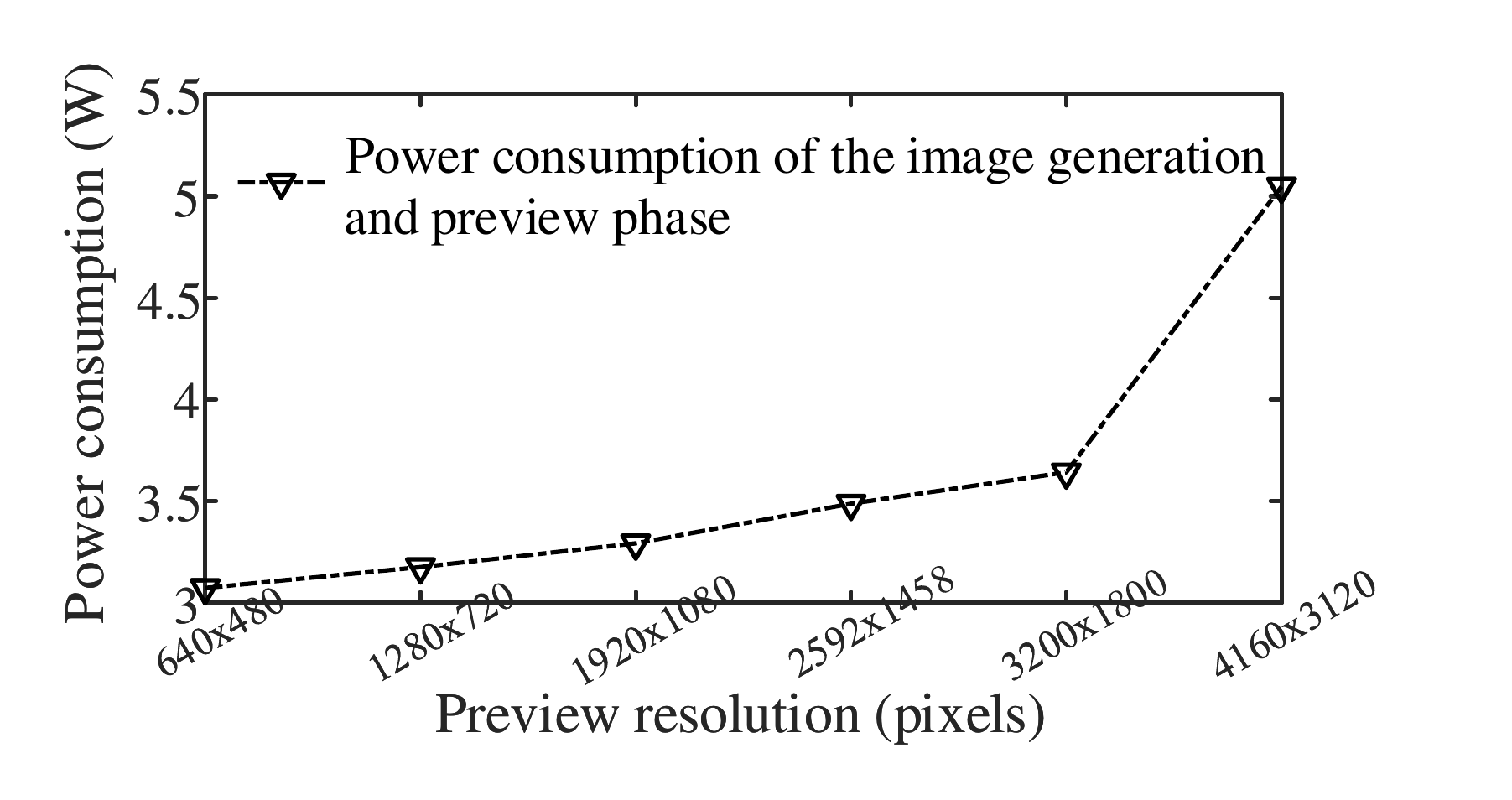}\label{fig:result19}}
\subfigure[Camera FPS vs. power consumption.] 
{\includegraphics[width=0.305\textwidth]{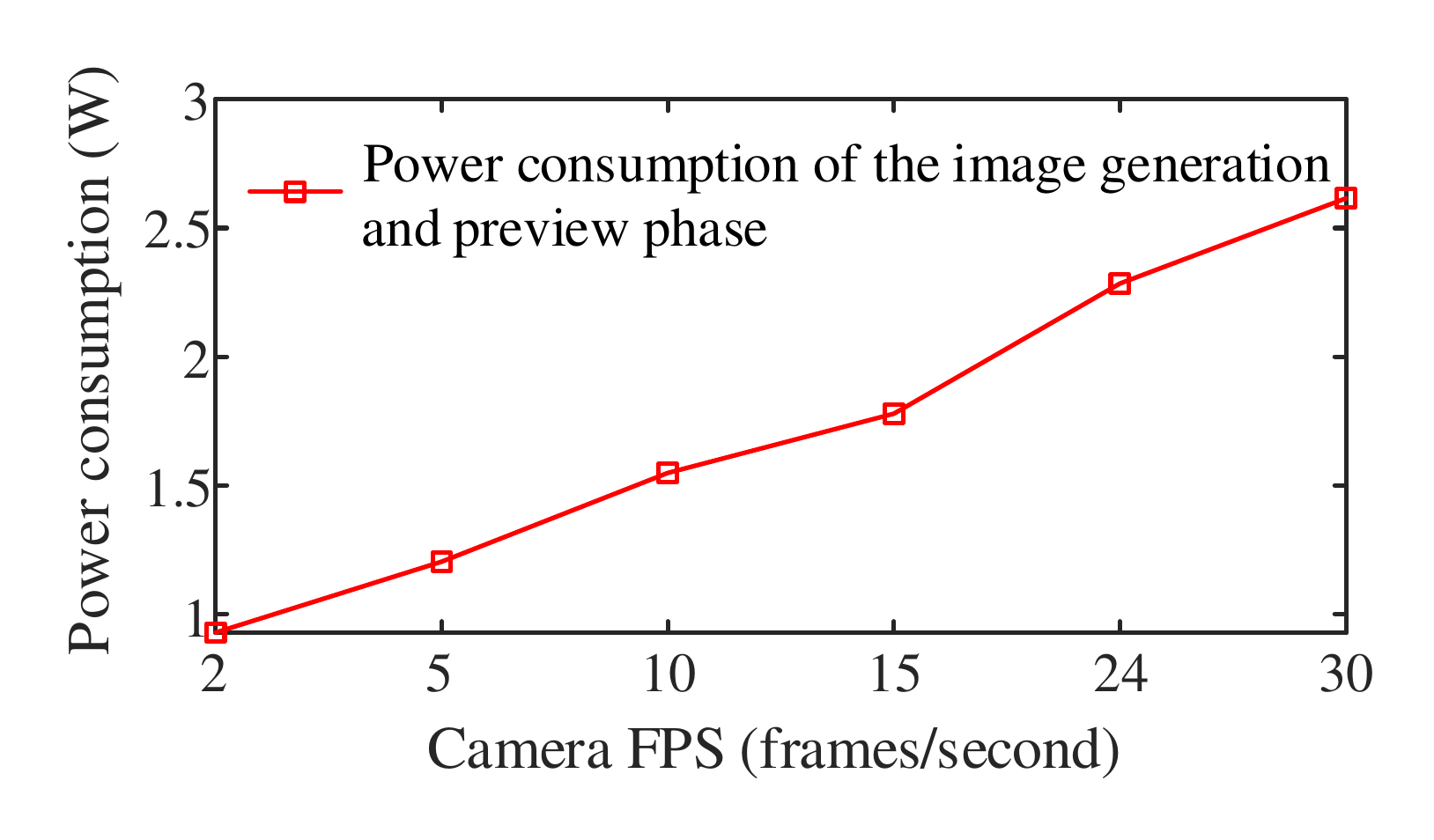}\label{fig:result111}}
\subfigure[Camera FPS vs. sampling efficiency.]
{\includegraphics[width=0.30\textwidth]{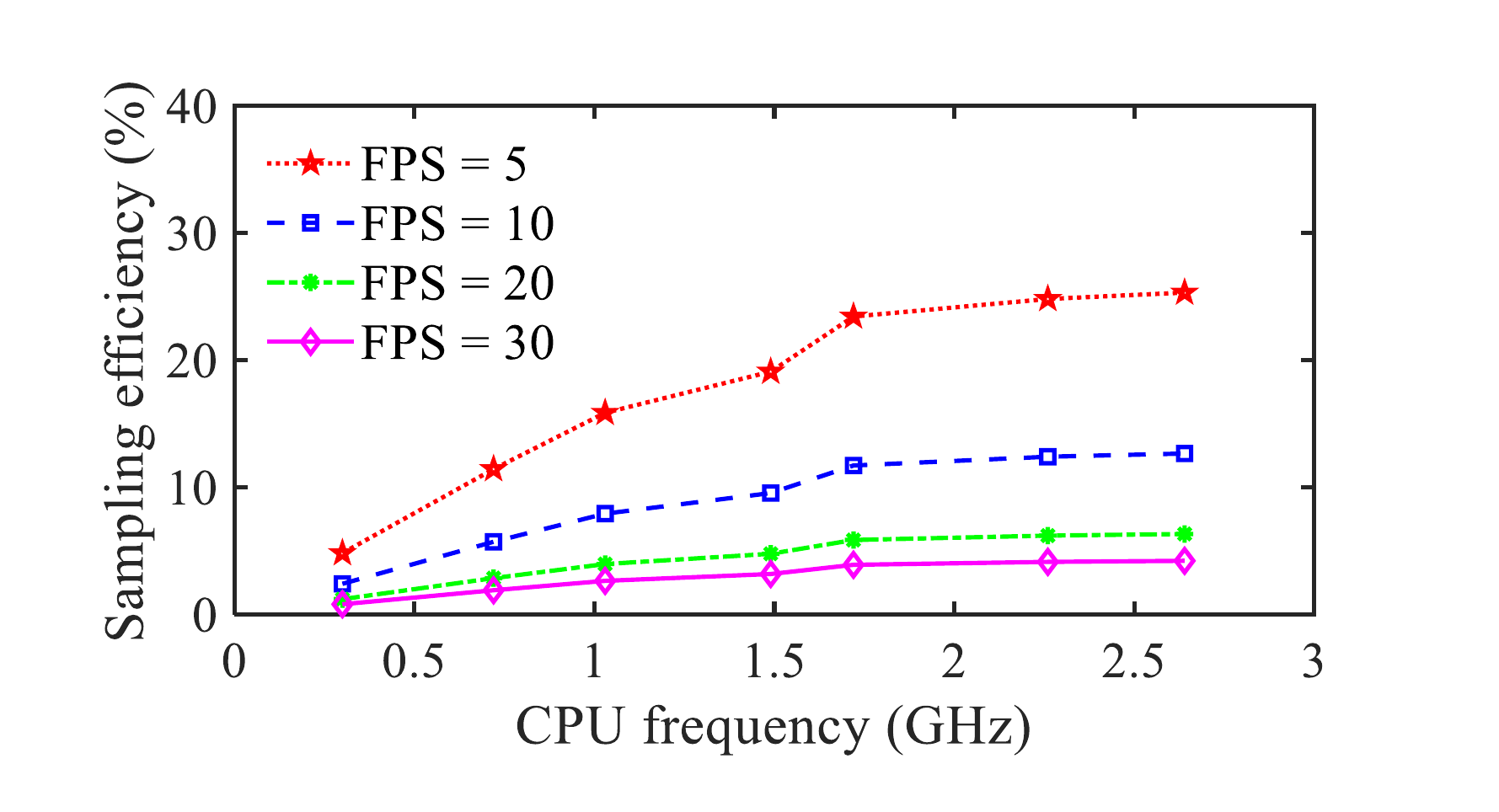}\label{fig:result1112}}
\subfigure[3A and image post processing algorithms vs. power consumption.] 
{\includegraphics[width=0.30\textwidth]{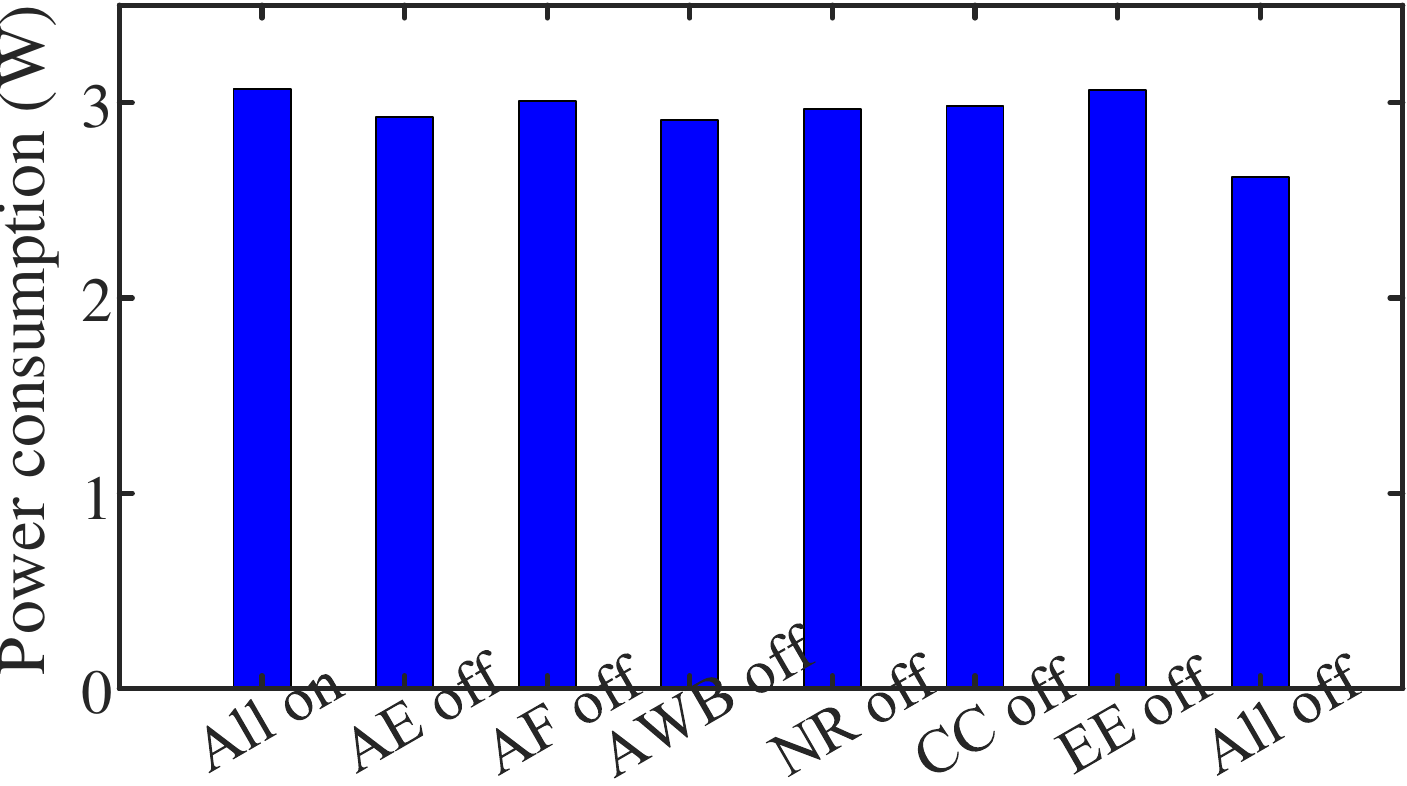}\label{fig:result110}}
\subfigure[Comparison of the per frame energy consumption.]
{\includegraphics[width=0.30\textwidth]{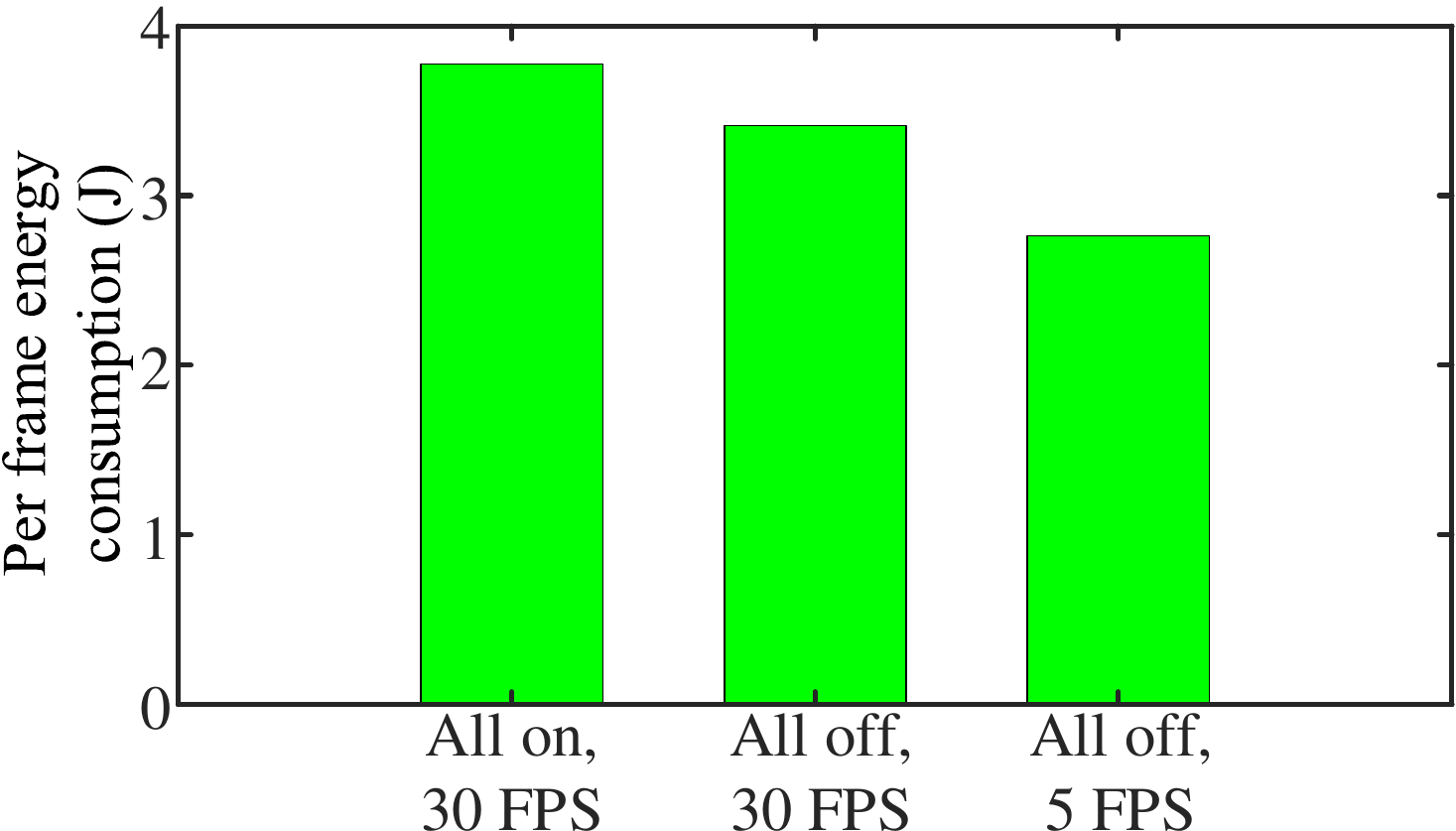}\label{fig:result112}}
\caption{Power consumption analyses of image generation and preview phases (remote execution).}
\label{fig:imagepro}   
\vspace{-0.1 in}
\end{figure*}

\textbf{Preview Resolution vs. Power Consumption.}
We first examine how the preview resolution influences the power consumption of image generation and preview phases, as shown in Fig. \ref{fig:result19}. We find that (29) as the preview resolution grows, the power consumption increases dramatically. \textit{Therefore, a preview with a higher frame resolution on the smartphone provides a better quality preview for users, but at the expense of battery drain, which is applicable for both local and remote execution cases.} 

\textbf{Camera FPS vs. Power Consumption.} We next vary the smartphone's camera FPS to explore how it impacts the device's power consumption, where the camera FPS is defined as the number of frames that the camera samples per second. Fig. \ref{fig:result111} shows that (30) a large camera FPS leads to a high power consumption. However, as shown in Fig. \ref{fig:pipeline}, not every camera captured image frame is sent to the edge server for detection. Because of the need (i) to avoid the processing of stale frames and (ii) to decrease the transmission energy consumption, only the latest camera sampled image frame is transmitted to the server. This may result in the smartphone expending significant reactive power for sampling non-detectable image frames. In Fig. \ref{fig:result1112}, we quantify the sampling efficiency with the variation of the camera FPS. As we expected, (31) a large camera FPS leads to a lower sampling efficiency (e.g., less than $2\%$ of the power is consumed for sampling the detectable image frames when the camera FPS is set to $30$). However, in most mobile AR applications, users usually request a high camera FPS for a smoother preview experience, which is critical for tracking targets in physical environments. Interestingly, (32) increasing CPU frequency can reduce the reactive power for sampling, as shown in Fig. \ref{fig:result1112}. \textit{These observations demonstrate that when a high camera FPS is requested, increasing CPU frequency can promote the sampling efficiency but may also boost the power consumption. Therefore, finding a CPU frequency that can balance this tradeoff is critical.}

\begin{figure}[t]
\centering
\subfigure[All enabled.]
{\includegraphics[width=0.2\textwidth]{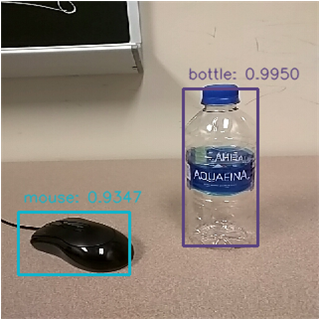}\label{fig:result113}}
\subfigure[All disabled.] 
{\includegraphics[width=0.2\textwidth]{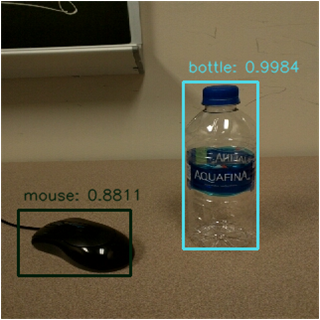}\label{fig:result114}}
\vspace{-0.1 in}
\caption{Comparison of the object detection results (remote execution).}
\label{fig:onandoff}   
\vspace{-0.1 in}
\end{figure}

\textbf{Image Post Processing and 3A Algorithms vs. Power Consumption.}
Lastly, we examine the effect of multiple image post processing and 3A algorithms on the power consumption of image generation and preview phases, as shown in Figs. \ref{fig:result110} and \ref{fig:result112}. Note that when the AE is disabled, we manually set the camera ISO and exposure time to $400$ and $20$ ms, respectively. We observe that (33) disabling the 3A, NR, CC, and EE algorithms decreases the power consumption by $14.8$\%. We conduct another experiment to understand if disabling these algorithms would impact the object detection performance. As shown in Fig. \ref{fig:onandoff}, (34) the detection performance does not degrade. Furthermore, we compare the per frame energy consumption among three cases, as depicted in Fig. \ref{fig:result112}: all enabled with camera capture frame rate $30$, all disabled with camera capture frame rate $30$, and all disabled with camera capture frame rate $5$. We find that (35) the per frame energy consumption of the second and the third cases decreases by approximately $10$\% and $27$\%, respectively, compared to the first case. \textit{Therefore, these three observations may answer the question that we presented in Section \ref{ssc:cpugovernor_local}: these energy-hungry image post processing algorithms may not be necessary for camera captured image frames to achieve successful object detection results.}

\subsection{The Impact of the Image Conversion Method}
\label{ssc:impconv_remote}

\begin{table*}[t!]
\newcommand{\tabincell}[2]{\begin{tabular}{@{}#1@{}}#2\end{tabular}}
 \begin{center}
    \caption{Image conversion latency results with different conversion methods.}
    \label{tb:comparison_imageconversion}
  \begin{tabular}{|l|l||c|c|c|c|c|c|c|}
    \hline
    \multicolumn{2}{|l||}{Preview Resolution (pixels)} & $320\times 240$ & $352\times 288$ & $640\times480$ & $720\times480$ & $800\times480$ & $800\times600$ & $1024\times768$\\\hline
    \multirow{2}{*}{\tabincell{l}{Conversion Latency (ms)}} & Java & 130.1  &  175.9  &  484.9 & 539.0 & 596.6 &  711.9  & 1157.1\\\cline{2-9}
                                                            &  C   & 14.7   &  16.4   & 40.3   & 48.3  & 48.4  &  54.4   & 82.6 \\\hline
    \multicolumn{2}{|l||}{Latency Reduction ($\%$)}                 & 88.7   &  90.7   & 91.7   & 91.0  & 91.9  &  92.4   & 92.9 \\\hline
  \end{tabular}
  \end{center}
\end{table*}

As depicted in Tables \ref{tb:local_cpu_latency} and \ref{tb:remote_cpu_latency}, the image conversion phase is one of the most time-consuming phases in both local and remote execution scenarios. This is because the image conversion method that we implemented in the testbed is developed based on Java, which is inefficient and slow. Thus, in order to improve the efficiency of the image conversion, we implement image conversion based on C in Android Native Development Kit (NDK). We compare these two methods by measuring their conversion latency with different preview resolutions. The measurement results are presented in Table \ref{tb:comparison_imageconversion}. We find that the image conversion method developed based on C decreases the image conversion latency by over $90\%$.

\subsection{Insights and Research Opportunities}
\label{ssc:sumdis_remote}
\textbf{Insights.}

\begin{itemize}

\item Offloading the object detection to the edge server does not always reduce the per frame latency and energy consumption of the mobile AR client compared to the local execution. For example, as we observed in our experiments, locally running a detection model with a size of $100\times100$ pixels achieves lower per frame latency and energy consumption than the remote execution that runs a CNN with a similar model size. In addition, the specific CNN model size when the local execution has better performance than the remote execution may vary with the computation capacities of the edge server and mobile AR clients, and even the wireless network bandwidth.

\item In the remote execution, the mobile AR client does not achieve the lowest per frame energy consumption when its CPU is set to the highest frequency, which is different from the local execution. For example, in our experiment, the lowest per frame energy consumption is obtained when the CPU frequency is around $2.26$GHz. Although this value may be different for diverse smartphones or wearable AR devices, this knowledge is important for designing the CPU scaling mechanism.

\end{itemize}

\textbf{Research Opportunities.}

\begin{itemize}

\item The energy consumption of the communication phase becomes the second largest portion of the per frame energy consumption when the frame resolution of the offloaded image is large (determined by the CNN model size). Thus, improving the image transmission energy efficiency is a potential research issue for the remote execution. For example, as we presented, when transmitting an image frame, the mobile AR client’s wireless interface experiences four phases: promotion, data transmission, tail, and idle. After completing the transmission, the wireless interface is forced to stay in the tail phase for a fixed duration and waits for other data transmission requests and the detection results. Therefore, developing a mechanism that can adaptively adjust the duration of the tail phase based on the predicted inference latency at the edge server and background activities of the mobile AR client may possibly improve the energy efficiency of the mobile AR client by allowing it to enter the idle phase faster. 

\item Although our experimental results indicate that some energy-consuming image post processing algorithms may not be necessary for mobile AR clients to achieve successful object detection results, more comprehensive studies are required to investigate this issue. For example, is this result influenced by other factors, such as the object category, frame resolution, and object detection algorithm?

\end{itemize}

\section{Threats to Validity}
\label{sc:validity}
\textbf{External validity.} External validity can be criticized by using a single version of Android OS on the tested Google Nexus $6$ smartphone. The threat is mitigated by running our benchmark applications on multiple smartphones with significantly different computation capacity (i.e., high-end and low-end), as described in Table \ref{tb:charoftest}. Furthermore, although the numerical values of our measurement with a specific experiment configuration (e.g., the per frame energy consumption of locally executing a $320\times320$ CNN model with Interactive CPU governor) cannot be generalized to all possible smartphones, such as iPhone 12 and Samsung Galaxy Note20, this paper focuses on investigating the trend of how the smartphone's energy consumption may vary when our benchmark applications are executed with different configurations, which will help predict variations on the energy consumption of other smartphones. In addition, an architecture-level comparison is not conducted in this paper (e.g., comparing the energy efficiency of running the benchmark applications on an iPhone with an Apple's bionic chip and an Android phone with a Samsung's Exynos chip). Because comparing different architectures is more challenging and needs more efforts on the hardware setup (e.g., selecting appropriate smartphones and using different ways to connect the power supply to each smartphone based on their different circuit designs) as well as the experiment design, we leave it to our future work. External validity may also be threatened by using a custom-designed object detection benchmark instead of real-world applications. However, our benchmark applications exercise most of the main functionalities of existing and potential mobile object detection applications, such as image generation, camera preview, image conversion, inference, data transmission, and virtual content rendering, which means that our custom-designed object detection benchmark applications are representative.

\textbf{Internal validity.} The collected power consumption and CPU frequency data might be influenced by the background activities. We mitigate this threat by terminating all other optional applications and services that can impact the smartphone's workload. One of the main contributions of this paper is comparing the energy consumption and latency of executing CNN-based object detections locally and remotely. To have a fair comparison between the local and remote executions, each comparison is conducted with the same configurations (e.g., CPU governor, CNN model size, preview resolution, and camera sampling rate) and under the same conditions. For instance, all the power measurements are conducted in a constant temperature laboratory. In addition, the temperature of the tested smartphone's CPU may increase when running the benchmark application, which may impact the power consumption of the smartphone. To mitigate this threat, a new measurement is launched only if the temperature of the CPU cools down to around $42$\textdegree{}C after the previous measurement.

\textbf{Construct validity.} Dissecting the energy consumption for each phase in an application is difficult. The accuracy of our evaluation is guaranteed by the energy measurement strategy we proposed in Section \ref{ssc:Measurement Strategy}, including the local clock synchronization and cooling-off period. Specifically, the precision of the local clock synchronization is in millisecond. In addition, we hypothesize that the power consumption is influenced by the workload accumulation, which is the assumption for breaking down the power consumption of the three parallel executions in our benchmark applications.

\section{Conclusion}
\label{sc:conclusion}
In this paper, we presented the first detailed experimental study of the energy consumption and the performance of a CNN-based object detection application. We examined both local and remote execution cases. We found that the performance of object detection is heavily affected by various factors, such as CPU governor, CPU frequency, and CNN model size. Although executing object detection on remote edge servers is one of the most commonly used approaches to assist low-end smartphones in improving their energy efficiency and performance, contrary to our expectation, local execution may consume less energy and obtain lower latency, as compared to remote execution. Overall, we believe that our findings provide great insights and guidelines to the future design of energy-efficient processing pipeline of CNN-based object detection.

\bibliographystyle{IEEEtran}
\bibliography{references}

\end{document}